\documentclass[letterpaper, 12pt]{article} 
\usepackage{amsmath}
\usepackage{amsfonts} 
\usepackage{amssymb} 
\usepackage{amsmath} 
\usepackage{epsfig} 
\usepackage{lscape} 
\usepackage{bbm}
\usepackage[dvips]{hyperref} 
\usepackage{color}
\usepackage[all,color]{xy} 
\usepackage{multirow} 
\usepackage{rotating}
\usepackage{longtable}
\usepackage{hypbmsec}

\usepackage{afterpage}
\usepackage[isu,bf]{caption}
\usepackage{cite} 
\usepackage[all]{xy} 
\usepackage[dvips]{geometry} 

\DeclareMathOperator{\ch}{ch} 
 
\DeclareMathOperator{\dP}{dP} 
 
\DeclareMathOperator{\Bl}{Bl}
\DeclareMathOperator{\tr}{tr}

\DeclareMathOperator{\Ext}{Ext}

\DeclareMathOperator{\Span}{span}
\DeclareMathOperator{\Pic}{Pic}
\DeclareMathOperator{\Td}{Td}
\DeclareMathOperator{\maximum}{max}

\newtheorem{theorem}{Theorem}

\def\clap#1{\hbox to 0pt{\hss#1\hss}}

\author{}

\newcommand{\drawsquare}[2]{\hbox{%
\rule{#2pt}{#1pt}\hskip-#2pt
\rule{#1pt}{#2pt}\hskip-#1pt
\rule[#1pt]{#1pt}{#2pt}}\rule[#1pt]{#2pt}{#2pt}\hskip-#2pt
\rule{#2pt}{#1pt}}
\newcommand{\fund}{\raisebox{-.5pt}{\drawsquare{6.5}{0.4}}}
\newcommand{\Ysymm}{\raisebox{-.5pt}{\drawsquare{6.5}{0.4}}\hskip-0.4pt%
         \raisebox{-.5pt}{\drawsquare{6.5}{0.4}}}
\newcommand{\Yasymm}{\raisebox{-3.5pt}{\drawsquare{6.5}{0.4}}\hskip-6.9pt%
        \raisebox{3pt}{\drawsquare{6.5}{0.4}}}
\newcommand{\antifund}{\overline{\fund}} 
\newcommand{\bYasymm}{\overline{\Yasymm}} 
\newcommand{\bYsymm}{\overline{\Ysymm}} 
\newcommand{\be}{\begin{equation}} 
\newcommand{\ee}{\end{equation}} 
\newcommand{\beq}{\begin{equation}} 
\newcommand{\eeq}{\end{equation}} 
\newcommand{\ba}{\begin{array}} 
\newcommand{\ea}{\end{array}} 
\newcommand{\bea}{\begin{eqnarray}} 
\newcommand{\eea}{\end{eqnarray}}

\newcommand{\ov}{\overline} 
\def\IR{\relax{\rm I\kern-.18em R}} 
 
\def\IP{\relax{\rm I\kern-.18em P}} 
\def\inbar{\vrule height1.5ex width.4pt depth0pt} 
\def\IC{\relax\,\hbox{$\inbar\kern-.3em{\rm C}$}} 

\newcommand{\Osheaf}{\ensuremath{\mathcal{O}}}
\newcommand{\OsheafY}{\ensuremath{\mathcal{O}_{Y}}}
\newcommand{\OsheafDa}{\ensuremath{\mathcal{O}_{D_a}}}
\newcommand{\OsheafDb}{\ensuremath{\mathcal{O}_{D_b}}}
\newcommand{\OsheafDc}{\ensuremath{\mathcal{O}_{D_c}}}
\newcommand{\OsheafF}{\ensuremath{\mathcal{O}_{F}}}

\newcommand{\cO}{\mathcal{O}}

\newcommand{\cC}{\mathcal{C}} 
 
\newcommand{\cL}{\mathcal{L}}

\newcommand{\cN}{\mathcal{N}}

\newcommand{\cB}{\mathcal{B}}

\newcommand{\cV}{\mathcal{V}}

\newcommand{\cM}{\mathcal M}

\newcommand{\R}{\text{Re}}

\newcommand{\bbZ}{\mathbb{Z}} 
 
\newcommand{\bbC}{\mathbb{C}} 
\newcommand{\bbP}{\mathbb{P}} 
 
\newcommand{\Z}{\mathbb{Z}} 
\newcommand{\Q}{\mathbb{Q}} 
\newcommand{\C}{\mathbb{C}} 
\newcommand{\CP}{\mathbb{P}}  

\def\K3{{\bf K3}}

\def\i{\iota}

\def\ov{\overline}

\def\n2d{\cN_{V^*}^{\otimes 2}}

\def\IC{\mathbb{C}} 

\def\IR{\mathbb{R}}

\def\IP{\mathbb{P}}

\def\cN{{\mathcal N}} 
\def\cM{{\mathcal M}}

\def\cO{{\mathcal O}} 
\def\cC{{\mathcal C}}

\def\cL{{\mathcal L}} 
\def\nn{\nonumber} 
 
\def\ch{\mbox{ch}}

\def\to{\rightarrow}

\begin{document} 
\baselineskip=15pt 
\parskip=3pt 
 
\title{ 
\begin{flushright} \vspace{-3.0cm} 
\small MPP-2008-144\\
\small DIAS-STP 08-15\\
\small Bonn-TH 2008-15\\
\small SLAC-PUB-13466 
\end{flushright} 
\vspace{1cm} 
GUTs in  Type IIB Orientifold Compactifications  
} 
\vspace{2.5cm} 
\author{\small 
  Ralph Blumenhagen$^{1}$, 
  Volker Braun$^{2}$, 
  Thomas W.~Grimm$^{3}$, 
  and Timo Weigand$^{4}$} 
 
\date{} 
 
\maketitle 
\vspace*{-.7cm}

\begin{center} 
\emph{$^{1         }$ Max-Planck-Institut f\"ur Physik, F\"ohringer Ring 6, \\ 
  80805 M\"unchen, Germany } \\ 
\vspace{0.1cm} 
 \emph{$^{2         }$ Dublin Institute for Advanced Studies, 10 Burlington
  Road, Dublin 4, Ireland} \\
\vspace{0.1cm}  
\emph{$^{3         }$ Bethe Center for Theoretical Physics and \\ 
                      Physikalisches Institut der Universit\"at Bonn, Nussallee 12, \\ 
                      53115 Bonn, Germany } \\ 
\vspace{0.1cm} 
\emph{$^{4}$ SLAC National Accelerator Laboratory, Stanford University, \\
2575 Sand Hill Road, Menlo Park, CA 94025, USA }  
\vspace{0.2cm}

\tt{blumenha@mppmu.mpg.de, vbraun@stp.dias.ie, grimm@th.physik.uni-bonn.de, timo@slac.stanford.edu} 
\vspace{0.0cm} 
\end{center} 
\vspace{.2cm} 
 
\begin{abstract} 
\noindent   
We systematically analyse globally consistent $SU(5)$ GUT models on
intersecting D7-branes in genuine Calabi-Yau orientifolds with O3- and
O7-planes. Beyond the well-known tadpole and K-theory cancellation
conditions there exist a number of additional subtle but quite
restrictive constraints.  For the realisation of $SU(5)$ GUTs with
gauge symmetry breaking via $U(1)_Y$ flux we present two classes of
suitable Calabi-Yau manifolds defined via del Pezzo transitions of the
elliptically fibred hypersurface $\IP_{1,1,1,6,9}[18]$ and of the
Quintic $\IP_{1,1,1,1,1}[5]$, respectively. To define an orientifold
projection we classify all involutions on del Pezzo surfaces. We work
out the model building prospects of these geometries and present five
globally consistent string GUT models in detail, including a
3-generation $SU(5)$ model with no exotics whatsoever.  We also
realise other phenomenological features such as the ${\bf 10 \, 10 \,
  5_H}$ Yukawa coupling and comment on the possibility of moduli
stabilisation, where we find an entire new set of so-called
swiss-cheese type Calabi-Yau manifolds.  It is expected that both the
general constrained structure and the concrete models lift to F-theory
vacua on compact Calabi-Yau fourfolds.

\end{abstract} 
 
\thispagestyle{empty} 
\clearpage 
 
\tableofcontents

\newpage


\section{Introduction}
The LHC experiment is widely expected not only to confirm  the existence of
the Higgs particle as the last missing ingredient of the Standard Model of Particle Physics, 
but also to reveal new structures going far
beyond.
As experiments are proceeding into this hitherto unexplored energy regime,
string theory, 
with its claim to represent the unified theory of all interactions, 
will have to render an account of its predictions
for physics beyond the Standard Model. 
Clearly, these depend largely on the value of the string scale $M_s$, the most 
dramatic outcome corresponding to $M_s$ 
close to the TeV scale. While this is indeed  a fascinating possibility,
in concrete string models it often  leads to severe cosmological
issues such as the cosmological moduli problem. 
In this light it might be fair to say that a more natural (but also more conservative) scenario involves a value of $M_s$ at the GUT, Planck or intermediate scale.

During the last years, various classes of four-dimensional string
compactifications with ${\cal N}=1$ spacetime supersymmetry have been
studied in quite some detail (see the reviews~\cite{Blumenhagen:2005mu,
  Douglas:2006es, Blumenhagen:2006ci,Denef:2007pq,Nilles:2008gq} for
references). From
the viewpoint of realising the Minimal Supersymmetric Standard Model
(MSSM) and some extension thereof the best understood such
constructions are certainly the perturbative heterotic string and Type
IIA orientifolds with intersecting D6-branes.  On the contrary, as far
as moduli stabilisation is concerned Type IIB orientifolds with O7-
and O3-planes look very promising. The combination of three-form
fluxes and D3-brane instantons can stabilise all closed string
moduli~\cite{Kachru:2003aw} even within the solid framework of
(conformal) Calabi-Yau manifolds where reliable computations can be
performed. Moreover, supersymmetry breaking via K\"ahler moduli
mediation and the resulting structure of soft terms bear some
attractive features and have been studied both for the LARGE volume
scenario~\cite{Balasubramanian:2005zx,Conlon:2005ki} with an
intermediate string scale and for a GUT scenario with the string scale
at the GUT scale~\cite{Aparicio:2008wh, Blumenhagen:2008kq}.

These considerations are reason enough to seriously pursue model
building within type IIB orientifolds.  The observation that the MSSM
gauge couplings appear to meet at the GUT scale furthermore suggests
the existence of some GUT theory at high energies. GUT gauge groups
such as $SU(5)$ and $SO(10)$ appear naturally in string theories based
on gauge group $E_8$ like the heterotic string.  On the other hand, it
has become clear that for perturbative orientifolds with D-branes,
exceptional gauge groups and features like the spinor representations
of $SO(10)$ do not emerge.  For $SU(5)$ D-brane models, by contrast,
the gauge symmetry and the desired chiral matter spectrum can be
realised, a fact welcome in view of the described progress in Type IIB
moduli stabilisation.  Still, at first sight there appears a serious
problem in the Yukawa coupling sector.  The ${\bf 10\, 10\, 5_H}$
Yukawa coupling violates global perturbative $U(1)$ symmetries which
are the remnants of former $U(1)$ symmetries rendered massive by the
St\"uckelberg mechanism~\cite{Blumenhagen:2001te}.  As a consequence
of these considerations it is sometimes argued that the natural
context for Type II GUT model building is the strong coupling limit,
where the crucial couplings in question are not ``perturbatively''
forbidden. The strongly coupled duals of type IIA and Type IIB
orientifolds are given by singular M-theory compactifications on $G_2$
manifolds and, respectively, by F-theory compactifications on
elliptically fibred Calabi-Yau fourfolds~\cite{Vafa:1996xn}.  The
local model building rules for such F-theory compactifications have
been worked out recently in~\cite{Donagi:2008ca, Beasley:2008dc,
  Hayashi:2008ba,
  Beasley:2008kw,Donagi:2008kj,Font:2008id,Heckman:2008qa}; For recent
studies of 7-branes from the F-theory perspective see
\cite{Lust:2005bd, Aluffi:2007sx, Braun:2008ua, Collinucci:2008pf,
  Braun:2008pz}.

On the other hand, investigations of non-perturbative corrections for
Type II orientifold models~\cite{Blumenhagen:2006xt, Ibanez:2006da,
  Florea:2006si} have revealed that the ${\bf 10\, 10\, 5_H}$ Yukawa
coupling
can be generated by Euclidean D-brane instantons wrapping suitable
cycles $\Gamma$ in the internal manifold with the right zero mode
structure~\cite{Blumenhagen:2007zk}.  Of course these couplings are
suppressed\footnote{Note that for a Georgi-Glashow $SU(5)$ GUT the
  ${\bf 10\, 10\, 5_H}$ Yukawa gives masses to up-type quarks, whereas
  for flipped $SU(5)$ it provides the down-type quark masses. In the
  latter case, the exponential suppression might explain the little
  hierarchy between up and down quarks~\cite{Blumenhagen:2007zk}.}  by
the exponential of the instanton action $\Re(T_{\rm inst})=g_s^{-1}
{\rm Vol}(\Gamma)$.
It is crucial to appreciate that this suppression is \emph{not} tied
to the inverse gauge coupling of the Standard Model, as would be the
case for effects related to gauge (as opposed to ``stringy'')
instantons, but can in principle take any value, depending on the
geometric details of our compactification manifold.  This feature,
which holds both for Type IIA and Type IIB orientifolds, opens up the
prospect of $SU(5)$ GUT model building already in the limit $T_{\rm
  inst}\to 0$ of perturbative Type II orientifolds.
Once we also take the nice features of moduli stabilisation
in Type IIB into account, one might seriously hope that
the strong coupling limit of Type IIB orientifolds, either in their genuinely 
F-theoretic disguise  or in their perturbative description as D-brane models 
with O7- and O3- planes and $T_{inst} \rightarrow 0$, may indeed provide
a promising starting point to construct realistic GUT models.

In the recent work~\cite{Beasley:2008dc,Beasley:2008kw}, the authors draw the first
conclusion (see also \cite{Donagi:2008ca,Donagi:2008kj}). Taking into account that F-theory models on elliptically
fibred fourfolds can admit degenerations of the elliptic fibre such
that exceptional gauge groups appear naturally, these references pursue
the program of studying GUT type F-theory compactifications.  As a
physical input, the authors of~\cite{Beasley:2008dc,Beasley:2008kw} propose the working hypothesis that the
Planck scale ought to be decouple-able from the GUT scale, even if
only in principle. {}From the Type IIB perspective, these F-theoretic
models do not only contain usual D-branes, but also so-called $(p,q)$
seven-branes which carry charge under both the R-R and the NS-NS
eight-forms. The new non-perturbative states, such as the gauge bosons
of exceptional groups or the spinor representations of $SO(10)$, are
given by $(p,q)$ string junctions starting and ending on these branes.
Unlike fundamental strings, these string junctions can have more than
two ends thus providing extra states.

{}From the guiding principle of decoupled gravity it is further argued
in~\cite{Beasley:2008kw} that the $(p,q)$ 7-branes should wrap
shrinkable four-cycles in the internal geometry. These are given by
del Pezzo surfaces\footnote{Local quiver type models on del Pezzo
  singularities have been studied, for example,
  in~\cite{Verlinde:2005jr, Buican:2006sn}.}. Lacking a global
description of Calabi-Yau fourfolds with the desirable degeneration,
the authors provide a local set-up of singularities or $(p,q)$
7-branes and line bundles so that the GUT particle spectrum is
realised. At a technical level there arises a challenge with GUT
symmetry breaking because a theory on a del Pezzo surface has neither
adjoints of $SU(5)$ nor discrete Wilson lines at its disposal to break
$SU(5)$ to $SU(3)\times SU(2)\times U(1)_Y$.  One option would be to
adopt the philosophy of heterotic compactifications and embed a
further non-trivial $U(1)$ line bundle, as discussed for the heterotic
string originally in~\cite{Witten:1985bz} and more recently
in~\cite{Blumenhagen:2005ga, Tatar:2006dc, Blumenhagen:2006ux,Blumenhagen:2006wj}.  For
line bundles non-trivial on the Calabi-Yau manifold,
the associated $U(1)$ generically becomes massive due to the
St\"uckelberg mechanism, but in presence of several line bundles special linear $U(1)$ 
combinations remain massless.  In the heterotic context of~\cite{Blumenhagen:2006ux}, 
in order to maintain gauge coupling unification without relying on large threshold corrections
it is  necessary to consider the large $g_s$ limit of heterotic M-theory~\cite{Tatar:2008zj}.

As a new and very central ingredient the authors
of~\cite{Beasley:2008kw} propose to break the GUT gauge group instead
by a line bundle embedded into $U(1)_Y$ such that it circumvents the
sort of no-go theorem mentioned above (see also \cite{Donagi:2008kj}). The idea is to support the
bundle on a non-trivial two-cycle inside the del Pezzo surface which
is trivial on the ambient four-fold base.  It was argued
in~\cite{Beasley:2008kw} that with this mechanism some of the
notorious problems of GUTs such as the doublet-triplet splitting
problem, dangerous dimension five proton decay operators and even
neutrino masses can be addressed and actually solved by appropriate
choices of matter localisations and line bundles on the del Pezzo
divisors.  Studies of supersymmetry breaking mechanisms for this class
of local models have appeared in~\cite{Heckman:2008es, Marsano:2008py,
  Marsano:2008jq}.

The Planck-scale decoupling principle might be a justification for a
local approach to string model building (and indeed a quite
constraining one), but in absence of a realisation of the described
mechanisms in globally consistent string compactifications it remains
an open question if these local GUT models do really consistently
couple to gravity.  In fact, it is the global consistency conditions
of string theory which decide whether a given construction is actually
part of the string landscape or merely of the ``swampland'' of gauge 
theories. At a technical level, it is therefore no wonder that they
constitute some of the biggest challenges in string model building,
and many interesting local constructions fail to possess a compact
embedding satisfying each of these stringy consistency conditions.
For example, whether or not a given $U(1)_Y$ flux actually leads to a
massless hypercharge depends on the global embedding of the divisor
supporting the 7-brane into the ambient geometry and cannot be decided
within a local context.

In F-theory the global consistency conditions, in particular
the D7-brane tadpole cancellation condition, are 
geometrised: They are contained in the statement
that indeed a compact elliptically fibred fourfold
exists such that the degenerations of the fibre 
realise the GUT model. This is a very top-down condition and given 
the complexity and sheer number of fourfolds  it is extremely hard to 
implement in practise.

It is the aim of this paper to address these global consistency
conditions by taking a different route.  As described above
Georgi-Glashow $SU(5)$ GUT models can naturally be realised on
two-stacks of D7-branes in a perturbative Type IIB orientifold.  Here
we have quite good control over the global consistency conditions as
they are very similar to the well-studied Type I or Type IIA
orientifolds.  Therefore, our approach is to first construct a GUT
model on a Type IIB orientifold, satisfy all consistency conditions,
check whether the top quark Yukawa is really generated by an
appropriate D3-instanton and then take the local $T_{\rm inst.}\to 0$
limit\footnote{Here we make the working assumption that a Type IIB
  vacuum satisfying all K-theory and supersymmetry constraints has an
  uplift to F-theory on a Calabi-Yau fourfold.  We are aware that this
  may not be so straightforward to show~\cite{wittenlecture}.}.

To follow this path, we start by partly newly deriving, partly summarising 
the model building rules for Type IIB orientifolds. We then
study how much of the appealing  structure proposed in the F-theoretic context, such as
the $U(1)_Y$ GUT gauge breaking, can already be realised 
in perturbative Type IIB orientifolds on Calabi-Yau threefolds with
intersecting D7-branes wrapping holomorphic surfaces
with non-trivial vector- or line-bundles.

In order for the $U(N)$ gauge factors on the D7-banes not to exhibit
chiral multiplets transforming in the adjoint representation, the
four-cycle wrapped by the 7-brane should be rigid.  A natural class of
such complex four-manifolds is again given by del Pezzo surfaces,
which therefore remain important ingredients from a phenomenological
point of view\footnote{Note, however, that the class of rigid surfaces
  is larger than that of shrinkable, that is, del Pezzo, surfaces. For
  example, the surface ${\rm dP}_9$, which is rigid, but not
  contractible, is still an interesting candidate for a GUT D7-brane
  in this respect.}.  This leads us to the study of a class of
suitable Calabi-Yau threefolds admitting complex divisors of del Pezzo
type whose non-trivial two-cycles are partially trivial when
considered as two-cycles of $Y$.  A large class of such Calabi-Yau threefolds
is given by del Pezzo transitions, as have been recently discussed in
a slightly different context in~\cite{Grimm:2008ed}. The idea is to
start for instance with the well-studied ``swiss cheese'' type
Calabi-Yau hypersurface $\IP_{1,1,1,6,9}[18]$, which is elliptically
fibred over $\IP_2$, and blow up points on the base.  As a second
starting point for such del Pezzo transitions we consider the quintic
$\IP_4[5]$. As we will see such Calabi-Yau manifolds are natural
candidates for GUT model building.

More concretely,
in \autoref{sec:OrientifoldBasics} we collect the string model building 
 rules for Type IIB orientifolds with O7- and O3-planes.
Since these involve intersecting D7-branes supporting
non-trivial gauge bundles, the relevant structure is
a combination  of Type I and Type IIA orientifold features
and as such is slightly more complicated and subtle. 
In particular, since the del Pezzo surfaces wrapped by the D7-branes
are not $Spin$, one has to take into account the
Freed-Witten anomaly~\cite{Freed:1999vc} shifting the
proper quantisation for  the gauge fluxes\footnote{An analogous
quantisation condition is expected to also arise for the
four-form flux in  F-theory compactifications on compact  as well as
on non-compact Calabi-Yau fourfolds.}. Moreover, for involutions
with $h^2_->0$ one  encounters dynamical $B$-field moduli, which have
to be appropriately dealt with.

In \autoref{sec:GUTsbreak} we discuss how we can realise a D-brane
sector supporting a Georgi-Glashow $SU(5)$ GUT model.  In many
respects this is very analogous to $SU(5)$ GUTs in intersecting
D6-brane scenarios (for concrete examples see, for
example,~\cite{Antoniadis:2000ena, Blumenhagen:2001te, Ellis:2002ci,
  Cvetic:2002pj,Gmeiner:2006vb,Cvetic:2006by,Antoniadis:2007jq}). We show that the GUT symmetry
breaking via $U(1)_Y$ flux can be realised in the perturbative Type IIB
orientifold framework and provide the conditions under which this flux
can solve the doublet-triplet splitting problem and the suppression of
dimension five proton decay operators. By studying which quantities
are affected by this flux we find that not only the gauge group is
broken but also the non-chiral spectrum and the D3-brane tadpole
generically changes. Very analogous features are expected
to arise also in F-theory compactifications on compact
Calabi-Yau fourfolds equipped with non-vanishing four-form
flux.

The slightly more mathematical \autoref{sec:class_of_examples} defines
a class of compact Calabi-Yau manifolds naturally containing del Pezzo
surfaces.  These compact manifolds have recently been considered
in~\cite{Grimm:2008ed} and contain the kind of holomorphic surfaces
allowing for the realisation of many of the GUT features we are
interested in.  They can be described as elliptic fibrations over del
Pezzo surfaces ${\rm dP}_n$, $n=1,\ldots, 8$ and their various connected
phases related via flop transitions.  In the elliptic fibration
itself, besides the ${\rm dP}_n$ basis, we find various ${\rm dP}_9$ surfaces,
which via the flop transitions become ${\rm dP}_8$ surfaces or $\IP_2$
surfaces with more than nine points blown up. In the course of this
section, to define the orientifold actions we have to investigate the
existence of appropriate involutions. In order not to interrupt the
physics elaborations too much this rather technical though central
discussion has mainly been shifted to appendix A. The mathematically
interested reader is encouraged to consult this appendix for more
details on the classification of involutions and the determination of
the fixed point loci.  As an important part of our analysis we will
prove the ``swiss cheese'' structure of those del Pezzo transitions
where the ${\rm dP}_9$ surfaces have all been flopped to ${\rm dP}_8$ surfaces. We
will show that as a consequence of this structure the D-term
supersymmetry conditions force the cycles supporting D-branes to take
a vanishing volume, that is, they are dynamically driven to the quiver
locus.

In Sections~\ref{sec:GutModelExample} and~\ref{sec:search} we present
some first concrete $SU(5)$ GUT models. These are the outcome of an
essentially manual search which has succeeded in implementing all
known global consistency conditions.  As a warm-up,
\autoref{sec:GutModelExample} discusses at length an $SU(5)$ model on
the Weierstra\ss{} model over ${\rm dP}_2$ with two chiral families of
$SU(5)$ GUT matter, one vector-like pair of Standard Model Higgs and
no chiral exotics. The GUT matter transforming in the ${\bf 10}$ is
localised in the bulk of the GUT branes, while the ${\bf \ov 5}$ and the
Higgs pair arise at matter curves. Upon breaking $SU(5)$ by means of
$U(1)_Y$ flux there arise extra vector-like pairs of Standard Model
matter.  As one of its phenomenologically appealing features, this
model contains a ${\bf 10\, 10\, 5_H}$ Yukawa coupling of order one
induced by a Euclidean D3-brane instanton in the limit $T_{inst.}
\rightarrow 0$, but the global consistency conditions do not allow for
the construction of a three-generation model on this particular
geometry. To remedy this we present, in \autoref{sec:search}, a string
vacuum of a similar type on the del Pezzo transition of this
Weierstra\ss{} model, but featuring three chiral families of Standard
Model matter, no chiral exotics and only two pairs of extra
vector-like states.  As a consequence of the swiss cheese structure of
the manifold, the D-term supersymmetry conditions drive the vacuum to
the boundary of the K\"ahler cone.  This can be avoided in a
three-generation GUT model on the Weierstra\ss{} model over ${\rm dP}_3$
as discussed in the remainder of this section. The key
phenomenological achievements and drawbacks of these three examples
are summarised in Tables~\ref{tab_gutmodel1}, \ref{tab_gutmodel2},
and~\ref{tab_gutmodel3}.

In \autoref{sec:quintic} we analyse yet another class of geometries
based on the quintic hypersurface which also contain del Pezzo
surfaces suitable for GUT model building by extending the examples
of~\cite{Grimm:2008ed}. In this type of geometries it is possible to
localise all GUT matter at matter curves, thus avoiding the appearance
of extra vector-like pairs in the bulk. The D-term supersymmetry
conditions can be realised for non-zero cycle volume. Once again due
to subtle global consistency constraints we only present a
two-generation GUT model exhibiting these features realised on ${\rm dP}_8$
surfaces, but remarkably we do find an interesting model with just
three generations of Standard Model matter and no exotics whatsoever
on a related geometry featuring ${\rm dP}_9$ surfaces. The detailed
properties of these two examples are summarised in
Tables~\ref{tab_gutmodelquint1} and~\ref{tab_gutmodelquint2},
respectively.

In \autoref{sec_moduli} we comment on the possibilities of moduli
stabilisation in our class of models. Taking our findings for GUT
model building into account, the most natural realisation of the LARGE
volume scenario of moduli stabilisation seems to place the GUT branes
and the instantons contributing to the superpotential on different,
non-intersecting ${\rm dP}_n$ surfaces. This is expected to lead to a
new pattern of soft terms at the GUT scale and consequently to
different collider signatures compared to the studies which have
appeared in the literature so far. Furthermore, we clearly need to
stabilise the string scale not at an intermediate, but at the GUT
scale, for example along the lines of~\cite{Blumenhagen:2008kq}. The
explicit elaboration of such aspects is, however, beyond the scope of
the present article.


\section{Orientifolds with Intersecting D7-Branes} 
\label{sec:OrientifoldBasics}

In this section we will introduce some preliminaries on Type IIB Calabi-Yau  
orientifold compactifications with space-time filling D7 branes.  
In order for such a scenario to be globally consistent and to  
preserve $\cN=1$ supersymmetry in the  four flat dimensions we consider  
an orientifold projection which allows for O3 and O7 planes and takes the form 
$\cO=(-1)^{F_L} \Omega_p \sigma$.  Here $\sigma$ is a holomorphic and 
isometric involution of the internal Calabi-Yau manifold $Y$. The action of 
$\sigma$ on the K\"ahler form $J$ of $Y$ is $\sigma^* J = J$, while the 
holomorphic $(3,0)$ form transforms as   $\sigma^*\Omega=-\Omega$. Similarly, 
for the other string fields to remain in the spectrum they have to transform with the  
appropriate parity under $\sigma$. 
 
To determine the four-dimensional effective theory one first needs to
examine the surviving bulk and brane fields.  At least locally, each
such field can be identified with an element of an appropriate bundle
valued cohomology group on the internal manifold and the cycles
wrapped by the D-branes. The involution $\sigma$ splits the cohomology
groups into eigenspaces, and allows one to identify the spectrum
preserved by the orientifold. Focusing on the bulk fields
corresponding to closed string excitations, one notes that
$H^{p,q}(Y,\bbC)$ splits as $H^{p,q}_+ \oplus H^{p,q}_-$ with
dimensions $h^{p,q}_\pm$ respectively.  One thus obtains the complex
dilaton $\tau = C_0 + i e^{-\phi}$, $h^{1,1}_+$ complexified K\"ahler
moduli $T_I$ and $h^{1,1}_-$ B-field moduli $G^i$ given
by~\cite{Grimm:2004uq, Grimm:2005fa}
\begin{equation} 
  \label{def-GT}
  T_I =   \int_{\Gamma^+_I} \Pi \ , \qquad  G^i = \int_{\gamma^i_-} \Pi\ , \qquad \Pi = e^{-B} \Big(e^{-\phi} \R \big[ e^{iJ} \big] + i C_{\rm RR}\Big)
\end{equation} 
where $C_{\rm RR} = C_0+C_2+C_4$. The cycles $\Gamma^+_I$ and
$\gamma^i_-$ form a basis of the homology groups $H_{2,2}^+$ and
$H_{1,1}^-$, respectively.  We will call the continuous moduli $G^i$
simply $B_-$ moduli, since they encode the variations of the NS-NS and
R-R two-forms.  While no dynamical moduli are associated with the
reduction of the B-field along the positive 2-cycles $ \Gamma^I_+ \in
H_{1,1}^+$ there can still be discrete non-zero B-flux $ \frac{1}{2
  \pi}\int_{\Gamma^I_+} B = \frac{1}{2}$.  This survives the
orientifold action due to the axionic shift symmetry
$\int_{\Gamma_+^I} B \rightarrow \int_{\Gamma_+^I} B+ 2 \pi$ and will
sometimes be referred to as $B_+$ flux. In the following we will
determine which quantities in the four-dimensional action depend on
which of these closed string moduli.

\subsection{Intersecting D7-Branes With Gauge Bundles}
\label{sec:generalities}
 
We first discuss the inclusion of space-time filling D7-branes in more
detail. Consider wrapping a stack of $N_a$ D7-branes around a
four-cycle $D_a$ in $Y$. The calibration condition for the D7-branes
requires $D_a$ to be a holomorphic divisor~\cite{Marino:1999af}. The
orientifold symmetry $\sigma$ maps $D_a$ to its orientifold image
$D_a'$ so that in the upstairs geometry each brane is accompanied by
its image brane.  There are three different cases to be distinguished:
\begin{enumerate}
\item \label{Dcase1} $[D_a] \neq [D_a']$,
\item \label{Dcase2} $[D_a] = [D_a']$ but $D_a \not= D_a'$ point-wise, and
\item \label{Dcase3} $D_a = D_a'$ point-wise, that is, D7-branes
  coincide with an O-plane.
\end{enumerate}
In the first two situations, the D7-brane may or may not intersect an
O7-plane.  For vanishing gauge flux, branes of the first type carry
unitary gauge groups, while those of the other two types yield
symplectic or orthogonal gauge
groups. 
In absence of CFT methods to uniquely distinguish SO vs. SP Chan-Paton
factors the rule of thumb is that a stack of $N_a$ branes plus their
$N_a$ image branes on top of an $O7^{-/+}$-plane\footnote{A
  $O7^{-/+}$-plane carries $-8/+8$ times the charge of a D7-brane in
  the upstairs geometry.} gives rise to a gauge group
$SO(2N_a)/SP(2N_a)$. The same configuration along a cycle of type~\ref{Dcase2}
with locally four Dirichlet-Neumann boundary conditions to the
$O7^{-/+}$-plane yields gauge group $SP(2N_a)/SO(2N_a)$.

\subsubsection*{Gauge fluxes on D7-branes}
 
Each stack of D7-branes can carry non-vanishing background flux for
the Yang-Mills field strength $F_a$.  Recall that the field strength
$F_a$ appears in the Chern-Simons and DBI action only in the gauge
invariant combination ${\cal F}_a = F_a + \i^* B \mathbf{1}$, where $\i: D_a
\rightarrow Y$ denotes the embedding of the divisor $D_a$ into the
ambient space. Therefore all physical quantities depend a priori only
on ${\cal F}_a$.  However, as we will describe in detail below, with
the exception of the D-term supersymmetry condition only the discrete
$B_+$-flux effectively enters the consistency conditions.

A consistent configuration of internal gauge flux is mathematically
described in terms of a stable holomorphic vector bundle\footnote{More
  generally, gauge flux is described by coherent sheaves, but for our
  purposes it suffices to consider vector bundles on the divisors.} by
identifying the curvature of its connection with the Yang-Mills field
strength.
For all concrete applications in this article it will be sufficient to
restrict  ourselves to the simplest case of line bundles $L_a$,
corresponding  to Abelian gauge flux.
For a single D-brane wrapping a simply connected divisor 
these are determined uniquely by 
their first Chern class $c_1(L_a)$ as an element of $H^2(D_a)$ or equivalently by a two-cycle $l_a$ with class in $H_2(D_a)$ as $L_a = {\mathcal O}(l_a)$. 
For stacks of several coincident branes wrapping the divisor $D_a$ we also have to specify the embedding of the $U(1)$ structure group of the line bundle into the original gauge group on the branes. 

Let us start with a stack of $N_a$ branes of type 1 and decompose the background value of the physical Yang-Mills field strength $\cal F$ as
\begin{equation}
{\cal F}_a = T_0 \, (   F^{(0)}_a  +  \i^* B ) + \sum_i T_i\, F^{(i)}_a.
\end{equation}
Here $T_0 = 1_{N_a \times N_a}$
refers to the diagonal $U(1)_a \subset U(N_a)$ while $T_i$ are the
traceless Abelian\footnote{This can be generalised to non-Abelian
  vector bundles. For example, on a stack of $N_a$ coincident branes
  one can define a holomorphic rank $n_a$ bundle (with $n_a$ dividing
  $N_a$) and embed its structure group $U(n_a)$ into the original
  $U(N_a)$ theory.   This breaks the
  four-dimensional gauge group down to the commutant $U(N_a/n_a)$ of
  $U(n_a)$ in $U(N_a)$. See ~\cite{Blumenhagen:2005pm, Blumenhagen:2005zh} for a general discussion in terms of D9-branes on Calabi-Yau manifolds.}  elements of $SU(N_a)$.  This defines the
line bundles $L_a^{(i)}$ as
\begin{equation}
\label{def_c1(L)}
c_1(L^{(0)}_a) = \frac{1}{2 \pi} (F^{(0)}_a + \i^*B) \in H^2(D_a), \quad c_1(L^{(i)}_a) = \frac{1}{2 \pi} F^{(i)}_a \in H^2(D_a).
\end{equation}
Note that
in view of the appearance of ${\cal F}_a$ in all physical equations 
the $B$-field is to be included in $c_1(L^0_a)$.

While the effect of $L_a^{(0)}$ is merely to split $U(N_a) \rightarrow SU(N_a) \times U(1)_a$, 
the other $L_a^{(i)}$ will break $SU(N_a)$ further. The relevant example we will be studying 
in detail is the breaking of $U(5)_a \rightarrow SU(5)\times U(1)_a \rightarrow 
SU(3) \times SU(2) \times U(1)_Y \times U(1)_a$ by means of diagonal flux and 
another line bundle corresponding to the hypercharge generator $T_Y$. Note that 
the Abelian gauge factors may become massive via the St\"uckelberg mechanism~\cite{erik}.
 
For a stack of $2N_a$ invariant branes of type 2 and 3 a non-trivial
bundle $L_a^{(0)}$ breaks $SO(2N_a)/ SP(2N_a) \rightarrow SU(N_a)
\times U(1)_a$ and the embedding of $L_a^{(i)}$ works in an analogous
manner.  We will be more specific in the context of the concrete setup
described in \autoref{subsec_GG}.

In general it is possible for some components of\footnote{In the
  sequel we will sometimes omit the superscripts in $L_a^{(i)}$ to
  avoid cluttering of notation.} $c_1(L_a)$ along $H^2(D_a)$ to be
trivial as elements of $H^2(Y)$.  Recall that the inclusion $\i: D_a
\rightarrow Y$ defines the pushforward $\i_*: H_2(D_a) \rightarrow
H_2(Y)$ and pullback $\i^*: H^2(Y) \rightarrow H^2(D_a)$.  Then one
can split $L_a$ as
\begin{equation}
  L_a = \i^* {\mathbb L}_a \otimes R_a
  ,
\end{equation}
with ${\mathbb L}_a = {\cal O}(\ell_a)$ defined as a line bundle on
the Calabi-Yau $Y$.  The part of $L_a$ trivial in $Y$, denoted as
${R}_a = {\cal O}(r_a)$, corresponds to a two-cycle $r_a$ which is
non-trivial on $D_a$ but a boundary in $Y$, that is, $[r_a] \in \ker(\i_*)$.
The possibility of considering such   gauge flux in the relative cohomology
of $D_a$ in $Y$  was first pointed out in~\cite{Jockers:2004yj,Jockers:2005zy} and its
relevance for model building was stressed in~\cite{Buican:2006sn,Grimm:2008ed}.

We need to understand which quantities are affected by a  relative flux 
$R$. In this context, we will make heavy use of the following 
integrals
\bea
\label{relativrela}
          &&\int_{D_a} c_1( \i^* {\mathbb L}_b )\wedge c_1( \i^* {\mathbb L}_c)=
          \int_Y [D_a] \wedge c_1({\cal O}({\ell}_b)) \wedge 
          c_1({\cal O}(   {\ell}_c))=\kappa_{abc}, \nonumber \\ 
           &&\int_{D_a} c_1( \i^* {\mathbb L}_b )\wedge c_1( {R}_c)=
          \int_{D_a\cap \ell_b}  c_1( R_c)  =0, \\ 
 &&\int_{D_a} c_1( R_b)\wedge c_1( R_c)=\eta_{abc}\; , \nonumber
\eea 
where $\kappa_{abc}$ and $\eta_{abc}$ are not necessarily zero.
We conclude that the integral over a divisor of a pull-back 
two form wedge a two-form which is trivial in $Y$ vanishes.
As will be detailed below, this implies that a bundle in the
cohomology which is trivial in $Y$ but non-trivial on $D_a$ does not
affect the chiral spectrum, the D-term supersymmetry conditions and the
D5-brane tadpole of the brane configuration.
However, it does affect the gauge symmetry, the D3-brane tadpole
and the non-chiral spectrum of the model.

\subsubsection*{Quantisation condition}

Essential both for consistency of the theory  and for concrete applications 
is to appreciate the correct quantisation conditions
on the gauge flux. Following~\cite{Freed:1999vc} they are determined by 
requiring that the worldsheet path integral for an open string wrapping 
the two-surface $\Sigma$ with boundary $\partial \Sigma$ along $D_a$ be single-valued. 
Consider first a single brane wrapping the divisor $D_a$ and carrying Abelian gauge flux $F_a$.
The quantity to be well-defined is given  by
\bea
\label{path}
{\rm Pfaff}(D) \,\,\,  {\rm exp} \, (i   \int_{\partial\Sigma}  A_a) \, \,\,  {\rm exp} \, (i\, \int_{\Sigma}  \,B )
\eea
in terms of the Pfaffian of the Dirac operator, the connection $A$ of the Abelian gauge bundle and the $B$-field. 
If $D_a$ is not {\it Spin}, that is, if 
$c_1(K_{D_a}) \ne 0 \mod 2$,
the Pfaffian picks up a holonomy upon transporting $\partial \Sigma$ around a loop on $D_a$~\cite{Freed:1999vc}.
This holonomy must be cancelled by the second factor in eq.~\eqref{path}. For internal line bundles this is guaranteed if the gauge flux obeys the condition
\begin{equation}
\label{quant1}
\int_\omega F_a  + \frac{1}{2}\int_{\omega}K_{D_a} \in \mathbb Z \quad \quad \forall \omega \in H_2(D_a,\mathbb Z), 
\end{equation}
or equivalently, using our convention eq.~\eqref{def_c1(L)}, 
\begin{equation}
\label{quant2}
c_1(L_a) - \i^* B + \frac{1}{2} c_1( K_{D_a})  \in H^2(D_a,\mathbb Z).
\end{equation}
Note in particular that for trivial $B$ flux along $D_a$, $\i^*B=0$,
the Abelian gauge bundle on the single brane $D_a$ has to be
half-integer\footnote{The quantisation condition eq.~\eqref{quant2} with
  non-trivial B-field is related to the concept of vector bundles
  without vector structure~\cite{Witten:1997bs} in Type I theory as
  studied recently, for example, in~\cite{Bachas:2008jv,Pesando:2008xt}.} quantised if the
divisor $D_a$ is not $Spin$.

This condition is readily generalised to line bundles on stacks of D-branes.
The probe argument of~\cite{Freed:1999vc} now implies that the path integral has to be well-defined for every disk worldsheet $\Sigma$ with boundary on each of the branes in the stack of $N_a$ D-branes wrapping $D_a$. This requires
\begin{equation}
\label{quant3}
T_0  \, (c_1(L^{(0)}_a) - \i^* B)   +  \sum_i T_i \, c_1(L^{(i)}_a)  +  \frac{1}{2} T_0\,  c_1(K_{D_a}) \in H^2(D_a,{\mathbb Z})_{N_a\times N_a}.
\end{equation}
where the notation on the right hand side means that all elements
of the $N_a\times N_a$ matrix on the left hand side are in 
$H_2(D_a,{\mathbb Z})$.
One concludes that depending on the precise from of $T_i$ the bundles $L^{(i)}_a$ can in general be fractionally quantised, a fact that will be very important for our applications.

A second constraint arises for the continuous $B_-$ moduli in $H^2_-$:
the restriction to $D_a$ of the characteristic class $\zeta$  of the 
$B_-$-field, introduced in~\cite{Freed:1999vc},  has to equal the 
third Stiefel-Whitney class of $D_a$. Recall from~\cite{Freed:1999vc} that modulo torsion, $\zeta$ is
given by the field strength $H= dB_-$
and that for complex divisors the third Stiefel-Whitney class 
is always zero. Moreover, for all surfaces considered
in this paper, we have $H^3(D_a, \mathbb Z)=0$ so that $H=dB_-$ always
restricts to zero on the divisor. Therefore no further condition
on the B-field moduli $G^i$ introduced in \eqref{def-GT} arises from these considerations.

\subsubsection*{Orientifold action} 
 
Let us now describe the orientifold action on the gauge flux.  To this
end note that the orientifold action $\sigma: D \rightarrow D'$
induces a map on cohomology, $\sigma^*: H^2(D_a,{\mathbb Z})
\rightarrow H^2(D_a',{\mathbb Z})$.  The full orientifold action on a
vector bundle on $D_a$ is given by the composition $\sigma^*
\Omega_p$.  Here $\Omega_p$ acts as dualisation, $L_a \rightarrow
L_a^{\vee}$. In particular, the Chern character of the image bundle is
\begin{equation} 
\label{L'} 
{\rm ch}_k (L_a') = (-1)^k \sigma^* {\rm ch}_k (L_a) = \sigma^* {\rm ch}_k(L_a^{\vee}). 
\end{equation} 

We now discuss the three cases introduced at the beginning of
\autoref{sec:generalities} in turn. In the first situation, where not
even the homology class of the brane is preserved, one can define two
divisors $D_a^\pm$ and two vector bundles $L^\pm_a$ by setting
\begin{equation} 
  \label{def-Dpm}
  D_a^\pm = D_a \cup (\pm D_a')\ , \qquad \quad L^\pm_a|_{D_a} = L_a\ ,\qquad
  L^\pm_a|_{D_a'} = \pm L_a'
  ,
\end{equation}
where $-D_a'$ is the cycle $D_a'$ with reversed orientation.  Upon
setting $H^2( D_a^+ ) = H^2(D_a ) \oplus H^2(D_a')$ and decomposing
the latter into positive and negative eigenspaces under $\sigma^*$,
$H^2(D_a^+) = H^2_+(D_a^+ ) \oplus H^2_-(D_a^+
)$~\cite{Jockers:2004yj}, it follows that $c_1(L^+_a) \in H^2_-(D^+_a
)$.

In the second case the homology class is preserved but the brane is
not point-wise fixed. Hence, the homology class of $D^-_a$ in
eq.~\eqref{def-Dpm} is trivial and we can use $[D_a] = \frac{1}{2}
[D_a^+]$.  The degree-2 cohomology group of $D_a$ thus splits again as
$H^2(D_a,{\mathbb Z} ) = H^2_+(D_a) \oplus H^2_-(D_a )$.  On the
covering space of the orientifold one requires an even number of
branes in the homology class of $D_a$ which are pairwise identified
under the involution $\sigma$. Clearly, this corresponds to an integer
number of branes on $D_a^+$.  The Chern class $c_1(L_a)$ on $D_a$ is
in the full $H^2(D_a)$ and $(D_a,L_a)$ is mapped to $(D_a,L_a')$ as in
eq.~\eqref{L'}.
 
In the third case, for $D_a$ on top of the orientifold, $H^2(D_a ) =
H^2_+(D_a )$ and $(D_a,L_a)$ is mapped to $(D_a,L_a^{\vee})$. This
case directly parallels the situation for D9-branes in Type I
compactifications.  An odd number of branes stuck on top of the
orientifold plane is not possible, as discussed recently
in~\cite{Collinucci:2008pf}. Formally we therefore work upstairs with
the system $2N_a \,D_a$ carrying the invariant bundle $L_a \oplus
L_a^{\vee}$.

\subsection{Tadpole Cancellation for Intersecting D7-Branes}

In consistent compactifications it is crucial to cancel the tadpoles  
of the space-time filling intersecting D7-branes.  
Satisfying the tadpole cancellation condition ensures that the  
spectrum is free of non-Abelian gauge anomalies. In general, D7-branes  
carry also induced D3- and D5- charges arising due to a non-trivial gauge-field  
configuration on the seven branes and through curvature corrections. All induced  
tadpoles for a compactification have to be cancelled.  

Throughout this article we will be working upstairs on the ambient Calabi-Yau manifold before taking the quotient by $\sigma$. 
Recall that the K-theoretic charges $\Gamma$ of a D7-brane and the O7-plane along divisors $D_a$ and $D_{O7}$ are encoded in the Chern-Simons coupling to the closed RR-forms $2 \pi\, \int_{ {\cal R}^{1,3} \times D_a} \sum_{2p} C_{2p} \Gamma$. 
Concretely these are given by 
\bea 
\label{CS_couplings} 
&& S_{D7}= 2 \pi \, \int_{{\cal R}^{1,3} \times D_a} \sum_{2p} C_{2p}\,    {\rm tr}\left[ \, e^{\frac{1}{2\pi} {\cal F}_a }\right]      \,  \sqrt{\frac{\hat A(TD)}{\hat A(ND)}}, \nonumber \\ 
&& S_{O7}= -16 \pi \int_{{\cal R}^{1,3} \times D_{O7}} \sum_{2p} C_{2p} \,  \sqrt{\frac{L(\frac{1}{4} \, TD_{O7})}{L(\frac{1}{4} \, ND_{O7})}} 
\eea 
in terms of the $\hat{\rm A}$-roof and the Hirzebruch genus associated with the tangent and normal bundles to the respective divisors. 
The D7-, D5-, and D3-brane charges follow upon straightforward
decomposition of eq.~\eqref{CS_couplings}.

Let us start by discussing  cancellation of the D7-charge. 
In a consistent orientifold compactification the D7-brane charge
 has to equal the charge   carried by the O7-planes, 
\begin{equation} 
\label{tadseven} 
  \sum_a  N_a\, ([D_a]+ [D'_a]) = 8\, [D_{\rm{O7}}]
  , 
\end{equation} 
where the sum is over all D7$_a$-branes.  Since supersymmetric
D7-branes together with their orientifold images wrap cycles $D_a+
D'_a$ in the homology classes of $H_4^+(Y)$, the orientifold invariant
charges are captured by the Poincar\'e dual cohomology $H^2_+(Y)$ on
the ambient Calabi-Yau manifold $Y$.
 
A net D5-brane charge can be induced by a gauge-field configuration
${\cal F}_a$ on $D_a$ (and the respective orientifold images) if there
exist non-trivial elements in $H^2_-(Y)$.  The D5-brane charge along
the element $\omega_{b} \in H^2(Y,{\mathbb Z})$ of a stack of $N_a$
branes carrying Abelian gauge flux as in eq.~\eqref{def_c1(L)} reads
\begin{equation} 
\Gamma^{D5}_{\omega} =  \frac{1}{2\pi} \, \int_Y  \omega \wedge [D_a] \wedge {\rm tr} {\cal F}_a\ ,
\end{equation} 
where
\begin{equation}
\label{trF}
\frac{1}{2\pi} {\rm tr} {\cal F}_a = \sum_I {\rm tr} \left[ T_I\right]  \,\, c_1(L_a^{(I)}), \quad\quad I=0,i\ . 
\end{equation}
The condition for cancellation of D5-brane charge therefore takes the form 
\begin{equation} 
\label{D5cancell} 
\sum_a  \int_Y \omega \wedge  \big(\, [D_a] \wedge  {\rm tr} {\cal F}_a + [D_a'] \wedge  {\rm tr} {\cal F}_{a'} \, \big) =0 \ ,
\end{equation} 
and has to be satisfied for all elements $\omega \in H^2_-(Y)$. 
Clearly this condition is vacuous if all branes are of the type $[D_a]=[D_a']$ and 
$c_1(L^{(I)}_{a'})= c_1\bigl( (L^{(I)}_a) ^{\vee} \bigr)$, but it may be quite restrictive in more general situations.

The general condition for cancellation of the D3-brane tadpole takes the form 
\begin{equation} \label{D3_tadpole} 
   (N_{\rm{D3}}+N_{\rm{D3'}}) + N_{\rm{flux}}  - 
\sum_a  (Q^a_{\rm D7} + Q'^a_{\rm D7})= \frac{N_{\rm O3}}{2} + Q_{\rm O7} \ . 
\end{equation} 
Here $N_{\rm{D3}}$ counts the number of $D3$-branes, each of which is
accompanied by its orientifold image\footnote{In particular, if
  $n_{D3}$ D3-branes lie inside an $O7$-plane they come together with
  their images and yield gauge group $Sp(2n_{D3})$.}. $N_{\rm flux}$
denotes the possible contributions from $G_3=F_3+\tau\, H_3$ form
flux, which is in particular important for complex structure moduli
stabilisation.  The induced D3-charge on the O7-planes is given by
\begin{eqnarray}
  Q_{\rm O7} =\frac{\chi(D_{\rm O7})}{6} &=& \frac{1}{6} \int_Y c_2(D_{\rm O7}) \wedge [D_{\rm O7}] \nonumber \\
  &=&\frac{1}{6} \int_Y [D_{\rm O7}]^3+ c_2(T_{Y}) \wedge [D_{\rm O7}] \; . 
\end{eqnarray}
If a stack of $N_a$ D7-branes wraps a smooth divisor of
type~\ref{Dcase1} or~\ref{Dcase3}, as defined on
page~\pageref{Dcase1}, their D3-charge reads
\begin{equation} 
  \label{D3_onD7} 
  Q^a_{\rm D7} = 
  N_a \, \frac{\chi(D_a)}{24} + 
  \frac{1}{8\pi^2}  \int_{D_a} \tr {\cal F}^2_a 
\end{equation}    
with   
\bea   
\frac{1}{8\pi^2} {\rm tr}{\cal F}^2_a   = \frac{1}{2}\,\,  \sum_{I,J} \tr [ T_I \, T_J] \,\, c_1 \bigl( L_a^{(I)} \bigr) \, c_1 \bigl( L_a^{(J)} \bigr) \  . 
\eea
More subtle is the case~\ref{Dcase2}, since eq.~\eqref{D3_tadpole}
will be modified as discussed in~\cite{Aluffi:2007sx,Collinucci:2008pf}.  One
replaces the Euler characteristic by
\begin{equation} 
  Q^a_{\rm D7} =N_a \,  \frac{\chi_o(D_a)}{24}+  \frac{1}{8\pi^2}  
  \int_{D_a}\!\! \tr{\cal F}^2_a  \ , \qquad \chi_o(D_a) = \chi(\Sigma_a) - n_{\rm pp}
  , 
\end{equation} 
where $\Sigma_a$ is an auxiliary surface  obtained by
blowing up the singular points in $D_a$, while $n_{\rm pp}$ counts the
number of pinch points in $D_a$.

The relation to the F-theory D3-brane tadpole condition becomes
obvious if one slightly reorders the terms in eq.~\eqref{D3_tadpole}
and divides by two,
\begin{equation} 
  \label{D3_tadpolezwei} 
  N_{\rm{D3}} + \frac{N_{\rm{flux}}}{2} 
  +N_{\rm gauge}
  = \frac{N_{\rm O3}}{4} + 
  {\chi(D_{\rm O7})\over 12} +\sum_a  N_a\, {\chi_o(D_a)\over 24} 
  \end{equation} 
with
\begin{equation}
N_{\rm gauge}=  - 
  \sum_a   \frac{1}{8\pi^2}  
  \int_{D_a}\!\! \tr{\cal F}^2_a = - \frac{1}{2} \, \sum_a N_a \int_{D_a}  \,  \sum_{I,J} \tr [ T_I \, T_J] \,\, c_1 \bigl( L_a^{(I)} \bigr) \, c_1 \bigl( L_a^{(J)} \bigr) \  . 
\end{equation}
The right-hand side of equation \eqref{D3_tadpolezwei} is precisely $\chi(Y_4)/24$ in the
F-theory lift of this Type IIB orientifold, where $Y_4$ denotes
the Calabi-Yau fourfold.  This implies that generically each
topologically different arrangement of ${\rm D7}$-branes
cancelling the RR eight-form tadpole constraints lifts
to a different Calabi-Yau fourfold with different
Euler characteristic. For the trivial solution with eight  ${\rm D7}$-branes
placed right on top of the smooth orientifold plane with $n_{\rm pp}=0$ 
the right
hand side of \eqref{D3_tadpolezwei} simplifies to
$\frac{N_{\rm O3}}{4}+{\chi(D_{\rm O7})\over 4}$. It is a consistency check
that this number is indeed an integer. 

Let us emphasise  that for the cancellation of anomalies only the D7 and D5-tadpole
constraints are important. The D3-brane tadpole is in some sense
only related to the non-chiral sector of the D-brane theory.
This is related to the fact that a D3-brane can never carry any chiral
modes, as it can in principle be moved to a position away from
the D7-branes.  The expectation is that a globally consistent
Type IIB orientifold model with a supersymmetric D7- and D5-brane sector
lifts up to F-theory on a compact Calabi-Yau fourfold. 
The cancellation of the D3-brane tadpole is an additional
attribute both in Type IIB orientfolds and in F-theory models.
Taking also into account that  for moduli stabilisation and the realisation
of inflation, the presence of (a small number of) anti-D3-branes 
is very welcome,
in this paper we take all the D7- and D5-brane supersymmetry constraints
very seriously but are a bit more relaxed about the existence of 
anti- D3-branes in the system. In fact, we will find that in our
semi-realistic GUT examples the D3-brane tadpole can easily be saturated by already
modest addition of gauge fluxes on the D7-branes.

\subsubsection*{Role of $\mathbf{B_-}$-moduli}

Before proceeding we would like to comment on the role
of the continuous $B_-$-moduli appearing in $\cal F$ and thus in the D-brane charges.  
It is natural to wonder how to reconcile their contribution with the discrete nature of a quantity such as the D5-~or D3-brane charge.

In fact the $B_-$ moduli
decouple from the tadpole equations by means of the
$D7$-brane tadpole cancellation condition \eqref{tadseven}
and the simple observation that the $B_-$ field restricts trivially to the 
O7-plane, 
\begin{equation} \label{restO7}
           \int_{D_{O7}} \omega \wedge B_- =0\ ,\qquad  \forall\, \omega \in H^2_-(Y)\; .
\end{equation}
Concretely, the $B_-$-contribution to the D5-brane tadpole condition \eqref{D5cancell} 
\begin{equation}
    \sum_a N_a \int_Y \omega \wedge \Bigl( [ D_a] \wedge B_- + [D_a']\wedge B_- \Bigr) 
    = 8 \int_Y \omega \wedge [ D_{O7} ]\wedge B_- =0\ 
\end{equation}
indeed vanishes due to \eqref{tadseven} and \eqref{restO7}. 

To isolate the $B_-$-moduli in the D3-brane tadpole let us introduce the quantity
\begin{equation} \label{B-splitoff}
c_1(\widetilde L_a) =  c_1(L_a) - B_-
\end{equation}
and rewrite the induced D3-brane tadpole as
\bea
    -\sum_a {N_a}  \left[\, \int_{D_a} \left( c_1(\widetilde L_a)+ B_- \right)^2
        + \int_{D_a'} \left( c_1(\widetilde L'_a)+ B_- \right)^2 \,\right].
\eea
For simplicity we are sticking to
gauge flux $\frac{1}{2\pi}{\cal F} =T_0 c_1(L_a)$. For the mixed term we find that 
\begin{equation}
     -\sum_a N_a \int_{Y} \Bigl( [D_a]\wedge c_1(\widetilde L_a) +  [D'_a]\wedge c_1(\widetilde L'_a)\Bigr)\wedge
         B_- = 0\ ,
\end{equation}
where we have used the D5-brane tadpole cancellation condition \eqref{D5cancell}.  
Finally, we evaluate
\begin{equation}
     -\sum_a N_a \int_{Y} \Bigl(  [D_a]\wedge  B_-^2 +  [D'_a]\wedge B_-^2
         \Bigr) =
        -8\, \int_{D_{O7}}  B_-^2 =0
\end{equation}
so that as anticipated the continuous $B_-$-moduli do not
appear in the tadpole cancellation conditions.

\subsubsection*{K-Theory charge cancellation}

Apart from cancellation of these homological charges, also all K-theoretic torsion charges have to sum up to zero. 
In general it is a very non-trivial  task to compute all in particular torsional
K-theory constraints.  
However, according to the probe brane argument of~\cite{Uranga:2000xp} cancellation of torsion charges is equivalent to absence of a global $SU(2)$ Witten anomaly on the worldvolume of every probe brane supporting symplectic Chan-Paton factors. 
In concrete compactifications this amounts to requiring an even number of
fundamental representations in the sector between the physical D7-branes and
each symplectic probe brane. Note that in a concrete model determining 
all symplectic four-cycles is also far from trivial.

\subsection{The Massless Spectrum} 
\label{masslessspec_gen} 
 
For applications to phenomenology it is essential to understand the
massless matter arising from open strings stretching between two
stacks of D7-branes.
 
Non-chiral matter transforming in the adjoint representation emerges
from strings with both endpoints on the same D-brane along $D_a$. It
consists of the vector multiplet together with $h^{1,0}(D_a)$ and
$h^{2,0}(D_a)$ chiral multiplets describing the Wilson line and
deformation moduli of the D7-branes.  Matter in the bifundamental
representation\footnote{For the general overview we only consider
  diagonal embeddings and postpone a discussion of more general line
  bundles to the applications in \autoref{sec:GutModelExample}.} $(\ov
N_a,N_b)$ and $(N_a, N_b)$ arises from open strings stretching between
two different D7-branes in the $(a,b)$ and $(a',b)$ sector,
respectively. Intersections between a brane and its image, that is, of
type $(a',a)$, yield matter in the (anti)symmetric representation.
For example, if all branes are on top of a $O7^{(-)}$-plane, then all
states in the $(a',a)$ sector are anti-symmetrised. On an invariant
brane with four Dirichlet-Neumann boundary conditions with an
$O7^{(-)}$-plane, the $(a',a)$ states are symmetrised. The chiral
spectrum is summarised in \autoref{tab_chir_spec}, see
also~\cite{Blumenhagen:2007sm}.
\begin{table}[htbp] 
  \renewcommand{\arraystretch}{1.5} 
  \begin{center} 
    \begin{tabular}{|c|c|c||c|} 
      \hline 
      \hline 
      sector & $U(N_a)$ & $U(N_b)$ & chirality   \\ 
      \hline \hline 
      $(a b)$         & $\antifund_{\, (-1)}$ & $\fund_{\, (1)}$  & $I_{ab}  $ \\
      \hline
      $(a' b)$         & $\fund_{\, (1)}$ & $\fund_{\, (1)}$   &  $ I_{a'b}$  \\
      \hline 
      $(a'a)$         & $\Yasymm_{\,(2)}$ & $1$ & $\frac{1}{2}( I_{a'a}+2I_{\rm{O7}a})   $  \\ 
      \hline
      $(a'a)$         & $\Ysymm_{\,(2)} $ & $1$ & $\frac{1}{2}( I_{a'a}-2I_{\rm{O7}a})   $ \\
      \hline 
      \hline  
    \end{tabular} 
    \caption{Chiral spectrum for intersecting D7-branes. The
      subscripts denote $U(1)$ charges.}
    \label{tab_chir_spec} 
  \end{center} 
\end{table}
For the concrete computation of the chiral index $I_{ab}$ and to
determine the vector-like spectrum we have to distinguish two
situations according to the localisation of matter on sub-loci of
different dimensions. For simplicity we again only discuss the case
where all D7-branes carry holomorphic line bundles.

\subsubsection{Matter Divisors} 
 
For two D7-branes wrapping the same divisor $D_a=D_b=D$ and carrying
line bundles $L_a$ and $L_b$, matter in the bifundamental
representation $(\ov N_a, N_b)$ is counted by the extension
groups~\cite{Katz:2002gh}
\begin{equation}
  \label{ext1} 
  \Ext^n (\i_* L_a, \i_* L_b), \qquad n=0,\ldots 3
  ,   
\end{equation}
where $i:D\to Y$ defines the embedding of $D$ in the Calabi-Yau $Y$.
The value $n=1$ refers to anti-chiral multiplets transforming as $(\ov
N_a, N_b)$, while $n=2$ corresponds to chiral multiplets in the same
representation.  For consistency, the states counted by the groups
corresponding to $n=0$ and $n=3$ have to absent. These states do not
correspond to matter fields but rather gauge fields and have been
named ghosts in~\cite{Donagi:2008ca}. We show in \autoref{SUSY} that
for supersymmetric configurations with the K\"ahler form inside the
K\"ahler cone these ghosts are automatically absent.  By running
through the spectral sequence, one can show that the sheaf extension
groups eq.~\eqref{ext1} are given by appropriate cohomology groups for
line bundles on the divisor $D$, concretely
\begin{eqnarray}
  \label{externcoh} 
  \Ext^0 (\i_* L_a, \i_* L_b)&=& H^0(D, L_a\otimes L^\vee_b), \nonumber \\ 
  \Ext^1 (\i_* L_a, \i_* L_b)&=& H^1(D, L_a\otimes L^\vee_b)+ 
  H^0(D, L_a\otimes L^\vee_b\otimes N_D), \nonumber \\  
  \Ext^2 (\i_* L_a, \i_* L_b)&=& H^2(D, L_a\otimes L^\vee_b)+ 
  H^1(D, L_a\otimes L^\vee_b\otimes N_D), \nonumber \\  
  \Ext^3 (\i_* L_a, \i_* L_b)&=& \phantom{aaaaaaaaaaaaaaa}\  
  H^2(D, L_a\otimes L^\vee_b\otimes N_D). 
\end{eqnarray}
By Serre duality and $N_D=K_D$ we can relate 
$ H^i(D, L_a\otimes L^\vee_b\otimes N_D)=H^{2-i}(D, L^\vee_a\otimes L_b)$. 
It straightforwardly follows that for the chiral index $I_{ab}$ counting bifundamental matter transforming as $(\ov N_a, N_b)$ one obtains 
\begin{equation}
  \label{chiralext} 
  \begin{split}
    I^{bulk}_{ab} =&\; 
    \sum_{n=0}^3 (-1)^n \dim \Ext^n (\i_* L_a, \i_* L_b)
    \\
    =&\; 
    \chi(D,  L_a\otimes L^\vee_b )-  \chi(D,  L_a\otimes 
    L^\vee_b\otimes N_D )
    \\
    =&\; 
    -\int_Y [D]\wedge [D]\wedge \left(\, c_1(L_a)-c_1(L_b)\, \right)
    .  
  \end{split}
\end{equation}
In these conventions $I_{ab}>0$ if there is an excess of chiral states
in the representation $(\ov N_a,N_b)$.  Note that this expression only
depends on the components of $c_1(L_i)$ which are non-trivial on the
ambient Calabi-Yau manifold, cf. eq.~\eqref{relativrela}.

We have the additional freedom to twist the line bundle $L_a$ on $D$
by a line bundle $R_a$ with $\i_* R_a={\cal O}$. This does not
change the chiral spectrum, though it can change the
vector-like one and will in general contribute to the D3-tadpole.

\subsubsection{Matter Curves} 

If the two D7-branes wrap different divisors $D_a$ and $D_b$ intersecting 
over a curve $C$ of genus $g$ and carrying line bundles $L_a$ and  
$L_b$, the cohomology groups counting matter in $(\ov N_a,N_b)$ are 
\begin{equation}
  H^i \left( C, L_a^\vee \otimes L_b \otimes K_C^\frac{1}{2} \right)
  . 
\end{equation}
Here $i=0$ and $i=1$ refer to chiral and anti-chiral states in the representation $(\ov N_a,N_b)$, respectively\footnote{Note the different assignment of chiral and anti-chiral multiplets with the extension groups of even and odd degree for bulk and localised matter. This can be derived, for example, by T-duality from the analogous phenomenon in D9-D5 systems~\cite{Blumenhagen:2005pm, Blumenhagen:2005zh}.}. 
The chiral index $I_{ab}$ counting the excess of chiral over anti-chiral states transforming as $(\ov N_a,N_b)$ follows from the Riemann-Roch-Hirzebruch theorem as 
\bea 
 I^{loc.}_{ab}=  \chi(C,  L_a^\vee \otimes L_b\otimes K_C^{1\over 2} )= 
  - \int_Y [D_a ]\wedge [D_b]\wedge \left(\, c_1(L_a)-c_1(L_b)\, \right)\; .  
\eea 
In terms of extension groups we therefore get
\begin{equation}
  \label{externcohzwei} 
    \Ext^i (\i_* L_a, \i_* L_b) =
    \begin{cases}
      0 & i=0 \\ 
      H^1(C, L^\vee_a\otimes L_b\otimes K_C^{1\over 2}) & i=1 \\  
      H^0(C, L^\vee_a\otimes L_b\otimes K_C^{1\over 2}) & i=2 \\  
      0 & i=3 
      .
    \end{cases}
\end{equation}
Finally, the index $I_{O7a}$ in \autoref{tab_chir_spec} is
\begin{equation}
  I_{O7a} =  \int_Y [D_{O7}]\wedge [D_a]\wedge c_1(L_a)
  ,
\end{equation}
reflecting the absence of a gauge bundle on top of the orientifold plane.

We leave it to the readers to convince themselves that, as in the context of the tadpole cancellation conditions, the $B_-$-moduli also drop out automatically from all cohomology groups describing the massless spectrum.

\subsection{F- and D-Term Supersymmetry Constraints} 
\label{SUSY} 

Let us discuss the constraints which need to be imposed in order for the brane
configuration to be supersymmetric. In the four-dimensional effective action
these constraints manifest themselves through the vanishing of F- and D-terms
in the vacuum. In the following we will discuss both sets of constraints in
turn. 

We first turn to the supersymmetry constraints imposed by the
vanishing of the D-terms. Recall that the D-term induced by a gauging
of a field-independent symmetry with Killing vector $X_a^L$ is of the
form $D_a = \bar X_a^L \partial_{\bar M^L} K$, where $K$ is the
K\"ahler potential. Let us recall the induced gauging for the complex
scalars $T_I$ and $G^i$ defined in \eqref{def-GT}. A gauging of $T_I$
can be induced for a non-trivial line bundle on a D7-brane
$(D_a,L_a)$, while $G^i$ can be gauged if there exists a $D'_a$ which
is not homologous to $D_a$, that is, if we are in the case~\ref{Dcase1}
defined at the beginning of \autoref{sec:generalities},
page~\pageref{Dcase1}. The Killing vectors for these gaugings are of
the form
\begin{equation} \label{KillingV}
  X_{a\,I} = \int_{D_a} [ \Gamma^I_+] \wedge c_1(\widetilde L_a)\ , \qquad X_a^i = \int_{D_a} [ \gamma_-^i]  \wedge B_-\ , 
\end{equation} 
where $\Gamma^I_+,\gamma_-^i$ are the two-cycles introduced
in eq.~\eqref{def-GT} and thereafter to define $T_I,G^i$. Note that there is no
continuous moduli dependence in $X_{a\,I}$ since we have explicitly
split off the $B_-$ field as in eq.~\eqref{B-splitoff}.  One next
notes that~\cite{Grimm:2004uq, Grimm:2004ua}
\begin{equation} \label{der_K}
   \partial_{T_I} K\ \sim\ r^I\ 
   ,\qquad 
   \partial_{G^i} K\ \sim\ s_i
   , 
\end{equation}
where $r^I,s_i$ arise in the expansions $J = r^I\, [\Gamma^+_I]$ and
$J \wedge B_- = s_i [\gamma^i_-]$.  It is important to note that the
expression for $\partial_{T_I} K$ in eq.~\eqref{der_K} is also valid
away from the large volume limit. For example, one of the $r^I$ can
become small while $\partial_{G^i} K$ will receive additional
corrections, for example, due to world-sheet instantons. We thus encounter a moduli
dependent Fayet-Iliopoulos term for the configuration of the
form~\cite{Marino:1999af, Jockers:2004yj}
\begin{equation} \label{FIterm}
     \xi_a\ \sim\ \int_{D_a} \i^* J \wedge (c_1(\widetilde L_a) + B_- ) = 
\int_{D_a} \i^* J \wedge c_1(L_a)\ . 
\end{equation}
Note that $\xi_a$ depends on the pullback of the K\"ahler form $J$ of $Y$  to
the D7-brane and as a consequence of eq.~\eqref{relativrela} only on the
components of $c_1(L_a)$ which are non-trivial on $Y$. 
Furthermore the $B_-$-moduli, encoded in $c_1(L_a)$, do not drop out of the D-terms.
For vanishing VEVs of the chiral fields charged under the $U(1)$ supported on
the D7-branes, the D-term supersymmetry condition requires these FI-terms to
vanish. This imposes conditions on the combined K\"ahler and $B_-$ moduli space.
As long as the K\"ahler moduli are chosen such that $J$ is indeed invariant under the orientifold action, the Fayet-Iliopoulos term for $(D_a,L_a)$ and $(D_a',L_a')$ coincide.
We will encounter that in a concrete example 
in \autoref{sec_offdiag}.

For line bundles $L_a\ne {\cal O}$ satisfying the D-term constraint
eq.~\eqref{FIterm}, we will now derive two important consequences:

\subsubsection*{No ghosts}

First, we realise that the FI-term is nothing else than the slope
$\mu(L_a)$ of the line bundle $L_a$.  Now to come back to the question
of ghosts in the massless spectrum in eq.~\eqref{externcoh}, it is
important to recall the general fact
that 
\begin{itemize}
\item For two vector bundles $V_a, V_b$ of equal slope and rank,
  $\mu(V_a)=\mu(V_b)$ and ${\rm rk}(V_a)={\rm rk}(V_b)$, the existence
  of a map $0\to V_a\to V_b$ implies that $V_a=V_b$.
\end{itemize}
We thus conclude for the extension groups between two supersymmetric
line bundles $L_a\ne L_b$ that ${\rm Ext}^0(\i_* L_a,\i_* L_b)={\rm
  Ext}^3(\i_* L_a,\i_* L_b)=0$. Indeed, if $H^0(D,L_a\otimes
L_b^\vee)$ were non-vanishing, we could define a map $0\to {\cal O}
\to L_a\otimes L_b^\vee$ where both ${\cal O}$ and, by hypothesis,
$L_a\otimes L_b^\vee$ have vanishing slope.  Therefore, $L_a=L_b$ in
contradiction to our assumption. The same reasoning for the dual
bundle $L^\vee_a\otimes L_b$ shows that $\Ext^3(\i_* L_a,\i_* L_b)=0$.

\subsubsection*{D3-tadpole contribution}

Second we note that for line bundles with vanishing slope
for a K\"ahler form inside the K\"ahler cone, the contribution
of the gauge flux to the D3-brane tadpole always has the same
sign
\begin{equation}
  -{N_a\over 2} \int_{D_a} c_1^2(L_a)\ge 0
  .
\end{equation}
Indeed, on a surface $D_a$  the set of $c_1(L_a)$ with vanishing slope
is given by $H_2(D_a)-\{ M\, \cup\, -M\}$, where $M$ denotes the
Mori cone of $D_a$. However, the Mori cone contains all classes $C$
with $C^2>0$ and $C\cdot K>0$. Therefore, $c^2_1(L_a)\le 0$.
This result implies that for supersymmetric brane configurations
the possible choices of line bundles are rather limited if
we do not want to introduce anti--D3-branes in the system.

Finally, let us mention that the other supersymmetry conditions,
namely the holomorphy of the divisor and the bundle, arise from a
superpotential
\begin{equation} 
  \label{def-superpot}
  W_{D7} = \int_{\cC(L_a,L_a')} \Omega
  ,
\end{equation}
where $\cC(L_a,L_a')$ is a chain ending on the two-cycle Poincar\'e
dual to $c_1(L_a^+)$ on the divisor $D_a+D_a'$.


\section{SU(5) GUTs and Their Breaking}
\label{sec:GUTsbreak}
 
After this discussion of the general model building rules for Type IIB
orientifolds with O7- and O3- planes we can now become more specific
about the realisation of $SU(5)$ Georgi-Glashow GUTs. Parts of the
logic are very similar to the implementation of GUT models in Type IIA
intersecting D-branes~\cite{Antoniadis:2000ena, Blumenhagen:2001te,
  Cvetic:2002pj} as described, for example, in~\cite{Blumenhagen:2001te,
  Blumenhagen:2007zk}.  Let us first transfer these rules to our Type
IIB setting.  Then we move forward to describe how the mechanism of
GUT symmetry breaking via $U(1)_Y$ flux, exploited
by~\cite{Beasley:2008kw} in the local F-theory context, can also be
realised in this perturbative orientifold limit.

\subsection{Georgi-Glashow SU(5) GUT} 
\label{subsec_GG} 

The starting point is the construction of a $U(5)\times U(1)$ gauge theory from a stack of five D7-branes wrapping a 
four-cycle $D_a$ and one additional brane wrapping $D_b$. 
These brane stacks  carry  holomorphic bundles 
$L_a$ and $L_b$, respectively.  
 
The orientifold action maps $(D_a,L_a) \rightarrow (D_{a'},L_a')$
(and similarly for $(D_b, L_b)$). As previously discussed, 
this includes the case that $D_a$ is invariant under $\sigma$.  
 First we diagonally embed two line bundles $L_a$ and $L_b$ by identifying 
their structure group with the diagonal $U(1)_a$ and $U(1)_b$, respectively. 
Each of the two Abelian factors $U(1)_a$ and $U(1)_b$ separately acquires a
 mass by the St\"uckelberg mechanism as long as $D_a$ and $D_b$ are 
non-trivial homology cycles~\cite{erik}. 

A more group theoretic way of describing the gauge group and its matter
content is to start with an $SO(12)$ gauge group. 
The embedding of two line bundles with structure groups $U(1)_{a,b}$ can break this
to $U(5)\times U(1)$, where the generators of the two $U(1)$s are embedded into 
$SO(12)$ as
\bea
\label{embedding1}
    U(1)_a&\in & {\rm diag}\left(\, 1_{5\times 5}, 0\, \vert\, 0, -1_{5\times
        5}\, \right),\nonumber \\
     U(1)_b&\in & {\rm diag}\left(\, 0_{5\times 5}, 1\, \vert\, -1, 0_{5\times
        5}\, \right).
\eea
The adjoint of $SO(12)$ decomposes into $SU(5)\times U(1)_a\times
 U(1)_b$ representations as
\bea
   [66]&=&[24]_{(0,0)} + [1]_{(0,0)} + [10]_{(2,0)} + [\ov{10}]_{(-2,0)}
     +[5]_{(1,-1)}+[\ov 5]_{(-1,1)} \nonumber \\
      &&+[5]_{(1,1)}+[\ov 5]_{(-1,-1)}.
\eea      
To ensure absence of a massless combination of $U(1)$ factors we require that 
$[D_a]$ and  $[D_b]$ be linearly independent in $H_4(Y, {\mathbb Z})$. 
Note that in the presence of further tadpole cancelling D7-branes it has to 
be ensured that the full mass matrix is of maximal rank.  

\subsubsection*{The MSSM spectrum}

The massless states transforming in the adjoint representation are
given by the deformations of the four-cycles, which are counted by
$H^1(D,{\cal O})$ (Wilson lines) and $H^2(D,{\cal O})$ (transversal
deformations).  In principle we could allow for precisely one such
adjoint of $SU(5)$, which might break the $SU(5)$ symmetry to the
Standard model by the Higgs mechanism.  An example of such a surface
with $h^{(2,0)}=1$ is $K3$.  However, a complete GUT model relying on
this mechanism would have to address the generation of a suitable
potential for the GUT Higgs field from string dynamics such that
$SU(5)$ is broken dynamically to the Standard Model. Since we will
rather be breaking the GUT symmetry by embedding $U(1)_Y$ flux, we
insist that the $SU(5)$ stack wraps a rigid four-cycle. This is
satisfied for del Pezzo surfaces, which have
$h^{1,0}(D)=h^{2,0}(D)=0$.  In view of the rules of
\autoref{tab_chir_spec} the charged GUT spectrum requires the chiral
intersection pattern listed in \autoref{tab_chir_modelgen}.

\begin{table}[htbp] 
\renewcommand{\arraystretch}{1.5} 
\begin{center} 
\begin{tabular}{|c|c||c|c|c|} 
\hline 
\hline 
state & number & sector & $U(5)$ & $U(1)$   \\ 
\hline \hline 
${\bf 10}$ & $3$ &  $(a'a)$         & $\Yasymm_{\,(2)}$ & $1$  \\ 
${\bf\ov 5}$ & $3$ & $(ab')$         & $\antifund_{\, (-1)}$ & $\antifund_{\, (-1)}$  \\
\hline
${\bf 1}_N$ & $3$ & $(b'b)$         & $1$ & $\Ysymm_{\,(2)}$  \\ 
${\bf 5}_H+ {\bf\ov 5}_H$ &  $1+1$ &  $(ab)$         & $\antifund_{\, (-1)}$ & $\fund_{\, (1)}$  \\
\hline 
\hline 
\end{tabular} 
\caption{Chiral spectrum for intersecting D7-brane model. The indices denote 
 the $U(1)$ charges. The last line gives the Higgs particles.}
\label{tab_chir_modelgen} 
\end{center} 
\end{table}

The first two lines contain the antisymmetric representation ${\bf 10}$ of
$SU(5)$ and the fundamental ${\bf{\overline 5}}$.
The states from the $b'b$ sector are necessary to satisfy
the ``formal'' $U(N_b)$ anomaly ($3\times(4+1) -3\times 5=0$)
and carry the charges of right-handed neutrinos.
States from the $(a'b)$ carry the right quantum numbers to be identified
with the pair of Higgs fields ${\bf{5}_H+\bf{\overline 5}_H}$. 
However, one can also realise the Higgs fields from intersections
$(ac)$  between  the $SU(5)$ brane stack and a third one.  
In contrast to $SO(10)$ GUTs, here all massless fields
are perturbatively realised by open string stretched between 
stacks of D7-branes.

The various fields are localised on the intersection of the various
divisors. As mentioned already in \autoref{sec:generalities}, these
are either curves or divisors. In the latter case one has to compute
cohomology classes over a del Pezzo surface, which in general gives
also vector-like matter. On the contrary, if two divisors intersect
over a curve vector-like states are much easier to suppress.  We will
exemplify this feature in the concrete examples to be discussed later.

\subsubsection*{Yukawa couplings} 

The Yukawa couplings which give masses to the MSSM fields
after GUT and electroweak symmetry breaking are
\bea 
\label{Yukawas}
       {\bf 10}^{(2,0)}\, {\bf 10}^{(2,0)}\, {\bf 5_H}^{(1,-1)}, \qquad
       {\bf 10}^{(2,0)}\, {\bf\overline 5}^{(-1,-1)}\, {\bf\overline
         5_H}^{(-1,1)}\qquad   
        {\bf 1}_N^{(0,2)}\, {\bf\overline 5}^{(-1,-1)}\, {\bf
         5_H}^{(1,-1)}  \; ,
\eea
where the upper indices denote the Abelian $U(1)_a\times U(1)_b$ charges.
If as indicated we realise the matter and Higgs fields as in \autoref{tab_chir_modelgen},
the last two Yukawa couplings, i.e. the ones generating masses
for the d-quarks and leptons, are allowed already at the perturbative level.
For them to be present the wave functions
of the massless modes have to overlap. If all states are localised on curves, this means that  
the three divisors have to meet at a point.
On a Calabi-Yau threefold, this is generically the case. 
Note that to first order the wavefunctions of the fields localise strictly along the matter curves and 
these perturbative Yukawa couplings are of rank one. Only higher order corrections to the wavefunction profile are responsible for a non-trivial family structure.
If on the other hand the ${\bf 10}$ and the ${\bf 1_N}$ arise from the bulk of the GUT brane and the $U(1)_b$ brane while the ${\bf \ov 5}$ and the Higgs are localised on curves, the perturbative Yukawa couplings involve the triple-product of the restriction of corresponding powers of $L_a$ and $L_b$ to the matter curve.

By contrast, it is obvious from their $U(1)$ charges in
eq.~\eqref{Yukawas} that the u-quark Yukawa couplings ${\bf 10\, 10\,
  5_H}$ are perturbatively forbidden. For quite some time this was
considered the main obstacle to the construction of open string SU(5)
GUT models.  This no-go was bypassed in~\cite{Blumenhagen:2007zk}
where it was pointed out that an isolated, rigid Euclidean D3-brane
instanton wrapping a four-cycle $D_{\rm inst}$ of $O(1)$ type (that is, in
particular invariant under orientifold action) can generate these
missing Yukawa couplings.  This requires extra fermionic charged
matter zero modes localised at the intersection of the instanton with
the two stacks of D7-branes.  Concretely, a necessary condition is
that the chiral intersection numbers
are 
\bea
\label{instzerocond}
       I_{a, {\rm inst}}=1, \qquad I_{b, {\rm inst}}=-1\; .
\eea
The resulting six chiral zero modes $\lambda^i_a, \lambda_b$, $i=1,\ldots 5$,
  can then be absorbed by the disc diagrams 
\bea
       {\bf 10}^{(2,0)}\, \lambda^i_a \lambda^j_a,\qquad
        {\bf 10}^{(2,0)}\, \lambda^k_a \lambda^l_a,\qquad
            {\bf 5}_H^{(1,-1)}\, \lambda^m_a \lambda_b\; .
\eea  
As detailed~\cite{Blumenhagen:2007zk} this results in non-perturbative couplings proportional to
\begin{equation}
\label{10_b}
Y^{\alpha}\, Y^{\beta} \, \epsilon_{ijklm} \, {\bf 10}_{\alpha}^{ij} \,  {\bf 10}_{\beta}^{kl} \, {\bf 5}_H^m
\end{equation}  
with $i,j,\ldots$ denoting $SU(5)$ group indices and $\alpha, \beta$
labelling families.  Note that the coupling eq.~\eqref{10_b} is of unit
rank in family space so that a single instanton gives rise to masses
only for one particular generation of u-quarks. This is a consequence
of the fact that the multiplicities of the $\lambda^i$-modes are only
due to their $SU(5)$ Chan-Paton factors. As stated above, this is a similar situation to the one for the perturbative couplings, which to first order are also of rank one. For non-perturbative couplings of the form (\ref{10_b}) the resolution has to involve several distinct instantons whose combined effect may be to give rise to higher rank couplings.

Of course the amplitude is suppressed by the instanton action
\bea
     \exp \left(- {1\over 2g_s}\,  \int_{D_{\rm inst}} J\wedge J \right).
\eea      
As such the scale of the coupling is independent of the GUT coupling, which is controlled by the cycle volume of the GUT brane. The instanton cycle entering the above suppression, however, is a priori unrelated to the GUT cycle.
Still in the perturbative regime $g_s\ll 1$ there is the danger that the coupling tends to be too
small.
In our approach we  will eventually take the
small  $T_{\rm inst}$  limit of the orientifold model and, besides
imposing the tadpole constraints,  will require that 
at least for the third family this Yukawa coupling is generated by a Euclidean  D3-brane instanton.
Of course we have to ensure that when $T_{\rm inst}\to 0$ not
the whole manifold degenerates.
In principle the large hierarchy in the u-quark masses between the third and the first two families can be engineered by different instantons with suitable suppression. 

Very similarly, if the Higgs fields originate from
the intersection of the $SU(5)$ branes with a third stack then
the bottom Yukawa couplings carry $U(1)^3$ charges 
\bea
 {\bf 10}^{(2,0,0)}\, {\bf\overline 5}^{(-1,-1,0)}\, {\bf\overline
         5_H}^{(-1,0,1)}\; 
\eea
and are therefore not gauge invariant any more.
In this case, also these couplings can only be generated
non-perturbatively by an appropriate D3-brane instanton, which 
intersects the $U(1)$ stacks $b$ and $c$ just once.

\subsubsection*{Dim=4 proton decay} 

In GUT theories there is the danger of generating dimension-four
operators violating
baryon or lepton number 
\bea
       U\, D\, D, \quad  Q\, D\, L, \quad   L\, L\, E\; .
\eea
Clearly if present they generate   unacceptably fast proton decay.
In Georgi-Glashow $SU(5)$ all these operators
descend from the 
\bea
 {\bf 10}^{(2,0)}\, {\bf\overline 5}^{(-1,-1)}\, {\bf\overline
         5}^{(-1,-1)}\; ,
\eea 
coupling, which is perturbatively forbidden due to $U(1)_b$ violation.
However, as just described, even perturbatively 
absent couplings can be generated non-perturbatively by
D3-brane instantons. 
In certain domains of the Ka\"ahler moduli space such instanton-induced dimension-four operators would be dangerous.
For an instanton to generate such a coupling three situations are in principle conceivable in view of the $U(1)_a$ and $U(1)_b$ charges: it
either carries the six charged matter zero modes
$\lambda_a^i,\ov\lambda_a^j$, $\ov\lambda_b^k$ with $i,j,k=1,2$ or alternatively the
four zero modes $\lambda_a,\ov\lambda_a$, $\ov\lambda_b^k$ with $k=1,2$.
The third possibility is that it carries just the two zero modes $\ov\lambda_b^k$ with $k=1,2$.
On the other hand, charged matter zero modes from intersections 
of the instanton
with the $SU(5)$ stack always appear in multiples of five.
We thus conclude that in absence of any known mechanism to absorb the
extra zero modes without pulling down more charged matter fields,
in the two first cases no  such dangerous 
dimension-four operators are  generated.
However, the third option is not a priori excluded.
Of course, if such an instanton exists the coupling is
exponentially suppressed, but we just learnt in the context of the ${\bf 10\, 10\, 5_H}$ Yukawa
coupling that this need
not be the case in the $T_{\rm inst} \rightarrow 0$ limit. Therefore, to be on the safe side we
require that such an instanton does not exist.

\subsubsection*{Neutrino masses} 

We have already seen that the Yukawa coupling ${\bf 1_N}^{(0,2)}\,
{\bf\ov 5}^{(-1,-1)}\, {\bf 5_H}^{(1,-1)}$ generates Dirac type
masses for the neutrinos. In order to realise for instance the see-saw
mechanism one also needs Majorana type masses of the order
$10^{12}-10^{15}$GeV. These can be either generated by higher
dimensional couplings involving some additional $SU(5)$ singlet fields
or by D3-brane instantons~\cite{Blumenhagen:2006xt, Ibanez:2006da,
  Cvetic:2007ku, Ibanez:2007rs}. Higher dimensional couplings are of
course suppressed by the string scale, so that one needs to explain
the high scale of these terms.

For directly generating a mass term 
\bea
S_{\rm mass}=M_N\, {\bf 1}_N^{(0,2)}\,{\bf 1}_N^{(0,2)}
\eea
via an instanton, it has to carry the four charged matter zero modes
$\lambda_b^{i}$, $i=1,\ldots, 4$. In this case the Majorana mass scale
is $M_s \exp(-S_{inst})$ which, that is,
depending on the size of the four-cycle, 
can still give a suppression by a few orders of magnitude.

\subsection
[Breaking SU(5) to SU(3) $\mathbf{\times}$ SU(2) $\mathbf{\times}$ U(1)]
[Breaking SU(5) to SU(3) x SU(2) x U(1)]
{Breaking SU(5) to SU(3) $\mathbf{\times}$ SU(2) $\mathbf{\times}$ U(1)}
\label{sec_breakbreak}

Let us describe how one can  break the $SU(5)$ GUT via a line bundle $L_Y$ 
whose structure group is embedded into $U(1)_Y$. This method was used in the context of local F-theory GUT models in~\cite{Beasley:2008kw}. Here we will discuss its implementation within the perturbative orientifold and find important quantisation constraints on the bundle $L_Y$. Clearly these have to be taken into account in a string theoretically consistent framework.

Suppose we  have designed the model such that the $SU(5)$ gauge symmetry 
is supported on a D7-brane wrapping a rigid divisor, which 
is a del Pezzo  surface ${\rm dP}_r\subset Y$ containing $r+1$ homological 2-cycles.
Therefore, even though we cannot turn on  
(discrete) Wilson lines (as $\pi^1(D)$ is vanishing), we have  
the chance to break the $SU(5)$ gauge symmetry to the 
Standard Model by turning on non-vanishing flux in $U(1)_Y$.  
This Abelian flux $F_Y$ is embedded into the fundamental
representation of $SU(5)$ as\, $F_Y\, T_{Y}\subset SU(5)$ with
\begin{equation}
  T_Y=   
  \begin{pmatrix} 
    -2 & & & & \\
    & -2 & & & \\
    & & -2 & & \\
    & & & 3 & \\
    & & & & 3 \\
  \end{pmatrix}
  .
\end{equation}
Such gauge flux through a non-trivial 2-cycle in $Y$ would lead, via
the Green-Schwarz mechanism, to a mass term by mixing with an
axion. However for flux supported on a two-cycle of the ${\rm dP}_r$ trivial
in $Y$, there is no axion to pair with and the $U(1)_Y$ remains
massless after gauge symmetry breaking~\cite{Buican:2006sn,
  Beasley:2008kw}.  As discussed in \autoref{sec:generalities}, this means
that for $U(1)_Y$ to remain massless we have to choose $U(1)_Y$ such
that its first Chern class $c_1(L_Y) \in H^2(D_a,{\mathbb Q})$ is
trivial on the ambient Calabi-Yau space $Y$, that is, the element $d_Y \in
H_2(D_a,{\mathbb Q})$ specifying $L_Y = {\cal O}_{D_a}(d_Y)$ must lie
in the kernel of the pushforward ${\i}_* H_2(D_a) \rightarrow H_2(Y)$.
{}From the relations eq.~\eqref{relativrela} it is immediately clear
that this flux does not change the chiral spectrum, the D7- and
D5-tadpole constraints and the Fayet-Iliopoulos terms.

Cancellation of the Freed-Witten anomaly again constrains the
quantisation of the bundles $L_a$ and $L_Y$.  In view of the diagonal
embedding of $L_a$ into $U(5)$ condition eq.~\eqref{quant3} becomes
\begin{equation}
\label{FW_stack}
T_0 \, (c_1(L_a) - \i^* B)  +  T_Y \, c_1(L_Y)  + \frac{1}{2} \,T_0  \,c_1(K_{D_a}) \in H^2(D_a, {\mathbb Z})_{5\times 5}
\end{equation}
with $T_0 = 1_{5 \times 5}$.  This equation has two
important consequences: First, $c_1(L_a)-\i^* B$ and $c_1(L_Y)$ can
take fractional values. For example, for a Spin divisor the choice
$c_1(L_a)-\i^* B = \frac{2}{5}$, $c_1(L_Y) = \frac{1}{5}$ would be
consistent. Second, the two independent conditions encoded in the
matrix valued equation~\eqref{FW_stack} cannot be satisfied
simultaneously for non-spin divisors without turning on non-trivial
flux $F_a$ as well\footnote{Note that this conclusion holds also in
  the context of F-theory compactifications. In a globally consistent
  setup, non-trivial $L_Y$ cannot be switched on at will, but only in
  combination with non-trivial and suitably quantised 4-form flux that
  takes the role of $L_a$.}.

The breaking of $SU(5)$ by means of $L_Y$ flux  induces the  
standard splitting of the GUT multiplets into MSSM representations, 
\bea 
\label{splitting} 
&&  {\bf 24} \rightarrow  ({\bf 8},{\bf 1})_{0_Y} +  
        ({\bf 1},{\bf 3})_{0_Y} + ({\bf 1},{\bf 1})_{0_Y} + 
          ({\bf 3},{\bf 2})_{5_Y}+  ({\bf \ov 3},{\bf 2})_{-5_Y},\nonumber\\ 
   && \ov {\bf 5}\ \, \rightarrow  (\ov{\bf 3},{\bf 1})_{2_Y} +  
      ({\bf 1},{\bf2 })_{-3_Y}, \\ 
&&  {\bf 10} \rightarrow ({\bf 3},{\bf 2})_{1_Y} +  
       ({\bf \ov 3},{\bf 1})_{-4_Y} + ({\bf 1},{\bf 1})_{6_Y}\; ,\nonumber\\ 
    && {\bf 5}_H\ \, \rightarrow  ({\bf 3},{\bf 1})_{-2_Y} +  
      ({\bf 1},{\bf2 })_{3_Y},\quad  
  \ov {\bf 5}_H\ \, \rightarrow  (\ov {\bf 3},{\bf 1})_{2_Y} +  
      ({\bf 1},{\bf2 })_{-3_Y}. \nonumber 
\eea 
As is familiar from the analogous embedding of U(1) bundles in heterotic compactifications~\cite{Blumenhagen:2005ga} the number of massless states after GUT symmetry breaking is computed by
dressing the bundles appearing in the cohomology groups
by a factor of $L_Y^q$. Here $q$ denotes the hypercharge of
the MSSM fields. 

{}From the decomposition of the adjoint of $SU(5)$ one deduces that $H^*
(D_a, L_Y^{\pm 5} )$ gives rise to extra massless states.  Clearly,
these vector-like exotics are phenomenologically unappealing, so we
require that these cohomology groups vanish.  This is a very strong
requirement and for fifth powers of {\emph{integer quantised}}
line-bundles on ${\rm dP}_r$ impossible to satisfy.

However, as discussed above, $c_1(L_Y)$ can really take values in $\mathbb Z/5$.
To illustrate this further one can modify the embedding as follows. Instead of embedding the line bundle $L_a$ entirely into the diagonal $U(1)_a$ of $U(5)$ as in eq.~\eqref{embedding1}, one defines 
two line bundles ${\cal L}_a$ and ${\cal L}_Y$ and identifies their field strengths with the following combinations of generators
$T_a$ and $T_Y$ of the diagonal $U(1)_a$ and hypercharge $U(1)_Y$,
\bea
&& {\cal L}_{a} \leftrightarrow T_a, \\ \nonumber
&& {\cal L}_Y\leftrightarrow \frac{2}{5} T_a + \frac{1}{5} T_Y.
\eea
The analogue of condition eq.~\eqref{FW_stack},
\bea
&& c_1({\cal L}_a) - \i^* B + \frac{1}{2} K_{D_a} \in \mathbb Z, \\ \nonumber
&&  c_1({\cal L}_a) +  c_1({\cal L}_Y) - \i^* B  + \frac{1}{2} K_{D_a} \in \mathbb Z, 
\eea
leads to $c_1({\cal L}_Y) \in \mathbb Z$ and in general half-integer quantised 
${\cal L}_a$ bundles. This agrees with the fact that all cohomology groups involve integer
powers of ${\cal L}_a$ and ${\cal L}_Y$.
It is important to realise that the gauge flux $U(1)_Y$, though
being trivial in the cohomology on $Y$ nevertheless does contribute
to the D3-tadpole condition. The contribution of the fluxes
$L_a$ and $L_Y$ reads
\bea
           N_{\rm gauge}=-{5\over 2} \int_{D_a} c^2_1(L_a) -15 \int_{D_a} c^2_1(L_Y),
\eea
where we have taken into account ${\rm tr}(T_{Y}^2)=30$.
Redefining as above $L_a={\cal L}_a\otimes {\cal L}_Y^{2\over 5}$
and  $L_Y={\cal L}_Y^{1\over 5}$ yields
\bea
\label{d3yflux}
  N_{\rm gauge}=-{5\over 2} \int_{D_a} c^2_1({\cal L}_a) - \int_{D_a} c^2_1({\cal L}_Y) - 2 \int_{D_a} c_1({\cal L}_a)\ c_1({\cal L}_Y) \; .
\eea$
{\cal L}_Y$ being trivial on the Calabi-Yau, the mixed term $\int c_1({\cal L}_a)\, c_1({\cal L}_Y)$ may be non-vanishing only 
if also ${\cal L}_a$ has components  trivial on $Y$.

Let us discuss the effect of ${\cal L}_Y$ in more detail:

\subsubsection*{Massless $\mathbf{U(1)_Y}$}

Since we have now embedded the line
bundle ${\cal L}_Y$ into a combination of $T_0$ and $T_Y$, one
might be worried that due to the Green-Schwarz mechanism
it is not directly $U(1)_Y$ which  remains massless.
To find the massive Abelian gauge symmetry, we have to
evaluate the relevant axion coupling
\bea
\label{greeni}
         \int_{\IR_{1,3}\times Y} C_4\wedge {\rm Tr}( F^2_{\rm GUT} ),
\eea
where $F_{\rm GUT}$ is the total $U(5)$ field strength supported on the
GUT brane stack. Splitting this into the four-dimensional 
parts $F^{4D}$ and the internal background values given by the
first Chern classes, we can write
\begin{equation}
\label{GS_stack}
F_{\rm GUT}= T_0 \, \Bigl(F^{4D}_0 + c_1({\cal L}_a) -\i^*B +{1\over 2}
               c_1(K_{D_a})+
          {2\over 5} c_1({\cal L}_Y)\Bigr)  +  
             T_Y \, \Bigl(F^{4D}_Y + {1\over 5} c_1({\cal L}_Y)\Bigr) \; .
\end{equation}
Inserting this into \eqref{greeni} and extracting the relevant term
with two legs of ${\rm Tr}( F^2_{\rm GUT})$ in the four-dimensional
Minkowski space and two legs on the GUT $D7$-brane, one immediately
realises that it is still the diagonal $F^{4D}_0$ which mixes
with the axions.

\subsubsection*{Exotics}

As already  described the decomposition of the adjoint of $SU(5)$ yields
massless states counted by the cohomology groups $H^* (D_a, {\cal L}_Y )$. 
For phenomenological reasons we require that  
these cohomology groups vanish. This gives already a very strong
constraint on the possible line bundles.

\subsubsection*{MSSM matter fields} 

Using the bundles ${\cal L}_a$ and ${\cal L}_Y$, we now express the
relevant cohomology groups counting the number quarks and lepton
fields. As mentioned these modes localise either on surfaces or on
curves.  To treat both cases simultaneously we express the number of
modes in terms of sheaf extension groups.  It is understood that for
the actual computation one uses the formulae collected in
\autoref{masslessspec_gen}.

The anti-fundamental matter representation of $SU(5)$ splits as 
\bea 
     &&(\ov{\bf 3},{\bf 1})_{2_Y}\  \, : \  
 \Ext^*(  {\cal L}_a , {\cal L}^{-1}_b ), \\
     &&({\bf 1},{\bf2 })_{-3_Y}:\  
   \Ext^*(  {\cal L}_a\otimes {\cal L}_Y , {\cal L}^{-1}_b ) \nonumber 
\eea 
with ${\cal L}_b=L_b$.
Similarly, for the anti-symmetric representation, we have to compute the three 
cohomology classes 
\bea 
  &&({\bf 3},{\bf 2})_{1_Y}\ \ \ :\  
   \Ext^*(  {\cal L}^{-1}_a \otimes  {\cal L}_Y^{-1}, {\cal L}_a ), \nonumber\\
  && ({\bf \ov 3},{\bf 1})_{-4_Y}\ : \ 
    \Ext^*(  {\cal L}^{-1}_a , {\cal L}_a ), \\
  && ({\bf 1},{\bf 1})_{6_Y}\ \ \, \, : \
   \Ext^*(  {\cal L}^{-1}_a \otimes  {\cal L}^{-2}_Y, {\cal L}_a ).  \nonumber
\eea 
Since ${\cal L}_Y$ is trivial in $Y$ it is guaranteed that the chiral
index of these representations does not change.  However, in general
the vector-like matter will change and will be different for MSSM
states descending from the same GUT multiplet.  This is avoidable if
all matter is localised on curves (and not on surfaces) such that the
restriction of ${\cal L}_Y$ to this matter curve vanishes.

\subsubsection*{Higgs field and 3-2 splitting}

The fundamental  representation for the Higgs field of $SU(5)$ splits as 
\bea 
\label{Higgs_Gen}
     &&({\bf 3},{\bf 1})_{-2_Y}\  \, : \  
  {\rm Ext}^*(  {\cal L}^{-1}_a , {\cal L}^{-1}_b ), \\
     &&({\bf 1},{\bf2 })_{3_Y}\quad :\  
  {\rm Ext}^*(  {\cal L}^{-1}_a \otimes {\cal L}_Y^{-1}, {\cal L}^{-1}_b ). \nonumber 
\eea 
Note that all these states are vector-like.
We need that the $SU(2)$ doublets remain massless and that the
$SU(3)$ triplets  get a mass of the GUT scale.
This translates into requiring that 
${\rm Ext}^*(  {\cal L}^{-1}_a , {\cal L}^{-1}_b )=(0,0,0,0)$
and $ {\rm Ext}^*(  {\cal L}^{-1}_a \otimes {\cal L}_Y^{-1}, {\cal L}^{-1}_b )=
(0,1,1,0)$.
For the Higgs fields localised on the intersection curve $C_{ab}$ of the
$U(5)$ divisor $D_a$ and the $U(1)$ divisor $D_b$, two possibilities
can occur.

The first option is that the intersection locus is a \emph{single} elliptic curve, that is, $C_{ab}=T^2$.
In this case the restriction of the line bundles to $C_{ab}$ have to be of degree zero so that
indeed no chiral matter is localised on $C_{ab}$.
Recall that a degree zero line bundle on the elliptic curve $C_{ab}$ can be written as
${\cal O}(p - q)$, where $p, q$ denote points different from the origin $0$ of the elliptic curve.
The trivial bundle $\cal O$ corresponds to $p - q =0$ and has cohomology $H^*(C_{ab}, {\cal O})=(1,1)$. If $p-q \neq 0$ the line bundle has a non-trivial Wilson line and the cohomology groups vanish.
It is therefore clear that for an appropriate choice of the line bundles appearing in eq.~\eqref{Higgs_Gen} it can be arranged that 
only the doublet remains massless while the triplets acquire string scale masses.
According to what we just said this happens provided the restriction of the line bundles appearing in eq.~\eqref{Higgs_Gen} to the genus one matter curve $C_{ab}$
take the form
\bea
\label{Wilsonlines_General}
&& {\cal L}_a^{-1} \otimes {\cal L}_b|_{C_{ab}} = {\cal O}(p -  q), \quad  p -  q \neq 0, \nonumber  \\
&& {\cal L}_a^{-1} \otimes {\cal L}_Y^{-1} \otimes {\cal L}_b|_{C_{ab}} = {\cal O}.
\eea
To see how to arrange for this, suppose one has found a model where the line bundles ${\cal L}_a$ and ${\cal L}_b$
can both be written as the pullback of line bundles from the Calabi-Yau,
\bea
\label{pulback_cond}
{\cal L}_a = \i_a^* \, {\mathbb L}_a, \quad \quad  {\cal L}_b = \i_b^* \, {\mathbb L}_b.
\eea
Since ${\cal L}_a^{-1} \otimes {\cal L}_b|_{C_{ab}}$ is of degree zero in this case ${\cal L}_a^{-1} \otimes {\cal L}_b|_{C_{ab}}= {\cal O}$ with trivial Wilson lines. The relations eq.~\eqref{Wilsonlines_General} can now simply be met by twisting ${\cal L}_a$ by a line bundle $R_a$ on $D_a$ which is trivial on the ambient manifold and satisfies
\bea
R_a = {\cal L}_Y^{-1}.
\eea
As with everything desirable in life this surgery is not for free, as the new contribution of the GUT brane to the D3-brane tadpole is increased from eq.~\eqref{d3yflux} to\footnote{Here we are using that ${\cal L}_a= \i_a^* \, {\mathbb L}_a$ prior to twisting by $R_a$ so that the cross-term in eq.~\eqref{d3yflux} vanishes.}
\bea
\label{d3yflux_b}
  N_{\rm gauge}^a=-{5\over 2} \int_{D_a} c^2_1({\cal L}_a) - \frac{3}{2} \int_{D_a} c^2_1({\cal L}_Y) .
\eea
We will however also encounter cases\footnote{This discussion is only relevant for the models proposed in Section 5 and Subsection 6.1.} in which eq.~\eqref{pulback_cond} is not the situation to begin with.
In particular it may be inconsistent to define the final GUT bundle as $\i_a^*
{\mathbb L}_a \otimes {\cal L}_Y^{-1}$ because this bundle might exhibit
ghosts in its spectrum. For the line bundle appearing in \eqref{Higgs_Gen},
in this case we have, supposing for definiteness that ${\cal L}_b=\i_b^* {\mathbb L}_b$, 
\bea
{\cal L}^{-1}_a  \otimes {\cal L}_Y^{-1} \otimes {\cal L}_b |_{C_{ab}} = {\cal O}(p_1 - q_1).
\eea
We then need to twist ${\cal L}_b$ by a line bundle $R_b$ trivial on the ambient space, and every such bundle
satisfies $R_b|_{C_{ab}} = {\cal O}(p_2 - q_2)|_{C_{ab}}$.
To ensure $p_1 + p_2 - (q_1 +q_2) =0$, as desired, might require adjusting some of the complex structure moduli of the manifold.

On the other hand, it was argued in~\cite{Beasley:2008kw} 
that in case $H_u$ and
$H_d$ are localised on a single curve, dimension five proton decay operators
$Q\,Q\,Q\, L$ can be generated by exchanging Kaluza-Klein modes of the 
Higgs-triplet.
To avoid such operators, it was suggested that the intersection
locus $D_a\cap D_b$ consists of two components $C_1\cup C_2$, such that
the ${\bf 5}_H$ originates from  $H^*(C_1, {\cal L}^{-1}_a\otimes {\cal L} _b
\otimes \sqrt{K_{C_1}})=(1,0)$ and
${\bf\ov 5}_H$ from $H^*(C_2, {\cal L}^{-1}_a\otimes {\cal L}_b
\otimes \sqrt{K_{C_1}})=(0,1)$. This is the second
option we have for the localisation of the Higgs field.

\subsubsection*{The top Yukawa couplings}

We have seen that in the $SU(5)$ GUT model the top quark
Yukawa coupling ${\bf 10\, 10\, 5_H}$ 
can be generated by a single rigid $O(1)$ instanton with the right 
charged matter zero mode structure.
We need to check whether this is compatible with the breaking
of the $SU(5)$ GUT group by the $U(1)_Y$ flux.

Recall that the five zero modes $\lambda^i_a$ transform in the
${\bf\ov 5}$ representation of the $SU(5)$ and the
single zero mode $\lambda_b$ in the singlet representation
of $SU(5)$ with $U(1)_b$ charge $q_b=+1$.
The $U(1)_Y$ flux splits the ${\bf\ov 5}$ representation
according to \eqref{splitting}, that is,  the $({\bf\ov 3},{\bf 1})_{2_Y}$
zero modes are counted by ${\rm Ext}^*({\cal L}_a,{\cal O})$
and the $({\bf\ov 1},{\bf 2})_{-3_Y}$ modes by
${\rm Ext}^*({\cal L}_a\otimes {\cal L}_Y,{\cal O})$.
As long as the $SU(5)$ stack of branes and the
instanton brane intersect over a 2-cycle non-trivial in the
Calabi-Yau manifold, the restriction of ${\cal L}_Y$ to the
intersection locus vanishes and we  get precisely 
three instanton zero modes  $\lambda^i_a$ transforming 
in the $({\bf\ov 3},{\bf 1})_{2_Y}$ representation
and two zero modes $\tilde\lambda^j_a$ transforming 
in the $({\bf\ov 1},{\bf 2})_{-3_Y}$ representation.
Since ${\cal L}_Y$ is supported on the $SU(5)$
stack, the single zero mode $\lambda_b$ also still
exists.
Then the Standard Model top-Yukawa couplings 
$({\bf 3},{\bf 2})\, ({\bf\ov 3},{\bf 1})\,
({\bf 1},{\bf 2})_H$ are generated
by the instanton via the following absorption of the six
charged instanton zero modes
\bea
 ({\bf 3},{\bf 2})_{1_Y} &\bigl\vert& ({\bf\ov 3},{\bf 1})^{\lambda_a}_{2_Y}\
                      ({\bf  1},{\bf 2})^{\tilde\lambda_a}_{-3_Y} \nonumber\\
 ({\bf\ov 3},{\bf 1})_{-4_Y} &\bigl\vert& ({\bf\ov 3},{\bf 1})^{\lambda_a}_{2_Y}\
                    ({\bf\ov 3},{\bf 1})^{\lambda_a}_{2_Y}\\
 ({\bf 1},{\bf 2})_{3_Y}  &\bigl\vert& ({\bf 1},{\bf
   2})^{\tilde\lambda_a}_{-3_Y}\ 
                    ({\bf  1},{\bf 1})^{\lambda_b}_{0_Y}\nonumber \; .
\eea

\subsubsection*{The $\mathbf{\mu}$-term}

The supersymmetric $\mu$ term clearly vanishes  at tree-level.
For having it directly generated non-perturbatively,
the rigid $O(1)$ instanton must carry the four charged
matter zero modes $\lambda_a,\ov\lambda_a$ and
$\lambda_b,\ov\lambda_b$. However, due to the 
$SU(2)$ Chan-Paton factor the $\lambda$ always come
in multiples of two, so that these simple non-perturbative
$\mu$-terms are absent. However, they can be generated
by higher dimensional operators involving $SU(5)$ gauge singlets, which
have to receive some non-vanishing vacuum expectation value.

\subsection{Summary of GUT Model Building Constraints}

In this section we have collected a number of perturbative
and non-perturbative stringy mechanisms to first realise
Georgi-Glashow $SU(5)$ GUTs and second to solve some
of their inherent problems.
The perfect string model, besides being globally consistent  
would of course satisfy all these
constraints. Eventually, one also has to address the issue
of moduli stabilisation by fluxes and instantons, some aspects of which 
we discuss in \autoref{sec_moduli}.
A more thorough and complete analysis is
beyond the main scope of this paper but truly on the agenda.

Let us summarise in \autoref{tab_gutfeatures} the main properties
a realistic string GUT model should have
\begin{table}[htbp] 
\renewcommand{\arraystretch}{1.5} 
\begin{center} 
\begin{tabular}{|c||c|c|} 
  \hline 
  \hline 
  property & mechanism & status    \\ 
  \hline \hline 
  globally consistent & tadpoles + K-theory & $\checkmark$ \\
  D-term susy & vanishing FI-terms inside K\"ahler cone & $\checkmark$ \\
  gauge group $SU(5)$ & $U(5)\times U(1)$ stacks & $\checkmark$ \\
  3 chiral generations & choice of line bundles ${\cal L}_{a,b}$ & $\checkmark$ \\
  no vector-like matter  & localisation on  curves & $\checkmark$ \\
  1 vector-like of  Higgs  & choice of line bundles & $\checkmark$ \\
  no adjoints & rigid 4-cycles $\leftarrow$  del Pezzo & $\checkmark$ \\
  GUT breaking   & $U(1)_Y$ flux ${\cal L}_{Y}$ on trivial 2-cycles  & $\checkmark$ \\
  3-2 splitting & Wilson lines on  $g=1$ curve   &
  $\checkmark$\\
  3-2 split + no dim=5 $p^+$-decay  & local. of $H_u, H_d$  on disjoint
  comp.   & $\checkmark$\\
  ${\bf 10\, \ov 5\, \ov 5}_H$ Yukawa & perturb. or D3-instanton
  & $\checkmark$ \\
  ${\bf 10\, 10\, 5}_H$ Yukawa & presence of appropriate D3-instanton
  & $\checkmark$ \\
  Majorana neutrino masses & presence of appropriate D3-instanton
  & $\checkmark$ \\
  \hline 
  \hline 
\end{tabular} 
\caption{
Summary of $SU(5)$ properties and their realisations
by different Type IIB orientifold mechanisms. The mark $\checkmark$
in the last column indicates that all features can in principle
be realised.}  
\label{tab_gutfeatures} 
\end{center} 
\end{table}
\vspace{-0.3cm}


\section
[Del Pezzo Transitions on $\mathbf{\IP_{1,1,1,6,9}[18]}$]
[Del Pezzo Transitions on {P(1,1,1,6,9)[18]}]
{Del Pezzo Transitions on $\mathbf{\IP_{1,1,1,6,9}[18]}$}
\label{sec:class_of_examples}
 
In this section we discuss a first class of Calabi-Yau orientifold
backgrounds which will later support our GUT models. The underlying
geometries are compact Calabi-Yau manifolds $M_n$ which can be either
constructed as elliptic fibrations over del Pezzo surfaces ${\rm dP}_n$, or
by performing del Pezzo transitions. To set the stage for our analysis
of the Calabi-Yau orientifolds we will first recall some basic
geometric facts about the surfaces ${\rm dP}_n$ in
\autoref{sec_delPezzobase} and classify all involutions on these
surfaces.  In \autoref{delPezzotrans} we argue that the elliptically
fibred Calabi-Yau threefolds over ${\rm dP}_n$ can also be obtained by del
Pezzo transitions starting from the degree $18$ hypersurface in
$\IP_{1,1,1,6,9}$ after flop transitions. Simple examples are obtained
if the base is one of the toric del Pezzo surfaces ${\rm dP}_2$ or ${\rm dP}_3$.
In Sections~\ref{sec:dP2fibration} and~\ref{dP3fibration} we construct
the corresponding Calabi-Yau threefolds and study their different
topological phases using toric geometry. In a second step we introduce
viable orientifold involutions $\sigma$ on these compact Calabi-Yau
manifolds and derive the induced tadpoles from O7 and O3 planes.
 
In the second part of this section we will have a closer look at the
topological phases of $M_n$ with $n$ ${\rm dP}_8$ surfaces. We show in
\autoref{swisskaas} that these are examples of so-called swiss-cheese
Calabi-Yau manifolds which can support large volume vacua with one
large and $n+1$ small four-cycles~\cite{Balasubramanian:2005zx,
  Conlon:2005ki, Cicoli:2008va}.  In \autoref{linebundles} we
discuss the D-term conditions arising from wrapping a D7-brane on the
small del Pezzo surfaces with orientifold invariant homology class.

\subsection{Del Pezzo Surfaces and Their Involutions}
\label{sec_delPezzobase}
 
The compact orientifold geometries for our GUT models will be obtained
from elliptic fibrations over del Pezzo surfaces ${\rm dP}_n$. In order to
study these threefolds it will be necessary to review some basic facts
about del Pezzo surfaces first. We will also discuss involutions on
these ${\rm dP}_n$.  We will determine their fix-point locus and action on
the exceptional curves of the del Pezzo surface. Since these
involutions on the base will descend to involutions on the entire
Calabi-Yau manifold, this will enable us to identify viable brane
configurations later on.

\subsubsection*{On the geometry of del Pezzo surfaces}

By definition, del Pezzo surfaces are the Fano surfaces, that is, the
algebraic surfaces with ample canonical bundle. These are either the
surfaces $\cB_n={\rm dP}_n$, which are obtained by blowing up $\bbP^2$
on $0\leq n \leq 8$ points\footnote{\label{fnote:generalpoints}The
  points must be general in the sense that no two points are
  infinitesimally close, no three are on one line, no six on a conic,
  no eight on a cubic with a node at one of them. In other words, one
  is not allowed to blow up points sitting on a $(-1)$-curve. If one
  were to blow up a point on a $(-1)$-curve, then the proper transform
  would be a $(-2)$-curve. So yet another characterisation of the
  allowed points is that there be no curves of self-intersection $-2$
  or less. Moreover, note that different sets of points can correspond
  to the same (complex structure on the) del Pezzo surface.}, or
$\bbP^1 \times \bbP^1$.  Their Hodge numbers are $h^{0,0}=h^{2,2}=1$,
$h^{1,1}\big( \cB_n \big)=n+1$ and $ h^{1,1}\big( \bbP^1 \times \bbP^1
\big)=2$, while all other $h^{p,q}$ vanish.  A basis of homologically
nontrivial two-cycles in $\cB_n$ consists of the class of a line $l$
in $\bbP^2$, and the $n$ exceptional curves $e_i$, one for each
blown-up point.  Their intersection numbers are $l^2=1$, $e_i\cdot e_j
= -\delta_{ij}$, $e_i\cdot l = 0$. Written in this basis, the first
Chern class is
\begin{equation} \label{def-K}
  c_1 \big( T\cB_n \big) 
  = 
  -K 
  = 
  3l - \sum_{i=1}^n e_i\  .
\end{equation}
The square of the canonical class 
\begin{equation} \label{K2delPezzo}
  K^2 
  =
  \int_{\cB_n} c_1^2
  =
  9-n 
\end{equation}
is also called the degree\footnote{To understand this notation, note
  that a degree $d=K^2$ del Pezzo surface with $d\geq 3$ can be
  realised as a degree-$d$ hypersurface in $\bbP^d$.} of the del
Pezzo surface. The second (top) Chern class is the Euler density,
hence
\begin{equation}
  \label{eulerdelpezzo} 
  \chi\big( \cB_n \big)
  =
  \int_{\cB_n} c_2
  =
  3+n \ .
\end{equation}
Let $C$ be a curve in the del Pezzo surface. Then its degree $\deg(C)$
and its arithmetic genus $g$ read
\begin{equation}
  \label{eq:deg_g} 
  \deg(C)=-K \cdot C\ ,\qquad
  g=\tfrac{1}{2}(C\cdot C+K \cdot C)+1\  . 
\end{equation} 
Of particular interest are the rigid genus-$0$ instantons, that is
rational curves of self-intersection $(-1)$. For convenience of the
reader we reproduce the classification of such $(-1)$-curves,
see~\cite{Demazure}, in \autoref{tab:DemazureTable3}.
\begin{table}[htbp] 
  \centering 
  \begin{displaymath}
    \renewcommand{\arraystretch}{1.3}
    \begin{array}{|c|rrrrrrrrr|} 
      \hline 
      \text{class}\in H_2(\cB_n,\Z)
      &\cB_1& \cB_2&\cB_3&\cB_4&\cB_5&\cB_6&\cB_7&\cB_8 &\\ 
      \hline
      (0;1)&1&2&3&4&5&6&7&8 &\\ 
      (1;-1^2)&&1&3&6&10&15&21&28&\\ 
      (2;-1^5)&&&&&1&6&21&56&\\ 
      (3;-2,-1^6)&&&&&&&7&56&\\ 
      (4;-2^3,-1^5)&&&&&&&&56&\\ 
      (5;-2^6,-1^2)&&&&&&&&28&\\ 
      (6;-3,-2^7)  &&&&&&&&8&\\ 
      \hline 
      \text{Total no.} &1&3&6&10&16&27&56&240&\\ 
      \hline 
    \end{array}     
  \end{displaymath}
  \caption{Number of $(-1)$-curves on the $\cB_n$ del Pezzo 
    surfaces. The coefficients $(a;b_1,\ldots b_n)$ are with respect
    to the standard basis $(l;e_1,\ldots,e_n)$. For example,
    $(1,-1^2)$ denotes all $\binom{n}{2}$ homology classes of the
    form $l-e_i-e_j$, $1\leq i<j\leq n$.
    Note that
    there are no $(-1)$-curves on $\cB_0=\CP^2$ and $\CP^1\times\CP^1$,
    which are omitted.} 
  \label{tab:DemazureTable3}
\end{table} 
 
The del Pezzo surfaces $\bbP^2=\cB_0$, $\bbP^1\times\bbP^1$, $\cB_1$,
$\cB_2$, and $\cB_3$ (that is, those of degree $K^2 \geq 6$) are toric
varieties. The remaining surfaces $\cB_4$, $\dots$, $\cB_8$ are not
toric varieties, but can of course be embedded into toric varieties.
In particular, the del Pezzo surfaces $\cB_5$, $\dots$, $\cB_8$ are
hypersurfaces or complete intersections in weighted projective
spaces. For this, let us denote by
$\bbP(d_1,\ldots,d_r|w_0,\ldots,w_m)$ the complete intersection of $r$
equations of homogeneous degree $d_1,\ldots,d_r$ in weighted
projective space with weights $w_0,\ldots,w_m$. These del Pezzo
surfaces are listed in \autoref{tab:delPinproj_space}.
One infers that the dimension of the complex deformation spaces for del Pezzo  
surfaces $\cB_n$ with $n\ge 5$ is ${\rm dim}\, H^1(T\cB_n)=2n-8$. 
\begin{table}[htbp] 
  \begin{displaymath} 
    \renewcommand{\arraystretch}{1.3}
    \begin{array}{|c|c|c|l|l|} 
      \hline 
      \text{del Pezzo} &
      K^2 &
      \text{hypersurface} &
      \text{coordinates} 
      \\
      \hline 
      \cB_5 & 4 & \bbP(2,2|1,1,1,1,1) & (x_1,x_2,x_3,x_4,x_5) \\ 
      \cB_6 & 3 & \bbP(3|1,1,1,1)&      (x_1,x_2,x_3,x_4)     \\ 
      \cB_7 & 2 & \bbP(4|2,1,1,1)&      (y,x_1,x_2,x_3)       \\ 
      \cB_8 & 1 & \bbP(6|3,2,1,1)&      (y,z,x_1,x_2)         \\ 
      \hline 
    \end{array} 
  \end{displaymath} 
  \caption{The del Pezzo surfaces of degree $K^2\leq 4$.}
  \label{tab:delPinproj_space}
\end{table}

\subsubsection*{Classification of involutions on del Pezzo surfaces} 

In order to systematically study GUT models on elliptically fibred
Calabi-Yau manifolds with del Pezzo base, one needs to classify all
different, non-trivial, holomorphic involutions on del Pezzo surfaces.
In the following we intend to discuss the final classification in
\autoref{tab:dPinv2} and give a first impression of the necessary
steps needed for this derivation. Most of the technical details and
geometric constructions are shifted into \autoref{sec:dPinvolutions}.
The classification in \autoref{tab:dPinv2} completes the list obtained in 
ref.~\cite{Grimm:2008ed}.

For a systematic classification of involutions we will look
at the pattern of rigid $\CP^1$ instantons, that is, the
$(-1)$-curves. Clearly, every involution induces a $\Z_2$ automorphism
of the $(-1)$-curves. Conversely, up to degree $6$, the automorphism
of the $(-1)$-curves determines the involution. In the remaining
degrees $\geq 7$ there either are no $(-1)$-curves, or they lie over a
line or point of the blown-up $\CP^2$. Hence, in the latter case they
do not ``fill out'' the space to uniquely determine the
involution. Technically, the $(-1)$-curves generate all of $H_2(S,\Z)$
if and only if the degree is $6$ or less. In a next step, one has to find all
conjugacy classes of involutions acting on the $(-1)$ curves and check that
these descend to actual geometric involutions on the corresponding 
del Pezzo surface. The details of this analysis can be found in 
\autoref{sec:dPinvolutions}. Here we will discuss the final classification 
summarised in \autoref{tab:dPinv2}.

In \autoref{tab:dPinv2} the complete list of del Pezzo surfaces
$\cB_n$ with involutions $\sigma_i$ is shown. For each pair
$(\cB_n,\sigma_i)$ it also includes detailed information about the
fix-point set.  We use the following notation:
\begin{itemize}
\item\label{item:Sigmadef} $\Sigma(\sigma)$ is the homology class of the genus-$g$ curve
  with $g\geq 1$ in the fixed point set of $\sigma$. As discussed
  in \autoref{sec:dPinvolutions}, there is at most one such curve.
\item $R(\sigma)$ are the homology classes of the rational genus
  $0$ curves in the fixed point set.
\item $B(\sigma)$ is the number of isolated fixed points that do not
  lie on $(-1)$-curves.
\item $P(\sigma)$ is the number of isolated fixed points that do lie on
  $(-1)$-curves, and hence may not be blown up further.
\item $(b_2^+,b_2^-)$ are the dimensions of the $\pm$ eigenspaces of $\sigma_i$ in $H_{2}(\cB_n)$.
\end{itemize}
In the last column we also displayed the explicit action of the involution on 
the basis $(l,e_1, \ldots,e_n)$ of $H_{2}(\cB_n)$ and basis $(l_1,l_2)$ of
$H_2(\bbP^1 \times \bbP^1)$. The $k$-dimensional identity matrix is simply
denoted by $\mathbf{1}_k$, while $H$ exchanges two elements $e_i
\leftrightarrow e_j$ or $l_1 \leftrightarrow l_2$. In addition there
are also five involutions $(I_{\cB_3}^{(2)}$, $I_{\cB_3}^{(3)}$, $I_{\cB_5}^{(5)}$, $I_{\cB_7}^{(9)}$, $I_{\cB_8}^{(9)})$ which should be
viewed as the additional building blocks for all non-trivial involutions on del Pezzo 
surfaces. We will introduce the explicit form of these involutions in turn.

To define the special involutions we will specify their 
action on the basis elements $(l,e_1,\ldots,e_n)$.
On the third del Pezzo surface we define the two involutions 
\begin{equation}
   I_{\cB_3}^{(2)} =  \left(\begin{smallmatrix}    
 2 &  1 &  1 &  1 \\
-1 & -1 &  0 & -1 \\
-1 &  0 & -1 & -1 \\
-1 & -1 & -1 &  0 \\
      \end{smallmatrix}\right) \ ,
      \qquad \quad I_{\cB_3}^{(3)}= 
       \left(\begin{smallmatrix}
 2 &  1 &  1 &  1 \\
-1 &  0 & -1 & -1 \\
-1 & -1 &  0 & -1 \\
-1 & -1 & -1 &  0 \\
      \end{smallmatrix}\right)\ .
\end{equation}
The remaining three involutions we need to introduce are well-known classical involutions.
They are minimal since they satisfy $\sigma(E)\neq E$ and $\sigma(E) \cap E \neq \emptyset$
for each $(-1)$-curve $E$. This implies that such an involution cannot be
obtained by blowing up a higher degree del Pezzo and extending an involution
defined on the del Pezzo surface before blow-up.
On $\cB_5$ there is a minimal involution known as the de Jonqui\`eres
involution which acts  as 
\begin{equation}
     I_{\cB_5}^{(5)} =
      \left(\begin{smallmatrix}
 3 &  2 &  1 &  1 &  1 &  1 \\
-2 & -1 & -1 & -1 & -1 & -1 \\
-1 & -1 & -1 &  0 &  0 &  0 \\
-1 & -1 &  0 & -1 &  0 &  0 \\
-1 & -1 &  0 &  0 & -1 &  0 \\
-1 & -1 &  0 &  0 &  0 & -1 \\
      \end{smallmatrix}\right)\ .
\end{equation}
There is one minimal involution for the del Pezzo surfaces of degree $1$ and
$2$, respectively. The del Pezzo surface 
$\cB_7$  admits the Geiser involution 
\begin{equation}
    I_{\cB_7}^{(9)} :\quad l\ \mapsto\ -l-3K \ , \quad e_i \ \mapsto\  -K - e_i\ .
\end{equation}
while on $\cB_8$ one has the Bertini involution acting as 
\begin{equation}
   I_{\cB_8}^{(9)} :\quad l\ \mapsto\ -l-6K \ , \quad e_i \ \mapsto\ -2K - e_i\ .
\end{equation}
Note that one can explicitly check that each involution on each del
Pezzo surface preserves its canonical class $K$ defined in
eq.~\eqref{def-K}.

With these definitions at hand, the involutions in \autoref{tab:dPinv2}
can be used for explicit computations. This will be particularly useful 
for the elliptically fibred threefolds over a del Pezzo base, since all 
involutions can be lifted to the corresponding Calabi-Yau threefold.
We are then in the position to construct explicit Calabi-Yau orientifolds
and compute the tadpoles induced by the O3- and O7-planes.
\begin{table}[htbp]
  \begin{center}
    \renewcommand{\arraystretch}{1.0}
    \small
    \begin{tabular}{|c|c|c|c|c|c|c|c|}
      \hline
      Involution & 
      $g$ & $\Sigma = [\Sigma_g]$ & $R$ & $B$ & $P$ & $(b_2^+,b_2^-)$ &
      action on $H_2$
      \\ \hline
      $(\CP^2,\sigma)$ & 
      & &
      $l$ & 
      $1$ & 
      $$ & 
      $(1,0)$ &
      $\textbf{1}_1$
      \\ \hline
      $(\CP^1\times\CP^1,\sigma_1)$ & 
      & &
      $(l_1)\cup (l_2)$ & 
      $$ & 
      $$ & 
      $(2,0)$ &
      $\textbf{1}_2$
      \\ 
      $(\CP^1\times\CP^1,\sigma_2)$ & 
      & &
      $$ & 
      $4$ & 
      $$ & 
     $(2,0)$  &
      $\textbf{1}_2$
      \\ 
      $(\CP^1\times\CP^1,\sigma_3)$ & 
      & &
      $l_1+l_2$ & 
      $$ & 
      $$ & 
      $(1,1)$ &
      $H$
      \\ \hline
      $(\cB_1,\sigma_1)$ & 
      & &
      $(l) \cup (e_1)$ & 
      $$ & 
      $$ & 
      $(2,0)$ &
      $\textbf{1}_2$
      \\ 
      $(\cB_1,\sigma_2)$ & 
      & &
      $l - e_1$ & 
      $1$ & 
      $1$ & 
      $(2,0)$ &
      $\textbf{1}_2$
      \\ \hline
      $(\cB_2,\sigma_1)$ & 
      & &
      $(l-e_1) \cup (e_2)$ & 
      $$ & 
      $1$ & 
      $(3,0)$ &
      $\textbf{1}_3$
      \\ 
      $(\cB_2,\sigma_2)$ & 
      & &
      $l - e_1-e_2$ & 
      $1$ & 
      $2$ & 
      $(3,0)$ &
      $\textbf{1}_3$
      \\ 
      $(\cB_2,\sigma_3)$ & 
      & &
      $l$ & 
      $$ & 
      $1$ & 
      $(2,1)$ &
      $\textbf{1}_1\oplus H$
      \\ \hline
      $(\cB_3,\sigma_1)$ & 
      & &
      $(l-e_1-e_2) \cup (e_3)$ & 
      $$ & 
      $2$ & 
      $(4,0)$ &
      $\textbf{1}_4$
      \\ 
      $(\cB_3,\sigma_2)$ & 
      & &
      $l - e_1$ & 
      $$ & 
      $2$ & 
      $(3,1)$&
      $\textbf{1}_2\oplus H$
      \\ 
      $(\cB_3,\sigma_3)$ & 
      & &
      $2l-e_1-e_2$ & 
      $$ & 
      $$ & 
      $(2,2)$ &
      $I_{\cB_3}^{(2)}$
      \\ 
      $(\cB_3,\sigma_4)$ & 
      & &
      $$ & 
      $4$ & 
      $$ & 
      $(3,1)$ &
      $I_{\cB_3}^{(3)}$
      \\ \hline
      $(\cB_4,\sigma_1)$ & 
      & &
      $l - e_1-e_2$ & 
      $$ & 
      $3$ & 
      $(4,1)$ &
      $\textbf{1}_3\oplus H$
      \\ 
      $(\cB_4,\sigma_2)$ & 
      & &
      $l$ & 
      $$ & 
      $1$ & 
      $(3,2)$ &
      $\textbf{1}_1\oplus 2H$
      \\ \hline
      $(\cB_5,\sigma_1)$ & 
      & &
      $l - e_1$ & 
      $$ & 
      $2$ & 
      $(4,2)$ &
      $\textbf{1}_2\oplus 2H$
      \\ 
      $(\cB_5,\sigma_2)$ & 
      & &
      $2l-e_1-e_2$ & 
      $$ & 
      $$ & 
      $(3,3)$ &
      $I_{\cB_3}^{(2)} \oplus H$
      \\ 
      $(\cB_5,\sigma_3)$ & 
      & &
      $$ & 
      $4$ & 
      $$ & 
      $(4,2)$ &
      $I_{\cB_3}^{(3)} \oplus H$
      \\ 
      $(\cB_5,\sigma_\text{dJ})$ & 
      $1$ &
      $3l-\sum_{i=1}^5 e_i$ &
      $$ &
      $$ &
      $$ &
      $(2,4)$ &
      $I_{\cB_5}^{(5)}$
      \\ \hline
      $(\cB_6,\sigma_1)$ & 
      & &
      $l - e_1-e_2$ & 
      $$ & 
      $3$ & 
      $(5,2)$ &
      $\textbf{1}_3\oplus 2H$
      \\ 
      $(\cB_6,\sigma_2)$ & 
      $1$ &
      $3l-\sum_{i=1}^6 e_i$ &
      $$ &
      $$ &
      $1$ &
      $(3,4)$ &
      $I_{\cB_5}^{(5)}\oplus \mathbf{1}_1$
      \\ \hline
      $(\cB_7,\sigma_1)$ & 
      & &
      $$ & 
      $4$ & 
      $$ & 
      $(5,3)$ &
      $I_{\cB_3}^{(3)} \oplus 2H$
      \\ 
      $(\cB_7,\sigma_2)$ & 
      $1$ & 
      $3l-\sum_{i=1}^7 e_i$ &
      $$ &
      $$ &
      $2$ &
      $(4,4)$ &
      $I_{\cB_5}^{(5)}\oplus \mathbf{1}_2$
      \\ 
      $(\cB_7,\sigma_3)$ & 
      $1$ &
      $3l-\sum_{i=1}^5 e_i$ &
      $$ &
      $$ &
      $$ &
      $(3,5)$ &
      $I_{\cB_5}^{(5)}\oplus H$
      \\ 
      $(\cB_7,\sigma_\text{G})$ & 
      $3$ & 
      $6 l - 2 \sum_{i=1}^7 e_i$ &
      $$ &
      $$ &
      $$ &
       $(1,7)$ &
      $I_{\cB_7}^{(9)}$
      \\ \hline
      $(\cB_8,\sigma_1)$ & 
      $1$ & 
      $3l-\sum_{i=1}^8 e_i$ &
      $$ &
      $$ &
      $3$ &
      $(5,4)$ &
      $I_{\cB_5}^{(5)}\oplus \mathbf{1}_3$
      \\ 
      $(\cB_8,\sigma_2)$ & 
      $1$ & 
      $3l-\sum_{i=1}^6 e_i$ &
      $$ &
      $$ &
      $1$ &
      $(4,5)$ &
      $I_{\cB_5}^{(5)}\oplus \mathbf{1}_1 \oplus H$
      \\ 
      $(\cB_8,\sigma_\text{B})$ & 
      $4$ & 
      $9 l - 3 \sum_{i=1}^7 e_i$ &
      $$ &
      $1$ &
      $$ &
      $(1,8)$ &
      $I_{\cB_8}^{(9)}$
      \\ \hline
    \end{tabular}
  \end{center}
  \caption{All involutions on del Pezzo surfaces. See
    page~\pageref{item:Sigmadef} for the definition of $\Sigma$, $R$, $B$,
    and $P$.}
  \label{tab:dPinv2}
\end{table}

\subsection
[The Geometry of Del Pezzo Transitions of {$\IP_{1,1,1,6,9}[18]$}]
[The Geometry of Del Pezzo Transitions of {P(1,1,1,6,9)[18]}]
{The Geometry of Del Pezzo Transitions of $\mathbf{\IP_{1,1,1,6,9}[18]}$} 
\label{delPezzotrans}
 
We are now in the position to construct compact Calabi-Yau 
threefolds $M_n$ associated to a del Pezzo base. 
The first construction is via an elliptic fibration over a 
del Pezzo base $\cB_n$, while the second construction is by 
employing del Pezzo transitions.

\subsubsection*{Elliptically fibred Calabi-Yau threefolds with del
  Pezzo base}

Let us construct Calabi-Yau threefolds $M_n$ as elliptic fibrations
over the del Pezzo base ${\rm dP}_n$. We consider elliptic fibrations which
are generically smooth with the worst degeneration of the fibre of
Kodaira type $I_1$~\cite{KodairaII, KodairaIII}. In the following we
will restrict further to elliptic fibrations with generic elliptic
fibres of type $E_8$ such that the generic elliptic fibres can be
represented by a degree $6$ hypersurface in $\bbP_{1,2,3}$ denoted by
$\bbP_{1,2,3}[6]$.  As shown, for example, in~\cite{Klemm:1996ts}, one
then finds that the Euler number of the elliptic fibration $M_n$ is
given by
\begin{equation}
  \label{general_chi}
  \chi(M_n) = - C_{(8)} \int_{\cB_n} c_1^2 = 60 (n-9)
  ,
\end{equation}
where $C_{(8)}=30$ is the dual Coxeter number of $E_8$ and we have
used \eqref{K2delPezzo} for the del Pezzo base $\cB_n$.  One can also
count the number of K\"ahler classes for these geometries.  One finds
that there are $n+1$ classes corresponding to the non-trivial
two-cycles of the del Pezzo base as well as the fibre class of the
elliptic fibration. This implies that $M_n$ has Hodge numbers
\begin{equation}
  \label{general_Hodge} 
  h^{1,1}(M_n) = n+2\ ,\qquad h^{2,1}(M_n) =
  272 - 29 n
  , 
\end{equation}
where we have used that $\chi = 2(h^{1,1}-h^{2,1})$.

The specification of $M_n$ as an elliptic fibration over the base
${\rm dP}_n$ will turn out to be particularly useful in the study of
orientifold involutions and brane configurations on $M_n$. Let us
introduce the map
\begin{equation}
   \pi: \ M_n \ \to \ \cB_n\ ,
\end{equation}
which is the projection from the threefold $M_n$ to the base. Note
that every $(-1)$ curve class $E$ in $\cB_n$ can be pulled back to a
divisor in $M_n$ using $\pi^*: E \mapsto \pi^* (E) \in
H_{4}(M_n,\bbZ)$. In fact, each such divisor is a ${\rm dP}_9$ surface. This
surface is defined as blow up of $\IP^2$ at nine points which arise at
the intersections of two cubic curves. Thus, ${\rm dP}_9$ is an elliptic
fibration over $\IP^1$ which has $12$ singular fibres\footnote{Roughly
  speaking, the ${\rm dP}_9$ is half a K3 surface which is an elliptic
  fibration over $\IP^1$ with $24$ singular fibres.}. The ${\rm dP}_9$ is
not strictly a del Pezzo surface, but the equations~\eqref{K2delPezzo}
and~\eqref{eulerdelpezzo} remain to be valid. Recall that the $(-1)$
curve in the base have been listed in \autoref{tab:DemazureTable3}. It
it thus straightforward to determine the intersections of these
curves. In case two curves $E_1,E_2$ intersect at a point the
corresponding two ${\rm dP}_9$ divisors $\pi^*(E_1)$ and $\pi^*(E_2)$ in
$M_n$ will intersect over a Riemann surface of genus $1$. Clearly each
$\pi^*(E)$ intersects the base $\cB_n$ in a $\bbP^1$. Already this
simple analysis allows us to infer the necessary information on the
triple intersections of the threefold $\cM_n$ in the elliptically
fibred phase.

We want to apply a similar logic also for the extension of an involution on the del
Pezzo base to an involution $\sigma$ on the threefold $\cM_n$. 
In fact, by appropriately defining $\sigma$ the action on the $(-1)$ 
curves of $\cB_n$ lifts to an action of their $\pi^*$ pull-backs. 
The fixed divisors wrapped by the O7-planes can then be determined using 
\autoref{tab:dPinv2}. Determining the number of O3-planes in the full 
set-up also depends on the precise form of the
involution on $M_n$. In particular, it is not generally the case that  
each isolated fix-point in the base lifts to a single fix-point in $M_n$.
Let us consider the case where the torus fibre over the fix-point is 
smooth, which will turn out to be the case in our explicit examples. In this simple 
situation, we have to distinguish three cases. Firstly, the fibre over the
isolated fix-point can be fixed itself. This extension of the involution should
be omitted, since this would imply the presence of O5-planes and violate the
condition $\sigma^* \Omega = - \Omega$. Secondly, the involution can act as 
shift on the torus fibre and hence have no fix-points in $M_n$. Thirdly,
$\sigma$ can act as the inversion of the torus fibre. Such an involution has 
$4$ fix-points, one on the zero section and three  on the
  tri-section
\begin{equation}  
  \label{def-tri}
  D_{T} = 3 \cB_n - 3 \pi^*(K)\ ,   
\end{equation}
where $K$ is the canonical class of the del Pezzo base. To define viable 
involutions on $M_n$ one needs to extend this analysis to the singular 
Kodaira fibres.

\subsubsection*{Del Pezzo transitions of $\mathbf{\IP_{1,1,1,6,9}[18]}$} 

An alternative way to construct the threefolds $M_n$ is to perform del
Pezzo transitions starting with the degree $18$ hypersurface in
weighted projective space $\IP_{1,1,1,6,9}$.  This Calabi-Yau manifold
$M_0 = \IP_{1,1,1,6,9}[18]$ is an elliptic $\IP_{1,2,3}[6]$ fibration
over the base $\cB=\IP_2$ and has $h^{1,1}=2$. In order to perform the
del Pezzo transition $M_0 \rightarrow M_1$ one generates a ${\rm dP}_8$
singularity by fixing $29$ complex structure
deformations~\cite{Morrison:1996pp, Grimm:2008ed}. This singularity is
then resolved by blowing up a del Pezzo surface ${\rm dP}_8$.  Clearly, in
this process the Hodge numbers precisely change as required in
eq.~\eqref{general_Hodge}. This process can be repeated to obtain the
manifolds $M_n$. However, it is important to note that the manifold
constructed via the elliptic fibration only coincide with the one
obtained by del Pezzo transitions after performing appropriate flop
transitions. In order to obtain del Pezzo surfaces ${\rm dP}_8$ out of the
${\rm dP}_9$ surfaces of the elliptic fibred phase, one has to perform a
flop transition for the $\IP^1$ intersecting the base.

\subsection
[Orientifold of An Elliptic Fibration Over ${\rm dP}_2$]
[Orientifold of An Elliptic Fibration Over dP2]
{Orientifold of An Elliptic Fibration Over $\mathbf{{\rm dP}_2}$}
\label{sec:dP2fibration}

\begin{figure}[htbp]
\centering
\hspace{40pt}
\includegraphics[width=0.9\textwidth]{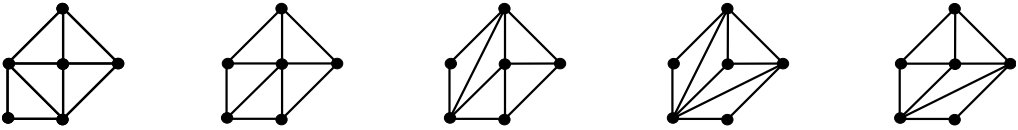}
\vspace{-10pt}
\caption{The five different triangulations of the toric ${\rm dP}_2$ base.}
\label{figtriang}
\end{figure}

The first Calabi-Yau threefold which we investigate in detail is the
manifold $M_2$ which corresponds to the elliptic fibration over
${\rm dP}_2$. The geometry of $M_2$ and its topological phases has been
studied from a different point of view in~\cite{Louis:1996mt}.  Note
that, using eqns.~\eqref{general_Hodge} and~\eqref{general_chi}, one
finds the Hodge numbers $h^{1,1}(M_2) = 4$, $h^{2,1}(M_2) = 214$ and
the Euler number $ \chi(M_2) = -420$.  It will be important that the
manifold $M_2 $ has actually $5$ distinct topological phases which are
connected by flop transitions.  They correspond to the five
triangulations of the toric ${\rm dP}_2$ base.  ${\rm dP}_2$ can be represented
torically by the points $(1,0)$, $(0,1)$, $(-1,0)$, $(0,-1)$,
$(-1,-1)$ in two dimensions.  The five triangulations of this
polyhedron are depicted in \autoref{figtriang}.

To study the Calabi-Yau space $M_2$ in detail, 
we will also invoke the methods of toric geometry. 
Note that $\IP_{1,1,1,6,9}$ is described by the 
six points $v^*_1=(1,0,0,0)$, $v^*_2=(0,1,0,0)$,  
$v^*_3=(0,0,1,0)$,  $v^*_4=(0,0,0,1)$, $v^*_5=(-9,-6,-1,-1)$,  
 $v^*_6=(-3,-2,0,0)$. Each of these points corresponds to 
 a divisor $D_i$ and the hypersurface $\IP_{1,1,1,6,9}[18]$ is defined to 
 admit the anti-canonical class $-\sum_{i=1}^6 D_i$. To obtain the manifold $M_2$ 
one introduces two blowing-up 
divisors $D_7$, $D_8$ corresponding to the points $v^*_7=(-6,-4,-1,0)$ 
and $v^*_8=(-6,-4,0,-1)$.  The compact hypersurface with anti-canonical 
class $-\sum_{i=1}^8 D_i$ is the manifold $M_2$. It admits 
five triangulations just as the ${\rm dP}_2$ base itself. 
In the following we will investigate two of them in more detail. 
The corresponding Calabi-Yau space will 
be denoted by $M_2^{({\rm dP}_9)^2}$ and $M_2^{({\rm dP}_8)^2}$. 
Here we indicate the type of the  divisors $D_7$, $D_8$ 
as we will check below\footnote{We are grateful to Albrecht Klemm for help
with the programs to perform the toric computations. The analysis of the
divisors and their intersection ring was carried out by using the Maple code Schubert.}.

\subsubsection*{The geometry of $\mathbf{M_2^{({\rm dP}_9)^2}}$} 
  
Let us first discuss the Calabi-Yau manifold $M_2^{({\rm dP}_9)^2}$.
The data of the associated linear sigma model is the following. We have eight 
complex coordinates  $x_i$. The divisors $D_i$ are defined by the constraints $x_i=0$. 
In addition there are four $U(1)$ symmetries. The corresponding charges are shown 
in \eqref{MoriA}. Note that we have chosen the charge vectors to correspond to 
the Mori cone generators for this triangulation. 
\bea 
  \label{MoriA} 
  \renewcommand{\arraystretch}{1.1} 
  \renewcommand{\arraycolsep}{7pt} 
  \begin{array}{|c|c|c|c|c|c|c|c|c||c|} 
  \hline 
  &x_1 & x_2 & x_3 & x_4 & x_5 & x_6 & x_7 &  x_8 & p \\ \hline\hline 
  \ell^{(1)} &3 & 2 & 0 & 0 & 0 & 1 & 0 & 0 & 6 \\ 
  \ell^{(2)} &0 & 0 & 1 & 0 & 1 & -1 & 0 & -1 & 0 \\ 
  \ell^{(3)} &0 & 0 & 0 & 1 & 1 & -1 & -1 & 0 & 0\\ 
  \ell^{(4)} &0 & 0 & 0 & 0 & -1 & -1 & 1 & 1 & 0\\ 
  \hline 
  \end{array} 
\eea 

The Mori cone for this triangulation is generated by four 
holomorphic curves $C^a$ which intersect the divisors $D_i$ 
as $D_i \cdot C^{a} = \ell^{(a)}_i$. Hence, the $\ell^{(a)}$
are the coordinates of the $C^{a}$ in the two-cycle basis dual to $D_i$. Since
there are as many Mori generators as K\"ahler parameters $h^{1,1}=4$, this
Mori cone is simplicial. 
Using the Mori generators it is also straightforward to determine a basis $K_i$ of 
four-cycles generating the K\"ahler cone. Expanding the K\"ahler form as
$J = r^i\, [K_i]$ the condition that all physical volumes are positive
translates into 
\bea
\int_{C^a} J = r^i\, \, C^a \cdot K_i> 0. 
\eea
This requires $r^i>0$ for    
\bea \label{Kaehlercone}
  K_1 &=& 3 D_5 + D_6 + 2 D_7 + 2 D_8\ ,\qquad \qquad 
  K_2 = D_5 + D_7\ ,\\  
  K_3 &=& D_5 + D_8\ ,\qquad \qquad \qquad \qquad \qquad  \ \, K_4 = D_5 + D_7 + D_8\ . \nn
\eea 
In this basis all triple intersections are positive, ensuring positivity 
of the divisor volumes and the total volume of $Y$. 
However, for our applications it is more useful to display the triple
intersection numbers in the basis $\{D_5,D_6,D_7,D_8\}$ as  
\begin{equation} 
\label{tripleint1} 
  I_3=D_6 \big( 7\,D_6^2 -D_5^2- D_7^2- D_8^2  - D_5\, D_6  -D_6\, D_7   -D_6\, D_8  
    +D_5 \, D_7 + D_5\, D_8 \big)\ .
\end{equation} 
Not surprisingly, there are both negative and positive intersections in this basis. 
 
To determine the geometry of the different divisors $D_i$ we now compute 
the Euler characteristic $\chi = \int_D c_2(T_D)$, as well as $\int_D c_1^2(T_D)$ 
for each divisor in $M_2^{({\rm dP}_9)^2}$.  We exemplify this for the divisor $D_8$ 
and restrict the intersection form \eqref{tripleint1} to this surface 
\begin{equation}  
   I_{D_8}=-D_6^2 - D_6\, D_8 +  D_5\, D_6 \ .
\end{equation} 
Using this intersection form one computes 
\begin{equation} \label{divisor_D8_indP9}
   \chi(D_8)= 12, \qquad \qquad \int_{D_8} c^2_1(T_{D_8})=0\; . 
\end{equation} 
{}From this we conclude that $D_8=\pi^*(E_2)$ is a ${\rm dP}_9$ surface. 
Analogously, we proceed for the remaining divisors $D_i$. We identify 
\begin{equation}
  \begin{array}{llll}   
    D_3=\pi^*(l-E_2) & K3\ ,\qquad  &  D_4=\pi^*(l-E_1) \quad & K3\ ,\\ 
   D_5=\pi^*(l-E_1-E_2) \quad & {\rm dP}_9\ ,\qquad \qquad & D_6= \cB \quad & {\rm dP}_2\ , \\    
   D_7=\pi^*(E_1) & {\rm dP}_9 \ , \qquad  & D_8=\pi^*(E_2) & {\rm dP}_9 \ .
  \end{array}
\end{equation}  
Let us note that indeed the exceptional divisors $D_7$ and $D_8$ are ${\rm dP}_9$
surfaces which justifies our notation $M_2^{({\rm dP}_9)^2}$.
Finally, using \eqref{MoriA} the divisor $D_1$ can be identified with  
\begin{equation} 
   D_1=3\, \cB + 9 \, \pi^*(l-E_1-E_2) +6\, \pi^*(E_1) + 6\, \pi^*(E_2) = D_T
   ,
\end{equation} 
that is, the tri-section $D_T$ defined in \eqref{def-tri} of the elliptic fibration over $\cB={\rm dP}_2$. 
With these identifications one can check that the triple intersection 
form \eqref{tripleint1} is indeed the one generated by
$\{\cB, \pi^*(E_i),\pi^*(l-E_1-E_2)\}$. All terms in \eqref{tripleint1}
contain the base $D_6$ and the expression in the brackets
corresponds to the intersection form of $E_i$ and $l-E_1-E_2$ as well as the
self-intersection $7$ of the anti-canonical class on ${\rm dP}_2$.

Let us now specify an orientifold projection $\Omega_p \sigma (-1)^{F_L}$. 
As a simple example consider the involution 
\begin{equation} \label{x3_map}
        \sigma: x_3\to -x_3 
\end{equation} 
and analyse the fixed point set. 
In order to do that we first note that the coordinates obey some scaling
relations dictated by the $U(1)$ weights $\ell^{(k)}$ displayed in \eqref{MoriA}.
This implies that in order to determine the fix-point set of \eqref{x3_map}, 
the coordinates $x_i$ need only to agree up to scaling such that 
\begin{equation} \label{xiscalings}
  x_j =\pm \prod_{k=1}^4 \lambda_{k}^{\ell^{(k)}_j } x_j \ ,
\end{equation}
where $\lambda_k \in \bbC^*$, and the minus sign should be used for $x_3$ while
the plus sign holds for all other coordinates. The value of the complex
scalars $\lambda_k$ is restricted by the Stanley-Reisner ideal of the toric
ambient space. More precisely, this ideal contains the information which
coordinates $x_i$ are not allowed to vanish simultaneously. For the case at
hand it reads 
\begin{equation}
  SR=\{ x_3\, x_5,\: x_3\, x_7,\: x_4\, x_5,\: x_4\, x_8,\:  x_7\, x_8,\:  x_1\, x_2\,x_6\}\ .
\end{equation} 
For example, since $x_1 x_5$ is in the Stanley-Reisner ideal, the
subspace $x_1 =x_5=0$ is not in the toric variety. Combining these
conditions with the scalings eq.~\eqref{xiscalings} fixes the
$\lambda_i$ to specific values and allows us to determine the
fix-point locus\footnote{Note that in general, the determination of
  the fix-point set in the Calabi-Yau hypersurface can be more
  tricky. This is due to fact that it will in general be non-generic
  hypersurface to admit the involution $\sigma$. In our examples, this
  will not introduce any new subtleties.}.

Let us apply this strategy explicitly to our example.  
The divisors $D_3=\{x_3=0\}$ and $D_7=\{x_7=0\}$ are fixed under $\sigma$
consistent with the scalings \eqref{MoriA}. As mentioned, $D_3$ 
is a $K3$ surface with $\chi=24$ and $D_7$ is a ${\rm dP}_9$ with $\chi=12$.
However, this is not the end of the story,  as there exist also fixed points
which give the location of $O3$-planes. Using the projective
identifications, we first get the two candidate fixed points 
\begin{equation} \label{one-fix}
   p_1= \{x_4=x_5=x_6=0\}, \qquad   p_2=\{x_5=x_6=x_8=0\}, 
\end{equation} 
where however, the first one $p_1$ is part of the Stanley-Reisner ideal 
and therefore discarded.  In addition there exist two fixed loci 
\begin{equation}  \label{three-fix}
   p_3= \{x_1=x_4=x_5=0\}, \qquad   p_4=\{x_1=x_5=x_8=0\}, 
\end{equation} 
where again the first is discarded and the second actually consists of
$3$ fixed points, which is essentially the space $\IP_{1,2}[6]$.  Note
that this is precisely the situation discussed in
\autoref{delPezzotrans}. {}From eq.~\eqref{one-fix} one infers that the
involution eq.~\eqref{x3_map} admits one isolated fix-point in the base
$D_6$. $\sigma$ acts on the smooth torus fibre over this point as
inversion, such that three fix-points eq.~\eqref{three-fix} arise in
the tri-section $D_1$.
 
To summarise, the fixed points locus consists of two non-intersecting  
$O7$ planes and four $O3$ planes. 
Therefore, the right-hand side of the $D7$-brane tadpole 
cancellation condition \eqref{tadseven} reads 
\bea 
     8 [D_{O7}]= 8\, \pi^*(l-E_2) + 8\, \pi^*(E_1) 
\eea 
and the right-hand side of the $D3$ brane tadpole condition 
\eqref{D3_tadpole} takes the form  
\begin{equation}  
  \label{D3tadpole_x3inv}
    \frac{\chi(K3) + \chi({\rm dP}_9)}{6} + \frac{N_{O3}}{2}= 8
  . 
\end{equation} 
Note that indeed we get a non-negative integer.  In order to cancel
the $D7$ brane tadpole we can only introduce branes wrapping entirely
the fibre. Candidates are of course $D_3$ and $D_7$ which are
point-wise invariant under the orientifold projection and therefore
belong to class 3.) introduced in \autoref{sec:generalities}.  A
natural candidate of class 2.) is $D_5=\pi^*(l-E_1-E_2)$.  This ${\rm dP}_9$
is not point-wise invariant.  In fact, using that $D_5$ intersects the
O7-plane in a genus $1$ curve and that all four O3 planes are inside
$D_5$ one uses the Lefschetz fix-point formula~\cite{Barth} to compute
for $D_5$ that $b^2_+ = 6$ and $b^2_-=4$. This involution indeed
corresponds to a viable involution of ${\rm dP}_9$.

Let us also consider the involution $x_1\to -x_1$, which 
is nothing else than the reflection of the torus fibre. 
The fixed point locus can be determined 
as $D_{O7}=D_1\cup D_6$, such that the right hand side of 
\eqref{tadseven} reads 
\begin{equation}
        8 [D_{O7}]=32\, \cB + 24 \,\pi^* c_1(\cB) 
\end{equation} 
and no fixed points. For the Euler characteristics we find 
$\chi(D_1)=435$ and $\chi(D_6)=\chi({\rm dP}_2)=5$, such that 
the right hand side of 
\eqref{D3_tadpole} takes the form
\begin{equation} 
    {\chi(D_{O7})  \over 6} + {N_{O3}\over 2}= \frac{220}{3}\; . 
\end{equation} 
Finally, let us determine how this involution acts on
$\pi^*(E_1)={\rm dP}_9$. The divisor $\cB$ intersects $\pi^*(E_1)$ of course
over a sphere and from $\chi(D_1\cap D_7)=D_1\cdot D_7\cdot
(-D_1-D_7)=-6$ we conclude that the Euler characteristic of the
fix-point set in ${\rm dP}_9$ is $\chi=-4$. This implies $b^2_+=2$ and
$b^2_-=8$ and corresponds to the blow-up of the Bertini involution
$(\cB_8,\sigma_{\rm B})$ of \autoref{tab:dPinv2} at one invariant
point.

\subsubsection*{The geometry of $\mathbf{M_2^{({\rm dP}_8)^2}}$} 

Let us now consider the triangulation which 
corresponds to a Calabi-Yau manifold with two exceptional ${\rm dP}_8$
divisors $D_7$ and $D_8$. Again we specify the data for the linear sigma-model 
such that the charge vectors correspond to the Mori cone generators for this
triangulation. 
 \bea 
  \label{MoriB} 
  \renewcommand{\arraystretch}{1.1} 
  \renewcommand{\arraycolsep}{7pt} 
  \begin{array}{|c|c|c|c|c|c|c|c|c||c|} 
  \hline 
  &x_1 & x_2 & x_3 & x_4 & x_5 & x_6 & x_7 &  x_8 & p \\ \hline\hline 
  \ell^{(1)} &3 & 2 & 0 & 1 & 1 & 0 & -1 & 0 & 6 \\ 
  \ell^{(2)} &3 & 2 & 1 & 0 & 1 & 0  & 0 & -1 & 0 \\ 
  \ell^{(3)} &0 & 0 & 0 & -1 & -1 & 1 & 1 & 0 & 0\\
  \ell^{(4)} &0 & 0 & 1 & 1 & 1 & -3 & 0 & 0 & 0\\ 
  \ell^{(5)} &0 & 0 & -1 & 0 & -1 & 1 & 0 & 1 & 0\\ 
  \hline 
  \end{array} 
\eea 

Note that the Mori cone eq.~\eqref{MoriB} is generated by five
holomorphic curves $C^a$ in this triangulation. Since this is more
than the $h^{1,1}=4$ K\"ahler deformations the Mori cone is
non-simplicial. Also for this case we have to determine the dual
K\"ahler cone spanned by four-cycles. By definition the K\"ahler cone
is generated by four-cycles $D$ which satisfy $D \cdot C^a \geq
0$. For our purposes it will be convenient to determine the K\"ahler
cone in coordinates $r^i$.  We therefore chose to discard one of the 5
Mori vectors \eqref{MoriB}, namely, the last one $\ell^{(5)}$. As in
the $M_2^{({\rm dP}_9)^2}$ phase it is then straightforward to
determine the K\"ahler cone generators \bea
\label{K5TrianB}
 K_1&=&3D_5+D_6+2D_7+3D_8\ , \qquad \quad K_2=-D_8,\nonumber \\
 K_3&=&3D_5+D_6+3D_7+3D_8\ , \qquad \quad K_4=D_5+D_7+D_8\ .
\eea
Note that for $J = r^i [K_i]$, the positivity of the curves $C^i,i=1,...,4$ is
ensured when $r^i>0$. In order that the last curve is positive we have to
additionally impose $\int_{C^5} J = r^i K_i \cdot C^5 = r^1 - r^2 + r^3 > 0$.
Clearly, a similar analysis can be carried out when discarding one of the other
$\ell^{(a)}$ which allows to define coordinates for the complete 
non-simplicial K\"ahler cone. 

For convenience we again display the triple intersection numbers for
$M_2^{({\rm dP}_8)^2}$  in the basis $\{D_5,D_6,D_7,D_8\}$,  
\begin{equation} 
\label{tripleint2} 
  I_3=9\,D_6^3 + D_7^3 +D_8^3 + 
 D_5\big(D_5\, D_6   + D_5\, D_7   +D_5\, D_8 -2\,
 D_5^2 - 3\, D_6^2 - D_7^2 - D_8^2 )\ .  
\end{equation} 
In particular, for the intersection form on the surface $D_8$  
we get  
\begin{equation} 
   I_{D_8}=D_5^2 +D_8^2 -D_5\, D_8 
\end{equation} 
so that 
\begin{equation} 
   \chi(D_8)= 11,\qquad \quad \int_{D_8} c^2_1(T_{D_8})=1\; , 
\end{equation}   
which we identify with the correct values for ${\rm dP}_8$. 
For $D_7$ we find the same result. Comparing this result with 
\eqref{divisor_D8_indP9} we note that in this cone of the complexified 
K\"ahler moduli space, one $\IP^1$ in each $\pi^*(E_i)$ of $M_2^{({\rm dP}_9)^2}$ 
has been flopped away so that ${\rm dP}_9\to {\rm dP}_8$. Indeed the two exceptional
divisors $E_i$ of the ${\rm dP}_2$ base in $M_2^{({\rm dP}_9)^2}$ are absent in
$M_2^{({\rm dP}_8)^2}$ since $\chi(D_6)=3$, and $\int c_2(T_{D_6}) = 9$.

Having performed the flop transitions, the exceptional 
$\IP^1$s have to reappear in other divisors. In fact, we compute  
\begin{equation}
    \chi(D_3)=25\ ,\quad  \chi(D_5)=14, \qquad \quad \int c^2_1(T_{D_3})=-1\ ,
     \quad \int c^2_1(T_{D_5})=-2\ .   
\end{equation} 
Again, the divisor $D_4$ has the same topology as $D_3$.
This implies that the divisor $D_3$, $D_4$ corresponding to the pull-back divisors
$\pi^*(l-E_i)$ in $M_2^{({\rm dP}_9)^2}$ now contain each one additional $\IP^1$.
The divisor $D_5$ corresponding to $\pi^*(l-E_1-E_2)$ contains two additional $\bbP^1$s.
 
Let us now investigate the involutions on $M_2^{({\rm dP}_8)^2}$. 
A simple involution exchanging the two ${\rm dP}_8$ surfaces has been 
employed in~\cite{Diaconescu:2007ah}. However, since these two del Pezzo
surfaces do not intersect, this involution will not be useful in constructing
GUT models. We therefore consider again the involution $\sigma: x_3\to -x_3$,
which still has the non-intersecting $O7$ planes $D_3$ and $D_7$. 
There exist $4$ fixed points so that the tadpole contribution is   
\bea 
    {\chi(D_{O7})  \over 6} + {N_{O3}\over 2}= 8\ . 
\eea 
This condition is identical to the condition eq.~\eqref{D3tadpole_x3inv} on 
$M_2^{({\rm dP}_9)^2}$ since the topology
change of the two O7 divisors precisely cancels. 
One checks that $3$ fix-points are located on the del Pezzo 8 defined 
by $x_8 = 0$ which intersects the orientifold locus $D_3$ on a genus $1$ curve. 
This implies using \autoref{tab:dPinv2} that $b_+^2=5$, $b_-^2=4$ for the  
del Pezzo $8$. 
 
The involution defined by the reflection 
$x_1\to -x_1$, again has the fixed point divisors 
$D_{O7} = D_1\cup D_6$. For the Euler characteristics we find 
$\chi(D_1)=435$ and $\chi(D_6)=3$. 
However, this time the involution also has the two 
fixed points $p_1=\{x_4=x_5=x_7=0\}$ and 
$p_2=\{x_3=x_5=x_8=0\}$. 
For the D3-tadpole contribution we therefore obtain 
\bea 
    {\chi(D_{O7})  \over 6} + {N_{O3}\over 2}=\frac{222}{3}\; . 
\eea 
Finally, let us determine how this involution acts on 
$D_7={\rm dP}_8$. The base $\cB=D_6$ does not 
intersect $D_7$ after the flop transition, while 
$D_1$ intersects $D_7$ over curve with $\chi=-6$. 
Moreover, only the fixed point $p_1$ lies on $D_7$ , 
while $p_2$ lies on $D_8$. Therefore, the Euler-characteristic of the 
fixed point set in $D_7$ is $\chi=-5$ and 
we obtain $b_2^+=1$, $b_2^-=8$. Comparing this result with 
\autoref{tab:dPinv2} we conclude that the involution 
$x_1\to -x_1$ acts on the ${\rm dP}_8$ as the Bertini involution $(\cB_8,\sigma_{\rm
  B})$.

\subsection
[Orientifold of An Elliptic Fibration Over ${\rm dP}_3$]
[Orientifold of An Elliptic Fibration Over dP3]
{Orientifold of An Elliptic Fibration Over $\mathbf{{\rm dP}_3}$}
\label{dP3fibration} 

\begin{figure}[htbp]
\centering
\hspace{40pt}
\includegraphics[height=1.8cm]{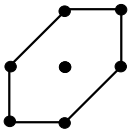}
\vspace{-10pt}
\caption{The points of the polyhedron of the ${\rm dP}_3$ base.}
\label{dP3figtriang}
\end{figure}

We can repeat the procedure just described also for the del Pezzo 
base ${\rm dP}_3$. This remains to be rather simple, 
since this del Pezzo is still toric and represented by the 
points $(1,0)$, $(0,1)$, $(-1,0)$, $(0,-1)$, $(-1,-1)$ and $(1,1)$ as 
shown in \autoref{dP3figtriang}.
In this case there are in fact $18$ triangulations of the ${\rm dP}_3$ base.
The toric ambient space for the corresponding Calabi-Yau hypersurface
$M_3$ is obtained by adding the point $v_9^* =(3,2,1,1)$ to the
polyhedron of $M_2$ above.  This polyhedron has $18$ triangulations
which yield different phases of the Calabi-Yau hypersurface $M_3$. As
determined by eqns.~\eqref{general_Hodge} and~\eqref{general_chi}, each
$M_3$ has the topological data $h^{1,1}(M_3) = 5$, $h^{2,1}(M_3) =
185$, and $\chi(M_3) =- 360$. In the following we discuss one out of
the $18$ phases in more detail. Namely, the Weierstra\ss{} phases
$M_3^{({\rm dP}_9)^3}$ where the divisors $D_7,D_8,D_9$ are ${\rm dP}_9$
surfaces. A second phase, $M_3^{({\rm dP}_8)^3}$ will be of importance in
\autoref{linebundles} where we discuss the issue of moduli
stabilisation for compactifications on $M_n$.  On $M_3$ we will also
be able to introduce interesting orientifold projections with a
non-trivial split $h^{1,1}=h^{1,1}_+ + h^{1,1}_-$.

\subsubsection*{The geometry of $\mathbf{M_3^{({\rm dP}_9)^3}}$}

Let us discuss the Weierstra\ss{} phase where all $(-1)$-curves 
lead to ${\rm dP}_9$ surfaces. Using \autoref{tab:DemazureTable3} we infer that 
${\rm dP}_3$ has six $(-1)$-curves which yield six ${\rm dP}_9$ surfaces.  
The Mori cone associated to this phase is shown in eq.~\eqref{MoridP3A}.
\bea 
  \label{MoridP3A} 
  \renewcommand{\arraystretch}{1.1} 
  \renewcommand{\arraycolsep}{7pt} 
  \begin{array}{|c|c|c|c|c|c|c|c|c|c||c|} 
  \hline 
  &x_1 & x_2 & x_3 & x_4 & x_5 & x_6 & x_7 &  x_8 & x_9 & p \\ \hline\hline 
  \ell^{(1)} & 3 & 2 & 0 & 0 & 0 &1 & 0 & 0 & 0 & 6  \\ 
  \ell^{(2)} &0 & 0 & 0 & 1 & 1 & -1 & -1 & 0 & 0 & 0 \\ 
  \ell^{(3)} &0 & 0 & 0 & 0 & -1 & -1 & 1 & 1 & 0 & 0\\ 
  \ell^{(4)} &0 & 0 & 1 & 1 & 0 & -1 & 0 & 0 & -1& 0\\
  \ell^{(5)} & 0 & 0 & 0&-1& 0 & -1 & 1 & 0 & 1 &0 \\
  \ell^{(6)} &0 & 0 & 1 & 0 & 1 & -1 & 0 & -1 & 0 & 0 \\
  \ell^{(7)} & 0 & 0 & -1 & 0 & 0 & -1 & 0 & 1 & 1 & 0\\
  \hline 
  \end{array} 
\eea 
Using the data for the Mori cone, it is straightforward to evaluate the
associated K\"ahler cone. However, we again display the triple intersection
numbers in the basis $D_6,\ldots, D_9$ 
\begin{equation}
  I_3=D_6 \big(6 D_6^2 -D_5^2 - D_7^2 - D_8^2 - D_9^2 - D_6 ( D_5 + D_7 + D_8
  + D_9) + D_5 D_8 + D_7 D_5  \big)\ .
\end{equation}
In a similar spirit as in the previous section 
we can also check that the divisors 
\bea
   &&D_3 = \pi^*(l-E_1 -E_3)\ ,\quad   D_4 = \pi^*(l-E_2 -E_3)\ , \quad  D_5 =
   \pi^*(l-E_1 -E_2)\ , \nn  \\
   &&D_7=\pi^*(E_2)\ ,\qquad \qquad \quad  D_8=\pi^*(E_1)\ ,\qquad \qquad
   \quad D_9=\pi^*(E_3)\ . 
\eea
are the six ${\rm dP}_9$ surfaces, while the ${\rm dP}_3$ base is the divisor $D_6$.

Let us now turn to the definition of the involution on $M_3^{({\rm dP}_9)^3}$. 
Using \autoref{tab:dPinv2}, we find four candidate involutions on the
${\rm dP}_3$ base. Our prime  focus will be on the exchange involution
$(\cB_3,\sigma_3)$, which admits the rational curve $2l-E_1-E_2$ as 
fix-point divisor. Lifted to the elliptically fibred threefold, this
involution descends to 
\begin{equation}
   \sigma:\quad x_7 \ \leftrightarrow \ x_3 \ , \qquad \quad x_8 \ \leftrightarrow \
   x_4 \ , \qquad \quad x_9 \ \leftrightarrow \ x_5\ .
\end{equation}
We can thus evaluate the action of $\sigma^*$ on the cohomology $H^{2}(M_3)$ spanned 
by $[D_5,D_6,D_7,D_8,D_9]$ to show that 
\begin{equation}
  h^{1,1}_+ = 3\ ,\qquad \quad h^{1,1}_- = 2\ .
\end{equation}
This implies that by using \eqref{def-GT} this orientifold compactification will 
admit three K\"ahler moduli $T_I$ and two B-field moduli $G^i$.

Again, we can determine the fixpoint set of $\sigma$ using toric geometry.
Taking into account the scaling relations \eqref{MoridP3A} and the
corresponding constraints from the Stanley-Reisner ideal, one finds the
fix-point divisor
\begin{equation} \label{fix_locdP3}
   x_5 x_8 x_3 x_9 - x_4 x_9 x_5 x_7 = 0 \ ,
\end{equation}
and no isolated fix-points meeting the hypersurface.  The isolated
fix-points and hence O3-planes can also be directly inferred from the
fact that there are no fix-points on the ${\rm dP}_3$ base for this
involution.  In accord with the general arguments presented in
\autoref{delPezzotrans}, the locus eq.~\eqref{fix_locdP3} corresponds
to an O7-plane on the divisor class $D_{O7} = D_4 + D_9 + D_5 +D_7 =
\pi^*(2l-E_1-E_2)$ and induces a tadpole
\begin{equation}
  8[D_{O7}] = 8\pi^*(2l-E_1-E_2)\ .
\end{equation}
For this O7-plane one computes $\chi(D_{O7})=48$, such that the induced
D3-tadpole is $\chi(D_{O7})/6 = 8$.

\subsection{The Swiss-Cheese Property}
\label{swisskaas}

Though we will deliver some more comments in \autoref{sec_moduli}, in
this paper we do not yet intend to combine the GUT model search with a
complete analysis of moduli stabilisation. However, we would like to
point out that some of the manifolds discussed so far provide new
examples of so-called swiss-cheese type Calabi-Yau manifolds. Here we
understand this term in the strong sense that the volume ${\cal V}$ of
the Calabi-Yau can be expressed as\footnote{It has been
  shown~\cite{Cicoli:2008va} that a much weaker condition is already
  sufficient to make the LARGE volume scenario work.}
\begin{equation}
\label{swisscheeseyeah}
   {\cal V}=\frac{\sqrt{8}}{6}
   \left[
     \tfrac{1}{\sqrt{3^5}} (\tau_0)^{3\over 2} - 
     \tfrac{1}{3} (\tau_\cB)^{3\over 2} - \sum_{i=1}^{h^{1,1}-2}  (\tau_i)^{3 \over
       2}
   \right]
   ,
\end{equation}
where the numerical coefficients are chosen for later convenience and 
$\tau_B,\tau_i$ are volumes of a basis of four-cycles
$\Gamma_0,\Gamma_\cB,\Gamma_i$ given by
\begin{equation} \label{def-taus}
   \tau_0 = \frac12 \int_{\Gamma_0} J \wedge J\ , \qquad  \tau_\cB = \frac12 \int_{\Gamma_\cB} J \wedge J\ , \qquad
   \tau_i = \frac12 \int_{\Gamma_i} J \wedge J\ .
\end{equation}
Our aim is to show that for the manifolds $M_n^{({\rm dP}_8)^n}$ 
one can always find such a basis $\Gamma_0,\Gamma_\cB,\Gamma_i$ 
such that $\cV$  is of the form \eqref{swisscheeseyeah}.

\subsubsection*{Swiss-Cheese property of $\mathbf{M_2^{({\rm dP}_8)^2}}$ and $\mathbf{M_3^{({\rm dP}_8)^3}}$}

We first consider the manifold $M_2^{({\rm dP}_8)^2}$.  Recall that this
Calabi-Yau is connected via a flop transition to the corresponding
Weierstra\ss{} model, which is an elliptic fibration over the ${\rm dP}_2$
base. In fact, the two exceptional two cycles $E_{1,2}$ in the base
${\rm dP}_2$ have been flopped away.  We will use the same notation as in
Sections~\ref{sec:dP2fibration} and~\ref{dP3fibration}.  The triple
intersection form on $M_2^{({\rm dP}_8)^2}$ in the basis $D_i$ was given in
eq.~\eqref{tripleint2}.  Expanding the K\"ahler form as $J=r_5\, D_5 +
r_6\, D_6 + r_7\, D_7 + r_8\, D_8$ and defining
\begin{equation}
  \Gamma_\cB = D_6\ ,\qquad \Gamma_1 = D_7\ ,\qquad \Gamma_2 = D_8\ ,
\end{equation}
we  compute the $\tau_i$ in \eqref{def-taus} as
\begin{equation}
     \tau_\cB= \tfrac12 (r_5-3\, r_6)^2\ ,\qquad
     \tau_1= \tfrac12 (r_5 - r_7)^2\ , \qquad
      \tau_2= \tfrac12 (r_5 - r_8)^2\ .
\end{equation}
Now, let us define the following divisor, which obviously is related
to the former tri-section $D_1$ in the Weierstra\ss{} phase
\begin{equation}
       \Gamma_0=D_1+3\, D_7+ 3\, D_8 = 3\, (3\, D_5 + D_6 + 3\, D_7+ 3\, D_8 )\ .
\end{equation}
The volume of this divisor is given by $\tau_0={3\over 2} r_5^2$.
We have found four divisors whose volumes can be written as  perfect squares
and it is now a simple calculation to show that the
total volume of the Calabi-Yau can be written as in \eqref{swisscheeseyeah},
showing the swiss cheese structure. Note that indeed
the three small cycles are of the type $\Gamma_\cB=\IP_2$, $\Gamma_{1,2}={\rm dP}_8$
and therefore all are shrinkable to a point.

Along the same lines also the swiss cheese structure of the
Calabi-Yau $M_3^{({\rm dP}_8)^3}$ can be shown. This manifold
is related to the Weierstra\ss{} model over ${\rm dP}_3$.
For completeness, let us list the relevant data.
The intersection form reads
\bea
  I_3&=&-2\, D_5^3 + 9\, D_6^3 + D_7^3 + D_8^3  + D_9^3 +D_5^2\, D_6 -
 3\, D_5\, D_6^2 +D_5^2\, D_7 -D_5\, D_7^2  \nonumber \\
   &&\quad +D_5^2\, D_8 -D_5\, D_8^2 \, .
\eea
For the volumes of the del Pezzo type divisors $\Gamma_\cB = D_6,\, \Gamma_i = D_{i+6}$ we get
\bea
      \tau_\cB= \tfrac12 (r_5-3\, r_6)^2\, , \quad
      \tau_1= \tfrac12(r_5 - r_7)^2\, , \quad
     \tau_2= \tfrac12 (r_5 - r_8)^2\, ,\quad
     \tau_3= \tfrac12  r_9^2\ ,
\eea
and with
\begin{equation}
      \Gamma_0=D_1+3\, D_7+ 3\, D_8 + 3\, D_9
          = 3\, (3\, D_5 + D_6 + 3\, D_7+ 3\, D_8 )\; .
\end{equation}
and $\tau_0={3\over 2} r_5^2$ the total volume of the Calabi-Yau can
again be written as eq.~\eqref{swisscheeseyeah}.

\subsubsection*{Proof of Swiss-Cheese property for $\mathbf{M_n^{
      ({\rm dP}_8)^n}}$}

What we have concretely confirmed for the latter two examples of
Calabi-Yau threefolds is, in fact, more generally true.  Starting with
the Weierstra\ss{} phase of an elliptic fibration over a ${\rm dP}_n$,
$n=0,\ldots, 8$ base, the phase related to this one by flopping away
all $n$ $\IP^1$-cycles in the base, is of the swiss-cheese type.  To
prove this, we show that we can define $n+2$ divisors such that the
triple intersection form is diagonal.  Before the flop transition we
have the pull-back divisors $\pi^*(E_i)={\rm dP}_9$. After the flop
transition, these lose the $\IP^1$ given by $\cB \cap \pi^*(E_i)$ and
we get $\Gamma_i:=\pi^*(E_i)_{\rm flop}={\rm dP}_8$ for $i=1,\ldots,
n$.  Clearly, these divisors satisfy
\begin{equation} \label{Gammai}
        \Gamma_i^3=1, \qquad \Gamma_i\cap \Gamma_j =0 \quad {\rm for}\ i\ne j\; .
\end{equation}
The former base ${\cal B}={\rm dP}_n$ is now just $\IP^2$ so that we define
$\Gamma_{\cB}:={\cal B}_{\rm flop}=\IP^2$. Since we have flopped away
the intersection locus with the $\pi^*(E_i)$, we can write
\begin{equation}
     \Gamma_{\cB}^3={\cal B}^3_{\rm flop}=c_1^2(\IP^2)\, {\cal B}_{\rm flop}=9,
\qquad
         \Gamma_{\cB}\cap \Gamma_i=0\; .
\end{equation}
For the remaining divisor we start with the former
tri-section $3({\cal B}+3\pi^*(l)-\sum_i \pi^*(E_i) )$ and realise
that, after the flop transition, this divisors gains an extra of
$3n$ $\IP^1$s. Therefore, this divisor cannot be diagonal to
the ones introduced so far.
However, we can define the new divisor
\bea
     \Gamma_0&=&3({\cal B}+3\pi^*(l)-\sum_i \pi^*(E_i) )_{\rm flop}
                +3\, \sum_i \pi^*(E_i)_{\rm flop} \nonumber \\
       &=& 3({\cal B}_{\rm flop} + 3\pi^*(l) ),
\eea
which satisfies
\begin{equation} \label{Gamma0}
    \Gamma_0^3=243 ,\qquad \Gamma_0\cap \Gamma_i=0, \qquad  \Gamma_0\cap \Gamma_{\cB}=0\; .
\end{equation}
Therefore, we have found a basis of $(n+2)$ divisors which diagonalise
the triple intersection form.  Taking into account that except
$\Gamma_0$ all four-cycles are shrinkable to a point, we expect that
inside the K\"ahler cone, we can write the volume of the Calabi-Yau in
the swiss cheese form eq.~\eqref{swisscheeseyeah}.  Apparently, the two
former toric Calabi-Yau manifolds are only two specific examples.

For realising the LARGE volume scenario it is not
necessary to have a Calabi-Yau having
the strong swiss-cheese type property~\cite{Cicoli:2008va} as in \eqref{swisscheeseyeah}.
Therefore, one can also discuss the case that from the $n$ initial
$\pi^*(E_i)$ divisors of the type ${\rm dP}_9$ only $r$ have been
flopped to the ${\rm dP}_8$ phase.
Since ${\rm dP}_9$ is not shrinkable to a point, one does not
expect a swiss-cheese type structure for them, but
along the same lines as above one can still write the volume as
\begin{equation} \label{cV_split}
   {\cal V}={\cal V}\left(M_{n-r}^{({\rm dP}_9)^{n-r}}\right) -\sum_{i=1}^r (\tau_i)^{3\over 2}\; .
\end{equation}
where $M_{n-r}^{({\rm dP}_9)^{n-r}}$ denotes the Weierstra\ss{} phase of the elliptic
fibration over $dP_{n-r}$.

\subsection{D-Term Conditions For D7-Branes on Del Pezzo Surfaces}
\label{linebundles}

In this section we have introduced a specific class of manifolds which
admit shrinkable del Pezzo surfaces as divisors. In the following we
like to address the question whether we can wrap D7-branes on these
surfaces and stabilise their volume at sizes significantly larger than
the string scale by demanding a vanishing D-term, eq.~\eqref{FIterm}.

Let us denote the del Pezzo surface with a wrapped D7-brane by
$(D_{dP},L_{dP})$ and its orientifold image by $(D_{dP}',L_{dP}')$. In
the following discussion it is crucial to again distinguish the three
cases defined at the beginning of \autoref{sec:generalities},
page~\pageref{Dcase1}. We will focus on the cases~\ref{Dcase2}
and~\ref{Dcase3} where $D_{dP}$ and $D'_{dP}$ are in the same homology
class. This implies that $D_{dP}$ cannot support $B_-$ moduli and the
D-term arises entirely from the gauging $X_{a\, I}$ given in
\eqref{KillingV}. The precise form of $X_{a\, I}$ depends on the
choice of four-cycles to define the coordinates $T_I$. Let us focus on
the swiss-cheese examples of \autoref{swisskaas}. We have argued in
eqns.~\eqref{Gammai}--\eqref{Gamma0} that one can choose a four-cycle
basis $\Gamma_0,\Gamma_\cB,\Gamma_i$ of $H_4(Y)$ such that the triple
intersection form reads
\begin{equation}
  I_3 = 243\ \Gamma_0^3 + 9\ \Gamma_\cB^3 + \sum_{i} \Gamma_i^3\ .
\end{equation} 
The $\Gamma_\cB$ and $\Gamma_i$ are del Pezzo surfaces, and hence are 
the candidate $D_{dP}$ for an appropriate orientifold projection.
We use this basis in the expansion of the K\"ahler form $J = - r_{dP}[D_{dP}]+
\ldots$, where $r_{dP}>0$ in the K\"ahler cone.  
The coordinate $T_{dP}$ associated to $D_{dP}$ is then
given by
\begin{equation}
   \R\, T_{dP} = \frac{1}{2} e^{-\phi} \int_{D_{dP}} J \wedge J \ \sim \  r^2_{dP}\ ,
\end{equation}
and the K\"ahler potential for the fields $G^i,T_I$ takes the form 
\begin{equation}
  K = - 2 \ln ({\cal V}_{\rm red} - (T_{dP} + \bar T_{dP})^{3\over 2})\ ,
\end{equation}
where we have used eq.~\eqref{swisscheeseyeah}. The important point is
that ${\cal V}_{\rm red} $ is independent of the moduli $T_{dP}$ and
only depends on the remaining $G^i,T_I$.  We can also evaluate the
Killing vector in the basis $\Gamma_0 , \Gamma_\cB, \Gamma_i$ and find
that it diagonalises in the $T_{dP}$ direction with the only
non-trivial contribution
\begin{equation} \label{KillingdP}
   X_{dP} = \int_{D_{dP}} [D_{dP}] \wedge c_1(\widetilde L_{dP})\ .
\end{equation}
Using these equations it is straightforward to evaluate the D-term 
\begin{equation} \label{xidP}
  \xi_{dP} \ \sim \ r_{dP}\ ,
\end{equation}
which thus has to vanish for a supersymmetric vacuum $\xi_{dP}=0$. This
implies that we are taken to the point $r_{dP}\rightarrow 0$, where the size
of the del Pezzo surface becomes of order string scale. 

Entering a small volume regime implies that our classical analysis is
no longer valid and additional corrections need to be included. In
particular, as in the underlying $\cN=2$ theory, world-sheet
instantons will correct the expressions. However, these corrections
will not alter the fact that $\partial_{T_{dP}}K \sim r_{dP}$ but
rather correct the definition of the $\cN=1$ coordinate
$T_{dP}$~\cite{Grimm:2004ua}.  Thus, if one still uses the
eq.~\eqref{KillingdP} for the Killing vector one is unavoidably driven
to the point where $r_{dP}$ is small. This is precisely the regime,
where for the local building the techniques of quiver gauge theories
on ${\rm dP}_r$ singularities are relevant~\cite{Douglas:1996sw,
  Aldazabal:2000sa, Verlinde:2005jr, Buican:2006sn, Conlon:2008wa}.
Thus, the global models we presented provide a concrete embedding of
these local constructions.

The question now is how general our findings are. One might naively
think that whenever one has shrinkable ${\rm dP}_r$, $r\le 8$ surfaces the
triple intersection form has the swiss-cheese form
eq.~\eqref{cV_split} so that for GUT branes on these cycles, one is
driven to the quiver locus. In \autoref{sec:quintic} we will present
another class of del Pezzo transitions, based on the Quintic, in which
we will instead find mutually intersecting ${\rm dP}_r$ surfaces, which
therefore do \emph{not} diagonalise the triple intersection form.
Therefore, it is not the shrink-ability of the del Pezzo surfaces but
rather the swiss-cheese form of the triple intersection form which is
responsible for the D-term minimisation at the quiver locus.


\section
[A GUT Model on $\mathbf{M_2^{({\rm dP}_9)^2}}$] 
[A GUT Model]
{A GUT Model on $\mathbf{M_2^{({\rm dP}_9)^2}}$}
\label{sec:GutModelExample}
 
In this section we investigate whether the simple geometries with
orientifold involutions introduced in the last section are already
sufficient to construct realistic, globally consistent intersecting
D7-brane GUT models. We work out one toy example in some detail which
exemplifies the necessary steps to build a realistic model. This will
illustrate the important role played by the structure of the manifold
to satisfy the constraints from \autoref{tab_gutfeatures}.  The
discussion of this section also serves as a preparation for the
construction of two three-generation GUT models in
\autoref{sec:search} on related geometries.

Concretely, we consider the Calabi-Yau manifold $M_2^{({\rm dP}_9)^2}$
of \autoref{sec:dP2fibration} and choose as the orientifold involution
$\Omega\sigma (-1)^{F_L}$ with $\sigma: x_3\to -x_3$.  As discussed
previously, the fixed point locus consists of the disjoint divisors
$\pi^*(E_1)$ and $\pi^*(l-E_2)$ and four additional fixed points. To
cancel the D7-brane tadpole eq.~\eqref{tadseven}, we introduce
D7-branes on the divisors
\begin{equation}
  D_a = \pi^*(E_1)
  ,\qquad
  D_b = \pi^*(l-E_1-E_2)
  ,\qquad 
  D_c = \pi^*(l-E_2)
  ,
\end{equation}
and denote the corresponding embeddings $\i_j: D_j \hookrightarrow Y$.
Recall from \autoref{sec:dP2fibration} that $D_a$, $D_b$ are ${\rm
  dP}_9$ surfaces while $D_c$ is a K3 surface. The O7-tadpole $8\big(
\pi^*(l-E_2)+\pi^*(E_1) \big)$ is cancelled by three stacks of
D7-branes with multiplicities
\begin{equation} 
  N\times D_a + (N-4)\times D_b + (8-N)\times D_c 
  , \qquad 
  N=4,5,6,7,8
\end{equation}
together with their orientifold images wrapping the same divisors.
The resulting gauge group is $SO(2N)\times SP(2N-8)\times
SO(16-2N)$. Note that the last stack has vanishing intersection with
the first two stacks, and will be hidden from the visible sector. The
next step is to break the first two gauge groups by turning on
non-trivial line-bundles in $U(N)\times U(N-4)$.

\subsection{The Chiral Model} 
 
For now, let us first focus on the chiral sector of the theory and
solely compute chiral indices. The computation of the entire
cohomology classes is postponed to \autoref{sec_globalcons}. As our
initial Ansatz for the line bundles on the divisors, we will pick all
three to be restrictions of global line bundles,
\begin{equation}
  \begin{split}
    L_a \;&= \i_a^* \; 
    \OsheafY \Big( k {\cal B} +\pi^*(\widetilde \eta_a) \Big)
    ,\\
    L_b \;&= \i_b^* \; 
    \OsheafY \Big( k {\cal B} +\pi^*(\widetilde \eta_b) \Big)
    ,\\
    L_c \;&= \i_c^* \; 
    \OsheafY \Big( m {\cal B} +\pi^*(\eta_c) \Big)
    .
  \end{split}
\end{equation}
In the following, we will be forced to modify this Ansatz by line
bundles that are trivial in $H^2(Y,\Z)$. However, this changes only
the vector-like pairs but not the chiral spectrum.

Since $D_a$ and $D_b$ are not {\it Spin}, it is convenient to
explicitly split off a factor $\sqrt{K}$ from $\widetilde
\eta_{a,b}$. Note that one can rewrite
\begin{equation}
  \sqrt{K_{D_a}} = \sqrt{\OsheafDa(-f)} = 
  \i_a^* \sqrt{\OsheafY(-D_a)} = 
  \i_a^*\; \OsheafY\big(- \tfrac{1}{2} \pi^*(E_1) \big)
\end{equation}
and similarly for $D_b$. Hence, let us set
\begin{equation}
  \widetilde \eta_a = \eta_a - \frac{1}{2} E_1
  , \quad 
  \widetilde \eta_b = \eta_b - \frac{1}{2} (l-E_1-E_2)
  . 
\end{equation}
With these definitions, we have parametrised the line bundles by 
\begin{equation}
  k,l,m \in \tfrac{1}{2}\Z
  ,\quad
  \eta_a \in H_2\big(D_a,\tfrac{1}{2}\Z\big)
  ,\quad
  \eta_b \in H_2\big(D_b,\tfrac{1}{2}\Z\big)
  ,\quad
  \eta_c \in H_2\big(D_c,\tfrac{1}{2}\Z\big)
  .
\end{equation}
Moreover, for vanishing $B_+$-field all have to be integral. For
non-zero discrete $B_+$-flux, they have to satisfy the quantisation
condition eq.~\eqref{quant2}.

In view of the rules form \autoref{tab_chir_spec} it is
straightforward to compute the resulting chiral spectrum, and we list
it in \autoref{tab_chir_modela}.
\begin{table}[htbp] 
\renewcommand{\arraystretch}{1.5} 
\begin{center} 
\begin{tabular}{|c||c|c|c|} 
\hline 
\hline 
number & $U(N)$ & $U(N-4)$ & $U(8-N)$  \\ 
\hline \hline 
$-2k$ & $\Yasymm_{\,(2)}$ & $1$ & $1$ \\ 
$-2k$ & $1$ & $\Ysymm_{\,(2)}$  & $1$ \\ 
$-2k$ & $\antifund_{\, (-1)}$ & $\antifund_{\, (-1)}$ & $1$ \\ 
\hline 
\hline 
\end{tabular} 
\caption{Chiral spectrum for intersecting D7-brane model. The indices denote 
 the $U(1)$ charges. } 
\label{tab_chir_modela} 
\end{center} 
\end{table} 
Let us make a couple of remarks concerning this spectrum: The third
stack does not carry any chiral modes and is completely hidden from
the first two. Moreover, the cubic $SU(N)$ and $SU(N-4)$ anomalies are
indeed cancelled.  Analysing the Abelian and mixed
Abelian--non-Abelian anomalies, we find that the linear combination
$U(1)_X={1\over N}U(1)_a + {1\over N-4}U(1)_b$ is
anomaly-free. However, due to the Green-Schwarz mass terms, it can
nevertheless receive a mass by mixing with an axionic mode. In fact,
this is the case as long as the first Chern classes of the line
bundles are independent as elements in $H^2(Y)$~\cite{erik}, and there
is no massless $U(1)$ prior to breaking $SU(5)$ to the Standard Model.
Intriguingly, for $N=5$ we get an $SU(5)$ model with $N_{\rm gen}=-2k$
generations of Standard Model particles. The $2k$ states in the
symmetric representation of $U(1)_b$ carry the quantum numbers of
right-handed neutrinos.  So far we have not required D-term
supersymmetry of the configuration. For this, one has to ensure that
one can choose the K\"ahler moduli inside the K\"ahler cone. We will
come to this in \autoref{secsusy}.

\subsection{D3-Brane Tadpole and K-Theory Constraints} 
\label{secktheo}

In this subsection, we now investigate the D3-brane tadpole
cancellation condition in some more detail. As we have seen, this
condition plays no role for the cancellation of the non-Abelian
anomalies. In absence of three-form flux the general equation \eqref{D3_tadpolezwei} is evaluated to be
\bea
N_{D3} + N_{\rm gauge} = 10,
\eea
where the contribution from the $U(1)$ fluxes on the
D7-branes 
\begin{eqnarray}
  \label{d3tadi} 
  N_{\rm gauge} &=& 
  - \frac12 \sum_i {N_i}\, \int_{D_i}  c^2_1(L_i) 
  \nonumber\\ 
  &=& 
  k\Bigl[ (N-2)\, k -N\, E_1\cdot \widetilde\eta_a -(N-4)\,
  (l-E_1-E_2)\cdot \widetilde\eta_b\Bigr] 
  \\ 
  && +m \Bigl[ (8-N)\, m -(8-N) \, (l-E_2) \cdot\eta_c \Bigr]
  \nonumber 
\end{eqnarray}
Clearly, for half-integer $k,m$ and pull-back classes
$\pi^*(\eta_{a,b,c})$, this is not always an integer. In this case we
simply cannot cancel this tadpole by introducing an integer number of
filler D3-branes. Only models with $N_\text{gauge}\in \Z$ can be
tad-pole free.

Let us now turn to the K-theory constraints. As discussed, these can
be determined by the SP-probe brane argument and cancellation of the
Witten anomaly.  In general, the identification of all potential
SP-branes is not an easy task and it is often hard to decide if one
has not missed a $\Z_2$ constraint.  As pointed out previously, since
we do not have a CFT description, it not straightforward to decide
whether a invariant four-cycle supports $SO$ or $SP$ Chan-Paton
factors.  Our strategy is to start from the branes on the O-planes,
that is, $\pi^*(E_1)$ and $\pi^*(l-E_2)$, from which we know that they
carry $SO$ Chan-Paton factors. A four-cycle with locally four
Neumann-Dirichlet boundary conditions relative to the O-planes is
expected to carry $SP$ Chan-Paton factors.  Consider the divisor
$\pi^*(l-E_1-E_2)$, which wraps the toroidal fibre and intersects
$\pi^*(E_1)$ in a point in the four-dimensional base ${\cal
  B}$. There, we expect that a brane wrapped on $\pi^*(l-E_1-E_2)$
carries $SP$ Chan-Paton factors. This identification is further
supported by the chiral spectrum in \autoref{tab_chir_modela}, where
the $a'a$ sector leads to fields in the anti-symmetric representation,
whereas the $b'b$ sector leads to symmetric ones.

This line of reasoning identifies three $4$-cycles supporting
symplectic Chan-Paton factors, namely
\begin{equation}
  \label{kcandidates}
  {\cal B}, \qquad \pi^*(E_2),\qquad  \pi^*(l-E_1-E_2)
  .
\end{equation}
All of these divisors are not {\it Spin} and, therefore, they have to
carry a half-integral bundle to comply with the Freed-Witten
quantisation condition. As usual, under the action of $\Omega$ the
field strength gets reflected, in which case the candidates in
eq.~\eqref{kcandidates} actually carry $U(N)$ Chan-Paton
factors. However, by turning on quantised B-flux through some cycles
in $H_2(Y,\Z)$, the quantisation conditions on the divisors can
change, in which case vanishing gauge flux is allowed resulting in a
non-trivial $\Z_2$ K-theory constraint.

\subsection{D-Flatness} 
\label{secsusy} 
 
According to the general discussion in \autoref{SUSY}, the FI-terms
for $U(1)_a$ and $U(1)_b$ are given by
\begin{eqnarray}
  \label{dterms} 
  \xi_a&\sim&  k (r_1-r_{\cal B}) +r_{\cal B}\, E_1\cdot \widetilde\eta_a, \nonumber \\ 
  \xi_b&\sim& k (r_l-r_1-r_2-r_{\cal B} ) +r_{\cal B}\, (l-E_1-E_2)\cdot \widetilde\eta_b,  \\ 
  \xi_c&\sim& m (r_l-r_2-2\, r_{\cal B} ) +r_{\cal B}\, (l-E_2)\cdot \eta_c \nonumber
  , 
\end{eqnarray}
where we have expanded 
\begin{equation}
  J=r_{{\cal B}}\, {\cal B} + r_l\, \pi^*(l) - r_1\, \pi^*(E_1)  - 
  r_2\, \pi^*(E_2)
  . 
\end{equation}
Since we do not want to give VEVs to Standard Model fields charged 
under $U(1)_a$ respectively $U(1)_b$, we have to require 
that the two FI-terms eq.~\eqref{dterms} vanish. 
We have to make sure that the resulting constraints define a plane 
inside the K\"ahler cone of this triangulation 
\begin{equation}
  r_{\cal B}>0,\qquad 
  r_1-{r_{\cal B}}>0, \qquad  
  r_2-{r_{\cal B}}>0, \qquad 
  r_l-r_1-r_2-r_{\cal B}>0
  . 
\end{equation}
However, we are actually interested in a slightly weaker
condition. Since the top Yukawa coupling ${\bf 10\, 10\, 5}_H$ is of
order one, the coupling-generating instanton necessarily has to be in
the non-perturbative regime. In fact, we would like to achieve
vanishing (classical) volume of the 4-cycle wrapped by the instanton,
corresponding to a particular boundary of the K\"ahler cone. Note that
world-sheet instanton corrections are expected to eventually fix the
size at the order of the string scale.  Of course, for the model to
make sense the gauge couplings of all space-time filling D7-branes
should stay finite, which can be achieved in presence of non-trivial
gauge flux.

Note that, even if one has satisfied the D7- and D5-brane tadpole cancellation
conditions guaranteeing already cancellation of non-Abelian anomalies,
the extra conditions of
\begin{itemize}
\item integer D3-brane tadpole contribution of the gauge fluxes on the
  D7-branes,
\item a number of $\Z_2$ K-theory constraints, and
\item satisfying the D-flatness conditions inside or at most on the
  boundary of the K\"ahler cone
\end{itemize}
provide further strong constraints. On the Calabi-Yau threefold
$M_2^{({\rm dP}_9)^2}$, we have not succeeded in realising a $SU(5)$
GUT model with an odd number of generations. One either does not
satisfy the first condition above or lands on an unacceptable boundary
of the K\"ahler cone in the sense that the volume of the Calabi-Yau
manifold vanishes. The best example we have found by our manual search
on $M_2^{({\rm dP}_9)^2}$ will be detailed in the next section. Having
said this, we will resolve the described problems for concrete
examples on the related manifold $M_2^{({\rm dP}_8)^2}$ in
\autoref{sec:search}.

\subsection{Globally Consistent Model} 
\label{sec_globalcons}
 
We now fix\footnote{That is, $N_a=5$, $N_b=1$, and $N_c=3$.} $N=5$,
leading to a gauge group $U(5)\times U(1)\times SO(6)$.  For an odd
number of generations we would have to consider an orientifold without
vector structure but with discrete NS-NS two-form flux $\int_F B
={1\over 2}$. As mentioned, in this case we did not succeed  in
satisfying integer D3-brane tadpole contribution of the gauge flux and
D-flatness inside the K\"ahler cone.
Therefore, we will settle for a model with an even number
of generations and choose 
\begin{equation}
  \label{bfluxbase}
  c_1(B) = \frac{1}{2} \pi^*(E_1)
  .
\end{equation}
In particular, this satisfies $\int_F B = 0$ and, therefore, $k \in
\Z$ must be integral.

Now we choose the line bundles on the three stacks of D7-branes to
be
\begin{equation}
  \label{eq:Labc}
  \begin{split}
    L_a &= \i_a^* \; 
    \OsheafY \Big( -{\cal B} -\pi^*(E_1) \Big)
    ,\\
    L_b &= \i_b^* \; 
    \OsheafY \Big( -{\cal B} \Big)
    ,\\
    L_c &= \i_c^* \; 
    \OsheafY 
    .
  \end{split}
\end{equation}
For explicitness, let us check the quantisation eq.~\eqref{quant2}. On
the $\dP_9$ surfaces, we will use the standard basis
\begin{equation}
  H_2\big( \dP_9, \Z \big) 
  =
  \Span_\Z\big\{ l, e_1, \dots, e_9 \big\}
  .
\end{equation}
Each such surface $D_a$, $D_b$ is elliptically fibred with fibre class
$f=3l-\sum e_i$ and zero-section
\begin{equation}
  {\cal B}\cap D_a = e_9 \in H_2\big(D_a,\Z\big)
  ,\quad
  {\cal B}\cap D_b = e_9 \in H_2\big(D_b,\Z\big)
  .
\end{equation}
Noting that the $B$-flux eq.~\eqref{bfluxbase} restricts trivially on
the four-cycle $D_c$, one finds that
\begin{equation}
  \renewcommand{\arraystretch}{1.5}
  \begin{array}{r@{:\qquad}c@{~-~}c@{~+~}c@{\quad}l}
    D_j& c_1(L_j) & B|_{D_j} & \frac{1}{2}c_1\big(K_{D_j}\big) &
    \in H_2\big(D_j,\Z\big)
    \\ \hline
    D_a = \pi^*(E_1)&
    -e_9+f &
    (-\frac{1}{2}f) &
    \frac{1}{2}(-f) &
    \in H_2\big(D_a,\Z\big)
    \\
    D_b = \pi^*(l-E_1-E_2)&
    -e_9 &
    \frac{1}{2}f &
    \frac{1}{2}(-f) &
    \in H_2\big(D_b,\Z\big)
    \\
    D_c = \pi^*(l-E_2)&
    0 & 0 & 0 & \in H_2\big(D_c,\Z\big)
    ,
  \end{array}
\end{equation}
and the bundles are, indeed, correctly quantised. For this choice, the
contribution to the D3-brane tadpole is
\begin{equation}
  \label{erstflux}
  N_\text{gauge}=   -\frac{1}{2} \sum_i N_i \int_{D_i} c_1^2\big(L_i\big)=
  8
\end{equation}
and we can cancel this tadpole, for example, by two dynamical D3-brane
carrying $SP(4)$ gauge group. However, according to
eq.~\eqref{d3yflux}, this cannot be the final answer as we will get an
additional contribution from the $U(1)_Y$ flux.

A D7-brane wrapping the divisor $\pi^*(l-E_1-E_2)$ now carries integer
quantised gauge flux. Therefore, the gauge bundle can be chosen to be
trivial leading to an $SP$-brane.  For such a probe brane we get a
Witten anomaly respectively expect a K-theory constraint. It is easy
to see that the above choice of branes and line bundles gives indeed a
even number of fundamental $SP$-representations.  The $B$-flux
eq.~\eqref{bfluxbase} does not restrict to the other candidate $SP$
cycles, ${\cal B}$ and $\pi^*(E_2)$ so that a trivial line bundle does
not exist. Consequently no further conditions arise.
 
The supersymmetry conditions become 
\begin{equation}
  \xi_a=r_1- {2}\, r_\sigma=0
  , \quad 
  \xi_b=r_l-r_1-r_2-  r_\sigma=0
  .
\end{equation}
The first condition can be satisfied inside the K\"ahler cone whereas
the second condition lies on the boundary of the K\"ahler cone where
the 2-cycle $C=l-E_1-E_2$ inside ${\rm dP}_2$ has zero size. As we
will explain in detail in the following, this is exactly as desired,
and there exists a D3-instanton wrapping $\pi^*(l-E_1-E_2)$ and
generating the top-Yukawa couplings. Due to the non-trivial line
bundle on the brane $D_b$, we find for the $U(1)$ gauge coupling on
this brane
\begin{equation}
  \frac{1}{g^2_{b}}
  \sim  
  -\int_{D_b} c^2_1(L_b)=O(1)
  ,
\end{equation}
which stays finite but, thankfully, leaves the perturbative regime.

\subsubsection*{Non-chiral SU(5) spectrum} 

We now move forward and compute the vector-like matter spectrum as
well. First, we turn to the GUT brane. To compute the relevant bundle
cohomology groups, we first evaluate the pull-back in the definition
of the line bundles eq.~\eqref{eq:Labc}. One obtains
\begin{equation}
  \label{eq:LabcD}
  \begin{split}
    L_a &= \i_a^* \; 
    \OsheafY \Big( -{\cal B} -\pi^*(E_1) \Big)
    = 
    \OsheafDa \Big( \big(-{\cal B} -\pi^*(E_1) \big) \cap D_a \Big)
    \\
    &=
    \OsheafDa \big( -e_9 + f \big)
    ,\\
    L_b &= \OsheafDb \big( -e_9 \big)
    ,\\
    L_c &= \OsheafDc
    .
  \end{split}
\end{equation}
For the $SU(5)$ matter, we now have to compute the cohomology of
powers of $L_a$. Using \autoref{sec:HdP9}, we easily find
\begin{equation}
  H^*\Big(\dP_9, L^\vee_a  \otimes L^\vee_a \Big)=(0,4,0)
  , \qquad 
  H^*\Big(\dP_9, L_a  \otimes L_a \Big)=(0,6,0)
  . 
\end{equation}
Thus we get two chiral and four vector-like pairs of matter fields 
in the anti-symmetric representation ${\bf 10}$ of $SU(5)$. 

Next, let us consider the states in the symmetric representation ${\bf
  1}$ of $U(1)$, that is, the right-handed neutrinos. Since the bundle
$(L_b^\vee)^2$ has sections, we get unwelcome non-trivial elements in
${\rm Ext}^0(\i_* L_b^\vee ,\i_* L_b)$ which would render the theory
inconsistent.  In order to get rid of these, we now use the freedom
to twist this pull-back bundle by a bundle $R_b$ whose push-forward is
trivial in $Y$. In particular, we pick
\begin{equation}
  R_a=\OsheafDa
  ,\quad
  R_b=\OsheafDb(-e_1+e_2\big)
  ,\quad
  R_c=\OsheafDc  
\end{equation}
and replace the bundle on $D_i$, $i=\{a,b,c\}$, with the tensor
product
\begin{equation}
  L_i
  ~\longrightarrow~
  \widetilde L_i  = L_i \otimes R_i.
\end{equation}
As explained previously, this modification does not change the chiral
matter content. However, the twisting with $R_b$ yields an extra
contribution to the D3-tadpole eq.~\eqref{erstflux}, which is now
\begin{equation}
  N_\text{gauge}=   
  -\sum_i N_i \int_{D_i} \ch_2\big(L_i\otimes R_i\big) = 9
  .
\end{equation}
The spectrum is, now, 
\begin{equation}
  H^*\Big(\dP_9, (L^\vee_b)^2 \otimes (R^\vee_b)^2
  \Big)=(0,4,0)
  , \qquad 
  H^*\Big(\dP_9, (L_a)^2 \otimes (R_b)^2 \Big)=(0,6,0)
  ,
\end{equation}
yielding two chiral and two vector-like pairs. 

Finally, consider the matter fields in the ${\bf \ov 5}$
representation, which are localised on the curve 
\begin{equation}
  D_a \cap D_b = F
  ,
\end{equation}
that is, on an elliptic fibre $\IP_{1,2,3}[6]\subset Y$.  We first
need to pull-back the bundles via the inclusion $\i_F: F\to D_a$ and
$D_b$, respectively. Note that the specific fibre $F$ is, viewed as a
curve in $D_a$ or $D_b$, in the fibre class $f$. In order to label its
intersection points with the sections $e_1$, $\dots$, $e_9$, let us
define
\begin{equation}
  \label{points}
  \begin{gathered}
    p_1 = \i_a(e_1) \cap F
    ,\quad
    p_2 = \i_a(e_2) \cap F
    ,\quad\dots,\quad
    p_8 = \i_a(e_8) \cap F
    ,\\
    p_1' = \i_b(e_1) \cap F
    ,\quad
    p_2' = \i_b(e_2) \cap F
    ,\quad\dots,\quad
    p_8' = \i_b(e_8) \cap F
    ,\\
    0 = \i_a(e_9) \cap F = \i_b(e_9) \cap F = {\cal B} \cap F
    .
  \end{gathered}
\end{equation}
Here we have implicitly fixed some of the complex structure moduli of the manifold such that indeed the sections of $D_a$ and $D_b$ intersect $F$ in the same points.
Using this notation, we obtain
\begin{equation}
  \label{eq:LabF}
  \begin{split}
    L_a|_F 
    &= \i_F^* L_a 
    = \OsheafF\Big( \big(-e_9+f\big) \cap F \Big) 
    = \OsheafF\big( -0 \big)
    ,\\
    \big(L_b\otimes R_b\big)|_F 
    &= \OsheafF(-0-p_1'+p_2'\big)
    .
  \end{split}
\end{equation}
With $K_F=\OsheafF$ we can easily compute the ${\bf \ov 5}$-spectrum,
and obtain
\begin{equation}
  \begin{split}
    H^*\Big( F, 
    \i_F^*\big( L_a^\vee\otimes (L_b\otimes R_b)^\vee \big) \Big)
    &= 
    H^*\Big( F, \OsheafF(0+0+p_1'-p_2') \Big)
    \\ &= 
    H^*\Big( F, \OsheafF(2 \text{pts.}) \Big)
    =
    (2,0)
    .
  \end{split}
\end{equation}
Hence, we get precisely $2$ chiral fields in the anti-fundamental
representation. The Higgs ${\bf 5}_H+{\bf \ov 5}_H$-spectrum, on the other hand, is
determined by 
\begin{equation}
  H^*\Big( F, \i_F^*\big( L_a^\vee\otimes (L_b\otimes R_b) \big) \Big)
  = 
  H^*\Big( F, \OsheafF(-p_1'+p_2') \Big)
  =
  (0,0)
  .
\end{equation}
One might worry that there is no candidate ${\bf 5}_H-{\bf \ov 5}_H$ pair
giving rise to the Higgs after symmetry breaking; However, as we will
see in the following, turning on a suitable $L_Y$ flux will generate
one vector-like pair as desired.

\subsubsection*{Instanton effects} 

A Euclidean D3-brane wrapping the divisor $\pi^*(l-E_1-E_2)$ with
trivial line bundle is of $O(1)$ type and clearly rigid. Therefore,
this instanton is a candidate to generate the ${\bf 10\, 10\, 5_H}$
Yukawa couplings. For the chiral charged matter zero modes, we indeed
get $I_{a,\text{inst}}=1$ and $I_{b,\text{inst}}=-1$. Therefore, the
necessary condition eq.~\eqref{instzerocond} for the generation of the
top Yukawa coupling is satisfied. Moreover, the line bundle on the
divisor $D_b=\pi^*(l-E_1-E_2)$ has first Chern class $c_1(L_b\otimes
R_b)=-e_9 -e_1+e_2$, and therefore
\begin{equation}
  H^*\big(D_b, L_b \otimes R_b  \big)=(0,1,0)
  , \quad 
  H^*\big(D_b, L^\vee_b \otimes R^\vee_b \big)=(0,0,0)
  .
\end{equation}
This implies $\Ext^*(L_b\otimes R_b,\OsheafDb)=(0,1,0,0)$ and shows
that there exists precisely one chiral zero mode $\lambda_b$ without
any additional vector-like pairs.  Since the instanton intersects the
brane $D_a$ over the fibre curve, we find exactly one chiral zero mode
$\lambda_a$.  Moreover, the D-term constraint for the brane $D_b$ has
fixed the K\"ahler moduli such that the instanton action goes to zero;
it follows that leaving (locally) the perturbative regime, the top
Yukawa couplings are really of order one.

To generate Majorana neutrino masses, one needs an $O(1)$ instanton
intersecting only the brane stack $D_b$. A candidate would be a
Euclidean D3-brane wrapping the divisor $\pi^*(E_2)$. However, since
this divisor is not {\it Spin}, it is not of $O(1)$ but $U(1)$ type.

\subsubsection*{Non-chiral $\mathbf{SU(3)\times SU(2)\times U(1)_Y}$ spectrum} 

Finally, let us break the $SU(5)$ gauge symmetry to the Standard
Model. To do so, we will turn on $U(1)_Y$ gauge flux supported on a
curve in $H_2(D_a,\Z)$ which is a boundary on $Y$. In particular, we
pick
\begin{equation}
  \cL_a = L_a
  ,\quad  
  \cL_Y= \OsheafDa(-e_1+e_2)
  . 
\end{equation}
The $\cL_Y$ bundle has vanishing cohomology classes on $D_a={\rm
  dP}_9$, and, therefore, no exotics are introduced.  The contribution
to the D3-tadpole from this flux is 
\begin{equation}
  N^Y_\text{gauge} = -\int_{D_a} c_1^2( \cL_Y ) = 2.
\end{equation}
The combined gauge flux contribution to the D3 tadpole is $N_{\rm
  gauge}=9+2=11$, overshooting by one unit. Therefore, to cancel this
tadpole, one needs to introduce one dynamical anti--D3-brane.
 
The relevant cohomology classes for the descendants of the
antisymmetric representation ${\bf 10}$ of $SU(5)$ are listed in
\autoref{tab_nonchiral}.
\begin{table}[htbp] 
  \renewcommand{\arraystretch}{1.5} 
  \begin{center} 
    \begin{tabular}{|c|c||c@{=}c|c|c|} 
      \hline 
      \hline 
      GUT &
      SM Field & 
      \multicolumn{2}{|c|}{Cohomology} & 
      chiral & 
      vector \\ 
      \hline \hline 
      \multirow{6}{*}{${\bf 10}$} &
      \multirow{2}{*}{$({\bf \ov 3},{\bf 1})_{-4_Y}$} 
      & $H^*(D_a, (\cL^\vee_a)^2 )$ & $(0,4,0)$ &
      \multirow{2}{*}{2} & 
      \multirow{2}{*}{4+4} \\
      && $H^*(D_a, \cL_a^2 )$ & $(0,6,0)$ && \\
      \cline{2-6}
      & \multirow{2}{*}{$({\bf 3},{\bf 2})_{1_Y}$} 
      & $H^*(D_a, (\cL^\vee_a)^2\otimes  \cL_Y^{-1})$ & $(0,5,0)$ &
      \multirow{2}{*}{2} & 
      \multirow{2}{*}{5+5} \\
      && $H^*(D_a, \cL_a^2\otimes  \cL_Y)$ & $(0,7,0)$ && \\
      \cline{2-6}
      & \multirow{2}{*}{$({\bf 1},{\bf 1})_{6_Y}$} 
      & $H^*(D_a, (\cL^\vee_a)^2\otimes  \cL^{-2}_Y)$ & $(0,8,0)$ &
      \multirow{2}{*}{2} & 
      \multirow{2}{*}{8+8} \\
      && $H^*(D_a, \cL_a^2\otimes  \cL_Y^2 )$ & $(0,10,0)$ && \\ 
      \hline
      \multirow{2}{*}{${\bf 1}$} &
      \multirow{2}{*}{$({\bf 1},{\bf 1})_{0_Y}$} 
      & $H^*(D_b, (L_b^\vee\otimes R_b^\vee)^2 )$ & $(0,4,0)$ &
      \multirow{2}{*}{2} & 
      \multirow{2}{*}{4+4} \\
      && $H^*(D_b, L_b^2\otimes R_b^2 )$ & $(0,6,0)$ && \\
      \hline
      \multirow{2}{*}{${\bf \ov 5}$} 
      & $({\bf \ov 3},{\bf 1})_{2_Y}$
      & eq.~\eqref{eq:CohLaLbRb}
      & $(2,0)$ & 2 & 0 
      \\
      & $({\bf 1},{\bf 2})_{-3_Y}$
      & eq.~\eqref{eq:CohLaLbRbLy}
      & $(2,0)$ & 2 & 0
      \\
      \hline
      \multirow{2}{*}{${\bf 5}_H+{\bf \ov 5}_H$} 
      & $({\bf 3},{\bf 1})_{-2_Y}$
      & $H^*( F, \OsheafF(-p_1'+p_2') )$
      & $(0,0)$ & 0 & 0 
      \\
      & $({\bf 1},{\bf 2})_{3_Y}$
      & $H^*( F, \OsheafF )$
      & $(1,1)$ & 0 & 1+1
      \\
      \hline 
      \hline 
    \end{tabular} 
  \end{center} 
  \caption{Spectrum for the orientifold model in
    \autoref{sec:GutModelExample}. The indices denote 
    the $U(1)$ charges.} 
  \label{tab_nonchiral} 
\end{table} 
Note that, of course, the chiral matter does not change; only extra
vector-like pairs of matter fields appear. We now turn toward the
matter fields in the anti-fundamental representation ${\bf \ov 5}$ as
well as the Higgs, both of which are localised on the intersection
curve $F$. In addition to eq.~\eqref{eq:LabF}, we have
\begin{equation}
  \i_F^* \cL_Y = \cL_Y|_F =
  \OsheafF\big( -p_1+p_2 \big)
  .
\end{equation}
First, note that the ${\bf \ov 5}$-spectrum is
unchanged since
\begin{subequations}
  \begin{align}
    \label{eq:CohLaLbRb}
    H^*\Big( F, 
    \i_F^*\big( \cL_a^\vee\otimes L_b^\vee \otimes R_b^\vee \big) \Big)
    =&
    H^*\Big( \OsheafF(0+0+p_1'-p_2') \Big)
    =
    (2,0)
    ,\\
    \label{eq:CohLaLbRbLy}
    H^*\Big( F, 
    \i_F^*\big( \cL_a^\vee\otimes L_b^\vee \otimes R_b^\vee \otimes
    \cL_Y^\vee \big) \Big)
    =&
    H^*\Big(\OsheafF(2\cdot 0+ p_1+p_1'-p_2-p_2') \Big)
    =
    (2,0)    
    .
  \end{align}
\end{subequations}
More interesting is the Higgs spectrum, which is determined by
\begin{equation}
  \i_F^*\big( \cL_a^\vee\otimes (L_b\otimes R_b) 
  \otimes \cL_Y^\vee \big)
  =
  \OsheafF(p_1-p_1'+p_2-p_2')
  =
  \OsheafF(0-q)
\end{equation}
for some point $q\in F$. The precise point can be computed using the
group law on the elliptic curve $F$, and will depend on the complex
structure of $D_a$, $D_b$ (and, therefore, $Y$). We assume that $q=0$,
which happens on a locus of codimension one in the complex structure
moduli space; See also the discussion around eq.~\eqref{points}. In
this case,
\begin{equation}
  H^*\Big( F, \i_F^*\big( \cL_a^\vee\otimes (L_b\otimes R_b) 
  \otimes \cL_Y^\vee \big) \Big)
  = 
  H^*\Big( F, \OsheafF \Big)
  =
  (1,1)
  .
\end{equation}
As desired, we then obtain one pair of Higgs-conjugate Higgs
fields. Moreover, the Higgs doublet is, in fact, split from the
dangerous colour triplet. The latter is still absent, thanks to
\begin{equation}
  H^*\Big( F, \i_F^*\big( \cL_a^\vee\otimes (L_b\otimes R_b) \big) \Big)
  = 
  H^*\Big( F, \OsheafF(-p_1'+p_2') \Big)
  =
  (0,0)
  .
\end{equation}
 
To summarise, we have defined a simple involution on $M_2^{({\rm
    dP}_9)^2}$ which allows for the introduction of a three-stack
intersecting D7-brane configuration cancelling the D7-, D5- and
D3-brane tadpoles, the latter at the cost of introducing one anti
D3-brane, as well as the K-theory tadpoles.  We have found an $SU(5)$
GUT-like model with two chiral generations of Standard Model particles
and one Higgs-conjugate Higgs pair. Moreover, we have been able to
realise the $U(1)_Y$ flux gauge symmetry breaking and computed the
resulting chiral and non-chiral matter spectrum.  The inevitable
appearance of the latter is one of the shortcomings of this example.
It can be traced back to the fact that our involution acts trivially
on the cohomology $H^2(Y, \Z)$.  As a consequence, antisymmetric
matter is localised not on a curve but on a whole divisor, widening
the sources of contributions to the cohomology.  By contrast, matter
and Higgs are localised on the elliptic fibre of $Y$. Since the Higgs
$H_u$ and $H_d$ are localised on the same curve, we cannot suppress
dimension-five proton decay operators. The D-term supersymmetry
conditions can be satisfied on the boundary of K\"ahler moduli space
such that a D3-brane instanton realises ${\bf 10\, 10\, 5}$ Yukawa
couplings of order $1$. In \autoref{tab_gutmodel1} we summarise
phenomenologically desirable features of this simple model.
\begin{table}[htbp] 
  \renewcommand{\arraystretch}{1.5} 
  \begin{center} 
    \begin{tabular}{|c||c|c|} 
      \hline 
      \hline 
      property & mechanism & status    \\ 
      \hline \hline 
      globally consistent & tadpoles + K-theory & $\checkmark^{*, **}$ \\
      D-term susy & vanishing FI-terms inside K\"ahler cone & $\checkmark^{***}$ \\
      gauge group $SU(5)$ & $U(5)\times U(1)$ stacks & $\checkmark$ \\
      3 chiral generations & choice of line bundles & $-$ \\
      no vector-like matter  & localisation on  curves & $-$ \\
      1 vector-like of  Higgs  & choice of line bundles & $\checkmark^{****}$ \\
      no adjoints & rigid 4-cycles, del Pezzo & $\checkmark$ \\
      GUT breaking   & $U(1)_Y$ flux on trivial 2-cycles  & $\checkmark$ \\
      3-2 splitting & Wilson lines on  $g=1$ curve   &
      $\checkmark$\\
      3-2 split + no dim=5 p-decay  & local. of $H_u, H_d$  on disjoint
      comp.   & $-$\\
      ${\bf 10\, \ov 5\, \ov 5}_H$ Yukawa & perturbative
      & $\checkmark$ \\
      ${\bf 10\, 10\, 5}_H$ Yukawa & presence of appropriate D3-instanton
      & $\checkmark$ \\
      Majorana neutrino masses & presence of appropriate D3-instanton
      & $-$$^{*****}$  \\
      \hline 
      \hline 
    \end{tabular} 
    \caption{Summary of $SU(5)$ properties realised 
      in the model of \autoref{sec:GutModelExample}. \newline 
      $^*$\ {\small overshooting in D3-tadpole $\to$ 1 $\overline{D3}$ brane}\newline
      $^{**}$\ {\small K-theory  to the best of our ability to detect $SP$ cycles} \newline
      $^{***}$\ {\small realised on acceptable boundary of K\"ahler moduli space}  \newline
      $^{****}$\ {\small for special choice of complex structure moduli}  \newline
      $^{*****}$  {\small at least not with O(1) instantons}   } 
    \label{tab_gutmodel1}
  \end{center} 
\end{table}


\section{GUT Model Search}
\label{sec:search}

The model presented in the previous section does not exhibit all
properties desirable for a nice string GUT model. However, as for each
single shortcoming it is quite clear how to improve on this.  The
non-trivial task is to achieve this in a globally consistent framework
and without loosing the good features already realised. To this end a
more systematic search is necessary and beyond the scope of this
paper.  In this section we provide a couple of manually found models
which incorporate some other desirable properties from
\autoref{tab_gutfeatures}, but come short on already realised
ones. The two features we focus on in this section are 3 chiral
families and the absence of vector-like matter fields.

\subsection
[A 3-Generation GUT Model on $M_{2}^{({\rm dP}_8)^2}$]
[A 3-Generation GUT Model]
{A 3-Generation GUT Model on $\mathbf{M_{2}^{({\rm dP}_8)^2}}$}
\label{sec:GutModelM2dP8}

We now present an example of a GUT model of the type described
previously which indeed gives rise to 3 chiral families of Standard
Model matter.  To this end we consider the manifold $M_2^{({\rm dP}_8)^2}$
introduced in \autoref{sec:dP2fibration}.  Since the intersection form
eq.~\eqref{tripleint2} differs considerably from the one of the
un-flopped Weierstra\ss{} phase, eq.~\eqref{tripleint1}, it seems
plausible that the no-go result for an odd number of generations can
be evaded.

We closely follow the philosophy spelt out in
\autoref{sec:GutModelExample} so that we can be brief. Concretely,
consider again a 3-stack model based on the divisors $D_a=D_7$,
$D_b=D_5$ and $D_c=D_5 + D_7$, wrapped by D7-branes with
multiplicities
\begin{equation}
  N \times D_7
  , \qquad 
  (N-4) \times D_5
  , \qquad
  (8-N) \times (D_5+D_7)
  ,
\end{equation}
plus their orientifold images, with $N=5$ corresponding to the GUT model we are interested in. The
above configuration satisfies the $D7$-brane tadpole constraint for
any $N$.

Let us first define the chiral $SU(5)$ GUT model by parametrising the part of line bundles $L_a$, $L_b$ and $L_c$ descending from the Calabi-Yau $Y$ as
\bea
&& L_a= \i_a^* \OsheafY \Big(a_1 D_5 + a_3 D_7\Big),  \quad  L_b= \i_b^* \OsheafY \Big(\sum_{i=1}^4 b_i D_{i+4}\Big),  \\\nonumber
&& L_c= \i_c^* \OsheafY \Big(c_1 D_{1} + c_2 D_{2} + c_4 D_{4} \Big).
\eea
For the correct definition of the
bundles it is essential to take into account that this time all three
divisors are not $Spin$.  According to the quantisation condition
eq.~\eqref{quant1} for vanishing B-field the parameters $b_1, a_3,
c_1$ are therefore half-integer while $a_1,b_2,b_3,b_4,c_2,c_4$
are integer.

As can be computed from the intersection form eq.~\eqref{tripleint2},
the divisors $D_a$ and $D_b$ still intersect along a genus 1 curve
where the $\ov \bf 5$ and the ${\bf 5}_H + { \bf \ov 5}_H$ are
localised.  Note also that the divisors $D_c$ and $D_a$ do not intersect, while $D_c$ and $D_b$ now intersect along a genus $0$ curve. Thus
there exists no massless exotic matter in the $D_a-D_c$ sector, even
at the vector-like level, while there might appear truly hidden sector
matter fields from the $D_b-D_c$ intersection.  This chiral matter is
displayed in \autoref{tab_chir_modelB}. Clearly all non-Abelian
anomalies vanish due to cancellation of D7- and absence of D5-brane
tadpoles. Note that for general bundles there exist both symmetric and
anti-symmetric states under $U(N-4)$.  In what follows we specialise
to the case $N=5$ corresponding to a GUT model with the first four
lines representing the ${\bf 10}$, ${\bf \ov 5}$, ${\bf 5_H} + {\bf
  \ov 5_H}$ and $N_R^c$.
\begin{table}[htbp] 
\renewcommand{\arraystretch}{1.5} 
\begin{center} 
\begin{tabular}{|c||c|c|c|} 
\hline 
\hline 
chirality & $U(N)$ & $U(N-4)$ & $U(8-N)$  \\ 
\hline \hline 
$-2a_1+2a_3$ & $\Yasymm_{\,(2)}$ & $1$ & $1$ \\ 
$-(a_1+b_1) + a_3 +b_3$                         &  $\antifund_{\, (-1)}$ & $\antifund_{\, (-1)}$  & $1$ \\ 
$-(a_1-b_1) + (a_3-b_3)$ & $\antifund_{\, (-1)}$ & $\fund_{\, (1)}$  & $1$ \\ 
$2 \left(  -b_1 +  b_3  \right)$ & $1$   & $\Ysymm_{\,(2)}$ &  $1$ \\ 
$2 \left(  -b_1 +  b_2 +  b_4  \right)$ & $1$   & $\Yasymm_{\,(2)}$ &  $1$ \\ 
\hline
\hline
$(b_1-c_1) - (b_2-c_2) - (b_4-c_4) $ & $1$ &    $\antifund_{\, (-1)}$ & $\fund_{\, (1)}$  \\
$ (b_1+c_1) -( b_2+c_2) - (b_4+c_4)       $ & $1$ &    $\antifund_{\, (-1)}$ & $\antifund_{\, (-1)}$  \\
$2(-c_1 + c_2 + c_4)$  &  $1$ & $1$ & $\Yasymm_{\,(2)}$  \\ 
\hline 
\end{tabular} 
\caption{Chiral spectrum for intersecting D7-brane model. The indices denote 
 the $U(1)$ charges.} 
\label{tab_chir_modelB} 
\end{center} 
\end{table}

\subsubsection*{Global consistency conditions}

The D3-brane tadpole condition is
\begin{equation}
  N_{D3} + N_\text{gauge}  = 10
  ,
\end{equation}
with
$N_{\rm{gauge}}$ given by 
\bea
-\frac{1}{2} \, \sum_a N_a \, c^2_1(L_a) &=& -\frac{5}{2} \left( a_1^2 + a_3^2 - 2a_1\,a_3\right ) -\frac{3}{2} \left(-c_1^2 + 2 c_1 c_2 - 3 c_2^2 + 2 c_1 c_4 - c_4^2 \right)    \nonumber \\
&-& \frac{1}{2}  \left(-2 b_1^2 + 2 b_1 b_2 - 3 b_2^2 + 2 b_1 b_3 - b_3^2 + 2 b_1 b_4 - b_4^2 \right) .
\eea
\noindent Note, however, that there will be additional contributions later on from the part of the line bundles trivial on the ambient space $Y$.
As in the previous section, due to the simple structure of the orientifold action,
the D5-brane tadpoles cancel automatically between the branes and their image.

On the other hand, there can arise K-theory constraints from the 3 invariant divisors $D_6$, $D_8$ and $D_5$ which may carry symplectic Chan-Paton factors.
The divisor $D_6$ is the former basis ${\rm dP}_2$ of the Weierstra\ss~model $M_{2}^{({\rm dP}_9)^2}$  with the two ${\mathbb P}^1$s
removed. As such it is the surface ${\mathbb P}^2$ with $K_{D6} = {\cal O}_{D6}(3)$ and obviously not {\it Spin}. 
This means that, according to eq.~\eqref{quant1},
the chiral part of a line bundle on $D_6$ 
\bea
L_{D_6}= \i^* \OsheafY \Big(x_1 D_5 + x_2 D_6\Big)
\eea
can be trivial only for $B = 0 \times D_5 + \frac{1}{2} D_6 + \ldots$. 
Otherwise the cycle does not carry symplectic gauge factors and therefore no K-theory constraint arises from $D_6$.

The K-theory constraint from $D_6$ therefore reads
\bea
b_1 + 3c_1 - 3(b_2 + 3c_2) \in
    2\mathbb Z  \quad {\rm if}   \quad   B= 0 \times D_5 + \frac{1}{2}  D_6.
\eea
A similar analysis for  $D_8$ and $D_5$ yields the two additional constraints
\bea
&& b_1 + 3c_1 - (b_4 + 3c_4) \in  2\mathbb Z \quad {\rm if} \quad   B= 0 \times D_5 + \frac{1}{2} D_8, \nonumber \\
&& 5a_1 - 5a_3 - 2(b_1+3c_1) + (b_2+3c_2) + b_3 + (b_4+3c_4) + 3 c_1 \in   2\mathbb Z  \\ 
&& \phantom{aaaaaaaaaaaaaaaaa}  \quad {\rm if}   \quad   B= \frac{1}{2} D_5 + 0 \times (D_6+D_7+D_8).  \nonumber 
\eea
In all other cases the K-theory constraints from these three divisors are trivial.

\subsubsection*{D-term supersymmetry constraints}

To determine the D-term supersymmetry conditions we have to expand the K\"ahler form $J$ in terms of the generators of the full K\"ahler cone and evaluate the Fayet-Iliopoulos terms.
For our purposes it will be sufficient to restrict our attention to the K\"ahler subcone corresponding to the vectors $K_i$ displayed in eq.~\eqref{K5TrianB} and take
\bea
J  = \sum_i r_i K_i, \quad\quad\quad r_i \in {\mathbb R}_+.
\eea
As discussed around eq.~\eqref{K5TrianB}, it suffices to check if the associated FI-terms
\bea
\xi_a &\simeq&  (a_1-a_3) r_1, \nonumber \\
\xi_b &\simeq&  b_2 r_4 + b_3 r_1 - b_1 r_2 + b_4 r_2 + b_1 r_3, \\
\xi_c &\simeq&  c_2 r_4 + c_4 r_2 + c_1 (r_1 - r_2 + r_3) \nonumber 
\eea
vanish for some values of $r_i>0$ for which in addition $r_1 - r_2 + r_3 >0$. 

\subsubsection*{ A 3-generation GUT model}

As a quick search reveals it is indeed possible to find globally consistent supersymmetric models with 3 chiral generations of SU(5) GUT matter.
As one example out of the {\cal O}(100) models we found we present the configuration with non-vanishing $B$-field
\bea
\label{choiceB2}
B= \frac{1}{2}D_5 + \frac{1}{2}D_7  
\eea
and line bundles
\bea
L_a&=& \i_a^* \OsheafY \Big( -\frac{7}{2} D_5 - 2 D_7 \Big), \nonumber \\
L_b&=& \i_b^* \OsheafY \Big(  - D_5 -D_6 + \frac{1}{2} D_7   \Big), \\
L_c&=& \i_c^* \OsheafY \Big( -D_5    \Big). \nonumber
\eea
According to \autoref{tab_chir_modelB} this choice yields precisely 3 chiral GUT families of ${\bf 10}$, ${\bf \ov 5}$ and $N_R^c$ with no chiral exotics. In addition there is chiral hidden matter as summarised in \autoref{tab_chir_modelB_conc}. 
\begin{table}[htbp] 
\renewcommand{\arraystretch}{1.5} 
\begin{center} 
\begin{tabular}{|c||c|c|c|} 
\hline 
\hline 
chirality & $U(5)$ & $U(1)$ & $U(3)$  \\ 
\hline \hline 
$3 $  & $\Yasymm_{\,(2)}$ & $1$ & $1$ \\ 
$3$                         &  $\antifund_{\, (-1)}$ & $\antifund_{\, (-1)}$  & $1$ \\ 
$3$ & $1$   & $\Ysymm_{\,(2)}$ &  $1$ \\ 
\hline
\hline
$1$  &  $1$ &    $\antifund_{\, (-1)}$ & $\fund_{\, (1)}$  \\
$1 $  &  $1$ &    $\fund_{\, (1)}$ & $\fund_{\, (1)}$  \\
$2 $  &  $1$ & $1$ & $\Yasymm_{\,(2)}$  \\ 
\hline 
\end{tabular} 
\caption{Chiral spectrum for intersecting D7-brane model with indices denoting 
 the $U(1)$ charges. The last three lines are completely hidden chiral matter.} 
\label{tab_chir_modelB_conc}
\end{center} 
\end{table}

The contributions to the D3-brane tadpole of this GUT model is
\bea
     N_{\rm gauge}=-{5\over 2}\cdot{9\over 4} + {1\over 2}\cdot {17\over 4} + {3 \over 2} \cdot {4 \over 4} =-2
\eea
so that at this stage we would need  to add $N_{D3}=12$ dynamical D3-branes.
Note that it is  suspicious that the contribution of the
gauge flux on the $SU(5)$ brane to the D3-brane tadpole is negative. Indeed, we will see
in a moment that this bundle leads to ghosts. We will avoid this conclusion by twisting it by an additional  
bundle $R_a$ which is trivial on $Y$.

For the above choice of B-field the K-theory constraints from $D_5,
D_6, D_8$ are vacuous.  As expected from the general consideration in
\autoref{linebundles}, the D-term constraint for $D_a$ drives us to
the boundary of K\"ahler moduli space in that it requires $r_1=0.$ The
general solution of the three D-term equations for the K\"ahler moduli
is
\bea
r_1=0, \quad r_2 =x, \quad  r_3=x,\quad  r_4=0.
\eea
For positive $x$ this solution  lies on the boundary of K\"ahler moduli space in that 
besides $r_1=0$ and $r_4=0$ also the volume of the generator $C^5$ of the Mori cone vanishes. In this regime
the classical volume of the divisor  $D_7$ vanishes, while all other brane volumes and the total volume of the Calabi-Yau are positive.
Note that for this model the classical  value of the GUT gauge coupling
is  $\alpha^{-1}_{GUT} \simeq -c_1^2(L_a)<0$, which we have just seen to be negative.
This is another indication that the  model is pathological in its present form and will be rectified momentarily by twisting $L_a$ further.

\subsubsection*{GUT breaking and $\mathbf{SU(3)\times SU(2)\times U(1)}$ spectrum}

To break the GUT symmetry to $SU(3)\times SU(2)\times U(1)_Y$ and to compute the vector-like MSSM spectrum we need 
the explicit pushforward and pullback maps $\i_*$ and $\i^*$ between the second (co)homology of $D_7$ and the ambient space $Y$.
Recall that $D_7$ is a ${\rm dP}_8$ surface with $H_2(D_7,\mathbb Z)$ spanned by $h,e_1,\ldots e_8$.  The pushforward $\i_*: H_2(D_7, {\mathbb Z})  \rightarrow H_2(Y, {\mathbb Z})$ 
follows immediately
once one takes into account that relative to $\pi^*(E_1)={\rm dP}_9$ 
the curve ${\cal B} E_1$ is flopped away in the present phase. Explicitly,
\bea 
\label{pushfor2} 
\i_*(e_i) = f,  \quad i=1, \ldots 8, \qquad
\i_*(h) = 3f,  
\eea 
where $-f \in H_2(Y, \mathbb Z)$ now denotes the class of the curve $D_7 \cap D_7$ in $Y$. This in turn follows from
$\i^*D_7 = K_{D_7} = -3h + \sum e_i$ together with the identity $\i_* \i^* = {\rm 1}$.
Finally one completes the pullback map to
\bea
\label{pullback2}
\i^*D_7= -3h +  \sum_{i=1}^8 e_i, \qquad \i^*D_5= 3h -  \sum_{i=1}^8 e_i
\eea
and all others vanishing.
Therefore, one finds for the pullback of ${\rm c}_1(L_a)$ to $D_7$
\bea
{\rm c}_1(L_a) = -\frac{3}{2} \left(3h - \sum_{i=1}^8 e_i \right).
\eea
For this bundle we can now compute
\bea 
      H^*\bigl({\rm dP}_8, L^\vee_a \otimes L^\vee_a   \bigr)=(7,0,0), \qquad 
      H^*\bigl({\rm dP}_8,  L_a \otimes L_a \bigr)=(0,0,4)\; ,
\eea
which implies  ${\rm Ext}^*(\i_* L^\vee_a,\i_* L_a)=(7,4,0,0)$.
Therefore, this line bundle on the $SU(5)$ stack leads to
ghosts in the spectrum. Since the K\"ahler form is on the boundary
of the K\"ahler cone this is not in contradiction with
the no-ghost theorem from \autoref{SUSY}.
However, we still have the freedom to tensor $L_a$ with a line bundle  $R_a$
 which is trivial on the ambient space $Y$.
This does not change the chiral spectrum but the non-chiral one.

This freedom can be used to  choose the bundles ${\cal L}_a = L_a \otimes R_a$ and ${\cal L }_Y$ as 
\bea
\label{LaYsec61}
{\rm c}_1({\cal L}_a) =  \frac{1}{2} \left(-h + \sum_{i=1}^4 e_i - \sum_{i=5}^8 e_i \right), \qquad {\rm c}_1({\cal L}_Y) = -e_1 + e_5.
\eea
This configuration leads to the multiplicities displayed in \autoref{tab_nonchiralB} for the decomposition of the ${\bf 10}$ into MSSM states. For its computation see \autoref{app_dPr}. Note the appearance of only two extra vector-like states.
\begin{table}[htbp] 
\renewcommand{\arraystretch}{1.5} 
\begin{center} 
\begin{tabular}{|c||c|c|} 
\hline 
\hline 
repr. & extension & spectrum    \\ 
\hline \hline 
$(\ov {\bf 3},{\bf 1})_{-4_Y}$ &  ${\rm Ext}^*( {\cal L}_a^{\vee}, {\cal L}_a  )$ & $(0,1,4,0)$ \\ 
 \hline 
$({\bf 3},{\bf 2})_{1_Y}$ &  ${\rm Ext}^*( {\cal L}^{\vee}_a \otimes {\cal L}_Y^{-1}, {\cal L}_a  )$ &$(0,0,3,0)$ \\ 
\hline 
$({\bf 1},{\bf 1})_{6_Y}$ & ${\rm Ext}^*( {\cal L}^{\vee}_a \otimes {\cal L}_Y^{-2}, {\cal L}_a  )$ & $(0,1,4,0)$ \\ 
 \hline 
\hline 
\end{tabular} 
\caption{Multiplicities of MSSM descendants from the SU(5) ${\bf 10}$.}  
\label{tab_nonchiralB} 
\end{center} 
\end{table}

To compute the MSSM descendants of the ${\bf \ov 5}$ we recall that the divisors $D_7$ and $D_5$ intersect again along a genus 1 curve $F$ with $c_1({\cal L}_Y|_F) = 0$. 
We will make use of our freedom to twist also $L_b$ by a line bundle $R_b$ which is trivial on the Calabi-Yau Y and define
\bea
{\cal L}_b = L_b \otimes R_b.
\eea
Since $R_b$ is trivial on $Y$ this does not change the chiral spectrum.
Irrespective of its form we  find $c_1\Big( {\cal L}^{\vee}_a \otimes {\cal L}^{\vee}_b|_{D_5 \cap D_7}\Big) = 3$, leading to  precisely $3$ multiplets of
$({\bf \ov 3},{\bf 1})_{2Y}$ and $({\bf 1},{\bf 2})_{-3Y}$ and no extra vector-like states.

On the other hand we will choose $R_b$ such that the Wilson lines for the bundles ${\cal L}^{\vee}_a \otimes {\cal L}_b|_{D_5 \cap D_7}$ and 
${\cal L}^{\vee}_a \otimes {\cal L}_Y^{\vee} \otimes {\cal L}_b|_{D_5 \cap D_7}$ give rise to precisely one pair of Higgs doublets and no Higgs triplet.
To this end we recall that $D_5$ is a ${\mathbb P}^2$ surface with 11 points blown up to a ${\mathbb P}^1$, and in analogy with the notation for del Pezzo surfaces $H^2(D_5,{\mathbb Z})$ is spanned by ${h, e_1,\ldots e_9,E_{10}, E_{11}}$´. Here $E_{10}$ and $E_{11}$ denote the extra two ${\mathbb P}^1$ which have been flopped into $D_5$ in the present phase. 
For a special choice of complex structure moduli  the elliptic curve ${D_5 \cap D_7}$ intersects $h, e_1, \ldots e_8$ in the same points  as the classes\footnote{This is actually more than we need since we only have to ensure that the sum of the Wilson lines add up to zero to engineer one Higgs pair.} $h, e_1, \ldots e_8$ in $H^2(D_7)$.
It is then clear that, given the choice eq.~\eqref{LaYsec61} for ${\cal L}_a$ and ${\cal L}_Y$, we have to pick
\bea
R_b = {\cal O}(4h - 2 e_1 - e_2 - e_3 -e_4 - e_5 - 2 e_6 - 2 e_7 - 2 e_8)
\eea
on $D_5$ to comply with the requirement stated in equation~\eqref{Wilsonlines_General}. 
 
The twist with the bundles $R_a$, $R_b$ and the addition of the bundle ${\cal L}_Y$ 
change the overall contributions
of the gauge fluxes to the D3-brane tadpole. These now read
\bea
   N_{\rm gauge}={5\over 2}\cdot{7\over 4} + {1\over 2}\cdot \Big( {17\over 4}+4 \Big) + {3 \over 2}  =10.
\eea
This time $N_{\rm gauge}$ precisely equals the D3-brane charge of the orientifold planes, and cancellation of the full D3-brane tadpole is possible in a supersymmetric manner without introducing further ${D3}$-branes.
Moreover, note that despite the vanishing volume  of the
GUT divisor $D_7={\rm dP}_8$, the gauge coupling now comes out positive.

Finally, one might wonder about the existence of ghosts on the $U(1)$ and $U(3)$ branes since we have not been able to satisfy all three D-term supersymmetry conditions inside the K\"ahler cone. On the other hand, one can convince oneself that it is possible to satisfy the supersymmetry conditions inside the K\"ahler cone for each of the line bundles $L_b$ and $L_c$ separately. This is already enough for our lemma in \autoref{SUSY} to guarantee absence of states in ${\rm Ext}^0$ and ${\rm Ext}^3$.

\subsubsection*{Instanton effects}

The existence of a model with three chiral generations of MSSM matter
rested upon the choice eq.~\eqref{choiceB2} for the B-field. The
downside of this choice is that none of the invariant divisors $D_5,
D_6, D_8$ allows for trivial line bundles. As a consequence there
exist no divisors that would give rise to symplectic gauge groups for
spacetime-filling branes, and thus no $O(1)$ instantons. To decide if
neutrino Majorana masses or the ${\bf 10 \, 10 \, 5_H}$ coupling are
generated non-perturbatively we would therefore have to study the
effects of D3-brane instantons wrapping non-invariant cycles, for
example along the lines of~\cite{Blumenhagen:2007bn}. This, however,
is beyond the scope of the present work.

We conclude this section by summarising the key phenomenological
properties of our model in \autoref{tab_gutmodel2}.
\begin{table}[htbp] 
\renewcommand{\arraystretch}{1.5} 
\begin{center} 
\begin{tabular}{|c||c|c|} 
  \hline 
  \hline 
  property & mechanism & status    \\ 
  \hline \hline 
  globally consistent & tadpoles + K-theory & $\checkmark^{*}$ \\
  D-term susy & vanishing FI-terms inside K\"ahler cone & $\checkmark^{**}$ \\
  gauge group $SU(5)$ & $U(5)\times U(1)$ stacks & $\checkmark$ \\
  3 chiral generations & choice of line bundles & $\checkmark$ \\
  no vector-like matter  & localisation on $\IP^1$ curves & $-$ \\
  1 vector-like of  Higgs  & choice of line bundles & $\checkmark^{***}$ \\
  no adjoints & rigid 4-cycles, del Pezzo & $\checkmark$ \\
  GUT breaking   & $U(1)_Y$ flux on trivial 2-cycles  & $\checkmark$ \\
  3-2 splitting & Wilson lines on  $g=1$ curve   &
  $\checkmark$\\
  3-2 split + no dim=5 p-decay  & local. of $H_u, H_d$  on disjoint
  comp.   & $-$ \\
  ${\bf 10\, \ov 5\, \ov 5_H}$ Yukawa & perturbative
  & $\checkmark$ \\
  ${\bf 10\, 10\, 5_H}$ Yukawa & presence of appropriate D3-instanton
  & $-$$^{****}$ \\
  Majorana neutrino masses & presence of appropriate D3-instanton
  & $-$$^{****}$ \\
  \hline 
  \hline 
\end{tabular} 
\caption{Summary of $SU(5)$ properties  realised 
in the model of \autoref{sec:GutModelM2dP8}. \newline
$^{*}$\ {\small K-theory to the best of our ability to detect $SP$ cycles}
\newline
$^{**}$\ {\small realised on acceptable boundary of K\"ahler cone}
\newline
$^{***}$\ {\small for special choice of complex structure moduli}
\newline
$^{****}$  {\small at least not with O(1) instantons}}  
\label{tab_gutmodel2}
\end{center} 
\end{table}

\subsection
[A  GUT Model on $M_3^{({\rm dP}_9)^3}$] 
[A  GUT Model]
{A  GUT Model on $\mathbf{M_3^{({\rm dP}_9)^3}}$} 
\label{sec_offdiag} 
 
In this section we investigate whether we can build a GUT model,
where the ${\bf 10}$ representation is also localised 
on a curve. This is expected to avoid the appearance
of extra vector-like states. 

Concretely, we consider the elliptic fibration over
the ${\rm dP}_3$ base in the Weierstra\ss{} phase with 
the section ${\cal B}={\rm dP}_3$ and the six ${\rm dP}_9$ pull-back
divisors $\pi^*(E_1)$, $\pi^*(E_2)$, $\pi^*(E_3)$,
$\pi^*(l-E_1-E_2)$,  $\pi^*(l-E_1-E_3)$ and  $\pi^*(l-E_2-E_3)$.
Moreover, we choose the involution acting as 
  \begin{equation}
    \begin{pmatrix}
      l \\ E_1 \\ E_2 \\ E_3
    \end{pmatrix}
    \mapsto
    \begin{pmatrix}
      2l-E_1-E_2-E_3 \\ l-E_1-E_3 \\ l-E_2-E_3 \\ l-E_1-E_2
    \end{pmatrix}    
  \end{equation}
on the ${\rm dP}_3$.
The orientifold $O7$-plane wraps the divisor
\bea
    D_{O7}=\pi^*(2l-E_1-E_2)
\eea
which has $\chi( D_{O7})=48$. Since there are no fixed points for
this involution, there are no O3-planes.
and the contribution of the curvature terms to
the $D3$ tadpole condition is $N_{D3}+N_{flux}=12$.

\subsubsection*{K\"ahler cone}

Expanding the K\"ahler form as
\bea   
            J=r_{{\cal B}}\, {\cal B} + r_l\, \pi^*(l) - r_1\, \pi^*(E_1)  - 
            r_2\, \pi^*(E_2)- r_3\, \pi^*(E_3)\; . 
\eea         
the K\"ahler cone is simply
\bea  
                 r_{\cal B}>0,\quad r_i-{r_{\cal B}}>0, 
               \quad r_l-r_i-r_j-r_{\cal B}>0,\qquad i<j\in\{1,2,3\}\; . 
\eea     
However, the involution $\sigma$ has $h^{1,1}_-=2$, so that
we expect that two of these five K\"ahler moduli are fixed.
Indeed, requiring that $J$ is invariant under $\sigma$ yields the two
relations
\bea
       r_1=r_2,\qquad   r_l=2\, r_2+r_3\; .
\eea
We are only left with three dynamical K\"ahler moduli.

In addition, in this case we have two $B_-$ moduli. With the general Ansatz
\bea   
            B_-=b_{{\cal B}}\, {\cal B} + b_l\, \pi^*(l) - b_1\, \pi^*(E_1)  - 
            b_2\, \pi^*(E_2)- b_3\, \pi^*(E_3)\;  
\eea  
subject to $\sigma(B_- )=-B_-$ we obtain the three constraints
\bea
      b_{{\cal B}}=0,\qquad  b_l=b_3\qquad   b_1=2\, b_3-b_2\; .
\eea

\subsubsection*{Tadpole cancellation}

To cancel the D7-brane tadpole eq.~\eqref{tadseven} 
we introduce three stacks of D7-branes on the divisors 
\bea
  &&D_a = \pi^*(E_2)\ ,\qquad \quad\quad\  \ D'_a = \pi^*(l-E_2-E_3), \nonumber \\
  &&D_b = \pi^*(l-E_1)\ ,\qquad \quad\, D'_b = \pi^*(l-E_2), \\
  &&D_c = \pi^*(E_3)\ ,\qquad\quad\quad\  \ D'_c = \pi^*(l-E_1-E_2). \nonumber
\eea  
As for the line bundles it is convenient to split off the continuous $B_-$-moduli by writing
$c_1(\widetilde L)= c_1(L)-B_-$.
Choosing the ${\cal B}$ part of the line bundles on these
divisors as
\bea
  c_1(\widetilde L_a)=3k{\cal B} + \pi^*(\eta_a), \quad
   c_1(\widetilde L_b)=-5k{\cal B} + \pi^*(\eta_b), \quad
    c_1(\widetilde L_c)=-3k{\cal B} + \pi^*(\eta_c), \quad
\eea
cancels also the D5-brane tadpole.   
The resulting chiral spectrum is listed in \autoref{tab_chir_modelc}.
\begin{table}[htbp] 
\renewcommand{\arraystretch}{1.5} 
\begin{center} 
\begin{tabular}{|c||c|c|c|} 
\hline 
\hline 
number & $U(5)$ & $U(3)$ & $U(5)$  \\ 
\hline \hline 
$6k$ & $\Yasymm_{\,(2)}$ & $1$ & $1$ \\ 
$2k$ & $\antifund_{\, (-1)}$ & $\antifund_{\, (-1)}$ & $1$ \\ 
$10k$ & $1$ & $\bYasymm_{\,(-2)}$  & $1$ \\ 
\hline 
\hline 
\end{tabular} 
\caption{Chiral spectrum for intersecting D7-brane model. The indices denote 
 the $U(1)$ charges. }  
\label{tab_chir_modelc}
\end{center} 
\end{table} 

\noindent
Some remarks are in order concerning this spectrum: One gets precisely
$6k$ generations of ${\bf \ov 10}$ and, taking into account the extra
multiplicities due to the $U(3)$ stack, $2k \times 3$ generations of
${\bf \ov 5}$, but without right-handed neutrinos. In fact we found
that none of the pull-back divisors carries symplectic Chan-Paton
factors.  In this model the flavour group is gauged.  Moreover, since
$E_2$ and $l-E_1$ do not intersect there are {\it no massless Higgs
  fields in the $(ab)$ sector}.  However in the $(ac')$ sector,
vanishing Wilson-lines along the elliptic fibre imply that we obtain
one vector-like matter field in the $ ({\bf 5},{\bf 1},{\bf 5})+({\bf
  \ov 5},{\bf 1},{\bf \ov 5})$ representation. These carry the GUT
quantum numbers to be identified with five pairs of Higgs fields. As
mentioned in \autoref{subsec_GG}, since the Higgs and ${\bf \ov
  5}$ matter fields are charged under different U(1) groups also the
bottom-type Yukawa couplings need to be realised by D3-brane
instantons.  Clearly, this is not a completely realistic model, but
some rough features are realised and in particular the massless matter
states in the anti-symmetric representation of $SU(5)$ are localised
on the fibre $F$ of the elliptic fibration.  The minimal choice
$k={1\over 2}$ gives already three generations.

\subsubsection*{Three generation model}

Taking now the quantisation conditions for the gauge fluxes
into account and turning on half-integer $B$-field flux 
through the fibre, that is, $c_1(B)={1\over 2}{\cal B}$, 
we choose
\bea
  c_1(\widetilde L_a)&=&{3\over 2}\,{\cal B} + {3\over 2} \pi^*(E_2), \quad
   c_1(\widetilde L_b)=-{5\over 2}\,{\cal B} + 5\, \pi^*(E_1), \nonumber \\
     c_1(\widetilde L_c)&=&-{3\over 2}\,{\cal B} -  {3\over 2} \pi^*(E_3) 
\eea
for the line bundles. 

The D-term constraints 
\bea
       \int_{D_a} J\wedge \Bigl( c_1(\widetilde L_a) + B_- \bigr)=0
\eea
for all three brane stacks give three constraints which can all be solved inside the K\"ahler cone provided
\bea
    r_{\cal B}>0\, \qquad -{3\over 2}< b_2 < {3\over 2}\; .
\eea

The contributions of each of the three stacks to the D3-tadpole condition are positive and
add up as
\bea
     N_{\rm gauge}={3\cdot 45\over 8} +{3\cdot 75\over 4} + 
            {3\cdot 45\over 8}=90\gg 12 \; .
\eea
Here we see explicitly that eventually the $B_-$ field
drops out so that one ends up really with an integer 
contribution.
Clearly this is a massive overshooting and requires the introduction
of anti D3-branes. 
Finally, for $c_1(B)={1\over 2}{\cal B}$ the divisor ${\cal B}$ 
can carry a trivial line bundle and is expected to have
$SP$ Chan-Paton factors.  The resulting global Witten anomaly
(K-theory) constraint is satisfied. 

\subsubsection*{Non-chiral spectrum and GUT breaking}

Nevertheless, the purpose was to demonstrate that by choosing
non-diagonal involutions, it is (in principle) possible to
have also the matter in the antisymmetric representation of $SU(5)$
localised on a curve. 
In fact here all matter is localised in the fibre elliptic curve
$C=F$ with trivial canonical line-bundle.
{}From that it immediately follows that there is no vector-like
matter in the ${\bf 10}+{\bf \ov 5}$ representation.
Since the $U(5)$ stack is a rigid ${\rm dP}_9$ surface, we can break
the $SU(5)$ GUT gauge symmetry to the Standard Model gauge group
by turning on  $U(1)_Y$ flux of the form ${\cal L}_Y={\cal O}(e_1-e_2)$.
In contrast to the first two examples we presented, this does not give any 
new vector-like matter. 
As for the Higgs sector, we are now in the favourable situation that the bundles ${\tilde L}_a$ and ${\tilde L}_b$ are both pullbacks from the ambient space. According to the discussion around eq.~\eqref{pulback_cond} it therefore suffices to twist $\tilde L_a$ by $R_a={\cal L}_Y^{-1}$ to arrange for precisely one Higgs doublet and no Higgs triplet \emph{without} further adjusting any complex structure moduli. Adding ${\cal L}_Y$ and $R_a$ results in an additional 3 units of D3-brane charge in the D3-tadpole equation.

\subsubsection*{Summary of features}

Let us summarise in \autoref{tab_gutmodel3} 
which of the desired features we were able to realise
in this simple model.
\begin{table}[htbp] 
\renewcommand{\arraystretch}{1.5} 
\begin{center} 
\begin{tabular}{|c||c|c|} 
  \hline 
  \hline 
  property & mechanism & status    \\ 
  \hline \hline 
  globally consistent & tadpoles + K-theory & $\checkmark^{*,**}$ \\
  D-term susy & vanishing FI-terms inside K\"ahler cone & $\checkmark$ \\
  gauge group $SU(5)$ & $U(5)\times U(3)$ stacks & $\checkmark$ \\
  3 chiral generations & choice of line bundles & $\checkmark$ \\
  no vector-like matter  & localisation on g=1 curves & $\checkmark$ \\
  5 vector-like   Higgs  & choice of line bundles & $\checkmark$ \\
  no adjoints & rigid 4-cycles, del Pezzo & $\checkmark$ \\
  GUT breaking   & $U(1)_Y$ flux on trivial 2-cycles  & $\checkmark$ \\
  3-2 splitting & Wilson lines on  $g=1$ curve   &
  $\checkmark$\\
  3-2 split + no dim=5 p-decay  & local. of $H_u, H_d$  on disjoint
  comp.   & $-$\\
  ${\bf 10\, \ov 5\, \ov 5_H}$ Yukawa & presence of appropriate D3-instanton
  & $-$$^{***}$ \\
  ${\bf 10\, 10\, 5_H}$ Yukawa & presence of appropriate D3-instanton
  & $-$$^{***}$ \\
  Majorana neutrino masses & presence of appropriate D3-instanton
  & $-$$^{***}$ \\
  \hline 
  \hline 
\end{tabular} 
\caption{
Summary of $SU(5)$ properties realised 
in the model of \autoref{sec_offdiag}. \newline 
$^*$\  {\small overshooting in D3-tadpole $\to$ $\ov D3$-branes}\newline
 $^{**}$\ {\small K-theory  to the best of our ability to detect $SP$ cycles} \newline
$^{***}$\ {\small at least not with $O(1)$ instantons}}  
\label{tab_gutmodel3}
\end{center} 
\end{table}


\section{GUTs on Del Pezzo Transitions of the Quintic}
\label{sec:quintic}

So far we have studied GUT models on descendants of
the elliptic fibration Calabi-Yau $\IP_{1,1,1,6,9}[18]$.
We have gone a long way to eventually arrive
at GUT like examples featuring many of the desired 
properties. One of the general aspects  of these models was
that for the $SU(5)$ GUT stack localised on a shrinkable
${\rm dP}_8$ surface the D-term conditions in conjunction  with
the swiss-cheese property of the triple intersection form
force the GUT four-cycle to collapse to string scale
size, that is, to the quiver locus. If this were a generic feature
of all Calabi-Yau orientifolds containing shrinkable
surfaces, it would clearly have strong implications for model
building.

The clarification of this point  is one of our   motivations
for studying another class of Calabi-Yau manifolds
containing del Pezzo surfaces. Instead of starting with
the elliptic fibration $\IP_{1,1,1,6,9}[18]$, we
take the simple Quintic $\IP_{1,1,1,1,1}[5]$ and perform
del Pezzo transitions. The mathematics of this construction
is collected in \autoref{sec_quintm} with the result
that here we can get intersecting ${\rm dP}_r$, $r\le 8$  surfaces,
so that the triple intersection forms  do not have the diagonal 
swiss-cheese type structure.
In \autoref{sec_quintex} we provide one more example
of a GUT model, for which all Standard Model matter 
is really localised on curves of genus zero and one,
respectively.

\subsection{Del Pezzo Transitions of the Quintic}
\label{sec_quintm}

In this section we introduce a class of compact Calabi-Yau manifolds 
which can be obtained from the quintic hypersurface by performing del Pezzo
transitions.
Again these spaces will be realised as hypersurfaces in an
ambient toric manifold. It will turn out that the del Pezzo surfaces arising after
the transitions can intersect and thus are ideal candidates for supporting
intersecting D7-branes.

\subsubsection*{The toric data and intersection forms}

Let us first give the points in the polyhedron for the toric ambient
spaces.  The hypersurface is then determined to have the
anti-canonical class given by the sum of all toric divisors as in
\autoref{delPezzotrans}.  The quintic hypersurface has the points
$v^*_1=(-1,0,0,0)$, $v^*_2=(0,-1,0,0)$ $v^*_3=(0,0,-1,0)$,
$v^*_4=(0,0,0,-1)$ and $v_5^*=(1,1,1,1)$.  Its Hodge numbers are
$h^{1,1}=1$ and $h^{2,1}=101$. By arranging some of the $h^{2,1}$
complex structure deformations in the hypersurface constraint one can
generate del Pezzo singularities and blow up del Pezzo surfaces. This
process increases $h^{1,1}$ by the number of K\"ahler moduli of the
del Pezzo four-cycles and lowers $h^{2,1}$ since one has to fix a
certain number of the complex structure moduli to generate a
singularity.

As a first transition we can blow up a ${\rm dP}_6$ surface by adding the
point $v^*_6=(1,0,0,0)$ to the polyhedron and consider the resulting
hypersurface $Q^{{\rm dP}_6}$.  In fact, the new Calabi-Yau manifold has
$h^{1,1}=2$ and $h^{2,1}=90$. There is only one triangulation for the
ambient toric space. Using the same methods as in
\autoref{delPezzotrans} one can explicitly check that the divisor
$D_6$ is a ${\rm dP}_6$ del Pezzo surface. The intersection form is simply
\begin{equation}
  I_3 = 3 D_6^3 + 3 D_5^2 D_6 - 3 D_6^2 D_5 + 2 D_5^3 \ .
\end{equation} 
This Calabi-Yau is also of the strong swiss-cheese type, since the
intersection form diagonalises for the transformation $(\tilde D_5 =
D_6 - D_5,\, \tilde D_6=D_6)$. This is also consistent with the fact that the ${\rm dP}_6$
surface in this manifold is generic which can be inferred from the fact
that in this transition $\Delta \chi = 24$. For generic transitions,
as the ones in \autoref{delPezzotrans} with Euler numbers
eq.~\eqref{general_chi}, one has $\Delta \chi = 2 C_{(n)}$, where
$C_{(n)}$ is the dual Coxeter number of the exceptional groups
associated to ${\rm dP}_n$.

To generate a second del Pezzo surface we add the point $v^*_7=(0,1,0,0)$
to the polyhedron. This blows up a point on the ${\rm dP}_6$ surface $D_6$ to a
$\bbP^1$ such that ${\rm dP}_6 \rightarrow {\rm dP}_7 \cong D_6$ 
and generates a second ${\rm dP}_7$ surface $D_7$. The ambient toric space  
has one Calabi-Yau triangulation and the hypersurface, denoted by
$Q^{({\rm dP}_7)^2}$, has Hodge numbers $h^{1,1}=3$ and $h^{2,1}=79$. One can then 
check that the two ${\rm dP}_7$ surfaces $D_6,D_7$ intersect on a $\bbP^1$. To do 
that one computes the triple intersection form 
\begin{equation}
  I_3 = 2 D_7^3 + 2 D_6^3 + 2 D_5^2 (D_7+D_6)  -  D_7^2 (2 D_5+D_6) 
 -  D_6^2 (2 D_5+D_7)+ D_5 D_6 D_7\ . 
\end{equation}
This intersection form cannot be diagonalised due to the intersection
of $D_6$ and $D_7$. In other words, the two ${\rm dP}_7$ surfaces are not
generic, but share a common $\bbP^1$. This can be also inferred from
the fact that $\Delta \chi =2 \times 24$ with respect to the quintic
hypersurface. This change is expected for a transition with two
generic ${\rm dP}_6$ surfaces and corresponds to the fact that the $E_6$
sublattices on the two ${\rm dP}_7$ surfaces are still trivial in the
Calabi-Yau threefold.  For two generic ${\rm dP}_7$ surfaces one would find
$\Delta \chi=2 \times 36$. That only an $E_6$ sublattice is trivial in
the Calabi-Yau can also be inferred by computing the BPS-instantons as
suggested in~\cite{Grimm:2008ed}. One finds that the representations
of $E_7$ appearing for a generic ${\rm dP}_7$ transition are split into
$E_6$ representations for our intersecting divisors $D_6$ and $D_7$.

We can perform a third transition by adding the point
$v^*_8=(0,0,1,0)$ to the polyhedron and denote the corresponding
hypersurface by $Q^{({\rm dP}_8)^3}$. In this case one additional
$\bbP^1$ is blown up on each ${\rm dP}_7$ surface $D_6,D_7$ such that
${\rm dP}_7 \rightarrow {\rm dP}_8$. The two ${\rm dP}_8$ surfaces
$D_6,D_7$ intersect a new ${\rm dP}_8$ surface $D_8$ in the blown-up
$\bbP^1$ curves. The new toric ambient space has one triangulation,
and the corresponding Calabi-Yau hypersurface has Hodge numbers
$h^{1,1}=4$ and $h^{2,1}=68$. The intersection form is given by
\begin{multline}
  I_3 = D_6^3 + D_7^3 + D_8^3 - D_5^3 +  D_5^2 (D_6 +D_7 + D_8)  
  \\
  -  D_6^2 (D_5+D_7+ D_8)
  - D_7^2 (D_5 + D_6 + D_8) - D_8^2 (D_5 +D_6+D_7) 
  \\
  +   D_5 (D_6 D_8 + D_7
  D_8 +  D_7 D_6)
  .
\end{multline}
Once again we note that the ${\rm dP}_8$ surfaces are non-generic, since they
intersect over $\bbP^1$ curve. This is an accord with the fact that $\Delta \chi =
3 \times 24$ with respect to the quintic. In fact, one concludes that in each del Pezzo
$8$ surface and $E_6$ sublattice is trivial in the Calabi-Yau space.

\begin{figure}[htbp]
  \centering
  \begin{picture}(400,80)
    \put(0,0){\includegraphics[width=0.95\textwidth]{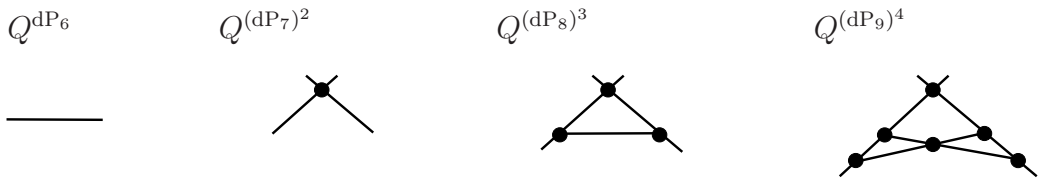}}
    \put(10,55){$Q^{{\rm dP}_6}$}
    \put(90,55){$Q^{({\rm dP}_7)^2}$}
    \put(195,55){$Q^{({\rm dP}_8)^3}$}
    \put(315,55){$Q^{({\rm dP}_9)^4}$}
  \end{picture}
  \vspace{-10pt}
  \caption{Schematics of the intersecting del Pezzo surfaces on
    transitions of the quintic.  Each intersection is a $\bbP^1$.}
  \label{fig_quintic_int}
\end{figure}

Finally, we can add the point
$v^*_9=(0,0,0,1)$ to the polyhedron and denote the corresponding hypersurface by
$Q^{({\rm dP}_9)^4}$. In fact, the new Calabi-Yau space has 
$h^{1,1}=5$ and $h^{2,1}=57$ and one checks that the toric divisors 
$D_6,D_7,D_8,D_9$ are ${\rm dP}_9$ surfaces. Clearly, this space cannot be obtained
by resolving a del Pezzo singularity, since ${\rm dP}_9$ can be only shrunk to a 
curve and not to a point. Nevertheless it is a viable Calabi-Yau background
with intersection form 
\begin{multline}
  I_3 = D_5 (D_6 D_8 + D_8 D_9 + D_6 D_9 + D_6 D_7
  + D_7 D_8 + D_7 D_9)  
  \\
  - D_6^2(D_7 + D_8 +D_9) 
        - D_7^2
  (D_6 + D_8 + D_9) - D_8^2(D_6 + D_7 + D_8)
  \\
  - D_9^2(D_6 + D_7 + D_8 )-  D_5^3 
  .
\end{multline}
One can check that each of the ${\rm dP}_9$ divisors intersects the other three
in a $\bbP^1$.  Schematically the intersecting del Pezzo surfaces in
the four manifolds $Q^{{\rm dP}_6}$, $Q^{({\rm dP}_7)^2}$, $Q^{({\rm dP}_8)^3}$ and
$Q^{({\rm dP}_9^4)}$ are shown in \autoref{fig_quintic_int}.  The Euler
number has changed again by $24$, such that $\Delta \chi = 4 \times
24$ with respect to the quintic hypersurface. Again, this corresponds
to the fact that there are four $E_6$ lattices on the ${\rm dP}_9$ surfaces
which are trivial in $Q^{({\rm dP}_9)^4}$.

\subsubsection*{K\"ahler cone and orientifold involution}

In Sections~\ref{sec_quintex} and~\ref{3G_quintic} we will construct
GUT models on $Q^{({\rm dP}_8)^3}$ and $Q^{({\rm dP}_9)^4}$ with matter and
Higgs localised on curves. However, in order to determine the spectrum
and check the supersymmetry conditions, we first need to calculate the
K\"ahler cone and the orientifold involution acting on it.

Let us begin by analysing the hypersurface $Q^{({\rm dP}_8)^3}$. The Mori
cone is simplicial for this Calabi-Yau phase, and generated by the
vectors
\begin{equation}
  \label{Mori_quinticdP83} 
  \renewcommand{\arraystretch}{1.1} 
  \renewcommand{\arraycolsep}{7pt} 
  \begin{array}{|c|c|c|c|c|c|c|c|c||c|} 
  \hline 
  &x_1 & x_2 & x_3 & x_4 & x_5 & x_6 & x_7 &  x_8 & p \\ \hline\hline 
  \ell^{(1)} & 1 & 0 & 0 & 1 & 1 & 0 & -1 & -1 & -1  \\ 
  \ell^{(2)} & 0& 0& 0& -1& -1& 1& 1& 1& -1 \\ 
  \ell^{(3)} &0& 0& 1& 1& 1& -1& -1& 0& -1 \\
  \ell^{(4)} &0& 1& 0& 1& 1& -1& 0& -1& -1 \\
  \hline 
  \end{array} 
\end{equation}
Its dual, the K\"ahler cone, is therefore again simplicial and
generated by
\begin{equation}
  \label{quintic_Kc}
  \begin{aligned}
    K_1 \;&= 2D_5+D_6+D_7+D_8  
    ,&\quad
    K_2 \;&= D_5+D_6           
    ,\\
    K_3 \;&= D_5+D_7           
    ,&
    K_4 \;&= D_5+D_8           
    .
  \end{aligned}
\end{equation}
As before, the K\"ahler cone is needed in order to evaluate the D-terms in
the physical region of the moduli space.

We next specify an orientifold involution $\sigma$ on
$Q^{({\rm dP}_8)^3}$. Explicitly, $\sigma$ is given by the exchange of
coordinates
\begin{equation}
  \label{quintic_involution}
  \sigma: \quad  
  x_2 \leftrightarrow x_3, 
  \qquad 
  x_7 \leftrightarrow x_8
  .
\end{equation}
This leads to a split $h^{1,1}_+ = 3$ and $h^{1,1}_- = 1$, yielding a
four-dimensional theory with $3$ K\"ahler moduli $T_I$ and one
$B_-$-modulus $G$ as defined in eq.~\eqref{def-GT}. The fixed point
locus of this involution contains one O7-plane wrapping the four-cycle
\begin{equation}
  \label{O7_quintic}
  D_{O7}=D_5+D_7+D_8
\end{equation}
and one fixed point, $N_{O3}=1$. Note that $\chi(D_{O7})=47$, and,
therefore, $\chi(D_{O7})+N_{O3}$ is indeed divisible by four.

Turning to the other Calabi-Yau phase $Q^{({\rm dP}_9)^4}$, the Mori cone is
now non-simplicial and generated by the vectors
\begin{equation}
  \label{Mori_quinticdP94} 
  \renewcommand{\arraystretch}{1.1} 
  \renewcommand{\arraycolsep}{7pt} 
  \begin{array}{|c|c|c|c|c|c|c|c|c|c||c|} 
    \hline 
    &x_1 & x_2 & x_3 & x_4 & x_5 & x_6 & x_7 &  x_8 & x_9 & p \\ 
    \hline\hline 
    \ell^{(1)} & 0 & 0 & -1 & 0 & -1 & 1 & 1 & 0 & 1 & -1 \\
    \ell^{(2)} & 0 & 1 & 0 & 1 & 1 & -1 & 0 & -1 & 0 & -1\\
    \ell^{(3)} & 1 & 0 & 0 & 1 & 1 & 0 & -1& -1 & 0 & -1 \\
    \ell^{(4)} &-1& 0 & 0 & 0 & -1& 0 & 1 & 1 & 1 & -1 \\
    \ell^{(5)} &0 & 1 & 1 & 0 & 1 & -1& 0 & 0 & -1& -1 \\
    \ell^{(6)} &0 & 0 & 0 & -1& -1& 1 & 1 & 1 & 0 & -1 \\
    \ell^{(7)} &0 & 0 & 1 & 1 & 1 & -1& -1& 0& 0& -1 \\
    \ell^{(8)} &1 & 1 & 0 & 0 & 1 & 0 & 0 & -1& -1& -1 \\
    \ell^{(9)} &1 & 0 & 1 & 0 & 1 & 0 & -1 & 0& -1& -1 \\
    \ell^{(10)} &0 & -1& 0 & 0 & -1& 1 & 0 & 1& 1& -1 \\
    \hline 
  \end{array} 
\end{equation}
In order to define coordinates for the K\"ahler cone, we first discard
the $\ell^{(\kappa)},\, \kappa=3,4,6,7,8$ and determine the dual basis of 
four-cycles 
\begin{equation}
  \begin{aligned}
    K_1 \;&= D_5 + D_6 + D_7
    ,&\qquad 
    K_2 \;&= D_5 + D_8 + D_9
    ,\\
    K_3 \;&= D_5 + D_6
    ,&
    K_4 \;&=  D_5 + D_7 + D_8
    ,\\
    K_5 \;&= D_5 + D_9
    .
  \end{aligned}
\end{equation}
Expanding $J = \sum_i r_i K_i$ we have to take $r_i\geq 0$ in the K\"ahler
cone. However, we note that the discarded $\ell^{(\kappa)}$ impose the 
additional conditions
\begin{equation}
  \begin{aligned}
    r_3 - r_4 + r_5 \geq&\; 0
    ,&\qquad  
    r_2 - r_3 + r_4 \geq&\; 0
    ,&\qquad
    r_1 + r_4 - r_5 \geq&\; 0
    ,\\
    r_2 -r_1  + r_5 \geq&\; 0
    ,&
    r_1 - r_2 + r_3 \geq&\; 0
    .
  \end{aligned}
\end{equation}
We have to ensure that these conditions are satisfied when evaluating the
D-terms in coordinates $r_i$.

Let us finally specify the involution on $Q^{({\rm dP}_9)^4}$. It is simply
given by the exchange of $x_2 \leftrightarrow x_3$ and $x_7
\leftrightarrow x_8$, the same as in
eq.~\eqref{quintic_involution}. This leads to a split $h^{1,1}_+ = 4$
and $h^{1,1}_- = 1$ and, hence, four $T_I$ K\"ahler moduli and one
$B_-$-modulus $G$. The fixed point locus of this involution contains
one O7-plane, wrapping the same linear combination as in
eq.~\eqref{O7_quintic} and three fixed points, $N_{O3}=3$.  Note that
now $\chi(D_{O7})=37$, and, therefore, $\chi(D_{O7})+N_{O3}$ is again
divisible by four.

\subsection{A GUT Model Without Vector-Like Matter}
\label{sec_quintex}

In this section we present a $SU(5)$ GUT model with all matter
realised on curves. One of our motivations to discuss this example is
to illustrate that the GUT brane can indeed wrap a shrinkable ${\rm dP}_r$,
$r \leq 8$ without being driven to the quiver locus by the D-terms.
Our starting point is the Calabi-Yau $Q^{({\rm dP}_8)^3}$, which contains
three intersecting ${\rm dP}_8$ surfaces.  Using the notation from the
previous section, we choose the involution
eq.~\eqref{quintic_involution}, which leads to an O7-plane
eq.~\eqref{O7_quintic} and one O3-plane.  In the following we will
specify the D7-branes which define a GUT model with two chiral
generations.

\subsubsection*{Two generation model}

To cancel the D7-brane tadpole \eqref{tadseven} 
we introduce three stacks of D7-branes on the divisors 
\begin{equation}
  \begin{aligned}
    U(5):&\quad&  D_a &= D_7,     & D'_a &= D_8,  & \chi(D_a)&=11, \\
    U(1):&&  D_b &= D_5,     & D'_b &= D_5, & \chi(D_b)&=25, \\
    SO(6):&& D_c &= D_5+D_7, & D'_c &= D_5+D_8, & \chi(D_c)&=36.
  \end{aligned}
\end{equation}
Here we note that $D_2 = D_5+D_7$ and $D_3 = D_5+D_8$ are toric divisors.
Then the contribution of the curvature terms to the D3-brane tadpole
cancellation condition reads
\bea
\frac{N_{\rm O3}}{4} + 
{\chi(D_{\rm O7})\over 12} +\sum_a  N_a\, {\chi_o(D_a)\over 24}=
  {1\over 4}+{47\over 12} + {5\cdot 11+ 1\cdot 25 + 3\cdot 36\over 24}=
   12\; . 
\eea   
Next, we choose $c_1(B)={1\over 2} D_5$ 
and split off the continuous $B_-$-moduli by writing
$c_1(\widetilde L)= c_1(L)-B_-$.
Taking into account the Freed-Witten quantisation
conditions, the following choice of line bundles 
\bea
  c_1(\widetilde L_a)={1\over 2}D_5 +{1\over 2}D_7 - D_8, \quad
   c_1(\widetilde L_b)=D_5 , \quad
    c_1(\widetilde L_c)=0 
\eea
cancels the D5-brane tadpole as well. Here we used the fact that 
$D_7$ restricts trivially to $D_c=D_5+D_7$.
The resulting chiral spectrum is listed in \autoref{tab_chir_modeld}.
\begin{table}[htbp] 
\renewcommand{\arraystretch}{1.5} 
\begin{center} 
\begin{tabular}{|c||c|c|c|} 
\hline 
\hline 
number & $U(5)$ & $U(1)$ & $SO(6)$  \\ 
\hline \hline 
$2$ & $\Yasymm_{\,(2)}$ & $1$ & $1$ \\ 
$2$ & $\antifund_{\, (-1)}$ & $\fund_{\, (1)}$ & $1$ \\ 
$2$ & $1$ & $\bYsymm_{\,(-2)}$  & $1$ \\ 
\hline 
\hline 
\end{tabular} 
\caption{Chiral spectrum for intersecting D7-brane model. The indices denote 
 the $U(1)$ charges.}  
\label{tab_chir_modeld}
\end{center} 
\end{table} 

\noindent
Let us make a couple of remarks concerning this spectrum:  
One obtains precisely two generations of MSSM particles
including the right-handed neutrinos. Moreover,
the states transforming in the ${\bf 10}$ representation
of $SU(5)$ are localised on the curve $D_7\cap D_8=\IP^1$, so
that there are no additional vector-like states. 
Similarly, the matter states in the ${\bf \ov{5}}$ representation
are localised on the curve $D_7\cap D_5=T^2$ so that
there are no vector-like states either.
Moreover, as in the examples before from the $(a'b)$ sector
we will get one vector-like pair of Higgs fields ${\bf 5}_H + {\bf \ov{5}}_H$ by twisting $L_b$ appropriately, see discussion at the end of this subsection.
Only the right-handed neutrinos are localised on a surface, namely
on $D_5$.

A D7-brane wrapped upon $D_5$ can carry a trivial line bundle so
that this brane is expected to carry symplectic Chan-Paton factors.
The resulting K-theory constraint
\begin{equation}
      \int_Y  [D_5]\wedge [D_7] \wedge c_1(L_a) + \int_Y [D_7]\wedge [D_5]\wedge  c_1(L_b) \in 2\mathbb Z
\end{equation}
is indeed satisfied for our model.
The D3-tadpoles induced by each single brane stack
are positive and
add up as
\bea
     N_{\rm gauge}={5\cdot 1\over 2} +{1\cdot 1\over 2} =3 \; .
\eea
Let us next evaluate the D-term constraints. 
The generators $K_i$ of the K\"ahler cone are given in \eqref{quintic_Kc}. 
We use these to expand the K\"ahler form as $J=\sum_i r_i  K_i$
with $r_i>0$. Note that due to the orientifold action
for the  last two K\"ahler cone generators we have $r_3=r_4$.
In addition there exists one $B_-$ modulus
\bea
             B_-=b\, (D_7-D_8)\; .
\eea
The D-term constraints 
\bea
       \int_{D_i} J\wedge \Bigl( c_1(\widetilde L_i) + B_- \bigr)=0
\eea
for all three brane stacks $i=a,b,c$ yield three conditions, 
which are solved on the boundary of the K\"ahler cone by
\bea
      r_1=0, \quad r_2=0, \quad  b=0.
\eea
However, the volumes of the three branes involved and the overall
volume are finite 
\bea
       &&\tau_7={1\over 2}\int_{D_7} J\wedge J= r_3^2,\qquad\quad
        \tau_5={1\over 2}\int_{D_5} J\wedge J=2\, r_3^2, \quad \nonumber \\ 
       &&{\rm Vol}(Y)={1\over 6}\int_Y J\wedge J\wedge J=2\, r_3^3,\; 
\eea
so that we still have parametric control over the $\alpha'$
expansion in the brane sector.

Finally, to break the $SU(5)$ GUT group to the MSSM, we
turn on the trivial line bundle ${\cal L}_Y$.
On $D_a=D_7={\rm dP}_8$ there are now three non-trivial two-cycles.
They include the two genus zero curves $\IP^1$s from the intersection
$D_7\cap D_6$ and $D_7\cap D_8$.
In addition there exists the genus one curve $D_7\cap D_5$
which is identical to $-D_7\cap D_7$.
Identifying $D_7\cap D_6=e_7$ and $D_7\cap D_6=e_8$
and  $D_7\cap D_5=3h-\sum_{i=1}^8 e_i\, $,  we realise
that the two-cycles on ${\rm dP}_8$ trivial in $Y$ are
the ones from ${\rm dP}_6$ orthogonal to $K$.
By definition this is the $E_6$ sublattice of $H_2({\rm dP}_6,\mathbb Z)$.
Therefore, choosing for instance $c_1({\cal L}_Y)=e_1-e_2$ 
breaks the $SU(5)$ gauge group  to the Standard  Model 
gauge group and contributes additional two units
to the D3-brane tadpole. 
To generate one pair of Higgs doublets and project out the triplet on the elliptic curve $C = D_5 \cap D_7$ we twist $\widetilde L_a$ by the bundle $R_a={\cal L}_Y^{-1}$ on $D_7$.
In essence this yields yet another unit of D3-brane charge in the D3-tadpole equation and
we need six dynamical
D3-branes to saturate it.
Since ${\cal L}_Y$ restricts trivially to $e_7$ and $e_8$,
there are precisely two generations of  charged quark and leptons
without any vector like states. 

As already mentioned a D7-brane on $D_5$ carries $SP$ Chan-Paton
factors, so that an Euclidean D3-brane instanton on the same cycle is
of type $O(1)$.  Indeed, such an instanton carries the right chiral
zero modes $I_{a,{\rm inst}}=1$ and\footnote{There is a change of sign
  compared to \eqref{instzerocond} as here the Higgs originates from
  the sector $(a'b)$ instead of $(ab)$. }  $I_{b,{\rm inst}}=1$ to
generate the top-Yukawa couplings. However, since the surface $D_5$
has $h^{(2,0)}(D_5)=1$, it is not rigid and there can be additional
vector-like zero modes from the intersection of the instanton with the
D7-brane wrapping $D_b=D_5$.

We summarise in \autoref{tab_gutmodelquint1} 
which of the desired features we were able to realise
in this simple model.
\begin{table}[htbp] 
\renewcommand{\arraystretch}{1.5} 
\begin{center} 
\begin{tabular}{|c||c|c|} 
  \hline 
  \hline 
  property & mechanism & status    \\ 
  \hline \hline 
  globally consistent & tadpoles + K-theory & $\checkmark^{*}$ \\
  D-term susy & vanishing FI-terms inside K\"ahler cone & $\checkmark^{**}$ \\
  gauge group $SU(5)$ & $U(5)\times U(1)$ stacks & $\checkmark$ \\
  3 chiral generations & choice of line bundles & $-$ \\
  no vector-like matter  & localisation on $g=0,1$ curves & $\checkmark$ \\
  5 vector-like   Higgs  & choice of line bundles & $\checkmark$ \\
  no adjoints & rigid 4-cycles, del Pezzo & $\checkmark$ \\
  GUT breaking   & $U(1)_Y$ flux on trivial 2-cycles  & $\checkmark$ \\
  3-2 splitting & Wilson lines on  $g=1$ curve   &
  $\checkmark$\\
  3-2 split + no dim=5 p-decay  & local. of $H_u, H_d$  on disjoint
  comp.   & $-$\\
  ${\bf 10\, \ov 5\, \ov 5_H}$ Yukawa & perturbative
    & $\checkmark$ \\
    ${\bf 10\, 10\, 5_H}$ Yukawa & presence of appropriate D3-instanton
    & $\checkmark^{***}$ \\
    Majorana neutrino masses & presence of appropriate D3-instanton
    & $-^{****}$ \\
    \hline 
    \hline 
\end{tabular} 
\caption{$SU(5)$ properties realised 
in the model of \autoref{sec_quintex}. \newline 
$^{*}$\ {\small K-theory to the best of our ability to detect SP-cycles} \newline
$^{**}$\ {\small realised on acceptable boundary of K\"ahler moduli space} \newline
$^{***}$\ {\small up to absorption of additional vector-like zero modes} \newline
$^{****}$\ {\small at least not with $O(1)$ instantons}} 
\label{tab_gutmodelquint1}
\end{center} 
\end{table}

\subsection
[A Three-Generation Model With Localised Matter on $Q^{({\rm dP}_9)^4}$]
[A Three-Generation Model With Localised Matter]
{A Three-Generation Model With Localised Matter on $\mathbf{Q^{({\rm dP}_9)^4}}$}
\label{3G_quintic}

While in on $Q^{({\rm dP}_8)^3}$ global consistency conditions are in
conflict with the construction of three-generation models, on its
cousin $Q^{({\rm dP}_9)^4}$ it turns out possible to find GUT models with
three generations and all GUT matter realised on matter curves, but
without running into half-integer D3 tadpoles.  The construction of
these models is almost identical to the two-generation example of the
previous subsection, and we can be quite brief.

We again cancel the D7-brane tadpole \eqref{tadseven} 
by introducing three stacks of D7-branes on the divisors 
\begin{equation}
  \begin{array}{llll}
    U(5): & D_a = D_7,  &  D'_a = D_8,  & \chi(D_7)=12, \nonumber \\[.1cm]
    U(1):& D_b = D_5,  & D'_b = D_5, & \chi(D_5)=13, \\[.1cm]
    U(3):\qquad & D_c, = D_5+D_7, \qquad &  D'_c  = D_5+D_8 \qquad & \chi(D_5 + D_7)=25. \nonumber
  \end{array}
\end{equation}
We consider the non-trivial $B_+$-flux
\bea
\label{B+9999}
c_1(B)= \frac{1}{2} (D_7+D_8 +D_9), 
\eea
which allows us to introduce the well-defined bundles
\bea
&&  c_1(\widetilde L_a) = 3 D_5 + 2 D_7 - \frac{1}{2} D_8 + \frac{1}{2} D_9,  \nonumber \\
&&  c_1(\widetilde L_b) =  \frac{5}{2} D_5  +  D_6 - \frac{1}{2} D_7 + \frac{5}{2} D_8 -\frac{1}{2} D_9, \\
&&  c_1(\widetilde L_c) =  \frac{1}{2} D_5   + \frac{3}{2} D_8 -\frac{1}{2} D_9.  \nonumber
\eea
This configuration cancels the D5-tadpole and achieves a pure three-generation spectrum as summarised in \autoref{tab_chir_model9999}.
\begin{table}[htbp] 
\renewcommand{\arraystretch}{1.5} 
\begin{center} 
\begin{tabular}{|c||c|c|c|} 
\hline 
\hline 
number & $U(5)$ & $U(1)$ & $U(3)$  \\ 
\hline \hline 
$3$ & $\Yasymm_{\,(2)}$ & $1$ & $1$ \\ 
$3$ & $\antifund_{\, (-1)}$ & $\fund_{\, (1)}$ & $1$ \\ 
$3$ & $1$ & $\bYsymm_{\,(-2)}$  & $1$ \\ 
\hline
$2$ & $1$ & $1$ & $\bYasymm_{\,(-2)}$ \\
$2$ & $1$ & $\antifund_{\, (-1)}$ & $\antifund_{\, (-1)}$ \\ 
\hline 
\hline 
\end{tabular} 
\caption{Chiral spectrum for intersecting D7-brane model. The indices denote 
 the $U(1)$ charges.}  
\label{tab_chir_model9999}
\end{center} 
\end{table} 
All phenomenological considerations detailed in the previous section regarding the spectrum of this model apply also for the present three-generation model since the intersection pattern of the divisors $D_a$, $D_b$ and $D_c$ is unchanged.

The only differences occur when analysing the global consistency and supersymmetry conditions, as we next discuss.
Prior to breaking $SU(5)$ via $U(1)_Y$ flux the D3-brane tadpole
\bea
\label{tad_9999}
N_{D3} + N_{\rm gauge} = 10
\eea
is just satisfied without room for extra $D3$-branes, as we verify by computing
\bea
N_{\rm gauge}=  \frac{5}{2} \cdot \frac{1}{2} +   \frac{1}{2} \cdot  \frac{31}{4} +  \frac{3}{2} \cdot  \frac{13}{4} =10.
\eea
Next we parametrise the K\"ahler form $J= \sum_{i=1}^5 K_i$ in terms of the five generators of the K\"ahler cone introduced previously, where compatibility with the orientifold action fixes $r_1 = r_2$.  One easily finds that the general solution to the D-term supersymmetry conditions relates the so-defined K\"ahler $r_1, r_2$ as well as the $B_-$-modulus $B_-= b \, (D_7 - D_8)$ to $r_4, r_5$ as
\bea
r_1 = \frac{2}{3} r_4 - r_5, \quad
r_3 = \frac{13}{6} r_4 - 2 r_5, \quad 
B_-= \frac{34 r_4 - 27 r_5}{38 r_4 - 30 r_5}.
\eea
Remarkably any such choice of K\"ahler form lies inside the K\"ahler cone as long as 
$r_5 > 0$ and  $\frac{3}{2} r_5 \leq r_4 \leq 2 \, r_5 $. The existence of solutions inside the K\"ahler cone is, together with the appearance of exactly three generations of Standard Model matter, one of the motivations to present this example.

However, two caveats require further attention.  As the reader is by
now very familiar with, we break $SU(5)$ further to the Standard Model
gauge group by turning on the line bundle ${\cal L}_Y$ on $D_a$ that
is trivial on the ambient Calabi-Yau space. As before the $E_6$
sublattice within $H^2(D_a)$ of the ${\rm dP}_9$ surface $D_a$ is
trivial on the Calabi-Yau, and the minimal choice for ${\cal L}_Y$
that avoids extra vector-like states is, for example, ${\cal L}_Y =
{\cal O}(e_1 - e_2)$. This, however, leads to an extra contribution of
$+2$ in $N_{\rm gauge}$ appearing in the D3-brane tadpole
equation~\eqref{tad_9999}. In addition, the by now familiar twist of $L_a$ by $R_a={\cal L}_Y^{-1}$ required to engineer one Higgs doublet and to project out the triplet contributes with $+1$ on the left-hand side of equation~\eqref{tad_9999}. In conclusion, there is a total overshooting by three units in the D3-brane tadpole equation, which requires the introduction of three
anti-$D_3$ branes.

A second subtlety is associated with cancellation of K-theory charge.
Applying our probe argument we note that the only invariant candidate
cycles for $SP$ Chan-Paton factors are $D_1, D_4, D_5, D_6$ and $D_9$,
each of which is non-Spin. In view of the $B_+$-flux
eq.~\eqref{B+9999} none of them can carry an invariant line bundle in
agreement with the Freed-Witten quantisation condition \emph{except
  $D_9$}.  In fact, any line bundle of the form $L^{(n)}_9=
\frac{n}{2} (D_7-D_8)$ with $n$ odd is a liable and invariant gauge
field configuration. As stressed several times by now, in absence of
unambiguous CFT techniques to establish the orientifold action in the
invariant sector it is hard to decide if $(D_9, L^{(n)}_9)$ carries
$SP$ or rather $SO$ Chan-Paton factors.  In the first case, its
worldvolume theory would suffer from a global Witten anomaly in our
three-generation model, as can be verified by computing that the
number of fundamental representations under the symplectic gauge group
is odd. In this case, the probe argument would suggest that model
would not be globally consistent due to the non-cancellation of
K-theory charge.  We do not attempt to settle this issue at this stage
but rather leave this model in the limbo of phenomenologically highly
appealing configurations whose liability as a genuine string vacuum
hinges upon as subtle and innocent a condition as the cancellation of
torsion K-theory charge.  With this warning in mind we conclude our
model building adventures with a summary of the phenomenological
properties of this model in \autoref{tab_gutmodelquint2}.
\begin{table}[htbp] 
\renewcommand{\arraystretch}{1.5} 
\begin{center} 
\begin{tabular}{|c||c|c|} 
  \hline 
  \hline 
  property & mechanism & status    \\ 
  \hline \hline 
  globally consistent & tadpoles + K-theory & $\checkmark^{*.**}$ \\
  D-term susy & vanishing FI-terms inside K\"ahler cone & $\checkmark$ \\
  gauge group $SU(5)$ & $U(5)\times U(1)$ stacks & $\checkmark$ \\
  3 chiral generations & choice of line bundles & $\checkmark$ \\
  no vector-like matter  & localisation on $g=0,1$ curves & $\checkmark$ \\
  5 vector-like   Higgs  & choice of line bundles & $\checkmark$ \\
  no adjoints & rigid 4-cycles, del Pezzo & $\checkmark$ \\
  GUT breaking   & $U(1)_Y$ flux on trivial 2-cycles  & $\checkmark$ \\
  3-2 splitting & Wilson lines on  $g=1$ curve   &
  $\checkmark$\\
  3-2 split + no dim=5 p-decay  & local. of $H_u, H_d$  on disjoint
  comp.   & $-$\\
  ${\bf 10\, \ov 5\, \ov 5_H}$ Yukawa & perturbative
  & $\checkmark$ \\
  ${\bf 10\, 10\, 5_H}$ Yukawa & presence of appropriate D3-instanton
  & $-$$^{***}$ \\
  Majorana neutrino masses & presence of appropriate D3-instanton
  & $-$$^{***}$ \\
  \hline 
  \hline 
\end{tabular} 
\caption{Summary of $SU(5)$ properties realised 
in the model of \autoref{3G_quintic}. \newline 
$^{*}$ {\small overshooting in D3-tadpole $\rightarrow$  3 $\ov{D3}$-branes } \newline
$^{**}$\ {\small K-theory to the best of our ability to detect SP-cycles and modulo the possible issue of $(D_9,L^{(n)}_9)$} \newline
$^{***}$ {\small at least not with $O(1)$ instantons}} 
\label{tab_gutmodelquint2}
\end{center} 
\end{table}


\section{Comments on Moduli Stabilisation}
\label{sec_moduli}

So far we have focused our attention on the construction of realistic
$SU(5)$ GUT models on intersecting D7-branes in  Type IIB
orientifold models. Eventually, to obtain a truly predictive
framework we have to address the central question of
moduli stabilisation.
Luckily, just for this kind of models very powerful techniques
for moduli stabilisation have been developed during
the last years. First to mention is  the possibility
of freezing the complex structure moduli and the dilaton
via three-form fluxes inducing a Gukov-Vafa-Witten
type superpotential. Combining this with D3-instanton induced
contributions depending on the K\"ahler moduli
very predictive scenarios with in principle all moduli stabilised
have been proposed. These include in particular  the original
KKLT scenario~\cite{Kachru:2003aw} with supersymmetry breaking via an
uplift potential\footnote{In this sense the presence of anti-D3 branes in some of our models might turn out to be some use.}. In some respects even better controlled is the
LARGE volume scenario with K\"ahler moduli dominated
supersymmetry breaking~\cite{Balasubramanian:2005zx,Conlon:2005ki}.

In this latter scenario it is essential to have a Calabi-Yau with
negative Euler characteristic (that is, $h^{2,1}>h^{1,1}$) and shrinkable,
rigid four-cycles supporting D3-brane instantons contributing to the
superpotential. Such cycles are given by del Pezzo surfaces ${\rm
  dP}_n$ with $n\le 8$.  Therefore, gauge coupling unification with
$SU(5)$ breaking via $U(1)_Y$ fluxes and controllable moduli
stabilisation with natural supersymmetry breaking both lead us to the
class of Type IIB orientifolds (with some four-cycles $T_{Yuk}\to 0$ )
on Calabi-Yau manifolds which contain ${\rm dP}_n$ surfaces supporting
two-cycles which are trivial in the Calabi-Yau manifold.

We have already discussed in \autoref{swisskaas} that the $M_n^{({\rm
    dP}_8)^n}$ manifolds exhibit the swiss-cheese structure of the
volume form.  Thus, the class of Calabi-Yau manifolds studied in
Sections~\ref{sec:class_of_examples}, \ref{sec:GutModelExample},
and~\ref{sec:search} as promising candidates for GUT model building
likewise exhibits some attractive features for LARGE volume moduli
stabilisation.  For the $M_n^{({\rm dP}_8)^n}$ manifolds we have also
shown that placing the $SU(5)$ GUT on a shrinkable ${\rm dP}_8$ cycle,
the D-terms force this cycle to shrink to string scale size.  If one
is not deterred by the appearance of quantum corrections, one can
consider this either as a global embedding of local quiver
theories\footnote{While this work was in its final stages the authors
  of~\cite{Conlon:2008wa} proposed a very similar scenario.}. If one
tries to avoid such corrections the above observation can alternatively serve as a
motivation to place the GUT branes on del Pezzo ${\rm dP}_9$ or other
non-shrinkable rigid surfaces instead.  In this latter case, the
D-term constraints can be solved in the large radius regime.  This results in a scenario where the GUT branes are localised on ${\rm dP}_9$
surfaces while instantons on ${\rm dP}_8$ or lower del Pezzo surfaces
can generate the superpotential contributions realising the LARGE
volume scenario.  
At this stage we also point out that for consistency in a GUT model
the string scale must be fixed not below the GUT scale, of
course. Thus in our context the original LARGE volume scenario, if
applied, has to be modified anyway as to stabilise the volume of the
manifold at not too LARGE values ${\cal V}\simeq 10^4$.
 For a scenario leading to $M_s =
M_{GUT}$ along these lines see~\cite{Blumenhagen:2008kq}.

The  arrangement just described  also resolves the constraints
pointed out in~\cite{Blumenhagen:2007sm} for the coexistence of a
chiral MSSM or GUT like intersecting D7-brane sector on the one hand and of a
D3-brane instanton sector contributing to the uncharged
superpotential on the other.  Since here the phenomenologically relevant sizes of
the D7-brane cycles are fixed by the D-term constraints instead of the
F-terms, the resulting soft-terms and low-energy signatures are
expected to be different from the ones computed so far in the
literature for the LARGE volume
scenario~\cite{Conlon:2005ki,Conlon:2006wz,
  Aparicio:2008wh,Blumenhagen:2008kq}.  There, it was mostly assumed
that the string scale is in the intermediate regime and that the MSSM
supporting D7-branes wrap the same cycle as the D3-brane instanton.

Very similar conclusions follow from the analysis of the quintic
descendants. In contrast to the models derived from
$\IP_{1,1,1,6,9}[18]$, here we have found intersecting ${\rm dP}_n$, $n\le
8$ surfaces, which therefore do not show a swiss cheese structure and
allow the D-terms to freeze the K\"ahler moduli such that the
volumes of these del Pezzo remain finite.  To arrange for the LARGE or rather GUT
volume scenario for this class of models additional points at generic
positions have to be blown up, presumably resulting in additional
del Pezzo divisors, which are orthogonal to ones supporting D7-branes.

The next logical step is to combine fluxes, instanton effects and GUT
D7-brane sectors such that a completely realistic and predictive model
arises. For this purpose, one first needs to study the coexistence of
three-form fluxes and D7-branes on the same Calabi-Yau, for which
additional consistency conditions arise.  Here to mention is both the
Freed-Witten condition $H_3\vert_D=0$ and a possible change for the
quantisation of the gauge fluxes due to the presence of $F_3$ form
flux~\cite{GarciaEtxebarria:2005qc}.  Moreover, also the coexistence
of Euclidean D3-brane instanton contributions to the superpotential
and the desired presence of a chiral GUT D7-brane sector implies
additional constraints~\cite{Blumenhagen:2007sm}. All this to be
evaluated and taken into account carefully to claim to have realised
the MSSM or a variation therefore, from a string compactification.

This is a formidable task, but not out of reach in the not too 
far future. We think the results reported in this paper on 
GUT realisations in Type IIB orientifolds provide an encouraging step
towards achieving this goal.


\section{Conclusions} 
 
In this paper we have started to systematically analyse the construction
of Georgi-Glashow like $SU(5)$ GUTs from Type IIB
orientifolds with D7- and D3-branes. First,  we formulated  
the quite restrictive global model building rules. 
Beyond the common tadpole and  K-theory constraints, there
arise a couple of additional subtle but quite restrictive constraints.
These include the  delicate quantisation rules
for the gauge flux on the D7-branes wrapping rigid del Pezzo
surfaces, which derive from the fact that  del Pezzo surfaces
are not {\it Spin}. In addition these gauge fluxes have to be chosen
such that the D-term constraints can be satisfied inside
the K\"ahler cone of the Calabi-Yau threefold. 
Applied to $SU(5)$ GUT models, in particular the
quantisation conditions cannot be satisfied with
only the GUT breaking line bundle ${\cal L}_Y$ supported
on the $SU(5)$ stack. The presence of a second bundle
${\cal L}_a$ embedded into the diagonal $U(1)\subset U(5)$
is essential.
It would be interesting to study the precise
lift of the consistency conditions 
to the F-theory description of these models.
While some details are known, we think it is fair to say that
the general picture is still not fully understood.

After outlining the general structure, we have provided a class of
concrete Calabi-Yau threefolds containing del Pezzo surfaces. Though
the construction is more general, we first considered examples
descending from the Calabi-Yau manifold $\IP_{1,1,1,6,9}[18]$ via del
Pezzo transitions. The resulting Calabi-Yau threefolds feature various
phases (triangulations) related via flop-transitions of curves in the
del Pezzo base.  To define orientifolds of Type IIB on these
manifolds, we have classified all their involutions resulting from
involutions of the del Pezzo base.  This provides already a large set
of models, which deserves a more systematic (statistical)
investigation than we could provide in this paper.  Clearly, there
exist more general involutions which also act on the elliptic fibre.
The prototype example is just the $y\to -y$ involution of the elliptic
fibre.  It would be interesting to study these more general
orientifolds, as well.  Another natural route to pursue is to start
with the related torus fibred Calabi-Yau manifolds
$\IP_{1,1,1,3,6}[12]$ and $\IP_{1,1,1,3,3}[9]$. More generally, one
could study systematically which Calabi-Yau manifolds in the known
class of hypersurfaces in toric varieties allow for similar del Pezzo
transitions.

Equipped with the general structure and appropriate concrete
Calabi-Yau manifolds, we have manually  searched for 
globally consistent examples.
We have presented three models in detail, each  
realising around 60-70\% of the desired GUT features, with 
almost every property being realised in at least one example.
Therefore, we do not see any conceptual obstacle to finding
GUT models exhibiting all features in a single configuration.

We have also introduced a second class of suitable  Calabi-Yau manifolds
defined via del Pezzo transitions of the simple
quintic hypersurface in $\IP_4$. In particular, these manifolds 
contain intersecting shrinkable del Pezzo surfaces, 
a property the first class based on $\IP_{1,1,1,6,9}[18]$ was
lacking due to the swiss-cheese structure of the 
triple intersection form. Finally, we have presented two
GUT models with all matter fields localised on curves
and therefore without any additional vector-like matter, the second of which realises just three families of Standard Model matter plus a Higgs pair.
Clearly, it would be interesting to generalise 
the here presented techniques for constructing (toric)
Calabi-Yau manifolds containing del Pezzo surfaces.

Our emphasis has been on the global string consistency conditions,
which in a first attempt seem to be easier to analyse in the IIB
orientifold framework than for F-theory compactifications on compact
elliptically fibred Calabi-Yau fourfolds.  The price one has to pay
for working in the orientifold phase is that some couplings such as
top-Yukawa couplings and Majorana neutrino masses are
non-perturbatively generated by D3-brane instantons.  With the recent
understanding of such instanton effects we have however been able 
to formulate a
criterion respectively constraint for their presence in concrete
set-ups.  The realistic corner in the moduli space of these models is
clearly where the 4-cycles wrapped by these instantons go to zero size.  In
this respect it would be very important to better understand the
relation of the orientifold construction to the F-theory uplift on
Calabi-Yau fourfolds.

Eventually, we have briefly discussed the issue of moduli stabilisation
for these models. We have shown that the manifolds $M_n^{({\rm dP}_8)^n}$ 
indeed feature  a swiss-cheese
 structure, which is a prerequisite for realising
the LARGE volume scenario. 
We think it is striking that both from 
the viewpoint of realising GUTs and from the viewpoint of
phenomenologically acceptable moduli stabilisation
one is led to the same class of string constructions, namely
Type IIB orientifolds (F-theory) on Calabi-Yau threefolds
with shrinkable four-cycles, that is, del Pezzo surfaces.

\section*{Acknowledgements}

We gratefully acknowledge helpful discussions with Andres Collinucci, Mirjam Cveti{\v c}, Hans Jockers,
Shamit Kachru, Albrecht Klemm, Dieter L\"ust, Sebastian Moster, Hans Peter Nilles, Erik
Plauschinn, Maximilian Schmidt-Sommerfeld and Gary Shiu. Furthermore we thank  the CERN Theory Institute for hospitality during parts of this work.
R.B., V.B. and T.G. acknowledge the hospitality of the Erwin-Schr{\"o}dinger-Institut, Vienna, and T.W. thanks the Aspen Center for Theoretical Physics and the Max-Planck-Institut, Munich, for hospitality during parts of this project.
This work
was supported in parts by the European Union 6th framework program
MRTN-CT-2004-503069 ``Quest for unification'', MRTN-CT-2004-005104
``ForcesUniverse'', MRTN-CT-2006-035863 ``UniverseNet'',
SFB-Transregio 33 ``The Dark Universe'' by the DFG and by the US Department of Energy under contract DE-AC02-76SF00515.

\newpage
\appendix 
\noindent {\bf \LARGE Appendices}

\section{Involutions on Del Pezzo Surfaces}
\label{sec:dPinvolutions}

\subsection{Del Pezzo Surfaces of High Degree}
\label{sec:dPinvolutionsHighDeg}

This appendix is the completion and continuation of
\autoref{sec_delPezzobase}.  We start with a more detailed examination
of the del Pezzo surfaces of high degree $\geq 6$. In the
Subsections~\ref{sec:P2involutions}--\ref{sec:B3involutions} we will
consider each such del Pezzo surface individually and classify their
involutions\footnote{That is, the different connected components of
  the moduli space of involutions.}. However, before we go into the
details of the different involutions let us recall
\autoref{tab:dPinv1}
\begin{sidewaystable}[htbp]
  \centering
  \renewcommand{\arraystretch}{1.3}
  \small
  \begin{tabular}[htbp]{|c|c|c|c|c|c|c|}
    \hline
    $S$ & 
    $\deg(S)$ &
    case &
    action &
    fixed point set $S^\sigma$ & 
    $[S^\sigma] \in H_2$  &
    action on $H_2(S,\Z)$
    \\ \hline 
    $\CP^2$ & 
    $9$ &
    $1$ &
    eq.~\eqref{eq:CP2involution} &
    $[0:*:*] \cup [1:0:0] \simeq \CP^1 \cup \{\text{pt.}\}$ &
    $l$ &
    \begin{math}
      \begin{pmatrix}
        1
      \end{pmatrix}      
    \end{math}
    \\ \hline
    $\CP^1\times \CP^1$ & 
    $8$ &
    \ref{item:P1P1caseA} &
    eq.~\eqref{eq:P1P1caseA} &
    $[1:1|*:*]\cup [-1:1|*:*]$ &
    $(l_2) + (l_2)$ &
    \begin{math}
      \begin{pmatrix}
        1 & 0 \\ 0 & 1
      \end{pmatrix}      
    \end{math}
    \\ \hline
    $\CP^1\times \CP^1$ & 
    $8$ &
    \ref{item:P1P1caseB} &
    eq.~\eqref{eq:P1P1caseB} &
    $[\pm 1:1|\pm 1:1] \simeq 4\text{ points}$ &
    $0$ &
    \begin{math}
      \begin{pmatrix}
        1 & 0 \\ 0 & 1
      \end{pmatrix}      
    \end{math}
    \\ \hline
    $\CP^1\times \CP^1$ & 
    $8$ &
    \ref{item:P1P1caseC} &
    eq.~\eqref{eq:P1P1caseC} &
    diagonal $\CP^1$ &
    $l_1+l_2$ &
    \begin{math}
      \begin{pmatrix}
        0 & 1 \\ 1 & 0
      \end{pmatrix}      
    \end{math}
    \\ \hline
    $\cB_1$ & 
    $8$ &
    \ref{item:B1caseA} &
    eq.~\eqref{eq:B1hypersurfCaseA} &
    $\pi^{-1}([0:*:*]) \cup e_1 \simeq 2\CP^1$ &
    $(l) + (e_1)$ &
    \begin{math}
      \begin{pmatrix}
        1 & 0 \\ 0 & 1
      \end{pmatrix}      
    \end{math}
    \\ \hline
    $\cB_1$ & 
    $8$ &
    \ref{item:B1caseB} &
    eq.~\eqref{eq:B1hypersurfCaseB} &
    $\CP^1 \cup \big\{2\text{ pts.}\big\}$ &
    $l-e_1$ &
    \begin{math}
      \begin{pmatrix}
        1 & 0 \\ 0 & 1
      \end{pmatrix}      
    \end{math}
    \\ \hline
    $\cB_2$ & 
    $7$ &
    \ref{item:B2caseA} &
    eq.~\eqref{eq:B2caseA} &
    $2 \CP^1 \cup \big\{\text{pt.}\big\}$ &
    $(l-e_1) + (e_2)$ &
    \begin{math}
      \left(
        \begin{smallmatrix}
          1 & 0 & 0 \\ 0 & 1 & 0 \\ 0 & 0 & 1
        \end{smallmatrix}      
      \right)
    \end{math}
    \\ \hline
    $\cB_2$ & 
    $7$ &
    \ref{item:B2caseB} &
    eq.~\eqref{eq:B2caseB} &
    $\CP^1 \cup \big\{3\text{ pts.}\big\}$ &
    $l-e_1-e_2$ &
    \begin{math}
      \left(
        \begin{smallmatrix}
          1 & 0 & 0 \\ 0 & 1 & 0 \\ 0 & 0 & 1
        \end{smallmatrix}      
      \right)
    \end{math}
    \\ \hline
    $\cB_2$ & 
    $7$ &
    \ref{item:B2caseC} &
    eq.~\eqref{eq:B2caseC} &
    $\CP^1 \cup \big\{\text{pt.}\big\}$ &
    $l$ &
    \begin{math}
      \left(
        \begin{smallmatrix}
          1 & 0 & 0 \\ 0 & 0 & 1 \\ 0 & 1 & 0
        \end{smallmatrix}      
      \right)
    \end{math}
    \\ \hline
    $\cB_3$ & 
    $6$ &
    \ref{item:B3caseA} &
    eq.~\eqref{eq:B3caseA} &
    $2 \CP^1 \cup \{2\text{ pts.}\}$ &
    $(e_1) + (l-e_2-e_3)$ &
    \begin{math}
      \left(
        \begin{smallmatrix}
          1 & 0 & 0 & 0 \\ 0 & 1 & 0 & 0 \\ 0 & 0 & 1 & 0 \\ 0 & 0 & 0 & 1
        \end{smallmatrix}      
      \right)
    \end{math}
    \\ \hline
    $\cB_3$ & 
    $6$ &
    \ref{item:B3caseB} &
    eq.~\eqref{eq:B3caseB} &
    $\CP^1 \cup \big\{2\text{ pts.}\big\}$ &
    $l-e_3$ &
    \begin{math}
      \left(
        \begin{smallmatrix}
          1 & 0 & 0 & 0 \\ 0 & 0 & 1 & 0 \\ 0 & 1 & 0 & 0 \\ 0 & 0 & 0 & 1
        \end{smallmatrix}      
      \right)
    \end{math}
    \\ \hline
    $\cB_3$ & 
    $6$ &
    \ref{item:B3caseC} &
    eq.~\eqref{eq:B3caseC} &
    $\CP^1$ &
    $2l -e_1-e_2$ &
    \begin{math}
      \left(
        \begin{smallmatrix}
           2 &  1 &  1 &  1 \\ 
          -1 & -1 &  0 & -1 \\ 
          -1 &  0 & -1 & -1 \\ 
          -1 & -1 & -1 &  0
        \end{smallmatrix}      
      \right)
    \end{math}
    \\ \hline
    $\cB_3$ & 
    $6$ &
    \ref{item:B3caseD} &
    eq.~\eqref{eq:B3caseD} &
    $4\text{ points}$ &
    $0$ &
    \begin{math}
      \left(
        \begin{smallmatrix}
           2 &  1 &  1 &  1 \\ 
          -1 &  0 & -1 & -1 \\ 
          -1 & -1 &  0 & -1 \\ 
          -1 & -1 & -1 &  0
        \end{smallmatrix}      
      \right)
    \end{math}
    \\ \hline
  \end{tabular}
  \caption{Involutions on del Pezzo surfaces of degree $\geq 6$.}
  \label{tab:dPinv1}
\end{sidewaystable}

\subsection{Involutions on the Projective Plane}
\label{sec:P2involutions}

Let us start with the simplest del Pezzo surface, $\CP^2$. There are
no $(-1)$-curves. Up to coordinate transformations, the unique
involution acts on the homogeneous coordinates as
\begin{equation}
  \label{eq:CP2involution}
  \sigma:\CP^2\to\CP^2
  ,\quad
  [z_0:z_1:z_2] \mapsto [-z_0:z_1:z_2]
  .
\end{equation}
The fixed point set of the involution $\sigma$ is 
\begin{equation}
  \Big( \CP^2 \Big)^\sigma = 
  \Big\{ [0:*:*] \Big\} \cup \Big\{ [1:0:0] \Big\} 
  \simeq \CP^1 \cup \{\text{pt}\} 
  ,
\end{equation}
and its homology class is $l \in H_2(\CP^2,\Z)$.

The projective plane is a toric variety, determined by the 
$2$-dimensional polytope shown in \autoref{fig:toricP2symm} as
follows. Associate one complex-valued variable to each point of the
polytope. Here, we label them $x_0$, $x_1$, and $x_2$. Whenever there
are two points that are not connected by a line, the two variables are
not allowed to vanish simultaneously. This does not happen here, but
will be important later on. Finally, for each linear relation amongst
the points we impose an equivalence under ``homogeneous''
rescaling. For example, the single linear relation
\begin{multline}
  \overbrace{\big( 0, 1\big)}^{x_1} + 
  \overbrace{\big(-1,-1\big)}^{x_0} + 
  \overbrace{\big( 1, 0\big)}^{x_2} = 0
  \\
  \Rightarrow \quad
  [x_0:x_1:x_2] = 
  [\lambda x_0: \lambda x_1: \lambda x_2] 
  \quad \forall \lambda \in \C^\times
\end{multline}
corresponds to the usual rescaling of the homogeneous
coordinates. Hence, the toric description of the projective plane is
\begin{equation}
  \frac{ \C^3 - \{\} }{ \C^\times }
  = 
  \CP^2
  .
\end{equation}
By construction, the ``algebraic torus'' $\big(\C^\times\big)^3$
acts\footnote{Clearly, the diagonal $\C^\ast$ is already modded out
  and acts trivially.} multiplicatively on the $3$ homogeneous
variables of $\CP^2$, hence the name toric variety. Any subgroup of
this action is called a toric group action. In particular, the
involution eq.~\eqref{eq:CP2involution} is a toric $\Z_2$ action.

Alternatively, the involution can be seen as a symmetry of the
polytope. The reflection symmetry
\begin{figure}[htbp]
  \centering
  \input{toricP2symm.pspdftex}
  \caption{Symmetry of the toric polytope defining $\CP^2$.}
  \label{fig:toricP2symm}
\end{figure}
shown in \autoref{fig:toricP2symm} generates the involution
\begin{equation}
  \label{eq:CP2involutionPermutation}
  \sigma:
  \frac{\C^3}{\C^\times} \to \frac{\C^3}{\C^\times}
  ,\quad
  [x_0:x_1:x_2] \mapsto [x_0:x_2:x_1]
  .
\end{equation}
This is the same group action as in eq.~\eqref{eq:CP2involution}, only
written in different coordinates\footnote{For future reference, we
  note that the fixed point set in the coordinates
  eq.~\eqref{eq:CP2involutionPermutation} is
  \begin{equation}
    \label{eq:CP2involutionPermutationInv}
    \Big( \CP^2 \Big)^\sigma = 
    \Big\{ [t_0:t_1:t_1] \Big| [t_0:t_1]\in \CP^1 \Big\} \cup \Big\{ [0:1:-1] \Big\} 
    \simeq \CP^1 \cup \{\text{pt}\} 
    .    
  \end{equation}
}.

\subsection{Involutions on the Product of Lines}
\label{sec:P1P1involutions}

\begin{figure}[htbp]
  \centering
  \input{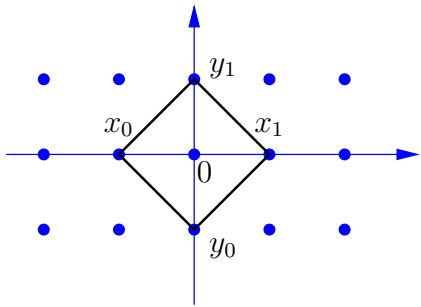}
  \caption{The toric polytope defining $\CP^1\times \CP^1$.}
  \label{fig:toricP1P1}
\end{figure}

There is only one non-trivial involution on $\CP^1$ acting as
$[z_0:z_1]\mapsto [-z_0:z_1]$, which can act on each factor of
$\CP^1\times\CP^1$. Together with the exchange of the two factors,
this generates all possible holomorphic involutions on
$\CP^1\times\CP^1$. All of these involutions arise from symmetries of
the toric polytope, see \autoref{fig:toricP1P1}.
The symmetry group of the toric polytope is $D_8$, the dihedral group
with $8$ elements. It has three conjugacy classes of $\Z_2$ subgroups,
namely:
\begin{subequations}
  \begin{enumerate}
  \item\label{item:P1P1caseA} Mirroring at vertical axis. In homogeneous
    coordinates, the induced action on $\CP^1\times\CP^1$ is
    \begin{equation}
      \label{eq:P1P1caseA}
      \sigma_1:\CP^1\times\CP^1\to\CP^1\times\CP^1
      ,\quad
      [x_0:x_1|y_0:y_1] \mapsto [x_1:x_0|y_0:y_1]
      .
    \end{equation}
  \item\label{item:P1P1caseB} Rotating by $\pi$, with induced action
    \begin{equation}
      \label{eq:P1P1caseB}
      \sigma_2:\CP^1\times\CP^1\to\CP^1\times\CP^1
      ,\quad
      [x_0:x_1|y_0:y_1] \mapsto [x_1:x_0|y_1:y_0]
      .
    \end{equation}
  \item\label{item:P1P1caseC} Mirroring at diagonal axis = Rotate by
    $\frac{\pi}{2}$ and mirror at vertical axis. The induced action is
    \begin{equation}
      \label{eq:P1P1caseC}
      \sigma_3:\CP^1\times\CP^1\to\CP^1\times\CP^1 
      ,\quad
      [x_0:x_1|y_0:y_1] \mapsto [y_0:y_1|x_0:x_1]
      .
    \end{equation}
  \end{enumerate}
\end{subequations}
According to the K\"unneth theorem, the homology group
$H_2(\CP^1\times\CP^1)=\Z^2$ is generated by the classes of the two
factors, which we call $l_1$ and $l_2$. The fixed point sets and their
homology classes are straightforward and listed in \autoref{tab:dPinv1}.

\subsection{Blow-up of the Projective Plane}
\label{sec:B1involutions}

We now come to the first case with a $(-1)$-curve, namely the blow-up $\cB_1$
of $\CP^2$ at one point. One possible realisation is the hypersurface
\begin{equation}
  \label{eq:blowupP2P1}
  \cB_1 =
  \Big\{ x_0 \cdot 0 + x_1 t_0 + x_2 t_1 = 0 \Big\}
  \subset \CP^2_{[x_0:x_1:x_2]} \times \CP^1_{[t_0:t_1]}
  .
\end{equation}
The obvious projection $\pi:\cB_1 \to \CP^2$,
$[x_0:x_1:x_2|t_0:t_1]\mapsto [x_0:x_1:x_2]$ is, in fact, the
corresponding blow-down map. To see this, consider the preimage:
\begin{itemize}
\item If $[x_0:x_1:x_2]\not=[1:0:0]$, then the preimage is the single
  point
  \begin{equation}
    \pi^{-1}\Big( [x_0:x_1:x_2] \Big) = [x_0:x_1:x_2|x_2:-x_1]
  \end{equation}
\item If $[x_0:x_1:x_2]=[1:0:0]$, then the preimage is
  \begin{equation}
    \pi^{-1}\Big( [1:0:0] \Big) = 
    \Big\{ [1:0:0|t_0:t_1] \Big| [t_0:t_1]\in \CP^1 \Big\}
    \simeq \CP^1
  \end{equation}
\end{itemize}
In other words, the hypersurface eq.~\eqref{eq:blowupP2P1} is the
blow-up at $[1:0:0]\in \CP^2$.

We now consider involutions of the hypersurface induced from the
ambient space $\CP^2\times\CP^1$. In fact, up to coordinate changes
there are two distinct possibilities, namely
\begin{subequations}
  \begin{equation}
    \label{eq:B1hypersurfCaseA}
    \sigma_1: \cB_1\to\cB_1, \quad
    \big[ x_0:x_1:x_2 \big| t_0:t_1 \big]
    \mapsto 
    \big[ -x_0:x_1:x_2 \big| t_0:t_1 \big]
  \end{equation}
  and
  \begin{equation}
    \label{eq:B1hypersurfCaseB}
    \sigma_2: \cB_1\to\cB_1, \quad
    \big[ x_0:x_1:x_2 \big| t_0:t_1 \big]
    \mapsto 
    \big[ x_0:x_2:x_1 \big| t_1:t_0 \big]
    .
  \end{equation}
\end{subequations}
In terms of the blown-up $\CP^2$, the two involutions can be
understood as follows. Recall from \autoref{sec:P2involutions} that
the fixed-point set on $\CP^2$ is the disjoint union of a line and a point.
\begin{enumerate}
\item\label{item:B1caseA} The first involution,
  eq.~\eqref{eq:B1hypersurfCaseA}, is the blow-up of the isolated
  fixed point on $\CP^2$. The corresponding fixed-point set in $\cB_1$
  is the whole exceptional $\CP^1$ as well as the fixed line in
  $\CP^1$. This $\Z_2$ group action is toric.
\item\label{item:B1caseB} The second involution,
  eq.~\eqref{eq:CP2involutionPermutationInv}, is the blow-up at a
  point on the fixed line on $\CP^2$. The exceptional $\CP^1$ is
  mapped to itself, but it is not point-wise fixed. Rather, the
  involution acts as a rotation by $\pi$ on this $\CP^1\simeq S^2$ and
  the north and south pole of the sphere end up being fixed. Looking
  at the whole $\cB_1$, the proper transform of the fixed line in
  $\CP^2$ passes through one of the fixed points in the exceptional
  curve. Hence, the fixed point set consists of this proper
  transform\footnote{\label{fnote:propertransform}The $\sigma_2$-fixed
    line $l_2$ in $\CP^2$ has the parametrisation
    \begin{equation}
      \xi \mapsto l_2(\xi) = [1:\xi:\xi]
      ,\quad
      \xi \in \C\cup\{\infty\} \simeq \CP^1
      .
    \end{equation}
    It passes through the point $l_2(0)=[1:0:0]$, which we
    are about to blow up. By definition, the proper transform is the
    curve
    \begin{equation}
      \tilde l =
      \quad 
      \pi^{-1}\circ l_2\Big( \CP^1 - \{0\} \Big) 
      ~\cup~ 
      \lim_{\xi \to 0} 
      \Big( \pi^{-1}\circ l_2\big(\xi\big) \Big)
      \quad \subset \quad
      \cB_1
      .
    \end{equation}
    Since $\tilde l \cdot l = 1 = \tilde l \cdot
    e_1$, the homology class of the proper transform must be $[\tilde
    l] = l - e_1$.}  $\tilde l \simeq \CP^1$ together with the
  remaining fixed point in the exceptional curve and the isolated
  fixed point that was already in $\CP^2$.
  
  In terms of toric geometry, this involution is induced from the
  reflection symmetry of the polyhedron shown in
  \autoref{fig:toricB2symm}
  \begin{figure}[htbp]
    \centering
    \input{toricB1symm.pspdftex}
    \caption{Symmetry of the toric polytope defining $\cB_1$.}
    \label{fig:toricB2symm}
  \end{figure}
\end{enumerate}
More abstractly, we can understand the two involutions from the
$(-1)$-curves. Since there is precisely one such curve, namely $e_1$,
this curve must necessarily be mapped to itself under any
involution. But whenever there is an \emph{invariant}\footnote{One can
  of course blow down any $(-1)$-curve and obtain a smooth surface,
  but the involution is lost (or, rather, becomes a birational map) if
  the $(-1)$-curve was not invariant.} exceptional curve on $\cB_n$,
then we can blow down this curve and obtain a involution on
$\cB_{n-1}$ (or $\CP^1\times\CP^1$ if $n=2$). This is why every
involution on $\cB_1$ is simply the (unique) involution on $\CP^2$
blown up at a fixed point. There are two connected components to the
fixed point set, and the choice of blow-up point coincides with the
two different ways that $e_1\simeq \CP^1$ can be mapped to itself:
\begin{enumerate}
\item If one blows up the isolated fixed point on $\CP^2$, then $e_1$
  is point-wise fixed under the induced involution on $\cB_1$.
\item If one blows up one point in the fixed line on $\CP^2$, then
  the induced involution on $\cB_1$ acts on $e_1\simeq S^2$ as
  rotation by $\pi$. 
\end{enumerate}

\subsection{Blow-up of the Projective Plane at Two Points}
\label{sec:B2involutions}

\begin{figure}[htbp]
  \centering
  \input{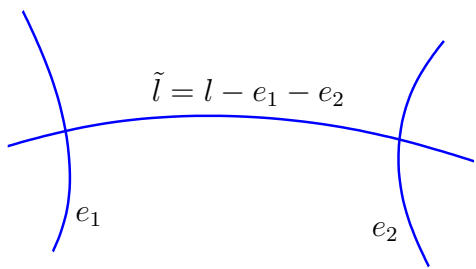}
  \caption{On the left, intersection pattern of the three
    $(-1)$-curves on $\cB_2$. The dual graph is shown on the right.}
  \label{fig:B2graph}
\end{figure}

The blow-up of $\CP^2$ at two points, $\cB_2$, is the first case with
an interesting pattern of $(-1)$-curves. Clearly, there are the
exceptional divisors $e_1$ and $e_2$. But there is also a third rigid
curve, namely the line on $\CP^2$ through the two blow-up points. As
we reviewed in \autoref{fnote:propertransform}, this line defines a
curve on the blow-up $\cB_2$. This so-called proper transform has
homology class
\begin{equation}
  \tilde l = l - e_1 -e_2
  .
\end{equation}
One can easily check that $\tilde{l}^2=-1$, as expected for a rigid
curve. We draw the intersection pattern of the three lines in
\autoref{fig:B2graph}.
In the following, we will always use the dual graph of the
$(-1)$-curves (and, by abuse of notation, drop the ``dual''). By
definition, this is the graph with
\begin{itemize}
\item One node for each $(-1)$-curve, and
\item One connecting line whenever two curves intersect. 
\end{itemize}
There are two different kinds of nodes, one of valence $2$ and two of
valence $1$. Blowing down the middle node yields $\CP^1\times \CP^1$,
while blowing down one of the nodes at the end yields $\cB_1$.

Clearly, the automorphism group of the graph is $\Z_2$ and the middle
node is always fixed. Hence, the easiest way to describe all 
involutions is as blow-up of $\CP^1\times\CP^1$, where there were
three distinct involutions. Just an in eq.~\eqref{eq:blowupP2P1}, we
will realise the blow-up at the point $[\xi_0:\xi_1|\eta_0:\eta_1]\in
\CP^1\times\CP^1$ as a degree-$(1,1,1)$ hypersurface
\begin{equation}
  \label{eq:B2hypersurface}
  \cB_2 = 
  \Big\{ 
  (\xi_1  x_0 - \xi_0  x_1) t_0 + 
  (\eta_1 y_0 - \eta_0 y_1) t_1  = 0 
  \Big\}
  \subset \CP^1_{[x_0:x_1]} \times \CP^1_{[y_0:y_1]} \times \CP^1_{[t_0:t_1]}
  .
\end{equation}
Using this construction, we can characterise the three different
involutions on $\cB_2$ as follows:
\begin{enumerate}
\item\label{item:B2caseA} First, let us start by blowing up
  ($\CP^1\times\CP^1,\sigma_1$) at a $\sigma_1$-fixed point. The fixed
  point set consists of two disjoint $\CP^1$, so one might think that
  there is a discrete choice. However, the two $\CP^1$ are exchanged
  by a remaining symmetry of $\CP^1\times\CP^1$, so they cannot be
  distinguished. Henceforth, we will pick the fixed point 
  \begin{equation}
    p = [1:1|0:1] 
    \in \CP^1\times\CP^1
  \end{equation}
  and define
  \begin{equation}
    \cB_2 = 
    \Big\{ 
    (x_0-x_1) t_0 + y_0 t_1  = 0 
    \Big\}
    \subset \CP^1_{[x_0:x_1]} \times \CP^1_{[y_0:y_1]} \times \CP^1_{[t_0:t_1]}
    .
  \end{equation}
  In order to make the hypersurface equation invariant under the
  involution, we must extend eq.~\eqref{eq:P1P1caseA} to
  \begin{equation}
    \label{eq:B2caseA}
    \sigma_1:\cB_2\to\cB_2
    ,\quad
    [x_0:x_1|y_0:y_1|t_0:t_1] \mapsto [x_1:x_0|y_0:y_1|-t_0:t_1]
    .
  \end{equation}
\item\label{item:B2caseB} Now we blow up one of the $4$ fixed points
  of $(\CP^1\times\CP^1,\sigma_2)$. Again, the fixed points are
  exchanged by residual symmetries, and cannot be
  distinguished. Hence, there is essentially only one choice which we
  take to be
  \begin{equation}
    \label{eq:B2fixedpointBC}
    p = [1:1|1:1] 
    \in \CP^1\times\CP^1
    .
  \end{equation}
  The blow-up with the induced involution is then
  \begin{gather}\
    \label{eq:B2hypersurfaceB}
    \cB_2 = 
    \Big\{ 
    (x_0-x_1) t_0 + (y_0-y_1) t_1  = 0 
    \Big\}
    \subset \CP^1_{[x_0:x_1]} \times \CP^1_{[y_0:y_1]} \times
    \CP^1_{[t_0:t_1]}
    ,
    \\ 
    \label{eq:B2caseB}
    \sigma_2:\cB_2\to\cB_2
    ,\quad
    [x_0:x_1|y_0:y_1|t_0:t_1] \mapsto [x_1:x_0|y_1:y_0|t_0:t_1]
    .
  \end{gather}
\item\label{item:B2caseC} Finally, we can blow-up one point on the
  fixed (diagonal) $\CP^1$ in $(\CP^1\times\CP^1,\sigma_3)$. For
  concreteness, let us take the point in
  eq.~\eqref{eq:B2fixedpointBC}, which is also fixed under
  $\sigma_3$. Hence, the hypersurface equation is the same as in
  eq.~\eqref{eq:B2hypersurfaceB}. However, the induced involution has
  to extend a different involution on $\CP^1\times\CP^1$, and must be
  \begin{equation}
    \label{eq:B2caseC}
    \sigma_3:\cB_2\to\cB_2
    ,\quad
    [x_0:x_1|y_0:y_1|t_0:t_1] \mapsto [y_0:y_1|x_0:x_1|t_1:t_0]
    .
  \end{equation}
\end{enumerate}
Equivalently, the three involutions can be described as blow-ups of
$\CP^2$ at two points. Let us quickly go over this equivalent point of
view.
\begin{enumerate}
\item The first two involutions correspond to the trivial automorphism
  of the graph of $(-1)$-curves. Hence, one must blow up two fixed
  points of $(\CP^2,\sigma)$. The first possibility is to blow up the
  isolated fixed point and one point on the fixed line in $\CP^2$. As
  we saw in \autoref{sec:B1involutions}, blowing up a point on the
  fixed $\CP^1$ leaves the proper transform $\tilde l$ (in the
  homology class $l-e_1$) and one isolated point fixed. Blowing up the
  fixed point leads to a point-wise fixed exceptional divisor $e_2$.
\item The second involution is the blow-up of two points on the fixed
  $\CP^1$ of $(\CP^2,\sigma)$. The fixed point set consists of
  \begin{itemize}
  \item The proper transform of the fixed line. Its homology class is
    $l-e_1-e_2$.
  \item One isolated fixed point on $e_1$ and one on $e_2$.
  \item The isolated fixed point that was already on $\CP^2$.
  \end{itemize}
\item The last involution corresponds to the non-trivial automorphism
  of the graph. There, the involution must exchange the two blow-up
  points. The fixed point set is the same as on $(\CP^2,\sigma)$.
\end{enumerate}
Clearly, the first two involutions act trivially on
$H_2(\cB_2,\Z)$. The third involution exchanges $e_1\leftrightarrow
e_2$ while leaving $l$ invariant.

\subsection{Blow-up of the Projective Plane at Three Points}
\label{sec:B3involutions}

The del Pezzo surface $\cB_3$ has $6$ rigid lines, the $3$ exceptional
divisors together with the $3$ lines connecting any pair of blow-up
points. Their homology classes are
\begin{equation}
  e_1
  ,~
  e_2
  ,~
  e_3
  ,~
  l - e_2-e_3
  ,~
  l - e_1-e_3
  ,~
  l - e_2-e_3
  .
\end{equation}
The graph of $(-1)$-curves is a hexagon, whose automorphism group is
$D_{12}$, the dihedral group with $12$ elements. It has $3$ conjugacy
\begin{figure}[htbp]
  \centering
  $\qquad$\input{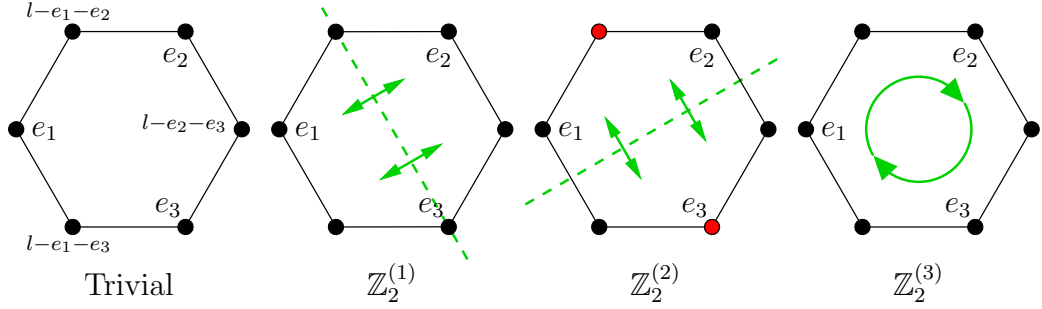}
  \caption{The trivial and the three order-$2$ automorphisms of the
    graph of $(-1)$-curves on $\cB_3$.}
  \label{fig:B3graph}
\end{figure}
classes of order $2$, which are depicted in \autoref{fig:B3graph}. We
now investigate which involutions on $\cB_3$ give rise to each graph
automorphism.
\begin{enumerate}
\item\label{item:B3caseA} Let us start with the trivial action on the
  graph. This is necessarily the blow-up of $(\CP^2,\sigma)$ at three
  fixed points. Note, however, that by construction no three point can
  lie on any one line and, in particular, not on the fixed
  line. Therefore, there is (up to coordinate changes) only one
  choice\footnote{Namely the isolated fixed point and two points on
    the fixed line in $\CP^2$.} of three fixed points to blow up. In
  the coordinates given by eq.~\eqref{eq:CP2involution}, these three
  points can be chosen to be
  \begin{equation}
    p_1 = [1:0:0]
    ,~
    p_2 = [0:1:0]
    ,~
    p_3 = [0:0:1]
  \end{equation}
  The first involution on $\cB_3$ is the one defined through the blow-up,
  \begin{equation}
    \label{eq:B3caseA}
    \big( \cB_3,\sigma_1) = 
    \Bl_{\{p_1,p_2,p_3\}}\big( \CP^2,\sigma \big)
    .
  \end{equation}
  We denote by $e_i$ the exceptional divisor of the blow-up at
  $p_i$. With this notation, the fixed point set consists of the
  exceptional divisor $e_1$, the proper transform of the fixed
  $\CP^1$, one isolated point on $e_2$, and one isolated point on
  $e_3$.
\item\label{item:B3caseB} We now consider the first non-trivial
  involution on the graph of $(-1)$-lines, which is denoted
  $\Z_2^{(1)}$ in \autoref{fig:B3graph}. There are two fixed
  $(-1)$-curves, one of which we already labelled $e_3$. Blowing down
  this exceptional divisor $e_3$, we clearly obtain the surface
  $\cB_2$ with the non-trivial action on its graph of
  $(-1)$-curves. There is only one such involution, namely
  $(\cB_2,\sigma_3)$. Therefore, the desired involution on $\cB_3$ is
  the blow-up of $(\cB_2,\sigma_3)$ at a fixed point. 

  Recall that $(\cB_2,\sigma_3)$ is the blow-up of $(\CP^2,\sigma)$ at
  a point--image point pair. We can pick coordinates such that
  \begin{equation}
    \big(\cB_2,\sigma_3\big) = 
    \Bl_{\big\{[1:1:0], [-1:1:0]\big\}} \big(\CP^2,\sigma\big)
    .
  \end{equation}
  The fixed point set of $(\cB_2,\sigma_3)$ has two connected
  components, an isolated point $[1:0:0]$ and
  $[0:*:*]\simeq\CP^1$. Note, however, that the isolated point is
  collinear with the previous blow-up points,
  \begin{equation}
    \big\{[1:1:0], [-1:1:0], [1:0:0]\big\}
    \in 
    \Big\{ [*:*:0] \Big\}
    .
  \end{equation}
  Hence, we cannot blow up $(\cB_2,\sigma_3)$ at the isolated fixed
  point. The only possibility is to pick a fixed point
  $q\not=[1:0:0]$. This point $q$ must lie on the fixed $\CP^1$ of
  $(\cB_2,\sigma_3)$. Hence
  we obtain the involution
  \begin{equation}
    \label{eq:B3caseB}
    \big(\cB_3,\sigma_2\big) = \Bl_q\big( \cB_2,\sigma_3 \big)
    .
  \end{equation}  
\item\label{item:B3caseC} Now, consider the $\Z_2^{(2)}$-automorphism
  shown in \autoref{fig:B3graph}. There is no fixed $(-1)$-curve, so
  we cannot describe it in terms of a blown-up $\cB_2$ del Pezzo. We
  can, however, blow down a pair of $(-1)$-curves that is exchanged by
  the involution \emph{and} does not intersect. There is only one such
  pair, marked in red. Blowing down this pair yields a del Pezzo
  surface of degree $8$ without any remaining $(-1)$-curves, that is,
  $\CP^1\times\CP^1$.

  There are three choices for the involution on $\CP^1\times\CP^1$. It
  turns out\footnote{On $(\CP^1\times\CP^1,\sigma_1)$ there is no
    suitable pair of non-fixed points to blow up into $\cB_3$. The
    second involution, $(\CP^1\times\CP^1,\sigma_3)$, will be used in
    \autoref{item:B3caseD}.} that $(\CP^1\times\CP^1,\sigma_3)$ gives
  rise to the right graph automorphism. Therefore, we set
  \begin{equation}
    \label{eq:B3caseC}
    \big( \cB_3,\sigma_3 \big) 
    = 
    \Bl_{\{p,q\}}\big( \CP^1\times\CP^1, \sigma_3 \big)
    ,
  \end{equation}
  where $p,q$ is a generic point and its image on
  $(\CP^1\times\CP^1,\sigma_3)$. The fixed point set 
  $\cB_3^{\sigma_4}$ is, by construction, the same as
  $(\CP^1\times\CP^1)^{\sigma_3}$. Since the $(-1)$-curves span
  $H_2(\cB_3,\Z)$, the action on the curve homology classes can be
  read off from \autoref{fig:B3graph}. One obtains
  \begin{equation}
    \begin{pmatrix}
      l \\ e_1 \\ e_2 \\ e_3
    \end{pmatrix}
    \mapsto
    \begin{pmatrix}
      2l-e_1-e_2-e_3 \\ l-e_1-e_3 \\ l-e_2-e_3 \\ l-e1-e_2
    \end{pmatrix}
  \end{equation}
\item\label{item:B3caseD} Finally, consider $\Z_2^{(3)}$. Again, there
  is no $(-1)$-curve, but we can blow down a pair of $(-1)$-curves and
  relate the involution to $\CP^1\times\CP^1$. In fact, only one
  involution on $\CP^1\times\CP^1$ gives rise to the desired graph
  automorphism. Hence, we set
  \begin{equation}
    \label{eq:B3caseD}
    \big( \cB_3,\sigma_4 \big) 
    = 
    \Bl_{\{p,q\}}\big( \CP^1\times\CP^1, \sigma_2 \big)
    ,
  \end{equation}
  where $p,q$ is a generic point and its image on
  $(\CP^1\times\CP^1,\sigma_2)$.
\end{enumerate}

\subsection{The Weyl Group and The Graph of Lines}
\label{sec:WeylLines}

Recall that, by definition, the degree of a curve is its intersection
with the canonical class $K = -3l + \sum_i e_i$.  
\begin{sidewaystable}[htbp]
  \centering
  \renewcommand{\arraystretch}{1.3}
  \begin{tabular}{|c|c|c|c|c|c|c|c|}
    \hline
    del Pezzo $S$ & 
    $\deg(S)$ &
    \renewcommand{\arraystretch}{1}
    \begin{tabular}{c}
      \# of\\ roots
    \end{tabular} &
    root lattice &
    Weyl group $W$ &
    Order $|W|$ &
    \renewcommand{\arraystretch}{1}
    \begin{tabular}{c}
      Number of $\Z_2$\\ conjugacy classes
    \end{tabular} &
    graph 
    \\ \hline
    $\CP^2$            & $9$ & $0$ & $\{\}$ & $1$ & $1$ & $0$ & $\{\}$ \\ \hline
    $\CP^1\times\CP^1$ & $8$ & $0$ &   $\{\}$ & $1$ & $1$ & $0$ & $\{\}$  \\ \hline
    $\cB_1$            & $8$ & $0$ &   $\{\}$ & $1$ & $1$ & $0$ & $\bullet$ \\ \hline
    $\cB_2$            & $7$ & $2$ &   $A_1$ & $\Z_2$ & $2$ & $1$ &
    \begin{math}
      \xygraph{
        !{<0cm,0cm>;<5mm,0mm>:<0mm,5mm>::}
        !{(-1,0)}*{\bullet}="a1"
        !{( 0,0)}*{\bullet}="a2"
        !{( 1,0)}*{\bullet}="a3"
        "a1"-"a2" "a2"-"a3" 
      } 
    \end{math}
    \\ \hline
    $\cB_3$            & $6$ & $8$ &   $A_2\oplus A_1$ & $D_6\times\Z_2 = D_{12}$ & $12$ & $3$ &
    ${
      \xygraph{
        !{<0cm,0cm>;<5mm,0mm>:<0mm,5mm>::}
        !{(0,-1)}*{\strut}
        !{(0, 1)}*{\strut}
        !{(0,0);a(0)**{}?(1.0)}*{\bullet}="a1"
        !{(0,0);a(60)**{}?(1.0)}*{\bullet}="a2"
        !{(0,0);a(120)**{}?(1.0)}*{\bullet}="a3"
        !{(0,0);a(180)**{}?(1.0)}*{\bullet}="a4"
        !{(0,0);a(240)**{}?(1.0)}*{\bullet}="a5"
        !{(0,0);a(300)**{}?(1.0)}*{\bullet}="a6"
        "a1"-"a2" "a2"-"a3" "a3"-"a4" "a4"-"a5" "a5"-"a6" "a6"-"a1"
      } 
    }$
    \\ \hline
    $\cB_4$ &  $5$ & $20$ &  $A_4$ & $S_5$ & $120$ & $2$ & 
    \renewcommand{\arraystretch}{1}
    \begin{tabular}{c}
      Petersen graph\\
      \autoref{fig:SpecialGraphs}
    \end{tabular} 
    \\ \hline
    $\cB_5$ &  $5$ & $40$ &  $D_5$ & $\text{Weyl}(D_5)$ & $1920$ & $5$ & 
    \renewcommand{\arraystretch}{1}
    \begin{tabular}{c}
      Clebsch graph\\
      \autoref{fig:SpecialGraphs}
    \end{tabular} 
    \\ \hline
    $\cB_6$ &  $5$ & $72$ &  $E_6$ & $\text{Weyl}(E_6)$ & $51840$ & $4$ &  $27$ nodes \\ \hline
    $\cB_7$ &  $5$ & $126$ &  $E_7$ & $\text{Weyl}(E_7)$ & $2903040$ & $9$ &  $56$ nodes \\ \hline
    $\cB_8$ &  $5$ & $240$ &  $E_8$ & $\text{Weyl}(E_8)$ & $696729600$ & $9$ & $240$ nodes \\ \hline
  \end{tabular}
  \caption{The Weyl groups and the number of $\Z_2$ conjugacy
    classes. The Weyl group equals the automorphism group of the
    graph of $(-1)$-curves on each del Pezzo
    surfaces.}
  \label{tab:linesautomorphisms}
\end{sidewaystable}
\begin{figure}[htbp]
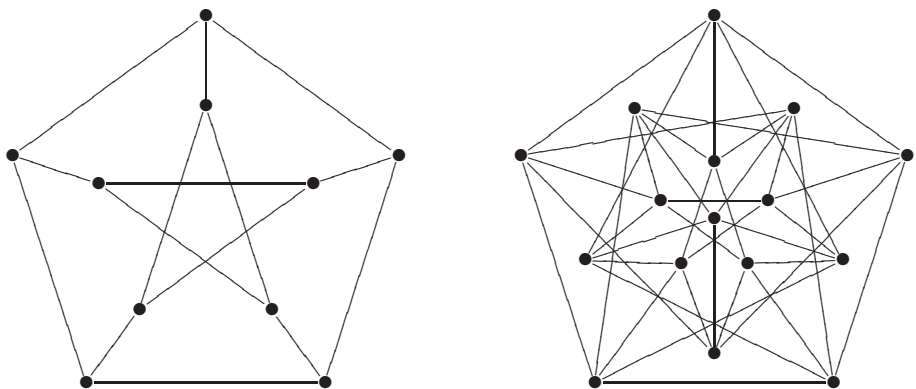

  \centering
  \mbox{
    \xygraph{
      !{<0cm,0cm>;<0cm,1.5cm>:<1.5cm,0cm>::}
      !{(0,0);a(0)**{}?(1.0)}*{\bullet}="a1"
      !{(0,0);a(72)**{}?(1.0)}*{\bullet}="a2"
      !{(0,0);a(144)**{}?(1.0)}*{\bullet}="a3"
      !{(0,0);a(216)**{}?(1.0)}*{\bullet}="a4"
      !{(0,0);a(288)**{}?(1.0)}*{\bullet}="a5"
      !{(0,0);a(0)**{}?(1.8)}*{\bullet}="b1"
      !{(0,0);a(72)**{}?(1.8)}*{\bullet}="b2"
      !{(0,0);a(144)**{}?(1.8)}*{\bullet}="b3"
      !{(0,0);a(216)**{}?(1.8)}*{\bullet}="b4"
      !{(0,0);a(288)**{}?(1.8)}*{\bullet}="b5"
      "a1"-"a3" "a3"-"a5" "a5"-"a2" "a2"-"a4" "a4"-"a1"
      "b1"-"b2" "b2"-"b3" "b3"-"b4" "b4"-"b5" "b5"-"b1"
      "a1"-"b1" "a2"-"b2" "a3"-"b3" "a4"-"b4" "a5"-"b5"
    } 
  }
  \hspace{1cm}
  \mbox{
    \xygraph{
      !{<0cm,0cm>;<0cm,1.5cm>:<1.5cm,0cm>::}
      !{(0,0);a(0)**{}?(1.8)}*{\bullet}="a1"
      !{(0,0);a(72)**{}?(1.8)}*{\bullet}="a2"
      !{(0,0);a(144)**{}?(1.8)}*{\bullet}="a3"
      !{(0,0);a(216)**{}?(1.8)}*{\bullet}="a4"
      !{(0,0);a(288)**{}?(1.8)}*{\bullet}="a5"
      !{(0,0);a(36)**{}?(1.2)}*{\bullet}="b1"
      !{(0,0);a(108)**{}?(1.2)}*{\bullet}="b2"
      !{(0,0);a(180)**{}?(1.2)}*{\bullet}="b3"
      !{(0,0);a(252)**{}?(1.2)}*{\bullet}="b4"
      !{(0,0);a(324)**{}?(1.2)}*{\bullet}="b5"
      !{(0,0);a(0)**{}?(0.5)}*{\bullet}="c1"
      !{(0,0);a(72)**{}?(0.5)}*{\bullet}="c2"
      !{(0,0);a(144)**{}?(0.5)}*{\bullet}="c3"
      !{(0,0);a(216)**{}?(0.5)}*{\bullet}="c4"
      !{(0,0);a(288)**{}?(0.5)}*{\bullet}="c5"
      !{(0,0);a(0)**{}?(0)}*{\bullet}="o"      
      "a1"-"c1" "a2"-"c2" "a3"-"c3" "a4"-"c4" "a5"-"c5"
      "b1"-"o" "b2"-"o" "b3"-"o" "b4"-"o" "b5"-"o"
      "a1"-"a2" "a2"-"a3" "a3"-"a4" "a4"-"a5" "a5"-"a1"
      "a1"-"b2" "a2"-"b3" "a3"-"b4" "a4"-"b5" "a5"-"b1"
      "a1"-"b4" "a2"-"b5" "a3"-"b1" "a4"-"b2" "a5"-"b3"
      "b1"-"c2" "c2"-"b2"
      "b2"-"c3" "c3"-"b3"
      "b3"-"c4" "c4"-"b4"
      "b4"-"c5" "c5"-"b5"
      "b5"-"c1" "c1"-"b1"
      "c1"-"c3" "c2"-"c4" "c3"-"c5" "c4"-"c1" "c5"-"c2"
    } 
  }
  \caption{The Petersen graph (left) and the Clebsch graph (right).}
  \label{fig:SpecialGraphs}
\end{figure}
As an alternative basis for the curve homology classes one can use the
canonical class and the degree zero sublattice
\begin{equation}
  K^\perp
  ~\subset H_2(\cB_n,\Z)
  .
\end{equation}
This sublattice contains a finite number of classes with
self-intersection $-2$. One can show~\cite{Manin} that these classes
span the root lattice\footnote{The lattice product on $K^\perp$ is
  \emph{minus} the intersection product in homology.} of a Lie algebra
for $n\ge 2$. By this identification, we will call the
$(-2)$-classes\footnote{Note that such a $(-2)$-homology class cannot
  be represented by a holomorphic curve on a del Pezzo surface.}  of
degree $0$ roots, as well. Explicitly, the simple roots can be taken
to be
\begin{equation}  
  \label{eq:delpezzolattice} 
  \begin{split}
    \alpha_i \;&= e_i-e_{i+1}
    , \quad 
    i=1,\ldots, n-1
    ,
    \\
    \alpha_n \;&= l-e_1-e_2-e_3  
    .
  \end{split}
\end{equation}  
The intersection matrix is given in terms of the Cartan matrix
$C_{ij}$ of the corresponding Lie algebra as
\begin{equation}
  \label{eq:int_matrix_delP} 
  \alpha_i \cdot \alpha_j = -C_{ij}
  ,\quad
  \alpha_i \cdot K = 0
  ,\quad
  K^2 = 9-n
  .
\end{equation}
The root lattices for the del Pezzo surfaces are listed in
\autoref{tab:linesautomorphisms} together with some information on the
corresponding Weyl groups.

One of the advantages of working with this root lattice is the
following characterisation of the symmetries of the graph of
$(-1)$-curves on a del Pezzo surface:
\begin{theorem}[Manin]
  The Weyl group of the root lattice associated to a del Pezzo surface
  equals the automorphism group of the graph of $(-1)$-curves.
\end{theorem}
In particular, we are interested in the conjugacy classes of
involutions acting on the graph of $(-1)$-curves, which we can easily
compute in terms of the conjugacy classes of $\Z_2$ subgroups in the
corresponding Weyl group. It is now a simple (but tedious) computation
to enumerate\footnote{We list the sizes of the conjugacy classes in
  \autoref{tab:H2action}.} all the Weyl group elements in each
conjugacy class.

Note that, on del Pezzo surfaces of degree
$6$ and higher ($\cB_n$ with $n\geq 3$), the canonical class and root
lattice span the whole homology group\footnote{In fact,
  $\Span\{K,\alpha_1,\dots,\alpha_n\}$ is a finite-index sublattice of
  $H_2(\cB_n,\Z)$ for $n\geq 3$. However, since there is no torsion in
  the homology of del Pezzo surfaces this does not matter in the
  following.} $H_2(\cB_n,\Q)$. {}From now on we will restrict ourselves
to this case, where we now have three equivalent bases for the
(rational) homology:
\begin{itemize}
\item The standard basis $l$, $e_1$, $\dots$, $e_n$.
\item The canonical class together with the $n$ simple roots.
\item Any maximal (that is, consisting of $n+1$) linearly independent set
  of $(-1)$-curves.
\end{itemize}
\afterpage{\clearpage
  \renewcommand{\arraystretch}{1.4}
  \begin{longtable}{|c|c|l|c|c|}
    \caption{The $\Z_2$ conjugacy classes in the Weyl groups 
      associated to del Pezzo surfaces. 
      These (together with the trivial group element) classify the 
      possible actions on the degree-$2$ homology of the 
      corresponding del Pezzo surface. 
      Note that in some cases the potential action cannot be realised
      on a del Pezzo surface. The $d\times d$ identity matrix
      is denoted by $\mathbf{1}_d$, the $2\times 2$ permutation 
      matrix by
      \protect\begin{math}
        H = 
        \left(
          \protect\begin{smallmatrix}
            0 & 1 \\ 1 & 0
            \protect\end{smallmatrix}  
        \right)
        \protect\end{math}.
    }
    \label{tab:H2action}
    \endfirsthead

    \multicolumn{5}{c}%
    {\textbf{\tablename\ \thetable{}} -- continued from previous page} 
    \\ \hline
    $\cB_n$ & 
    Weyl &
    Action on $H_2\big(\cB_n,\Z\big)=\Span\big\{l,e_1,\dots,e_n\big\}$ &
    $\big( b_2^+,b_2^- \big)$ &
    $|I_{\cB_n}^{(\cdot)}|^\text{Weyl}$
    \endhead

    \hline \multicolumn{5}{r}{Continued on the next page} 
    \\
    \endfoot

    \endlastfoot

    \hline
    \rotatebox{90}{del Pezzo} & 
    \rotatebox{90}{Weyl group} &
    Action on $H_2\big(\cB_n,\Z\big)=\mathrm{Span}\big\{l,e_1,\dots,e_n\big\}$ &
    $\big( b_2^+,b_2^- \big)$ &
    \rotatebox{90}{
      \renewcommand{\arraystretch}{1}
      \begin{tabular}{c}
        Size of\\ conj. class
      \end{tabular} 
    }
    \\ \hline
    \multirow{3}{*}{$\cB_3$} &  
    \multirow{3}{*}{$A_2\oplus A_1$} &  
    $I_{\cB_3}^{(1)} =\mathbf{1}_2 \oplus H$ 
    & $(3,1)$ & $3$ 
    \\[1pt]\nopagebreak
    & & 
    \begin{math}
      I_{\cB_3}^{(2)} =
      \left(\begin{smallmatrix}    
 2 &  1 &  1 &  1 \\
-1 & -1 &  0 & -1 \\
-1 &  0 & -1 & -1 \\
-1 & -1 & -1 &  0 \\
      \end{smallmatrix}\right)
    \end{math}
    & $(2,2)$ & $3$
    \\\nopagebreak
    & & 
    \begin{math}
      I_{\cB_3}^{(3)} =
      \left(\begin{smallmatrix}
 2 &  1 &  1 &  1 \\
-1 &  0 & -1 & -1 \\
-1 & -1 &  0 & -1 \\
-1 & -1 & -1 &  0 \\
      \end{smallmatrix}\right)
    \end{math}
    & $(3,1)$ & $1$
    \\ \hline
    \multirow{2}{*}{$\cB_4$} &  
    \multirow{2}{*}{$A_4$} &  
    $I_{\cB_4}^{(1)} =\mathbf{1}_3 \oplus H$ 
    & $(4,1)$ & $10$ 
    \\\nopagebreak
    & & 
    $I_{\cB_4}^{(2)} =\mathbf{1}_1 \oplus 2H$ 
    & $(3,2)$ & $15$ 
    \\ \hline
    \multirow{5}{*}{$\cB_5$} &  
    \multirow{5}{*}{$D_5$} &  
    $I_{\cB_5}^{(1)} =\mathbf{1}_4 \oplus H$
    \hfill\textcolor{red}{No such $\cB_5$ involution!}
    & $(5,1)$ & $20$
    \\
    & &
    $I_{\cB_5}^{(2)} =\mathbf{1}_2 \oplus 2H$
    & $(4,2)$ & $60$
    \\
    & &
    $I_{\cB_5}^{(3)} = I_{\cB_3}^{(2)} \oplus H$
    & $(3,3)$ & $60$
    \\
    & &
    $I_{\cB_5}^{(4)} = I_{\cB_3}^{(3)} \oplus H$
    & $(4,2)$ & $10$
    \\[1pt]
    & &
    \begin{math}
      I_{\cB_5}^{(5)} =
      \left(\begin{smallmatrix}
 3 &  2 &  1 &  1 &  1 &  1 \\
-2 & -1 & -1 & -1 & -1 & -1 \\
-1 & -1 & -1 &  0 &  0 &  0 \\
-1 & -1 &  0 & -1 &  0 &  0 \\
-1 & -1 &  0 &  0 & -1 &  0 \\
-1 & -1 &  0 &  0 &  0 & -1 \\
      \end{smallmatrix}\right)
    \end{math}
    & $(2,4)$ & $5$
    \\ \hline
    \multirow{4}{*}{$\cB_6$} &  
    \multirow{4}{*}{$E_6$} &  
    $I_{\cB_6}^{(1)} = \mathbf{1}_5 \oplus H$
    \hfill\textcolor{red}{No such $\cB_6$ involution!}
    & $(6,1)$ & $36$
    \\\nopagebreak
    & &
    $I_{\cB_6}^{(2)} = \mathbf{1}_3 \oplus 2H$
    & $(5,2)$ & $270$
    \\\nopagebreak
    & & 
    $I_{\cB_6}^{(3)} = \mathbf{1}_1 \oplus 3H$
    \hfill\textcolor{red}{No such $\cB_6$ involution!}
    & $(4,3)$ & $540$
    \\\nopagebreak
    & &
    $I_{\cB_6}^{(4)} = I_{\cB_5}^{(5)} \oplus \mathbf{1}_1$
    & $(3,4)$ & $45$
    \\ \hline
    & &
    $I_{\cB_7}^{(1)} = \mathbf{1}_6 \oplus H$
    \hfill\textcolor{red}{No such $\cB_7$ involution!}
    & $(7,1)$ & $63$
    \\\nopagebreak
    & &
    $I_{\cB_7}^{(2)} = \mathbf{1}_4 \oplus 2H$
    \hfill\textcolor{red}{No such $\cB_7$ involution!}
    & $(6,2)$ & $945$
    \\\nopagebreak
    & &
    $I_{\cB_7}^{(3)} = \mathbf{1}_2 \oplus 3H$
    \hfill\textcolor{red}{No such $\cB_7$ involution!}
    & $(5,3)$ & $3780$
    \\\nopagebreak
    & &
    $I_{\cB_7}^{(4)} = I_{\cB_3}^{(2)} \oplus 2H$
    \hfill\textcolor{red}{No such $\cB_7$ involution!}
    & $(4,4)$ & $3780$
    \\\nopagebreak
    & &
    $I_{\cB_7}^{(5)} = I_{\cB_3}^{(3)} \oplus 2H$
    & $(5,3)$ & $315$
    \\\nopagebreak
    & &
    $I_{\cB_7}^{(6)} = I_{\cB_5}^{(5)} \oplus \mathbf{1}_2$
    & $(4,4)$ & $315$
    \\\nopagebreak
    $\cB_7$ & $E_7$ &
    $I_{\cB_7}^{(7)} = I_{\cB_5}^{(5)} \oplus H$
    & $(3,5)$ & $945$
    \\[1pt]\nopagebreak
    & &
    \begin{math}
      I_{\cB_7}^{(8)} =
      \left(\begin{smallmatrix}
 4 &  3 &  1 &  1 &  1 &  1 &  1 &  1 \\
-3 & -2 & -1 & -1 & -1 & -1 & -1 & -1 \\
-1 & -1 & -1 &  0 &  0 &  0 &  0 &  0 \\
-1 & -1 &  0 & -1 &  0 &  0 &  0 &  0 \\
-1 & -1 &  0 &  0 & -1 &  0 &  0 &  0 \\
-1 & -1 &  0 &  0 &  0 & -1 &  0 &  0 \\
-1 & -1 &  0 &  0 &  0 &  0 & -1 &  0 \\
-1 & -1 &  0 &  0 &  0 &  0 &  0 & -1 \\
       \end{smallmatrix}\right)
    \end{math}
    \hfill\llap{\textcolor{red}{
        \begin{minipage}{2cm}
          \flushright
          No such $\cB_7$\\ involution!    
        \end{minipage}
      }
      \hspace{-5mm}
    }
    & $(2,6)$ & $63$    
    \\
    & &
    \begin{math}
      I_{\cB_7}^{(9)} =
      \left(\begin{smallmatrix}
 8 &  3 &  3 &  3 &  3 &  3 &  3 &  3 \\
-3 & -2 & -1 & -1 & -1 & -1 & -1 & -1 \\
-3 & -1 & -2 & -1 & -1 & -1 & -1 & -1 \\
-3 & -1 & -1 & -2 & -1 & -1 & -1 & -1 \\
-3 & -1 & -1 & -1 & -2 & -1 & -1 & -1 \\
-3 & -1 & -1 & -1 & -1 & -2 & -1 & -1 \\
-3 & -1 & -1 & -1 & -1 & -1 & -2 & -1 \\
-3 & -1 & -1 & -1 & -1 & -1 & -1 & -2 \\
    \end{smallmatrix}\right)
    \end{math}
    & $(1,7)$ & $1$    
    \\ \hline
    & &
    $I_{\cB_8}^{(1)} = \mathbf{1}_7 \oplus H$
    \hfill\textcolor{red}{No such $\cB_8$ involution!}
    & $(8,1)$ & $120$
    \\\nopagebreak
    & &
    $I_{\cB_8}^{(2)} = \mathbf{1}_5 \oplus 2H$
    \hfill\textcolor{red}{No such $\cB_8$ involution!}
    & $(7,2)$ & $3780$
    \\\nopagebreak
    & &
    $I_{\cB_8}^{(3)} = \mathbf{1}_3 \oplus 3H$
    \hfill\textcolor{red}{No such $\cB_8$ involution!}
    & $(6,3)$ & $37800$
    \\\nopagebreak
    & &
    $I_{\cB_8}^{(4)} = \mathbf{1}_1 \oplus 4H$
    \hfill\textcolor{red}{No such $\cB_8$ involution!}
    & $(5,4)$ & $113400$
    \\\nopagebreak
    & &
    $I_{\cB_8}^{(5)} = I_{\cB_5}^{(5)} \oplus \mathbf{1}_3$
    & $(5,4)$ & $3150$
    \\\nopagebreak
    $\cB_8$& $E_8$ &
    $I_{\cB_8}^{(6)} = I_{\cB_5}^{(5)} \oplus \mathbf{1}_1 \oplus H$
    & $(4,5)$ & $37800$
    \\\nopagebreak
    & &
    $I_{\cB_8}^{(7)} = I_{\cB_7}^{(8)} \oplus \mathbf{1}_1$
    \hfill\textcolor{red}{No such $\cB_8$ involution!}
    & $(3,6)$ & $3780$
    \\\nopagebreak
    & &
    $I_{\cB_8}^{(8)} = I_{\cB_7}^{(9)} \oplus \mathbf{1}_1$
    \hfill\textcolor{red}{No such $\cB_8$ involution!}
    & $(2,7)$ & $120$
    \\[1pt]\nopagebreak
    & &
    \begin{math}
      I_{\cB_8}^{(9)} =
      \left(\begin{smallmatrix}
17 &   6 &   6 &   6 &   6 &   6 &   6 &   6 &   6 \\
-6 &  -3 &  -2 &  -2 &  -2 &  -2 &  -2 &  -2 &  -2 \\
-6 &  -2 &  -3 &  -2 &  -2 &  -2 &  -2 &  -2 &  -2 \\
-6 &  -2 &  -2 &  -3 &  -2 &  -2 &  -2 &  -2 &  -2 \\
-6 &  -2 &  -2 &  -2 &  -3 &  -2 &  -2 &  -2 &  -2 \\
-6 &  -2 &  -2 &  -2 &  -2 &  -3 &  -2 &  -2 &  -2 \\
-6 &  -2 &  -2 &  -2 &  -2 &  -2 &  -3 &  -2 &  -2 \\
-6 &  -2 &  -2 &  -2 &  -2 &  -2 &  -2 &  -3 &  -2 \\
-6 &  -2 &  -2 &  -2 &  -2 &  -2 &  -2 &  -2 &  -3 \\
       \end{smallmatrix}\right)
    \end{math}
    & $(1,8)$ & $1$    
    \\
    \hline
  \end{longtable}
  \clearpage } 
The second basis is especially useful to determine the homology action
of an involution on a del Pezzo surface. By definition, the canonical
class is invariant under the action of a holomorphic map and the Weyl
group acts on its orthogonal complement $K^\perp$. In
\autoref{tab:H2action}, we use this to find the action on the homology
of each possible involution of the graph of $(-1)$-curves. Note that
conjugate involutions can have different matrix expressions; We pick
particularly ``nice'' representatives by picking block-diagonal ones
such that the bottom right blocks are either $2\times 2$ permutation
matrices, or identity matrices.

Note that no two $(-1)$-curves are homologous, that is, the homology
classes of the $(-1)$-curves are all distinct. Hence, the homology
class of a $(-1)$-curve can only be invariant under an involution if
the $(-1)$-curve is geometrically invariant. Similarly, a pair of
homology classes of self-intersection $-1$ is exchanged under an
involution if and only if the actual $(-1)$-curves are exchanged by
the involution. Hence, 
\begin{itemize}
\item If an involution on $\cB_n$ leaves one $(-1)$-class invariant,
  then said involution is the blow-up of an involution on $\cB_{n-1}$
  at one invariant point. In the standard basis, the group action on
  $H_2(\cB_n,\Z)$ is block diagonal, consisting of the action on
  $H_2(\cB_{n-1},\Z)$ and one $\mathbf{1}_1$ block.
\item If an involution on $\cB_n$ exchanges two $(-1)$-classes that do
  not intersect, then said involution is the blow-up of an involution
  on $\cB_{n-2}$ at a pair of points (that is, a non-fixed point and
  its image point). In the standard basis, the group action on
  $H_2(\cB_n,\Z)$ is block diagonal, consisting of the action on
  $H_2(\cB_{n-2},\Z)$ and one $H$ block.
\end{itemize}

\clearpage

\subsection{Minimal Involutions}
\label{sec:minimalmodels}

Analysing the list of possible actions on $H_2(S,\Z)$ in
\autoref{tab:dPinv1}, one easily sees that many are related by adding
$\mathbf{1}_1$ or $H$ blocks. If the action is induced from a del
Pezzo surface, then these operations correspond to blowing up a fixed
point and blowing up a point--image point pair, respectively. The
minimal group actions, which cannot be built from simpler
ones, are
\begin{itemize}
\item $\mathbf{1}_1$ acting on $H_2(\CP^2,\Z)$ 
  \hfill Realised by $(\CP^2,\sigma)$, eq.~\eqref{eq:CP2involution}.
\item $\mathbf{1}_2$ acting on $H_2(\CP^1\times\CP^1,\Z)$ 
  \begin{flushright}
  Realised by $(\CP^1\times\CP^1,\sigma_1)$ and
  $(\CP^1\times\CP^1,\sigma_2)$, eqns.~\eqref{eq:P1P1caseA}
  and~\eqref{eq:P1P1caseB}.    
  \end{flushright}
\item $H$ acting on $H_2(\CP^1\times\CP^1,\Z)$ 
  \hfill Realised by $(\CP^1\times\CP^1,\sigma_3)$, 
  eq.~\eqref{eq:P1P1caseA}.  
\item $I_{\cB_5}^{(5)}$ acting on $H_2(\cB_5,\Z)$ 
\item $I_{\cB_7}^{(8)}$ acting on $H_2(\cB_7,\Z)$ 
\item $I_{\cB_7}^{(9)}$ acting on $H_2(\cB_7,\Z)$ 
\item $I_{\cB_8}^{(9)}$ acting on $H_2(\cB_8,\Z)$ 
\end{itemize}
Moreover, the first three (and only the first three) are related by
adding \emph{and subtracting} blocks of the form $\mathbf{1}_1$ and
$H$. In terms of geometric involutions, this means that they are
birational. Hence, there are five disconnected cases of involutions
modulo blow-up/down. One obvious way to understand these cases is to
look at the minimal involution, that is, those that cannot be further
blown down. We already analysed the minimal models $(\CP^2,\sigma)$,
$(\CP^1\times\CP^1,\sigma_1)$, $(\CP^1\times\CP^1,\sigma_2)$, and
$(\CP^1\times\CP^1,\sigma_3)$, all of which are birationally
equivalent\footnote{Evidently, a minimal model can be a ``local
  minimum''}. The remaining four minimal involutions are classically
known involutions:
\begin{itemize}
\item[$I_{\cB_8}^{(9)}$:] The Bertini involution on $\cB_8$, which we
  denote by $(\cB_8,\sigma_\text{B})$. Its fixed point set see
  eq.~\eqref{eq:ExplicitInvolutions}, consists of a genus-$4$ curve
  and one isolated fixed point. Since $b_2^+=1$, all invariant
  homology classes must be proportional to the canonical class. The
  constant of proportionality determines the genus of a corresponding
  holomorphic curve via eq.~\eqref{eq:deg_g}. Using this, the homology
  class of the fixed point set must be
  \begin{equation}
    [\Sigma_4] = 
    - 3 K = 9 l - 3 \sum_{i=1}^8 e_i
    ~\in H_2\big( \cB_8,\Z \big)^{\sigma_\text{B}}
    = \Z K
    .
  \end{equation}
\item[$I_{\cB_7}^{(9)}$:] The Geiser involution on $\cB_7$, which we
  denote by $(\cB_7,\sigma_\text{G})$.  The fixed point set, see
  eq.~\eqref{eq:ExplicitInvolutions}, consists of a genus-$3$ curve
  with homology class
  \begin{equation}
    [\Sigma_3] = 
    - 2 K = 6 l - 2 \sum_{i=1}^7 e_i
    ~\in H_2\big( \cB_7,\Z \big)^{\sigma_\text{G}}
    = \Z K
    .
  \end{equation}
\item[$I_{\cB_7}^{(8)}$:] The de Jonqui\`eres involution of degree $4$
  on the blow-up of $\CP^2$ at $7$ points. In this case, $6$ of the
  points necessarily lie on a conic\cite{dolgachev-2006b}, so this
  involution cannot be realised on a del Pezzo surface. The invariant
  homology classes are the rank-$2$ lattice generated by $K$ and
  $l-e_1$. The fixed point set consists of a single genus-$2$ curve
  $\Sigma_2$. Its homology class is
  \begin{multline}
    [\Sigma_2] = 
    - K + (l-e_1) = 4 l - 2 e_1 - \sum_{i=2}^7 e_i
    \\
    \in H_2\big( \cB_7,\Z \big)^{\sigma_\text{dJ}}
    = \Span_\Z \big\{ K, l-e_1 \big\}
    .
  \end{multline}
\item[$I_{\cB_5}^{(5)}$:] The de Jonqui\`eres involution of degree $3$
  on $\cB_5$, which we denote by $(\cB_5,\sigma_\text{dJ})$. The fixed
  point set, see eq.~\eqref{eq:ExplicitInvolutions}, consists of a
  single genus-$1$ curve (that is, an elliptic curve) $\Sigma_1$ in
  the class
  \begin{equation}
    [\Sigma_1] = 
    - K = 3 l - \sum_{i=1}^5 e_i
    ~ \in H_2\big( \cB_5,\Z \big)^{\sigma_\text{dJ}}
    = \Span_\Z \big\{ K, l-e_1 \big\}
    .
  \end{equation}
\end{itemize}
Each of these involutions comes with moduli, which can be identified
with the moduli of the fixed point curve $\Sigma_g$~\cite{bayle-1999,
  defernex-2004-174}. For example, the moduli space of
$(\cB_5,\sigma)$ is the usual moduli space of elliptic
curves\footnote{That is, the upper half plane modulo
  $PSL(2,\Z)$.}. Since these moduli spaces are all connected, we learn
that there is (up to continuous deformation) a unique involution
giving rise to the actions ($I_{\cB_5}^{(5)}$, $I_{\cB_7}^{(8)}$,
$I_{\cB_7}^{(9)}$, and $I_{\cB_8}^{(9)}$, respectively) in homology.

Looking at the fixed point sets, we can understand the five different
birationally connected components of involutions as follows. Note that
blowing up a point on a surface always generates an exceptional
divisor $\simeq \CP^1$ and does not change the genus of other
curves. Hence, all fixed points in del Pezzo involutions birationally
connected to $(\CP^2,\sigma)$ are either isolated fixed points or of
genus $0$. The fixed point set in the other $4$ disconnected
components all contain a single curve of genus $1$, $2$, $3$, and $4$,
respectively. Again, since the genus of these divisors is a birational
invariant, these $5$ cases cannot be connected by a chain of
blow-ups/downs.

\subsection{Blow-ups of Minimal Models}
\label{sec:blowupMM}

In order to list all involutions on del Pezzo surfaces, we now just
have to start with the $5$ minimal involutions and perform successive
blow-ups at fixed points or point--image point pairs. Recall that, by
definition, the blow-up point must not lie on $(-1)$ curves, see
\autoref{fnote:generalpoints}. Together with our local analysis of the
fixed-point set after blow-up, \autoref{sec:B1involutions}, we can
summarise the blow-up procedure as follows:
\begin{itemize}
\item Blowing up an isolated fixed point is only allowed if this fixed
  point is not part of a $(-1)$-curve. After the blow-up, the isolated
  fixed point is replaced with the point-wise fixed exceptional
  divisor $e_n \simeq \CP^1$ on $\cB_n$. No points of $e_n$ may be
  blown up any further.
\item On a point-wise fixed curve $C$ one may blow up points as long
  as $C^2\geq 0$ and the point is not one of the finitely many
  intersection points with $(-1)$-curves. After the blow up to
  $\cB_n$, the fixed curve $C$ is replaced by its proper transform
  $\tilde C$ and one isolated point $\tilde p$. The new exceptional
  divisor $e_n$ is not point-wise fixed, but intersects $\tilde C
  \cdot e_n = 1$ and contains $\tilde p \in e_n$. In particular,
  $\tilde p$ may not be blown up further. The homology class of the
  new fixed-point set is
  \begin{equation}
    [\tilde C] = [C-e_n] \in H_2\big( \cB_n, \Z \big)
  \end{equation}
  and $\tilde{C}^2=C^2-1$.
\item A generic point $p$ is neither fixed nor part of the finitely
  many $(-1)$-curves. Hence, one can always find such a point $p$ and
  image point $q$. However, since the two points are not independent,
  one must check that all points are still in general position, see
  \autoref{fnote:generalpoints}.  The fixed point set does not change
  when blowing up such a pair of points.
\end{itemize}
Clearly, the important data about the fixed points set are the curves
and the isolated fixed points that can/cannot be blown up further.

The successive blow-ups of minimal involutions are then subject to the
requirement that the blow-up points are in general position. This
yields the following restrictions:
\begin{itemize}
\item Consider a generic point--image point pair on $(\CP^2,\sigma)$, 
  \begin{equation}
    p = [y:x_0:x_1]
    ,\quad
    q = [-y,x_0:x_1] 
    ~\in \CP^2
    .
  \end{equation}
  $p$, $q$, and the isolated fixed point $[1:0:0]\in (\CP^2)^\sigma$
  are collinear. Hence, one may either blow up this isolated fixed
  point or blow up a point--image point pair, but not all three.
\item Amongst $4$ invariant points on $(\CP^2,\sigma)$ there are $3$
  collinear ones. This excludes actions containing $\mathbf{1}_5$ and
  higher. 
\item Any $3$ point--image point pairs on $(\CP^2,\sigma)$ lie on a
  conic. This excludes actions containing $3H$ and $4H$.
\item Any $3$ point--image point pairs on
  $(\CP^1\times\CP^1,\sigma_3)$ lie on a conic. This excludes the
  action $I_{\cB_3}^{(2)}\oplus 2H$.
\item The de Jonqui\`eres involution of degree $4$
  on the blow-up of $\CP^2$ at $7$ points is excluded since there are
  $6$ points lying on a conic.
\item Recall the action of the Geiser involution on
  $\cB_7=\Bl_{p_1,\dots,p_7}\CP^2$, the image of a point $q\in \cB_7$
  is the remaining basepoint of the pencil of cubics going through the
  $8$ points $\{p_1,\dots,p_7,q\}$. If this remaining basepoint
  coincides with $q$, then the cubic has a node at $q$. Hence, blowing
  up a fixed point of $(\cB_7,\sigma_\text{G})$ is excluded as it
  would result in a nodal cubic going through the $8$ points.
\end{itemize}

Starting from the minimal involutions, this lets us enumerate all
involutions on del Pezzo surfaces. The result is presented in
\autoref{tab:dPinv2}.
\begin{figure}[htbp]
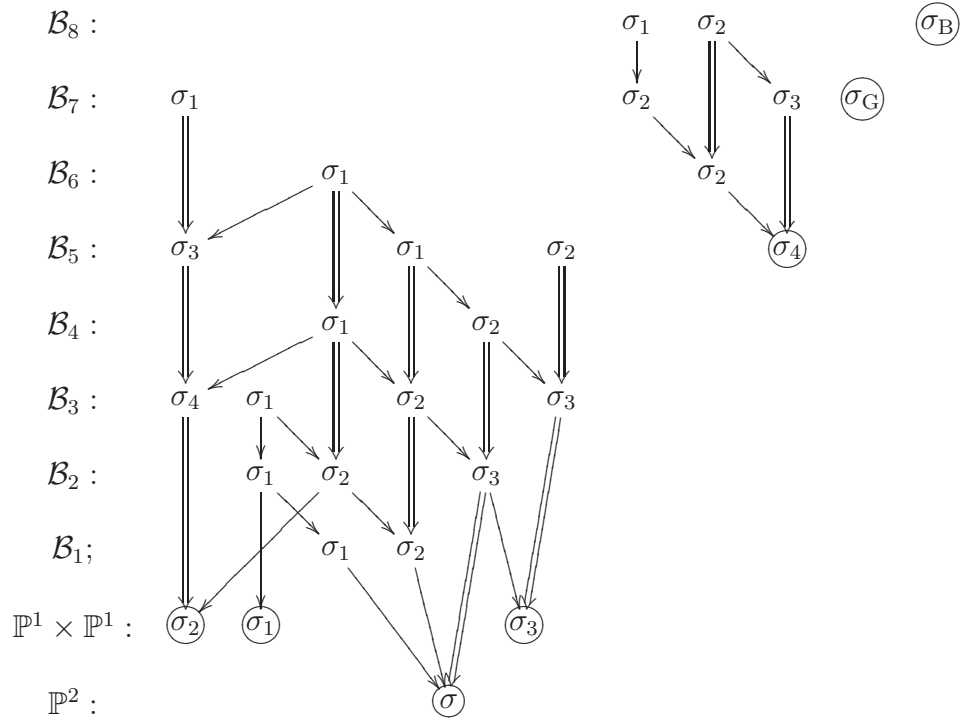

  \centering
  \rotatebox{0}{
    \begin{math}
    \xygraph{
      !{<0cm,0cm>;<10mm,0mm>:<0mm,10mm>::}
      !{(-4.5, 8)}*{\cB_8:}
      !{(-4.5, 7)}*{\cB_7:}
      !{(-4.5, 6)}*{\cB_6:}
      !{(-4.5, 5)}*{\cB_5:}
      !{(-4.5, 4)}*{\cB_4:}
      !{(-4.5, 3)}*{\cB_3:}
      !{(-4.5, 2)}*{\cB_2:}
      !{(-4.5, 1)}*{\cB_1;}
      !{(-4.5, 0)}*{\CP^1\times\CP^1:}
      !{(-4.5,-1)}*{\CP^2:}
      !{( 7, 8)}*+[o][F-]{\sigma_\text{B}}="B"      
      !{( 6, 7)}*+[o][F-]{\sigma_\text{G}}="G"      
      !{( 4, 8)}*+{\sigma_2}="I5_1_H"
      !{( 5, 7)}*+{\sigma_3}="I5_H"
      !{( 3, 8)}*+{\sigma_1}="I5_3"
      !{( 3, 7)}*+{\sigma_2}="I5_2"
      !{( 4, 6)}*+{\sigma_2}="I5_1"
      !{( 5, 5)}*+[o][F-]{\sigma_4}="I5"
      !{(-2, 3)}*+{\sigma_1}="4"
      !{(-1, 6)}*+{\sigma_1}="3_2H"
      !{(-1, 4)}*+{\sigma_1}="3_H"
      !{(-2, 2)}*+{\sigma_1}="3_s1"
      !{(-1, 2)}*+{\sigma_2}="3_s2"
      !{( 0, 5)}*+{\sigma_1}="2_2H"
      !{( 0, 3)}*+{\sigma_2}="2_H"
      !{(-1, 1)}*+{\sigma_1}="2_s1"
      !{( 0, 1)}*+{\sigma_2}="2_s2"
      !{( 1, 4)}*+{\sigma_2}="1_2H"
      !{( 1, 2)}*+{\sigma_3}="1_H"
      !{( 0.5,-1)}*+[o][F-]{\sigma}="1"
      !{( 2, 5)}*+{\sigma_2}="I3_2_H"
      !{( 2, 3)}*+{\sigma_3}="I3_2"
      !{( 1.5, 0)}*+[o][F-]{\sigma_3}="P11_H"
      !{(-3, 7)}*+{\sigma_1}="I3_3_2H"
      !{(-3, 5)}*+{\sigma_3}="I3_3_H"
      !{(-3, 3)}*+{\sigma_4}="I3_3"
      !{(-2, 0)}*+[o][F-]{\sigma_1}="P11_2_s1"
      !{(-3, 0)}*+[o][F-]{\sigma_2}="P11_2_s2"
%
      "I5_1_H"-@{->}"I5_H"
      "I5_3"-@{->}"I5_2"
      "I5_2"-@{->}"I5_1"
      "I5_1"-@{->}"I5"
      "I5_1_H"-@{=>}"I5_1"
      "I5_H"-@{=>}"I5"
      "1_H"-@{=>}"1" "1_2H"-@{=>}"1_H" 
      "2_H"-@{=>}"2_s2" 
      "2_2H"-@{=>}"2_H" 
      "3_H"-@{=>}"3_s2" 
      "3_2H"-@{=>}"3_H" 
      "3_2H"-@{->}"2_2H" "2_2H"-@{->}"1_2H"
      "3_H"-@{->}"2_H" "2_H"-@{->}"1_H"
      "4"-@{->}"3_s1" 
      "4"-@{->}"3_s2" 
      "3_s1"-@{->}"2_s1" "3_s2"-@{->}"2_s2" "2_s1"-@{->}"1" "2_s2"-@{->}"1"
      "3_s1"-@{->}"P11_2_s1" 
      "3_s2"-@{->}"P11_2_s2" 
      "1_H"-@{->}"P11_H"
      "I3_2_H"-@{=>}"I3_2" "I3_2"-@{=>}"P11_H"
      "1_2H"-@{->}"I3_2"
      "I3_3_2H"-@{=>}"I3_3_H" "I3_3_H"-@{=>}"I3_3" 
      "I3_3"-@{=>}"P11_2_s2"
      "3_2H"-@{->}"I3_3_H"
      "3_H"-@{->}"I3_3"
    } 
  \end{math}
  }
  \caption{Relations between the involutions on del Pezzo  
    surfaces. The single arrows ($\rightarrow$) are the blow-downs of
    a fixed $(-1)$-curve, the double arrows ($\Rightarrow$) are the
    blow-downs of a $(-1)$-curve and its image curve. The minimal
    involutions are encircled.}
  \label{fig:dPinvRelations}
\end{figure}
These involutions are related through blowing up fixed points or
point--image point pairs. This is depicted in
\autoref{fig:dPinvRelations}.

\subsection{Explicit Realisations}
\label{sec:explicit_involutions}

Finally, let us list explicit examples for the involutions on del
Pezzo surfaces of degree $5$ and less. These can be written as
hypersurfaces in weighted projective spaces. For simplicity, we pick a
particular simple point in the complex moduli space in each case:
\begin{equation}
  \begin{aligned}
    \cB_6 =&\; 
    \Big\{ 
    x_0^3 + x_1^3 + x_2^3 + x_3^3 = 0 
    \Big\}
    &
    & \subset \CP^{1,1,1,1}
    \\
    \cB_7 =&\; 
    \Big\{ 
    y^2 + x_0^4 + x_1^4 + x_2^4 = 0 
    \Big\}
    &
    & \subset \CP^{2,1,1,1}
    \\
    \cB_7 =&\; 
    \Big\{ 
    y^2 + z^3 + x_0^6 + x_1^6 = 0 
    \Big\}
    &
    & \subset \CP^{3,2,1,1}
  \end{aligned}
\end{equation}
The involutions listed in \autoref{tab:dPinv2} then act as follows on
the hypersurfaces:
\begin{equation}
  \label{eq:ExplicitInvolutions}
  \begin{aligned}
    \sigma_1: &\;\cB_6\to\cB_6
    ,&
    \big[ x_0:x_1:x_2:x_3 \big] \;&\mapsto
    \big[ x_1:x_0:x_3:x_2 \big]
    ,\\
    \sigma_2: &\;\cB_6\to\cB_6
    ,&
    \big[ x_0:x_1:x_2:x_3 \big] \;&\mapsto
    \big[ x_1:x_0:x_2:x_3 \big]
    ,\\ 
    \sigma_1: &\;\cB_7\to\cB_7
    ,&
    \big[ y:x_0:x_1:x_2 \big] \;&\mapsto
    \big[ -y:-x_0:x_1:x_2 \big]
    ,\\
    \sigma_2: &\;\cB_7\to\cB_7
    ,&
    \big[ y:x_0:x_1:x_2 \big] \;&\mapsto
    \big[ y:-x_0:x_1:x_2 \big]
    ,\\
    \sigma_3: &\;\cB_7\to\cB_7
    ,&
    \big[ y:x_0:x_1:x_2 \big] \;&\mapsto
    \big[ y:x_1:x_0:x_2 \big]
    ,\\
    \sigma_\text{G}: &\;\cB_7\to\cB_7
    ,&
    \big[ y:x_0:x_1:x_2 \big] \;&\mapsto
    \big[ -y:x_0:x_1:x_2 \big]
    ,\\ 
    \sigma_1: &\;\cB_8\to\cB_8
    ,&
    \big[ y:z:x_0:x_1 \big] \;&\mapsto
    \big[ y:z:-x_0:x_1 \big]
    ,\\
    \sigma_2: &\;\cB_8\to\cB_8
    ,&
    \big[ y:z:x_0:x_1 \big] \;&\mapsto
    \big[ y:z:x_1:x_0 \big]
    ,\\
    \sigma_\text{B}: &\;\cB_8\to\cB_8
    ,&
    \big[ y:z:x_0:x_1 \big] \;&\mapsto
    \big[ -y:z:x_0:x_1 \big]
    .
  \end{aligned}
\end{equation}

\section{Cohomology of Line Bundles over del Pezzo Surfaces} 
\label{app_dPr} 

On a del Pezzo surface $\cB_n =
\Bl_{p_1,p_2,\dots,p_n}\big(\CP^2\big)$, $n\leq 8$, the line
bundles\footnote{We are going to label the exceptional divisors such
  that $e_k\subset \cB_n$ corresponds to the blow-up point $p_k\in
  \CP^2$, $k=1,\dots,n$.}
\begin{equation}
  \Pic(\cB_n)=H^2(\cB_n,\Z) 
  = \Span_\Z\big\{ l, e_1, e_2, \dots, e_n \big\}
  \simeq
  \Z^{n+1}
\end{equation}
are classified by their first Chern class. We will parametrise the
first Chern class of any line bundle $L$ as
\begin{equation}
   L=
   \cO \Big(
     a l + \sum_{i\in I} b_i e_i - \sum_{j\in J} c_j e_j
   \Big)
   ,\quad
   a \in \Z
   ,~
   b_i \in \Z_\geq
   ,~
   c_j \in \Z_>
   ,
\end{equation}
where we split the index range $\{1,2,\dots,n\} = I\cup J$ into two
disjoint index sets. 

In order to compute the bundle cohomology groups of $L$, we fist
recall the following two facts:
\begin{itemize}
\item The index of $L$ is
  \begin{equation}
    \begin{split}
      \chi(L) 
      \;&= 
      \sum_{k=0}^2 (-1)^k h^k(\cB_n,L) = 
      \int_{\cB_n} \ch(L) \Td(T\cB_n) = 
      \\ 
      &=
      \binom{a+2}{2} 
      - \sum_{i\in I} \frac{b_i (b_i-1)}{2}
      - \sum_{j\in J} \frac{c_j (c_j+1)}{2}
      .
    \end{split}
  \end{equation}
\item Serre duality relates
  \begin{equation}
    \begin{split}
      H^k\big(\cB_n,L\big)^\vee
      \;&=
      H^{2-k}\big(\cB_n,L^\vee \otimes K\big)      
      \\
      \;&=
      H^{2-k}\big(\cB_n,L^\vee \otimes \cO(-3l+e_1+\cdots+e_n) \big)      
    \end{split}
  \end{equation}
\end{itemize}
Using Serre duality if necessary, it therefore suffices to calculate
the cohomology groups of $L$ for $a \geq -2$. In the following, we
will always assume this to be the case.

First, the cohomology of $\cO(al)$ is clearly identical to the
cohomology of $\cO_{\CP^2}(a)$, which is
\begin{equation}
  H^*\Big( \cB_n,\cO(al) \Big)
  =
  \begin{cases}
    0 & \ast=2 \\
    0 & \ast=1 \\
    \binom{a+2}{2} & \ast=0 
    .
  \end{cases}
\end{equation}
The $\tbinom{a+2}{2}$ global sections are nothing but the degree-$a$
homogeneous polynomials in the $3$ homogeneous variables. Similarly,
the global sections of $\cO(al - \sum c_i e_i)$ can be identified with
the degree-$a$ homogeneous polynomials that vanish at the blow-up
point $p_j$ to the degree $c_j$, $j\in J$.  Note that the homogeneous
polynomials are a linear space spanned by the monomials, and,
therefore, counting the dimension of the space of such sections is a
simple linear algebra problem. We denote the dimension of the sections
vanishing at $p_1$, $\dots$, $p_n$ by\footnote{The expected dimension
  is
  \begin{equation}
    A_{\sum c_i p_i}^\text{expect}(a) = 
    \maximum\left\{
      0,~ \binom{a+2}{2}- \sum \frac{c_j (c_j+1)}{2} 
    \right\}
    ,
  \end{equation}
  and this is often the actual value. However, for example,
  $\cO_{\cB_3}(3-3 e_1 -2 e_2 -e_3)$ has
  $A_{3e_1+2e_2+e_1}^\text{expect}(3)=0$ while the actual value is
  $A_{3e_1+2e_2+e_1}(3)=1$. Moreover, for special values of the
  complex structure moduli of $\cB_n$, the dimension of the cohomology
  group jumps and $A_{\sum c_i p_i}(a) > A_{\sum c_i
    p_i}^\text{expect}(a)$. Note that one can always pick coordinates
  on $\CP^2$ such that
  \begin{equation}
    p_1 = [1:0:0]
    ,~
    p_2 = [0:1:0]
    ,~
    p_3 = [0:0:1]
    ,~
    p_4 = [1:1:1]
    .
  \end{equation}
  In other words, only $\cB_n$ with $n\geq 5$ have complex structure
  moduli, parametrised by the position of the points $p_5$, $\dots$,
  $p_n$. 
}
\begin{equation}
  A_{\sum c_i p_i}(a) = 
  \dim
  \big\{ 
  P_a(x,y,z)
  \big| 
  P_a(p_i) = 
  0
  \text{ to order }
  c_i
  \big\}
  .
\end{equation}
To obtain the higher-degree cohomology
groups, consider the standard short exact sequence
\begin{equation}
  \vcenter{\xymatrix{
      0 \ar[r] &
      \cO_{\cB_n}\big( al - \sum c_j e_j \big) \ar[r] &
      \cO_{\cB_n}\big( al \big) \ar[r] &
      \oplus_{j\in J}
      \cO_{c_j e_j} \ar[r] &
      0
      .
    }}
\end{equation}
{}From the corresponding long exact sequence\footnote{By abuse of
  notation, we just write $k$ instead of a $k$-dimensional vector space
  in long exact sequences.},
\begin{equation}
  \vcenter{\xymatrix@R=10pt@M=4pt@H+=22pt{
      0 \ar[r] & 
      H^0\Big( \cO_{\cB_n}\big( al - \sum c_j e_j \big) \Big)
      \ar[r] &
      \binom{a+2}{2}
      \ar[r] &
      \sum \frac{c_j (c_j+1)}{2}
      \ar`[rd]`[l]`[dlll]`[d][dll] & 
      \\
      & 
      H^1\Big( \cO_{\cB_n}\big( al - \sum c_j e_j \big) \Big)
      \ar[r] &
      0
      \ar[r] &
      0
      ,
      &
    }}
\end{equation}
it follows that the cohomology is concentrated in degrees $0$ and $1$
only. Hence, using the index, one obtains that
\begin{equation}
    H^2\Big( \cB_n,\cO(al - \sum c_j e_j) \Big) = 0
    .
\end{equation}
Finally, consider the short exact sequence
\begin{equation}
  \vcenter{\xymatrix{
      0 \ar[r]  &
      \cO_{\cB_n}\big( a l - \sum c_j e_j \big) \ar[r] &
      L \ar[r] &
      \oplus_{i\in I}
      \cO_{b_i e_i}(-b_i) \ar[r] &
      0
    }}
  .
\end{equation}
The corresponding long exact sequence reads
\begin{equation}
  \vcenter{\xymatrix@R=10pt@M=4pt@H+=22pt{
      0 \ar[r] & 
      H^0\Big( \cO_{\cB_n}\big( al - \sum c_i e_i \big) \Big)
      \ar[r] &
      H^0(L)
      \ar[r] &
      0
      \ar`[rd]`[l]`[dlll]`[d][dll] & 
      \\
      & 
      H^1\Big( \cO_{\cB_n}\big( al - \sum c_i e_i \big) \Big)
      \ar[r] &
      H^1(L)
      \ar[r] &
      \sum \frac{b_i (b_i+1)}{2}
      \ar`[rd]`[l]`[dlll]`[d][dll] & 
      \\
      & 
      0
      \ar[r] &
      H^2(L)
      \ar[r] &
      0 \ar[r]
      &
      0
      .
    }}
\end{equation}
One immediately notices that the cohomology of $L$ is concentrated in
degrees $0$ and $1$, and therefore is determined by the index and $A$.
Therefore, the cohomology of $L=\cO(al + \sum b_i e_i - \sum c_j e_j)$
with $a\geq -2$ is given by
\begin{equation}
  \label{eq:CohdPrResult}
  H^*\Big( \cB_n,L \Big)
  =
  \begin{cases}
    0 & \ast=2 \\
    -\chi(L)
    - A_{\sum c_i p_i}(a) & \ast=1 \\
    A_{\sum c_i p_i}(a) & \ast=0 
    .
  \end{cases}
\end{equation}

\section{Cohomology of Line Bundles On Rational Elliptic Surfaces} 
\label{sec:HdP9} 

By definition, a $\dP_9$ surface is a rational elliptic surface,
meaning that it is simultaneously a blow-up of $\CP^2$ and
elliptically fibred. Given this definition, one can show that there
must be precisely $9$ blow-up points\footnote{In contrast to the del Pezzo
  case, the blow-up points can be ``infinitesimally close''. This
  generates $(-2)$-curves in the blown-up surface, which are allowed
  on a $\dP_9$ but not on a del Pezzo surface. In fact, the
  irreducible components of most Kodaira fibres (all except $I_0$,
  $I_1$, and $III$) have self-intersection $-2$.}
\begin{equation}
  \dP_9 = \Bl_{\{p_1,\dots,p_9\}}\big( \CP^2 \big)
  ,
\end{equation}
that the base of the fibration must be $\CP^1$ and can\footnote{And we
  will make this choice always in the following.} be taken to be
$e_9$), and that the fibre class is
\begin{equation}
  f = 3 l - \sum_{i=1}^9 e_i
  \quad
  \in H_2\big( \dP_9,\Z\big)
  .
\end{equation}
However, the position of the $9$ points are not arbitrary. In other
words, not every blow-up of $\CP^2$ at $9$ points is elliptically
fibred, but, rather, the blow-up points have to be in the right
position. The immediate consequence for line bundles is that often the
actual cohomology groups are larger than what one would obtain from
the naive application of \autoref{app_dPr}. For
example~\cite{Braun:2005zv},
\begin{equation}
  H^*\big( \dP_9, \Osheaf_{\dP_9}(f) \big) = (2,1,0)
  ,
\end{equation}
while eq.~\eqref{eq:CohdPrResult} would have yielded\footnote{Of
  course, the index is the same since it is a topological quantity.}
$(1,0,0)$.

However, there are two special cases\footnote{Here, and in the
  following, we will assume that we are on a sufficiently generic
  $\dP_9$ surface where all blow-up points $p_1$, $\dots$, $p_9$ can
  be chosen to be distinct.} where we can, in fact, simply apply the
results of \autoref{app_dPr} for $\cB_r$, $r=0,\dots,8$:
\begin{itemize}
\item Consider the case where one of the exceptional divisors (say,
  $e_9$) does not appear in the line bundle, that is,
  \begin{equation}
    L=
    \cO \Big(
    a l + \sum_{i\in I} b_i e_i  - \sum_{j\in J} c_j e_j + 0 \cdot e_9
    \Big)
    ,\quad
    a \in \Z
    ,~
    b_i \in \Z_\geq
    ,~
    c_j \in \Z_>
    ,
  \end{equation}
  with $I \cup J = \{1,\dots,8\}$ disjoint index
  sets. Then, we can identify the cohomology with the analogue 
  \begin{equation}
    H^*\big( \dP_9, L\big) = 
    H^*\Big(\cB_8, 
    \Osheaf_{\cB_8}\big(
    {
      \textstyle
      a l + \sum_{i\in I} b_i e_i  - \sum_{j\in J} c_j
      e_j
    }
    \big)
    \Big)
  \end{equation}
  on $\cB_8$, where eq.~\eqref{eq:CohdPrResult} holds.
\item The canonical bundle is
  \begin{equation}
    K_{\dP_9} = \Osheaf_{\dP_9}(-f)
    ,
  \end{equation}
  and, therefore, Serre duality allows us to identify
  \begin{equation}
    H^*\big( \dP_9, L \otimes \Osheaf_{\dP_9}(-f) \big) =     
    H^{2-*}\big( \dP_9, L^\vee \big)^\vee
    .
  \end{equation}
\end{itemize}

However, in general one has to use the use the program detailed
in~\cite{Braun:2005zv}, and use the Leray-Serre spectral sequence
for the elliptic fibration $\pi: \dP_9 \to \CP^1$. This revolves
around the three steps:
\begin{enumerate}
\item Compute the push-down $R^q \pi_* L$ for linear
  combinations of sections, adding and subtracting one section $s$ at
  a time. This amounts to repeated applications of the short exact
  sequence
  \begin{equation}
    0 
    \longrightarrow
    \Osheaf_{\dP_9}(D-s)
    \longrightarrow
    \Osheaf_{\dP_9}(D)
    \longrightarrow
    \Osheaf_s(D\cdot s)
    \longrightarrow
    0
  \end{equation}
  and the corresponding long exact sequence 
  \begin{equation}
    \vcenter{\xymatrix@R=10pt@M=4pt@H+=22pt{
        0 \ar[r] & 
        \pi_*\big( \Osheaf_{\dP_9}(D-s) \big)
        \ar[r] &
        \pi_*\big( \Osheaf_{\dP_9}(D) \big)
        \ar[r] &
        \Osheaf_{\CP^1}(D\cdot s)
        \ar`[rd]`[l]`[dlll]`[d][dll] & 
        \\
        & 
        R^1 \pi_*\big( \Osheaf_{\dP_9}(D-s) \big)
        \ar[r] &
        R^1 \pi_*\big( \Osheaf_{\dP_9}(D) \big)
        \ar[r] &
        0
        ,
      &
    }}
\end{equation}
  for push-downs. 
\item Use the \emph{projection formula} 
  \begin{equation}
    \label{eq:projectionformula}
    R^q\pi_\ast \Big( L \otimes \Osheaf_{\dP_9}(n f) \Big)
    = 
    R^q\pi_\ast \big( L \big) \otimes \Osheaf_{\CP^1}(n)
  \end{equation}
  to shift fibre classes to the base $\CP^1$.
\item Compute the cohomology groups of $R^q \pi_* (L)$ (on $\CP^1$)
  and use the Leray-Serre spectral sequence to conclude
  \begin{equation}
    H^*\big( \dP_9, L \big) = 
    \begin{cases}
      H^1\big( \CP^1, R^1\pi_*(L) \big)
      & *=2 
      ,\\
      H^1\big( \CP^1, \pi_*(L) \big)
      \oplus
      H^0\big( \CP^1, R^1\pi_*(L) \big)
      & *=1 
      ,\\
      H^0\big( \CP^1, \pi_*(L) \big)
      & *=0 
      .
    \end{cases}
  \end{equation}
\end{enumerate}

As an example, let us consider $L = \cL^{-2}\otimes \cL_Y^{-1} =
\Osheaf_{\dP_9}(2e_9+e_1-e_2-2f)$ in \autoref{tab_nonchiral}. Adding
and subtracting sections yields (vanishing $R^1\pi_*$s are skipped)
\begin{equation}
  \begin{split}
    \pi_*\big( \Osheaf_{\dP_9} \big) 
    &=
    \Osheaf_{\CP^1}
    ,\quad
    R^1 \pi_*\big( \Osheaf_{\dP_9} \big) 
    =
    \Osheaf_{\CP^1}(-1)
    ,\\
    \pi_*\big( \Osheaf_{\dP_9}(e_9) \big) 
    &=
    \Osheaf_{\CP^1}
    ,\\
    \pi_*\big( \Osheaf_{\dP_9}(2e_9) \big) 
    &=
    \Osheaf_{\CP^1} \oplus 
    \Osheaf_{\CP^1}(-2)
    ,\\
    \pi_*\big( \Osheaf_{\dP_9}(2e_9+e_1) \big) 
    &=
    \Osheaf_{\CP^1} \oplus 
    \Osheaf_{\CP^1}(-2) \oplus
    \Osheaf_{\CP^1}(-1)
    ,\\
    \pi_*\big( \Osheaf_{\dP_9}(2e_9+e_1-e_2) \big) 
    &=
    \Osheaf_{\CP^1}(-2) \oplus
    \Osheaf_{\CP^1}(-1)
    .
  \end{split}
\end{equation}
Hence, using eq.~\eqref{eq:projectionformula},
\begin{equation}
  \begin{split}
    \pi_*\big( L \big) 
    &= 
    \Osheaf_{\CP^1}(-4) \oplus
    \Osheaf_{\CP^1}(-3)
    \\
    R^1 \pi_*\big( L \big) 
    &= 
    0
    .
  \end{split}
\end{equation}
Finally, the Leray-Serre spectral sequence yields 
\begin{equation}
  H^*\big( \dP_9, L \big) = (0, 5, 0)
  .
\end{equation}

\bibliographystyle{utphys} 
\renewcommand{\refname}{Bibliography}
\addcontentsline{toc}{section}{Bibliography} 
\bibliography{rev}

\end{document}